\documentclass{article}
\usepackage{latexsym,amsfonts,amssymb,exscale,enumerate}
\usepackage[centertags,sumlimits,intlimits,namelimits,reqno]{amsmath}
\usepackage{hyperref}
\hypersetup{final=true}

\usepackage{pstricks}
\usepackage{pst-grad}

\psset{linewidth=0.3pt,dimen=middle}
\psset{xunit=.70cm,yunit=0.70cm}

\usepackage{graphicx}
\usepackage[all, knot]{xy}
\SelectTips{cm}{} \xyoption{arc}
\usepackage{epsfig}
\usepackage{diagram}

\renewcommand{\to}{\rightarrow}
\newcommand{\To}{\Rightarrow}
\def\stackto #1 { \, {\stackrel{#1}{\longrightarrow}}\, }

\newcommand{\Set}{{\rm Set}}
\newcommand{\Cat}{{\rm Cat}}
\newcommand{\Top}{{\rm Top}}
\newcommand{\Tang}{{\rm Tang}}
\newcommand{\Braid}{{\rm Braid}}

\newcommand{\Grp}{{\rm Grp}}

\newcommand{\Rep}{{\rm Rep}}
\newcommand{\Vect}{{\rm Vect}}
\newcommand{\Hilb}{{\rm Hilb}}
\newcommand{\Fin}{{\rm Fin}}

\newcommand{\Cob}{{\rm Cob}}
\newcommand{\Bimod}{{\rm Bimod}}

\newcommand{\tensor}{\otimes}
\newcommand{\maps}{\colon}

\newcommand{\iso}{\cong}

\newcommand{\tr}{{\rm tr}}

\renewcommand{\hom}{{\rm hom}}

\newcommand{\Path}{{\cal P}}
\newcommand{\Trans}{{\rm Trans}}

 \newcommand{\Ob}{{\rm Ob}}

\newcommand{\SU}{{\rm SU}}
\newcommand{\U}{{\rm U}}
\renewcommand{\O}{{\rm O}}
\newcommand{\SO}{{\rm SO}}
\newcommand{\SL}{{\rm SL}}

\renewcommand{\P}{{\bf P}}

\newcommand{\g}{{\mathfrak g}}

\newcommand{\ssl}{{\mathfrak{sl}}}

\newcommand{\R}{{\mathbb R}}
\newcommand{\C}{{\mathbb C}}

\newcommand{\Z}{{\mathbb Z}}

\newcommand{\scs}{\scriptstyle}
\newcommand{\ten}{\otimes }

\newcommand{\smult}{\lambda}

\newcommand{\revisions}[1] {}

\newcommand{\cxymatrix}[1]{\vcenter{\xymatrix{#1}}}

\input xy \xyoption{all} \xyoption{poly} 
\labelmargin-{1.2pt}  

\newcommand{\thetagraph}[3]  
{ \xymatrix{ *{\bullet} \ar@{-} @/^1.5pc/ [r] ^{#1} \ar@{-}
@/_1.5pc/ [r] _{#3} \ar@{-} [r]^{#2} & *{\bullet}
\\}
}

\newcommand{\fourtheta}[4]  
{ \cxymatrix{ *{\bullet} \ar@{-} @/^1.5pc/ [r] ^{#1} \ar@{-}
@/_1.5pc/ [r] _{#4} \ar@{-} @/^/ [r]^{#2} \ar@{-} @/_/ [r]_{#3} &
*{\bullet}
\\}
}

\newcommand{\TetJ}[6]{
\def\lab{\ifcase\xypolynode\or #1 \or #2 \or #3 \fi}
\begin{xy}
\xygraph{!{<3.2pc,0pc>:}
 *{\bullet}
 !P3"A"{~><{@{{}{-}*{\bullet}}} ~>>{_{\lab}}}
 "A0" -@-_{#4} "A1"
 "A0" -@-_{#5} "A2"
 "A0" -@-^{#6} "A3"
}
\end{xy}
}

\newcommand{\TenJ}{
\def\lab{\ifcase\xypolynode\or 12 \or 23 \or 34 \or 45 \or 15 \fi}
\begin{xy}
\xygraph{!{<4pc,0pc>:}
 !P5"A"{~><{@{{}{-}*{\bullet}}} ~>>{_{j_{\lab}}}}
 "A1" -@-_{j_{13}} "A3"
 "A2" -@-_{j_{24}} "A4"
 "A3" -@-_{j_{35}} "A5"
 "A4" -@-_{j_{14}} "A1"
 "A5" -@-_{j_{25}} "A2"
}
\end{xy}
}

\newcommand{\TenA}{
\def\lab{\ifcase\xypolynode\or 12 \or 23 \or 34 \or 45 \or 15 \fi}
\begin{xy}
\xygraph{!{<4pc,0pc>:}
 !P5"A"{~><{@{{}{-}*{\bullet}}} ~>>{_{a_{\lab}}}}
 "A1" -@-_{a_{13}} "A3"
 "A2" -@-_{a_{24}} "A4"
 "A3" -@-_{a_{35}} "A5"
 "A4" -@-_{a_{14}} "A1"
 "A5" -@-_{a_{25}} "A2"
}
\end{xy}
}

\newcommand{\Ten}{%
\def\lab{\ifcase\xypolynode\or 1,2 \or 2,3 \or 3,4 \or 4,5 \or 5,1 \fi}
\begin{xy}
\xygraph{!{<2pc,0pc>:}
 !P5"A"{~><{@{{}{-}*{\bullet}}} ~>>{_{}}}
 "A1" -@-_{} "A3"
 "A2" -@-_{} "A4"
 "A3" -@-_{} "A5"
 "A4" -@-_{} "A1"
 "A5" -@-_{} "A2"
}
\end{xy}
}

\newcommand{\TenL}{
\def\lab{\ifcase\xypolynode\or 1,2 \or 2,3 \or 3,4 \or 4,5 \or 5,1 \fi}
\begin{xy}
\xygraph{!{<4pc,0pc>:}
 !P5"A"{~><{@{{}{-}*{\bullet}}} ~>>{_{\smult}}}
 "A1" -@-_{\smult} "A3"
 "A2" -@-_{\smult} "A4"
 "A3" -@-_{\smult} "A5"
 "A4" -@-_{\smult} "A1"
 "A5" -@-_{\smult} "A2"
}
\end{xy}
}

\newcommand{\tenl}{
\def\lab{\ifcase\xypolynode\or 1,2 \or 2,3 \or 3,4 \or 4,5 \or 5,1 \fi}
\begin{xy}
\xygraph{!{<2pc,0pc>:}
 !P5"A"{~><{@{{}{-}*{\bullet}}} ~>>{_{\smult}}}
 "A1" -@-_{\smult} "A3"
 "A2" -@-_{\smult} "A4"
 "A3" -@-_{\smult} "A5"
 "A4" -@-_{\smult} "A1"
 "A5" -@-_{\smult} "A2"
}
\end{xy}
}

\newcommand{\periodictable}{
\begin{tabular}{|c|c|c|c|}  \hline
        & $n = 0$ & $n = 1$ & $n = 2$\\ \hline
$k = 0$ & sets & categories & 2-categories     \\     \hline
$k = 1$  & monoids   & monoidal   & monoidal         \\
        &           & categories & 2-categories     \\     \hline
$k = 2$  &commutative& braided    & braided          \\
        & monoids   & monoidal   & monoidal         \\
        &           & categories & 2-categories     \\     \hline
$k = 3$  &`'         & symmetric  & sylleptic \\
        &           & monoidal   & monoidal         \\
        &           & categories & 2-categories     \\     \hline
$k = 4$  &`'         & `'         & symmetric \\
        &           &            & monoidal         \\
        &           &            & 2-categories     \\     \hline
$k = 5$  &`'         &`'          & `'               \\
        &           &            &                  \\
        &           &            &                  \\     \hline
$k = 6$  &`'         &`'          & `'               \\
        &           &            &                  \\
        &           &            &                  \\     \hline
\end{tabular}
\vskip 0.5em
}

\newcommand{\periodictableII}{
{
\[ %
\xy (-75,60)*{k=0}; (-75,30)*{k=1}; (-75,0)*{k=2};
(-75,-30)*{k=3}; (-75,-60)*{k=4}; (-40,75)*{n=0}; (0,75)*{n=1};
(40,75)*{n=2};
(0,0)*{ 
  \xy 0;/r.20pc/: 
  (25,25)*{};
(-2,12)*{\bullet}="1"+(-2,1)*{ \scriptstyle x};
(1,16)*{\bullet}="2"+(-2,2)*{\scriptstyle x^{\ast}};
(10,14)*{\bullet}="3"+(1,3)*{\scriptstyle x};
(1,-7)*{\bullet}="4"+(-1,3)*{\scriptstyle x}; "3";"4" **\crv{}
\POS?(.3)*{\hole}="J"; ?(0)*\dir{>}; "1";"J" **\crv{(15,-5)};
?(.15)*\dir{>}; "2";"J" **\crv{} ?(.28)*\dir{<};
(-14,10)*{}="TL"; (14,10)*{}="TR"; (14,-10)*{}="BR";
(-14,-10)*{}="BL"; (-6,20)*{}="xTL"; (22,20)*{}="xTR";
(22,0)*{}="xBR"; (-6,0)*{}="xBL";
    "TL";"TR" **\dir{-};
    "TR";"BR" **\dir{-};
    "BR";"BL" **\dir{-};
    "BL";"TL" **\dir{-};
    "xTL";"xTR" **\dir{-};
    "xTR";"xBR" **\dir{-};
    "xBR";"xBL" **\dir{.};
    "TL";"xTL" **\dir{-};
    "TR";"xTR" **\dir{-};
    "BL";"xBL" **\dir{.};
    "BR";"xBR" **\dir{-};
    "xTL";"xBL" **\dir{.};
\endxy
   };
   (0,30)*{
    \xy 0;/r.20pc/: 
(-7,10)*{\bullet}="1"+(0,3)*{x};
(-1,10)*{\bullet}="2"+(0,3)*{x^{\ast}};
(6,10)*{\bullet}="3"+(0,3)*{x}; (1,-10)*{\bullet}="4"+(0,-3)*{x};
 "1"; "2" **\crv{(-7,2) & (-1,2)}; ?(.2)*\dir{>}; ?(.9)*\dir{>};
 "3"; "4" **\crv{(9,4) & (-5,-1)}; ?(.5)*\dir{>};
(-10,10)*{}; (10,10)*{} **\dir{-}; (10,-10)*{}; (10,10)*{}
**\dir{-}; (-10,-10)*{}; (-10,10)*{} **\dir{-}; (-10,-10)*{};
(10,-10)*{} **\dir{-};
\endxy};
(40,0)*{
\xy 0;/r.20pc/:
(8.25,-1.25)*\ellipse(2,.65){-}; (6,18)*{}="b";
  (1,16)*{}="a";
 \vunder~{(1,17.5)}{(6,18)}{(3.5,15.5)}{(6,16)};
 (1,17.5)*{}; (6,16)*{} **\crv{(-5,15)& (6,13) }; \POS?(.75)*{\hole}="J1";
 "J1";(6,18)*{} **\crv{(10,14)&(11,18)};
  (6,-2)*{}="b";
  (1,-4)*{}="a";
 \vunder~{(1,-2.5)}{(6,-2)}{(3.5,-4.5)}{(6,-4)};
 (1,-2.5)*{}; (6,-4)*{} **\crv{(-5,-5)& (6,-7) }; \POS?(.75)*{\hole}="J1";
 "J1";(6,-2)*{} **\crv{(10,-6)&(11,-2)};
 (-1,16)*{}="TL";
 (9,16)*{}="TR";
 (-1,-4)*{}="BL";
 (9.25,-3.5)*{}="BR";
 (18.5,-2.5)*{}="BRR";
 (14.5,-2.5)*{}="BRR2";
 "TL";"BL" **\crv{(-2,13) & (1,6)};
 "TR";"BRR" **\crv{(9,12) & (10,6)};
 "BR";"BRR2" **\crv{(7,12) & (12.5,0)};
(-14,10)*{}="TL"; (14,10)*{}="TR"; (14,-10)*{}="BR";
(-14,-10)*{}="BL"; (-6,20)*{}="xTL"; (22,20)*{}="xTR";
(22,0)*{}="xBR"; (-6,0)*{}="xBL";
    "TL";"TR" **\dir{-};
    "TR";"BR" **\dir{-};
    "BR";"BL" **\dir{-};
    "BL";"TL" **\dir{-};
    "xTL";"xTR" **\dir{-};
    "xTR";"xBR" **\dir{-};
    "xBR";"xBL" **\dir{.};
    "TL";"xTL" **\dir{-};
    "TR";"xTR" **\dir{-};
    "BL";"xBL" **\dir{.};
    "BR";"xBR" **\dir{-};
    "xTL";"xBL" **\dir{.};
(5,22)*{\textbf{4d}}; (25,25)*{};
\endxy};
(-40,0)*{ 
\xy  0;/r.20pc/:
(-5,5)*{\bullet}+(1,3)*{x}; (5,4)*{\bullet}+(1,3)*{x^{\ast}};
(1,-7)*{\bullet}+(1,3)*{x}; (-10,10)*{}; (10,10)*{} **\dir{-};
(10,-10)*{}; (10,10)*{} **\dir{-}; (-10,-10)*{}; (-10,10)*{}
**\dir{-}; (-10,-10)*{}; (10,-10)*{} **\dir{-};
\endxy
};(0,-30)*{ 
\xy 0;/r.20pc/: 
(-4,14)*{\bullet}="1"+(,2)*{ \scriptstyle x};
(0,14)*{\bullet}="2"+(,2)*{\scriptstyle x^{\ast}};
(10,14)*{\bullet}="3"+(1,3)*{\scriptstyle x};
(1,-7)*{\bullet}="4"+(-1,3)*{\scriptstyle x}; "3";"4"
**\crv{(10,4) & (-1,-6)}; ?(.6)*\dir{>};
(-14,10)*{}="TL"; (14,10)*{}="TR"; (14,-10)*{}="BR";
(-14,-10)*{}="BL"; (-6,20)*{}="xTL"; (22,20)*{}="xTR";
(22,0)*{}="xBR"; (-6,0)*{}="xBL";
    "TL";"TR" **\dir{-}; \POS?(.35)*{\hole}="J"; \POS?(.5)*{\hole}="J1";
    "TR";"BR" **\dir{-};
    "BR";"BL" **\dir{-};
    "BL";"TL" **\dir{-};
    "xTL";"xTR" **\dir{-};
    "xTR";"xBR" **\dir{-};
    "xBR";"xBL" **\dir{.};
    "TL";"xTL" **\dir{-};
    "TR";"xTR" **\dir{-};
    "BL";"xBL" **\dir{.};
    "BR";"xBR" **\dir{-};
    "xTL";"xBL" **\dir{.};
 "1";"J" **\crv{}?(.9)*\dir{>};
 "2";"J1" **\crv{} ?(.3)*\dir{<};
 "J";"J1" **\crv{(-4,5) & (0,5)};
 (5,22)*{\textbf{4d}};
 (25,25)*{};
\endxy
};
 (0,-60)*{
   \xy 0;/r.20pc/: 
(-4,14)*{\bullet}="1"+(,2)*{ \scriptstyle x};
(0,14)*{\bullet}="2"+(,2)*{\scriptstyle x^{\ast}};
(10,14)*{\bullet}="3"+(1,3)*{\scriptstyle x};
(1,-7)*{\bullet}="4"+(-1,3)*{\scriptstyle x}; "3";"4"
**\crv{(10,4) & (-1,-6)}; ?(.6)*\dir{>};
(-14,10)*{}="TL"; (14,10)*{}="TR"; (14,-10)*{}="BR";
(-14,-10)*{}="BL"; (-6,20)*{}="xTL"; (22,20)*{}="xTR";
(22,0)*{}="xBR"; (-6,0)*{}="xBL";
    "TL";"TR" **\dir{-}; \POS?(.35)*{\hole}="J"; \POS?(.5)*{\hole}="J1";
    "TR";"BR" **\dir{-};
    "BR";"BL" **\dir{-};
    "BL";"TL" **\dir{-};
    "xTL";"xTR" **\dir{-};
    "xTR";"xBR" **\dir{-};
    "xBR";"xBL" **\dir{.};
    "TL";"xTL" **\dir{-};
    "TR";"xTR" **\dir{-};
    "BL";"xBL" **\dir{.};
    "BR";"xBR" **\dir{-};
    "xTL";"xBL" **\dir{.};
 "1";"J" **\crv{}?(.9)*\dir{>};
 "2";"J1" **\crv{} ?(.3)*\dir{<};
 "J";"J1" **\crv{(-4,5) & (0,5)};
 (5,22)*{\textbf{5d}};
 (25,25)*{};
\endxy
};
(-40,30)*{ 
\xy   
(0,0)*{\bullet}+(1,3)*{x^{\ast}}; (5,0)*{\bullet}+(1,3)*{x};
(-5,0)*{\bullet}+(1,3)*{x}; (-10,0)*{}; (10,0)*{} **\dir{-};
\endxy
};
(-40,60)*{ 
\xy (0,0)*{\bullet}+(2,3)*{x^{\ast}};
\endxy
};
(-40,-30)*{ 
  \xy 0;/r.20pc/: 
(-2.5,4)*{\bullet}+(1,2)*{\scriptstyle x}; (5,7)*{
\bullet}+(3,1)*{\scriptstyle x^{\ast}};
(6,0)*{\bullet}+(1,2)*{\scriptstyle x};
(-14,10)*{}="TL"; (14,10)*{}="TR"; (14,-10)*{}="BR";
(-14,-10)*{}="BL"; (-6,20)*{}="xTL"; (22,20)*{}="xTR";
(22,0)*{}="xBR"; (-6,0)*{}="xBL";
    "TL";"TR" **\dir{-};
    "TR";"BR" **\dir{-};
    "BR";"BL" **\dir{-};
    "BL";"TL" **\dir{-};
    "xTL";"xTR" **\dir{-};
    "xTR";"xBR" **\dir{-};
    "xBR";"xBL" **\dir{.};
    "TL";"xTL" **\dir{-};
    "TR";"xTR" **\dir{-};
    "BL";"xBL" **\dir{.};
    "BR";"xBR" **\dir{-};
    "xTL";"xBL" **\dir{.};
(25,25)*{};
\endxy
}; (-40,-60)*{
  \xy 0;/r.20pc/: 
(-2.5,4)*{\bullet}+(1,2)*{\scriptstyle x}; (5,7)*{
\bullet}+(3,1)*{\scriptstyle x^{\ast}};
(6,0)*{\bullet}+(1,2)*{\scriptstyle x};
(-14,10)*{}="TL"; (14,10)*{}="TR"; (14,-10)*{}="BR";
(-14,-10)*{}="BL"; (-6,20)*{}="xTL"; (22,20)*{}="xTR";
(22,0)*{}="xBR"; (-6,0)*{}="xBL";
    "TL";"TR" **\dir{-};
    "TR";"BR" **\dir{-};
    "BR";"BL" **\dir{-};
    "BL";"TL" **\dir{-};
    "xTL";"xTR" **\dir{-};
    "xTR";"xBR" **\dir{-};
    "xBR";"xBL" **\dir{.};
    "TL";"xTL" **\dir{-};
    "TR";"xTR" **\dir{-};
    "BL";"xBL" **\dir{.};
    "BR";"xBR" **\dir{-};
    "xTL";"xBL" **\dir{.};
(25,25)*{}; (5,22)*{\textbf{4d}};
\endxy
}; (0,60)*{
 \xy  0;/r.20pc/:  
(0,10)*{\bullet}="a"+(2.5,1)*{x};
(0,-10)*{\bullet}="b"+(2.5,1)*{x}; "a";"b" **\dir{-};
?(.45)*\dir{>};
\endxy
}; (40,60)*{
 \xy 0;/r.20pc/: 
(-10,10)*{\bullet}="TL"+(-1,3)*{x}; (10,10)*{}="2"="TR"+(1,3)*{x};
(10,-10)*{\bullet}="BR"+(1,-3)*{x};
(-10,-10)*{\bullet}="BL"+(-1,-3)*{x}; "TL";"TR" **\dir{-};
?(.5)*\dir{>}; "BL";"BR" **\dir{-}; ?(.5)*\dir{>}; "TL";"BL"
**\dir{-}; "TR";"BR" **\dir{-}; (0,3)*{}="a"; (0,-3)*{}="a'";
{\ar@{=>} "a";"a'"};
\endxy
}; (40,30)*{
 \xy 0;/r.20pc/: 
(-9,10)*{\bullet}="1"+(-1,-3)*{x^{\ast}};
(-1,10)*{\bullet}="2"+(-1,-3)*{x};
 (-9,-10)*{\bullet}="b1"+(-1,-3)*{x^{\ast}};
(-1,-10)*{\bullet}="b2"+(-1,-3)*{x};
(13,20)*{\bullet}="4"+(1,3)*{x};
(7,20)*{\bullet}="3"+(1,3)*{x^{\ast}}; (13,0)*{\bullet}="b4";
(7,0)*{\bullet}="b3"+(1,3); "1";"2" **\crv{(-4,15) & (3,15)};
?(.15)*\dir{<}; "3";"4" **\crv{(3,15) & (10,15)}; ?(.85)*\dir{>};
 (-1,-5.25)*{}="M";
"1";"b1"  **\dir{-}; "2";"b2"  **\dir{-}; "b1";"M"  **\dir{-};
?(.76)*\dir{>}; "b3";"M"  **\dir{.}; "b2";"b4"  **\dir{-};
?(.4)*\dir{<}; "4";"b4"  **\dir{-}; "3";"b3"  **\dir{.};
(0,13.25)*{}="z1"; (7,16.15)*{}="z2"; "z1";"z2" **\crv{(2,-3) &
(7,3)};
(-14,10)*{}="TL"; (14,10)*{}="TR"; (14,-10)*{}="BR";
(-14,-10)*{}="BL"; (-6,20)*{}="xTL"; (22,20)*{}="xTR";
(22,0)*{}="xBR"; (-6,0)*{}="xBL";
    "TL";"TR" **\dir{-};
    "TR";"BR" **\dir{-};
    "BR";"BL" **\dir{-};
    "BL";"TL" **\dir{-};
    "xTL";"xTR" **\dir{-};
    "xTR";"xBR" **\dir{-};
    "xBR";"xBL" **\dir{.};
    "TL";"xTL" **\dir{-};
    "TR";"xTR" **\dir{-};
    "BL";"xBL" **\dir{.};
    "BR";"xBR" **\dir{-};
    "xTL";"xBL" **\dir{.};
(25,25)*{};
\endxy
}; (40,-30)*{
 \xy 0;/r.20pc/:
(-1.5,-2)*\ellipse(3,1){-}; (3.5,-2)*\ellipse(3,1){-};
(6,18)*{}="b";
  (1,16)*{}="a";
 \vunder~{(1,17.5)}{(6,18)}{(3.5,15.5)}{(6,16)};
 (1,17.5)*{}; (6,16)*{} **\crv{(-5,15)& (6,13) }; \POS?(.75)*{\hole}="J1";
 "J1";(6,18)*{} **\crv{(10,14)&(11,18)};
 (-1,16)*{}="TL";
 (9,16)*{}="TR";
 (-6,-4)*{}="BLL";
 (0,-4)*{}="BL";
 (10,-4)*{}="BRR";
 (4,-4)*{}="BR";
 (3,5)*{}="C";
 (4.25,6.25)*{}="C2";
 (4.25,15)*{}="C3";
   "TL";"BLL" **\crv{(-2,13) & (-1,6)};
   "TR";"BRR" **\crv{(9,13) & (10,6)};
   "C";"BL" **\crv{};
   "C";"BR" **\crv{(6,10)};
    "C3";"C2" **\dir{.};
(-14,10)*{}="TL"; (14,10)*{}="TR"; (14,-10)*{}="BR";
(-14,-10)*{}="BL"; (-6,20)*{}="xTL"; (22,20)*{}="xTR";
(22,0)*{}="xBR"; (-6,0)*{}="xBL";
    "TL";"TR" **\dir{-};
    "TR";"BR" **\dir{-};
    "BR";"BL" **\dir{-};
    "BL";"TL" **\dir{-};
    "xTL";"xTR" **\dir{-};
    "xTR";"xBR" **\dir{-};
    "xBR";"xBL" **\dir{.};
    "TL";"xTL" **\dir{-};
    "TR";"xTR" **\dir{-};
    "BL";"xBL" **\dir{.};
    "BR";"xBR" **\dir{-};
    "xTL";"xBL" **\dir{.};
(5,22)*{\textbf{5d}}; (25,25)*{};
\endxy
}; (40,-60)*{
\xy 0;/r.20pc/:
(-1.5,-2)*\ellipse(3,1){-}; (3.5,-2)*\ellipse(3,1){-};
(6,18)*{}="b";
  (1,16)*{}="a";
 \vunder~{(1,17.5)}{(6,18)}{(3.5,15.5)}{(6,16)};
 (1,17.5)*{}; (6,16)*{} **\crv{(-5,15)& (6,13) }; \POS?(.75)*{\hole}="J1";
 "J1";(6,18)*{} **\crv{(10,14)&(11,18)};
 (-1,16)*{}="TL";
 (9,16)*{}="TR";
 (-6,-4)*{}="BLL";
 (0,-4)*{}="BL";
 (10,-4)*{}="BRR";
 (4,-4)*{}="BR";
 (3,5)*{}="C";
 (4.25,6.25)*{}="C2";
 (4.25,15)*{}="C3";
   "TL";"BLL" **\crv{(-2,13) & (-1,6)};
   "TR";"BRR" **\crv{(9,13) & (10,6)};
   "C";"BL" **\crv{};
   "C";"BR" **\crv{(6,10)};
    "C3";"C2" **\dir{.};
(-14,10)*{}="TL"; (14,10)*{}="TR"; (14,-10)*{}="BR";
(-14,-10)*{}="BL"; (-6,20)*{}="xTL"; (22,20)*{}="xTR";
(22,0)*{}="xBR"; (-6,0)*{}="xBL";
    "TL";"TR" **\dir{-};
    "TR";"BR" **\dir{-};
    "BR";"BL" **\dir{-};
    "BL";"TL" **\dir{-};
    "xTL";"xTR" **\dir{-};
    "xTR";"xBR" **\dir{-};
    "xBR";"xBL" **\dir{.};
    "TL";"xTL" **\dir{-};
    "TR";"xTR" **\dir{-};
    "BL";"xBL" **\dir{.};
    "BR";"xBR" **\dir{-};
    "xTL";"xBL" **\dir{.};
(5,22)*{\textbf{6d}}; (25,25)*{};
\endxy
}:
\endxy 
\]
} }

\newcommand{\feynmandiagram}{
  \xy 0;/r.22pc/:      
 (-3,10)*{}="TL"; (3,10)*{}="TR";
 (0,-2)*{}="B";
 (0,-12)*{}="BB";
    "TL";"B" **\dir{-}?(.5)*\dir{>};
    "TR";"B" **\dir{~};
    "B";"BB" **\dir{-}?(.5)*\dir{>};
 \endxy
                    \; \; \; + \; \;
\xy 0;/r.22pc/:
 (-3,10)*{}="TL"; (3,10)*{}="TR";
 (0,3)*{}="B";
 (0,-12)*{}="BB";
    "TL";"B" **\dir{-}?(.5)*\dir{>};
    "TR";"B"+(0,-.6) **\dir{~};
    "B";"BB" **\dir{-}?(.5)*\dir{>} ?(.24)*\dir{}="1" ?(.84)*\dir{}="2";
    {\ar@/^.35pc/@{~} "1"+(1,0);"2"+(1,0)};
 \endxy
                    \; \; + \; \;
\xy
 (-3,10)*{}="TL"; (4,10)*{}="TR";
 (0,-2)*{}="B";
 (0,-12)*{}="BB";
    "TL";"B" **\dir{-}?(.5)*\dir{>}?(.17)*\dir{}="1";
    "TR";"B" **\dir{~} ?(.5)*\dir{}="mid";
    "B";"BB" **\dir{-}?(.5)*\dir{>} ?(.7)*\dir{}="2";
    {\ar@/^1pc/@{~}|<<<<{\hole} "1"+(1,0);"2"+(1,0)};
 \endxy
                     \; \; + \; \cdots \; + \;
 \xy
 (-3,10)*{}="TL"; (3,10)*{}="TR";
 (0,3)*{}="B";
 (0,-12)*{}="BB";
    "TL";"B" **\dir{-}?(.35)*\dir{>} ?(.75)*\dir{>};
    "TR";"B"+(0,-.6) **\dir{~};
    "B";"BB" **\dir{-}?(.4)*\dir{>}?(.7)*\dir{>} ?(.24)*\dir{}="1" ?(.85)*\dir{}="2";
    {\ar@/^.35pc/@{~} "1"+(.9,0);"2"+(.9,0)};
     {\ar@{~} (-5.2,-2.7);(0,-8.1)};
   {\ar@{~} (-1.9,8);(-5,2.1)};
     (-5,0)*\xycircle(2,2.8){-};
     {\ar@{~} (-3,0);(0,-5)};
 \endxy
    \; \; + \; \cdots
}

\newcommand{\pachnerII}{
 \vcenter{\xy 0;/r.28pc/:
 (-10,0)*{}="L";
 (10,0)*{}="R";
 (0,16)*{}="T";
 (0,6)*{}="M";
 (-10,12)*{}="TL";
 (10,12)*{}="TR";
    "T";"L" **\dir{-};
    "R";"T" **\dir{-};
    "L";"R" **\dir{-};
 \endxy}
\qquad = \qquad
 \vcenter{\xy 0;/r.28pc/:
 (-10,0)*{}="L";
 (10,0)*{}="R";
 (0,16)*{}="T";
 (0,6)*{}="M";
    "L";"T" **\dir{-};
    "R";"T" **\dir{-};
    "L";"R" **\dir{-};
    "T";"M" **\dir{-};
    "R";"M" **\dir{-};
    "L";"M" **\dir{-};
 \endxy}
 }

\newcommand{\bigpentagon}{
\xy 0;/r.20pc/:
    (-24.73,8.03)*+{\scriptstyle
(w \otimes (x \otimes y)) \otimes z}="l";
    (0,26)*+{\scriptstyle ((w \otimes x) \otimes y) \otimes z}="t";
    (24.73,8.03)*+{\scriptstyle (w \otimes x) \otimes (y \otimes z)}="r";
    (15.28,-21.03)*+{\scriptstyle w \otimes (x \otimes (y \otimes z))}="br";
    (-15.28,-21.03)*+{\scriptstyle w \otimes ((x \otimes y) \otimes z)}="bl";
     {\ar^{} "t";"l"};
     {\ar_{} "l";"bl"};
     {\ar_{} "bl";"br"};
     {\ar_{} "r";"br"};
     {\ar^{} "t";"r"};
(-51.35,16.68)*+{
    \xy 0;/r.16pc/:
     (-10.6,-14.63)*{}="t1";
     (-17.1,5.56)*{}="t2";
     (0,18)*{}="t3";
     (17.1,5.56)*{}="t4";
     (10.6,-14.63)*{}="t5";
   {\ar@{-}^{w} "t1";"t2"};
   {\ar@{-}^{x} "t2";"t3"};
   {\ar@{-}^{y} "t3";"t4"};
   {\ar@{-}^{z} "t4";"t5"};
   {\ar@{-} "t1";"t5"};
   {\ar@{-} "t2";"t4"};
   {\ar@{-} "t1";"t4"};
  \endxy};
(0,54)*+{
    \xy 0;/r.16pc/:
     (-10.6,-14.63)*{}="t1";
     (-17.1,5.56)*{}="t2";
     (0,18)*{}="t3";
     (17.1,5.56)*{}="t4";
     (10.6,-14.63)*{}="t5";
    {\ar@{-}^{w} "t1";"t2"};
    {\ar@{-}^{x} "t2";"t3"};
    {\ar@{-}^{y} "t3";"t4"};
    {\ar@{-}^{z} "t4";"t5"};
    {\ar@{-} "t1";"t5"};
    {\ar@{-} "t1";"t3"};
    {\ar@{-} "t1";"t4"};
   \endxy};
(51.36,16.68)*+{
    \xy 0;/r.16pc/:
     (-10.6,-14.63)*{}="t1";
     (-17.1,5.56)*{}="t2";
     (0,18)*{}="t3";
     (17.1,5.56)*{}="t4";
     (10.6,-14.63)*{}="t5";
   {\ar@{-}^{w} "t1";"t2"};
   {\ar@{-}^{x} "t2";"t3"};
   {\ar@{-}^{y} "t3";"t4"};
   {\ar@{-}^{z} "t4";"t5"};
   {\ar@{-} "t1";"t5"};
   {\ar@{-} "t1";"t3"};
   {\ar@{-} "t3";"t5"};
  \endxy};
(31.73,-43.68)*+{
    \xy 0;/r.16pc/:
     (-10.6,-14.63)*{}="t1";
     (-17.1,5.56)*{}="t2";
     (0,18)*{}="t3";
     (17.1,5.56)*{}="t4";
     (10.6,-14.63)*{}="t5";
    {\ar@{-}^{w} "t1";"t2"};
    {\ar@{-}^{x} "t2";"t3"};
    {\ar@{-}@{-}^{y} "t3";"t4"};
    {\ar@{-}^{z} "t4";"t5"};
    {\ar@{-} "t1";"t5"};{\ar@{-} "t2";"t5"};
    {\ar@{-} "t3";"t5"};
  \endxy};
(-31.73,-43.68)*+{
    \xy 0;/r.16pc/:
     (-10.6,-14.63)*{}="t1";
     (-17.1,5.56)*{}="t2";
     (0,18)*{}="t3";
     (17.1,5.56)*{}="t4";
     (10.6,-14.63)*{}="t5";
    {\ar@{-}^{w} "t1";"t2"};
    {\ar@{-}^{x} "t2";"t3"};
    {\ar@{-}^{y} "t3";"t4"};
    {\ar@{-}^{z} "t4";"t5"};
    {\ar@{-} "t1";"t5"};
    {\ar@{-} "t2";"t4"};
    {\ar@{-} "t2";"t5"};
   \endxy};
\endxy
}

\newcommand{\tinytwothreemove}{
 \xy 0;/r.09pc/:
 (6.18,19)*{}="t1"; 
 (-16.18,11.74)*{}="t2";
 (-16.18,-11.74)*{}="t3";
(6.18,-19)*{}="t4";
 (20,0)*{}="t5";
   {\ar@{-}"t1";"t2"};
   {\ar@{-} "t2";"t3"};
   {\ar@{-} "t3";"t4"};
   {\ar@{-} "t4";"t5"};
   {\ar@{-} "t1";"t5"};
   {\ar@{-} "t1";"t3"}; {\ar@{-} "t3";"t5"};
   {\ar@{-}|>>>>>>{ \hole \; \hole} "t1";"t4"};
   {\ar@{-}|<<<<<<{\hole}|<<<<<<<<<<{ \hole} "t2";"t4"};
\endxy
\quad   = \quad
 \xy 0;/r.09pc/:
 (6.18,19)*{}="t1"; 
 (-16.18,11.74)*{}="t2";
 (-16.18,-11.74)*{}="t3";
(6.18,-19)*{}="t4";
 (20,0)*{}="t5";
   {\ar@{-}"t1";"t2"};
   {\ar@{-} "t2";"t3"};
   {\ar@{-} "t3";"t4"};
   {\ar@{-} "t4";"t5"};
   {\ar@{-} "t1";"t5"};
   {\ar@{-} "t1";"t3"}; {\ar@{-} "t3";"t5"};
   {\ar@{-}|<<<<<<{ \hole \; \hole}|>>>>>>{ \hole \; \hole} "t1";"t4"};
   {\ar@{-}|<<<<<<{ \hole } "t2";"t5"};
   {\ar@{-}|<<<<<<{\hole}|<<<<<<<<<<{ \hole} "t2";"t4"};
\endxy
}

\newcommand{\bigtwothreemove}{
\xy 
 (6.18,19)*{}="t1"; 
 (-16.18,11.74)*{}="t2";
 (-16.18,-11.74)*{}="t3";
(6.18,-19)*{}="t4";
 (20,0)*{}="t5";
   {\ar@{-}"t1";"t2"};
   {\ar@{-} "t2";"t3"};
   {\ar@{-}|<<<<<<<<<<<<<<<{ \hole \; \hole}|>>>>>>>>>>>>>>>{ \hole \; \hole} "t2";"t4"};
   {\ar@{-} "t3";"t4"};
   {\ar@{-} "t4";"t5"};
   {\ar@{-} "t1";"t5"};
   {\ar@{-} "t1";"t3"}; {\ar@{-} "t3";"t5"};
   {\ar@{-}|>>>>>>>>>>>>>>>{ \hole \; \hole}  "t1";"t4"};
\endxy
\qquad \qquad  = \qquad  \qquad
 \xy 
 (6.18,19)*{}="t1"; 
 (-16.18,11.74)*{}="t2";
 (-16.18,-11.74)*{}="t3";
(6.18,-19)*{}="t4";
 (20,0)*{}="t5";
   {\ar@{-}"t1";"t2"};
   {\ar@{-} "t2";"t3"};
   {\ar@{-} "t3";"t4"};
   {\ar@{-} "t4";"t5"};
   {\ar@{-} "t1";"t5"};
   {\ar@{-} "t1";"t3"}; {\ar@{-} "t3";"t5"};
   {\ar@{-}|<<<<<<<<<<<<<<<{ \hole \; \hole}|>>>>>>>>>>>>>>>{ \hole \; \hole} "t1";"t4"};
   {\ar@{-}|<<<<<<<<<<<<<<{ \hole \; \hole} "t2";"t5"};
   {\ar@{-}|<<<<<<<<<<<<<<<{ \hole \; \hole}|>>>>>>>>>>>>>>>{ \hole \; \hole} "t2";"t4"};
\endxy
}

\newcommand{\mediumtwothreemove}{
 \xy 0;/r.17pc/:
 (6.18,19)*{}="t1"; 
 (-16.18,11.74)*{}="t2";
 (-16.18,-11.74)*{}="t3";
(6.18,-19)*{}="t4";
 (20,0)*{}="t5";
   {\ar@{-}"t1";"t2"};
   {\ar@{-} "t2";"t3"};
   {\ar@{-} "t3";"t4"};
   {\ar@{-} "t4";"t5"};
   {\ar@{-} "t1";"t5"};
   {\ar@{-} "t1";"t3"}; {\ar@{-} "t3";"t5"};
   {\ar@{-}|>>>>>>>>>>>{ \hole \; \hole} "t1";"t4"};
   {\ar@{-}|<<<<<<<<<<<{ \hole \; \hole}|>>>>>>>>>>>{ \hole \; \hole} "t2";"t4"};
\endxy
\qquad   \longleftrightarrow   \qquad
 \xy 0;/r.18pc/:
 (6.18,19)*{}="t1"; 
 (-16.18,11.74)*{}="t2";
 (-16.18,-11.74)*{}="t3";
(6.18,-19)*{}="t4";
 (20,0)*{}="t5";
   {\ar@{-}"t1";"t2"};
   {\ar@{-} "t2";"t3"};
   {\ar@{-} "t3";"t4"};
   {\ar@{-} "t4";"t5"};
   {\ar@{-} "t1";"t5"};
   {\ar@{-} "t1";"t3"}; {\ar@{-} "t3";"t5"};
   {\ar@{-}|<<<<<<<<<<<{ \hole \; \hole}|>>>>>>>>>>>{ \hole \; \hole} "t1";"t4"};
   {\ar@{-}|<<<<<<<<<<<<{ \hole \; \hole} "t2";"t5"};
   {\ar@{-}|<<<<<<<<<<<{ \hole \; \hole}|>>>>>>>>>>>{ \hole \; \hole} "t2";"t4"};
\endxy
}

\newcommand{\smalltwothreemove}{
 \xy 0;/r.10pc/:
 (6.18,19)*{}="t1"; 
 (-16.18,11.74)*{}="t2";
 (-16.18,-11.74)*{}="t3";
(6.18,-19)*{}="t4";
 (20,0)*{}="t5";
   {\ar@{-}"t1";"t2"};
   {\ar@{-} "t2";"t3"};
   {\ar@{-} "t3";"t4"};
   {\ar@{-} "t4";"t5"};
   {\ar@{-} "t1";"t5"};
   {\ar@{-} "t1";"t3"}; {\ar@{-} "t3";"t5"};
   {\ar@{-}|>>>>>>>{ \hole \; \hole} "t1";"t4"};
   {\ar@{-}|<<<<<<<{  \hole}|>>>>>>{  \hole} "t2";"t4"};
\endxy
\qquad   = \qquad
 \xy 0;/r.10pc/:
 (6.18,19)*{}="t1"; 
 (-16.18,11.74)*{}="t2";
 (-16.18,-11.74)*{}="t3";
(6.18,-19)*{}="t4";
 (20,0)*{}="t5";
   {\ar@{-}"t1";"t2"};
   {\ar@{-} "t2";"t3"};
   {\ar@{-} "t3";"t4"};
   {\ar@{-} "t4";"t5"};
   {\ar@{-} "t1";"t5"};
   {\ar@{-} "t1";"t3"}; {\ar@{-} "t3";"t5"};
   {\ar@{-}|<<<<<<<{ \hole \; \hole}|>>>>>>>{ \hole \; \hole} "t1";"t4"};
   {\ar@{-}|<<<<<<<{ \hole } "t2";"t5"};
   {\ar@{-}|<<<<<<<{  \hole}|>>>>>>{  \hole} "t2";"t4"};
\endxy
}

\newcommand{\smallfouronemove}{
\xy 0;/r.15pc/:
 (-10,-5 )*{}="1";
 (8,-10)*{}="2";
 (15,0)*{}="3";
 (1,12)*{}="4";
    {\ar@{-} "1";"2" };
    {\ar@{-}"2";"3" };
    {\ar@{-} "4";"3" };
    {\ar@{-} "1";"4" };
    {\ar@{-} "4";"2" };
    {\ar@{.}|>>>>>>>>>>{\hole \hole} "1";"3"};
 \endxy
\qquad =\qquad
 \xy 0;/r.15pc/:
 (-10,-5 )*{}="1";
 (8,-10)*{}="2";
 (15,0)*{}="3";
 (1,12)*{}="4";
 (0,3)*{}="m";
    {\ar@{-} "1";"m" };
    {\ar@{-} "2";"m" };
    {\ar@{-} "3";"m" };
    {\ar@{-} "4";"m" };
    {\ar@{-} "1";"2" };
    {\ar@{-}"2";"3" };
    {\ar@{-} "4";"3" };
    {\ar@{-} "1";"4" };
    {\ar@{-} "4";"2" };
    {\ar@{.}|>>>>>>>>>>{\hole \hole} "1";"3"};
 \endxy
 }

\newcommand{\mediumfouronemove}{
 \xy
 (-10,-5 )*{}="1";
 (8,-10)*{}="2";
 (15,0)*{}="3";
 (1,12)*{}="4";
    {\ar@{-} "1";"2" };
    {\ar@{-}"2";"3" };
    {\ar@{-} "4";"3" };
    {\ar@{-} "1";"4" };
    {\ar@{-} "4";"2" };
    {\ar@{-}|>>>>>>>>>>{\hole \hole} "1";"3"};
 \endxy
\qquad \longleftrightarrow \qquad
 \xy
 (-10,-5 )*{}="1";
 (8,-10)*{}="2";
 (15,0)*{}="3";
 (1,12)*{}="4";
 (0,3)*{}="m";
    {\ar@{-} "1";"m" };
    {\ar@{-} "2";"m" };
    {\ar@{-} "3";"m" };
    {\ar@{-} "4";"m" };
    {\ar@{-} "1";"2" };
    {\ar@{-} "2";"3" };
    {\ar@{-} "4";"3" };
    {\ar@{-} "1";"4" };
    {\ar@{-} "4";"2" };
    {\ar@{-}|>>>>>>>>>>{\hole \hole}|>>>{\hole \hole} "1";"3"};
 \endxy
}

\newcommand{\stickassociator}{
\xy 0;/r.13pc/:
 (-30,0)*{
  \xy 0;/r.15pc/:
    (-8,10)*{}="TL";
    (8,10)*{}="TR";
    (-2,10)*{}="X'";
    (-5,7)*{}="XM'";
    (2,10)*{}="X";
    (5,7)*{}="XM";
    (0,2)*{}="M";
    (0,-10)*{}="B";
    "TL";"M" **\dir{-};
    "X";"XM" **\dir{-};
    "X'";"XM'" **\dir{-};
    "TR";"M" **\dir{-};
    "B";"M" **\dir{-};
    \endxy
    }="1";
 (30,0)*{
    \xy 0;/r.15pc/:
    (-8,10)*{}="TL";
    (8,10)*{}="TR";
    (-2,10)*{}="C";
    (2,10)*{}="X";
    (0,8)*{}="XM";
    (3,5)*{}="CM";
    (0,2)*{}="M";
    (0,-10)*{}="B";
    "TL";"M" **\dir{-};
    "X";"XM" **\dir{-};
    "C";"CM" **\dir{-};
    "TR";"M" **\dir{-};
    "B";"M" **\dir{-};
    \endxy
    }="5";
 (15,-30)*{
    \xy 0;/r.15pc/:
    (-8,10)*{}="TL";
    (8,10)*{}="TR";
    (2,10)*{}="C";
    (-2,10)*{}="X";
    (0,8)*{}="XM";
    (-3,5)*{}="CM";
    (0,2)*{}="M";
    (0,-10)*{}="B";
    "TL";"M" **\dir{-};
    "X";"XM" **\dir{-};
    "C";"CM" **\dir{-};
    "TR";"M" **\dir{-};
    "B";"M" **\dir{-};
    \endxy
    }="4";
 (0,20)*{
    \xy 0;/r.15pc/:
    (-8,10)*{}="TL";
    (8,10)*{}="TR";
    (2,10)*{}="C";
    (-2,10)*{}="X";
    (-5,7)*{}="XM";
    (-3,5)*{}="CM";
    (0,2)*{}="M";
    (0,-10)*{}="B";
    "TL";"M" **\dir{-};
    "X";"XM" **\dir{-};
    "C";"CM" **\dir{-};
    "TR";"M" **\dir{-};
    "B";"M" **\dir{-};
\endxy
    }="2";
 (-15,-30)*{
 \xy 0;/r.15pc/: 
    (8,10)*{}="TL";
    (-8,10)*{}="TR";
    (-2,10)*{}="C";
    (2,10)*{}="X";
    (5,7)*{}="XM";
    (3,5)*{}="CM";
    (0,2)*{}="M";
    (0,-10)*{}="B";
    "TL";"M" **\dir{-};
    "X";"XM" **\dir{-};
    "C";"CM" **\dir{-};
    "TR";"M" **\dir{-};
    "B";"M" **\dir{-};
\endxy
    }="3";
    {\ar@{=>} "2";"1"};
    {\ar@{=>} "2";"5"};
    {\ar@{=>} "1";"3"};
    {\ar@{=>} "4";"3"};
    {\ar@{=>} "5";"4"};
\endxy}

\newcommand{\LTQTtable}{
\begin{tabular}{|c|c|}
  \hline
  \textbf{2D Lattice Field Theory} & \textbf{3D Lattice Field Theory} \\
  \hline
  $\vcenter{\xy (0,0)*{\LARGE \bullet}; \endxy}$  unlabeled
&
  $\vcenter{\xy (0,0)*{\LARGE \bullet}; \endxy}$  unlabeled
\\
  $\vcenter{\xy {\ar (-6,0)*+{\bullet};
  (6,0)*+{\bullet}}; (0,5)*{};(0,-5)*{};\endxy}$  $A \in \Vect$
 & $\vcenter{\xy {\ar (-6,0)*+{\bullet};
  (6,0)*+{\bullet}}; (0,5)*{};(0,-5)*{};\endxy}$  $A \in$ 2-$\Vect$ \\
  $ \vcenter{  \xy 0;/r.16pc/:
  (0,6)*{\bullet};
 (-10,0)*{}="L";
 (10,0)*{}="R";
 (0,16)*{}="T";
 (0,6)*{}="M";
 (0,-4)*{}="B";
 (-10,12)*{}="TL";
 (10,12)*{}="TR";
    "T";"L" **\dir{.};
    "R";"T" **\dir{.};
    "L";"R" **\dir{.};
    "TL";"M" **\dir{-}?(.5)*\dir{>};
    "TR";"M" **\dir{-}?(.5)*\dir{>};
    "M";"B" **\dir{-}?(.6)*\dir{>};
 \endxy}$ $m \maps A \tensor A \to A$ &   $ \vcenter{  \xy 0;/r.16pc/:
  (0,6)*{\bullet};
 (-10,0)*{}="L";
 (10,0)*{}="R";
 (0,16)*{}="T";
 (0,6)*{}="M";
 (0,-4)*{}="B";
 (-10,12)*{}="TL";
 (10,12)*{}="TR";
    "T";"L" **\dir{.};
    "R";"T" **\dir{.};
    "L";"R" **\dir{.};
    "TL";"M" **\dir{-}?(.5)*\dir{>};
    "TR";"M" **\dir{-}?(.5)*\dir{>};
    "M";"B" **\dir{-}?(.6)*\dir{>}; (0,-12)*{};
 \endxy}$ $m \maps A \tensor A \to A$ \footnote{Care must be taken in the definition
 of this tensor product but using a basis this can be done without to much difficulty.} \\
  $\xy 0;/r.16pc/:
 (0,10)*{}="mt";
 (0,-10)*{}="mb";
 (-16,0)*{}="l";
 (16,0)*{}="r";
  (6,0)*{}="xr";
  (-6,0)*{}="xl";
  (-12,10)*{}="xlt";
  (-12,-10)*{}="xlb";
    (12,10)*{}="xrt";
  (12,-10)*{}="xrb";
  (6,0)*{\bullet};
  (-6,0)*{\bullet};
    "mt";"mb" **\dir{.};
    "mb";"l" **\dir{.};
    "mt";"l" **\dir{.};
    "mb";"r" **\dir{.};
    "mt";"r" **\dir{.};
    "xl";"xr" **\dir{-}?(.5)*\dir{<};
    "xl";"xlt" **\dir{-}?(.5)*\dir{<};
    "xl";"xlb" **\dir{-}?(.65)*\dir{>};
    "xr";"xrt" **\dir{-}?(.5)*\dir{<};
    "xr";"xrb" **\dir{-}?(.5)*\dir{<};
 \endxy
 \quad = \quad
 \xy 0;/r.16pc/:
 (0,10)*{}="mt";
 (0,-10)*{}="mb";
 (-16,0)*{}="l";
 (16,0)*{}="r";
  (0,-4)*{}="xr";
  (0,4)*{}="xl";
  (10,10)*{}="xlt";
  (-10,10)*{}="xlb";
    (10,-10)*{}="xrt";
  (-10,-10)*{}="xrb";
  (0,4)*{\bullet};
  (0,-4)*{\bullet};
    "l";"r" **\dir{.};
    "mb";"l" **\dir{.};
    "mt";"l" **\dir{.};
    "mb";"r" **\dir{.};
    "mt";"r" **\dir{.};
    "xl";"xr" **\dir{-}?(.58)*\dir{>};
    "xl";"xlt" **\dir{-}?(.5)*\dir{<};
    "xl";"xlb" **\dir{-}?(.5)*\dir{<};
    "xr";"xrt" **\dir{-}?(.5)*\dir{<};
    "xr";"xrb" **\dir{-}?(.65)*\dir{>};
 \endxy$ &  $\xy 0;/r.16pc/:
 (0,10)*{}="mt";
 (0,-10)*{}="mb";
 (-16,0)*{}="l";
 (16,0)*{}="r";
  (6,0)*{}="xr";
  (-6,0)*{}="xl";
  (-12,10)*{}="xlt";
  (-12,-10)*{}="xlb";
    (12,10)*{}="xrt";
  (12,-10)*{}="xrb";
  (6,0)*{\bullet};
  (-6,0)*{\bullet};
    "mt";"mb" **\dir{.};
    "mb";"l" **\dir{.};
    "mt";"l" **\dir{.};
    "mb";"r" **\dir{.};
    "mt";"r" **\dir{.};
    "xl";"xr" **\dir{-}?(.5)*\dir{<};
    "xl";"xlt" **\dir{-}?(.5)*\dir{<};
    "xl";"xlb" **\dir{-}?(.65)*\dir{>};
    "xr";"xrt" **\dir{-}?(.5)*\dir{<};
    "xr";"xrb" **\dir{-}?(.5)*\dir{<};
 \endxy
 \quad \xy {\ar@{=>}^{\scs \alpha} (-3,0);(3,0)}; \endxy \quad
 \xy 0;/r.16pc/:
 (0,10)*{}="mt";
 (0,-10)*{}="mb";
 (-16,0)*{}="l";
 (16,0)*{}="r";
  (0,-4)*{}="xr";
  (0,4)*{}="xl";
  (10,10)*{}="xlt";
  (-10,10)*{}="xlb";
    (10,-10)*{}="xrt";
  (-10,-10)*{}="xrb";
  (0,4)*{\bullet};
  (0,-4)*{\bullet};
    "l";"r" **\dir{.};
    "mb";"l" **\dir{.};
    "mt";"l" **\dir{.};
    "mb";"r" **\dir{.};
    "mt";"r" **\dir{.};
    "xl";"xr" **\dir{-}?(.58)*\dir{>};
    "xl";"xlt" **\dir{-}?(.5)*\dir{<};
    "xl";"xlb" **\dir{-}?(.5)*\dir{<};
    "xr";"xrt" **\dir{-}?(.5)*\dir{<};
    "xr";"xrb" **\dir{-}?(.65)*\dir{>};
 \endxy$\\
  $\vcenter{\xy 0;/r.16pc/:
  (0,6)*{\bullet};
(0,24)*{}; 
 (-10,0)*{}="L";
 (10,0)*{}="R";
 (0,16)*{}="T";
 (0,6)*{}="M";
 (0,-4)*{}="B";
 (-10,12)*{}="TL";
 (10,12)*{}="TR";
    "T";"L" **\dir{.};
    "R";"T" **\dir{.};
    "L";"R" **\dir{.};
    "TL";"M" **\dir{-};
    "TR";"M" **\dir{-};
    "M";"B" **\dir{-};
 \endxy}
\quad = \quad
 \vcenter{\xy 0;/r.16pc/:
 (0,24)*{}; 
(-10,0)*{}="L";
 (10,0)*{}="R";
 (0,16)*{}="T";
 (0,6)*{}="M";
    "L";"T" **\dir{.};
    "R";"T" **\dir{.};
    "L";"R" **\dir{.};
    "T";"M" **\dir{.};
    "R";"M" **\dir{.};
    "L";"M" **\dir{.};
 (0,-4)*{}="B";
 (-10,12)*{}="TL";
 (10,12)*{}="TR";
 (-3.5,8)*{}="tl";
 (3.5,8)*{}="tr";
 (0,2.5)*{}="b";
    "TL";"tl" **\dir{-};
    "TR";"tr" **\dir{-};
    "b";"B" **\dir{-};
    "tl";"tr" **\dir{-};
    "tr";"b" **\dir{-};
    "tl";"b" **\dir{-}?(.6)*\dir{};
 \endxy}$ &  $\vcenter{\xy 0;/r.16pc/:
  (0,6)*{\bullet};
(0,24)*{}; 
 (-10,0)*{}="L";
 (10,0)*{}="R";
 (0,16)*{}="T";
 (0,6)*{}="M";
 (0,-4)*{}="B";
 (-10,12)*{}="TL";
 (10,12)*{}="TR";
    "T";"L" **\dir{.};
    "R";"T" **\dir{.};
    "L";"R" **\dir{.};
    "TL";"M" **\dir{-};
    "TR";"M" **\dir{-};
    "M";"B" **\dir{-};
 \endxy}
\quad \xy {\ar@{=>}^{} (-3,1);(3,1)};
          {\ar@{=>}^{} (3,-4);(-3,-4)};\endxy \quad
 \vcenter{\xy 0;/r.16pc/:
 (0,24)*{}; 
(-10,0)*{}="L";
 (10,0)*{}="R";
 (0,16)*{}="T";
 (0,6)*{}="M";
    "L";"T" **\dir{.};
    "R";"T" **\dir{.};
    "L";"R" **\dir{.};
    "T";"M" **\dir{.};
    "R";"M" **\dir{.};
    "L";"M" **\dir{.};
 (0,-4)*{}="B";
 (-10,12)*{}="TL";
 (10,12)*{}="TR";
 (-3.5,8)*{}="tl";
 (3.5,8)*{}="tr";
 (0,2.5)*{}="b";
    "TL";"tl" **\dir{-};
    "TR";"tr" **\dir{-};
    "b";"B" **\dir{-};
    "tl";"tr" **\dir{-};
    "tr";"b" **\dir{-};
    "tl";"b" **\dir{-}?(.6)*\dir{};
 \endxy}$ \\
   & $\stickassociator$ \\
   &  \\
  \hline
\end{tabular}
}

\newcommand{\LTQTtableII}{
\begin{tabular}{|c|c|}
  \hline
  \textbf{2D Lattice Field Theory} & \textbf{3D Lattice Field Theory} \\
  \hline
  $\vcenter{\xy (0,0)*{\LARGE \bullet}; \endxy}$  unlabeled
&
  $\vcenter{\xy (0,0)*{\LARGE \bullet}; \endxy}$  unlabeled
\\
  $\vcenter{\xy {\ar (-6,0)*+{\bullet};
  (6,0)*+{\bullet}}; (0,5)*{};(0,-5)*{};\endxy}$  $A \in \Vect$
 & $\vcenter{\xy {\ar (-6,0)*+{\bullet};
  (6,0)*+{\bullet}}; (0,5)*{};(0,-5)*{};\endxy}$  $A \in$ 2-$\Vect$ \\
  $ \vcenter{  \xy 0;/r.16pc/:
  (0,6)*{\bullet};
 (-10,0)*{}="L";
 (10,0)*{}="R";
 (0,16)*{}="T";
 (0,6)*{}="M";
 (0,-4)*{}="B";
 (-10,12)*{}="TL";
 (10,12)*{}="TR";
    "T";"L" **\dir{.};
    "R";"T" **\dir{.};
    "L";"R" **\dir{.};
    "TL";"M" **\dir{-}?(.5)*\dir{>};
    "TR";"M" **\dir{-}?(.5)*\dir{>};
    "M";"B" **\dir{-}?(.6)*\dir{>};
 \endxy}$ $m \maps A \tensor A \to A$ &   $ \vcenter{  \xy 0;/r.16pc/:
  (0,6)*{\bullet};
 (-10,0)*{}="L";
 (10,0)*{}="R";
 (0,16)*{}="T";
 (0,6)*{}="M";
 (0,-4)*{}="B";
 (-10,12)*{}="TL";
 (10,12)*{}="TR";
    "T";"L" **\dir{.};
    "R";"T" **\dir{.};
    "L";"R" **\dir{.};
    "TL";"M" **\dir{-}?(.5)*\dir{>};
    "TR";"M" **\dir{-}?(.5)*\dir{>};
    "M";"B" **\dir{-}?(.6)*\dir{>}; (0,-12)*{};
 \endxy}$ $m \maps A \tensor A \to A$ \footnote{Care must be taken in the definition
 of this tensor product but using a basis this can be done without to much difficulty.} \\
  $\xy 0;/r.16pc/:
 (0,10)*{}="mt";
 (0,-10)*{}="mb";
 (-16,0)*{}="l";
 (16,0)*{}="r";
  (6,0)*{}="xr";
  (-6,0)*{}="xl";
  (-12,10)*{}="xlt";
  (-12,-10)*{}="xlb";
    (12,10)*{}="xrt";
  (12,-10)*{}="xrb";
  (6,0)*{\bullet};
  (-6,0)*{\bullet};
    "mt";"mb" **\dir{.};
    "mb";"l" **\dir{.};
    "mt";"l" **\dir{.};
    "mb";"r" **\dir{.};
    "mt";"r" **\dir{.};
    "xl";"xr" **\dir{-}?(.5)*\dir{<};
    "xl";"xlt" **\dir{-}?(.5)*\dir{<};
    "xl";"xlb" **\dir{-}?(.65)*\dir{>};
    "xr";"xrt" **\dir{-}?(.5)*\dir{<};
    "xr";"xrb" **\dir{-}?(.5)*\dir{<};
 \endxy
 \quad = \quad
 \xy 0;/r.16pc/:
 (0,10)*{}="mt";
 (0,-10)*{}="mb";
 (-16,0)*{}="l";
 (16,0)*{}="r";
  (0,-4)*{}="xr";
  (0,4)*{}="xl";
  (10,10)*{}="xlt";
  (-10,10)*{}="xlb";
    (10,-10)*{}="xrt";
  (-10,-10)*{}="xrb";
  (0,4)*{\bullet};
  (0,-4)*{\bullet};
    "l";"r" **\dir{.};
    "mb";"l" **\dir{.};
    "mt";"l" **\dir{.};
    "mb";"r" **\dir{.};
    "mt";"r" **\dir{.};
    "xl";"xr" **\dir{-}?(.58)*\dir{>};
    "xl";"xlt" **\dir{-}?(.5)*\dir{<};
    "xl";"xlb" **\dir{-}?(.5)*\dir{<};
    "xr";"xrt" **\dir{-}?(.5)*\dir{<};
    "xr";"xrb" **\dir{-}?(.65)*\dir{>};
 \endxy$ &  $\xy 0;/r.16pc/:
 (0,10)*{}="mt";
 (0,-10)*{}="mb";
 (-16,0)*{}="l";
 (16,0)*{}="r";
  (6,0)*{}="xr";
  (-6,0)*{}="xl";
  (-12,10)*{}="xlt";
  (-12,-10)*{}="xlb";
    (12,10)*{}="xrt";
  (12,-10)*{}="xrb";
  (6,0)*{\bullet};
  (-6,0)*{\bullet};
    "mt";"mb" **\dir{.};
    "mb";"l" **\dir{.};
    "mt";"l" **\dir{.};
    "mb";"r" **\dir{.};
    "mt";"r" **\dir{.};
    "xl";"xr" **\dir{-}?(.5)*\dir{<};
    "xl";"xlt" **\dir{-}?(.5)*\dir{<};
    "xl";"xlb" **\dir{-}?(.65)*\dir{>};
    "xr";"xrt" **\dir{-}?(.5)*\dir{<};
    "xr";"xrb" **\dir{-}?(.5)*\dir{<};
 \endxy
 \quad \xy {\ar@{=>}^{\scs \alpha} (-3,0);(3,0)}; \endxy \quad
 \xy 0;/r.16pc/:
 (0,10)*{}="mt";
 (0,-10)*{}="mb";
 (-16,0)*{}="l";
 (16,0)*{}="r";
  (0,-4)*{}="xr";
  (0,4)*{}="xl";
  (10,10)*{}="xlt";
  (-10,10)*{}="xlb";
    (10,-10)*{}="xrt";
  (-10,-10)*{}="xrb";
  (0,4)*{\bullet};
  (0,-4)*{\bullet};
    "l";"r" **\dir{.};
    "mb";"l" **\dir{.};
    "mt";"l" **\dir{.};
    "mb";"r" **\dir{.};
    "mt";"r" **\dir{.};
    "xl";"xr" **\dir{-}?(.58)*\dir{>};
    "xl";"xlt" **\dir{-}?(.5)*\dir{<};
    "xl";"xlb" **\dir{-}?(.5)*\dir{<};
    "xr";"xrt" **\dir{-}?(.5)*\dir{<};
    "xr";"xrb" **\dir{-}?(.65)*\dir{>};
 \endxy$\\
  $\vcenter{\xy 0;/r.16pc/:
  (0,6)*{\bullet};
(0,24)*{}; 
 (-10,0)*{}="L";
 (10,0)*{}="R";
 (0,16)*{}="T";
 (0,6)*{}="M";
 (0,-4)*{}="B";
 (-10,12)*{}="TL";
 (10,12)*{}="TR";
    "T";"L" **\dir{.};
    "R";"T" **\dir{.};
    "L";"R" **\dir{.};
    "TL";"M" **\dir{-};
    "TR";"M" **\dir{-};
    "M";"B" **\dir{-};
 \endxy}
\quad = \quad
 \vcenter{\xy 0;/r.16pc/:
 (0,24)*{}; 
(-10,0)*{}="L";
 (10,0)*{}="R";
 (0,16)*{}="T";
 (0,6)*{}="M";
    "L";"T" **\dir{.};
    "R";"T" **\dir{.};
    "L";"R" **\dir{.};
    "T";"M" **\dir{.};
    "R";"M" **\dir{.};
    "L";"M" **\dir{.};
 (0,-4)*{}="B";
 (-10,12)*{}="TL";
 (10,12)*{}="TR";
 (-3.5,8)*{}="tl";
 (3.5,8)*{}="tr";
 (0,2.5)*{}="b";
    "TL";"tl" **\dir{-};
    "TR";"tr" **\dir{-};
    "b";"B" **\dir{-};
    "tl";"tr" **\dir{-};
    "tr";"b" **\dir{-};
    "tl";"b" **\dir{-}?(.6)*\dir{};
 \endxy}$ &  $\vcenter{\xy 0;/r.16pc/:
  (0,6)*{\bullet};
(0,24)*{}; 
 (-10,0)*{}="L";
 (10,0)*{}="R";
 (0,16)*{}="T";
 (0,6)*{}="M";
 (0,-4)*{}="B";
 (-10,12)*{}="TL";
 (10,12)*{}="TR";
    "T";"L" **\dir{.};
    "R";"T" **\dir{.};
    "L";"R" **\dir{.};
    "TL";"M" **\dir{-};
    "TR";"M" **\dir{-};
    "M";"B" **\dir{-};
 \endxy}
\quad \xy {\ar@{=>}^{} (-3,1);(3,1)};
          {\ar@{=>}^{} (3,-4);(-3,-4)};\endxy \quad
 \vcenter{\xy 0;/r.16pc/:
 (0,24)*{}; 
(-10,0)*{}="L";
 (10,0)*{}="R";
 (0,16)*{}="T";
 (0,6)*{}="M";
    "L";"T" **\dir{.};
    "R";"T" **\dir{.};
    "L";"R" **\dir{.};
    "T";"M" **\dir{.};
    "R";"M" **\dir{.};
    "L";"M" **\dir{.};
 (0,-4)*{}="B";
 (-10,12)*{}="TL";
 (10,12)*{}="TR";
 (-3.5,8)*{}="tl";
 (3.5,8)*{}="tr";
 (0,2.5)*{}="b";
    "TL";"tl" **\dir{-};
    "TR";"tr" **\dir{-};
    "b";"B" **\dir{-};
    "tl";"tr" **\dir{-};
    "tr";"b" **\dir{-};
    "tl";"b" **\dir{-}?(.6)*\dir{};
 \endxy}$ \\
   $\xy (0,15)*{};(0,-15)*{}; \endxy$ & $\smalltwothreemove$ \\
   $\xy (0,-10)*{}; \endxy$& $\smallfouronemove$ \\
  \hline
\end{tabular}
}

\newcommand{\bubblenattrans}{
  \vcenter{
 \xy 
    (-6,4)*{};(6,4)*{};
      **\crv{(5,-10) & (-5,-10)};
\endxy}
\qquad  \xy {\ar@{=>} (-3,0);(3,0)} \endxy  \qquad
 \vcenter{\xy 0;/r.14pc/:
  (0,6)*{};
 (0,6)*{}="M";
 (-6,-4)*{}="B";
 (-9.5,-10)*{}="b";
 (-1.5,-16)*{}="b'";
 (-4,14)*{}="TL";
 (4,12)*{}="TR";
 (12,6)*{}="tr";
    "TL";"M" **\dir{-} ;
    "TR";"M" **\dir{-};
    "M";"B" **\dir{-};
(-16,14)*{}="a";
    "a";"B" **\dir{-};
   "b";"B" **\dir{-};
   "b'";"tr" **\dir{-} ;
  "TR";"tr" **\crv{(8,18) & (16,12)};
  "b";"b'" **\crv{ (-14,-16)&(-6,-22) };
 \endxy}
 }

\newcommand{\dualbubblenattrans}{
  \vcenter{
 \xy 
    (-6,4)*{};(6,4)*{};
      **\crv{(5,-10) & (-5,-10)};
\endxy}
\qquad  \xy {\ar@{=>} (3,0);(-3,0)} \endxy  \qquad
 \vcenter{\xy 0;/r.14pc/:
  (0,6)*{};
 (0,6)*{}="M";
 (-6,-4)*{}="B";
 (-9.5,-10)*{}="b";
 (-1.5,-16)*{}="b'";
 (-4,14)*{}="TL";
 (4,12)*{}="TR";
 (12,6)*{}="tr";
    "TL";"M" **\dir{-} ;
    "TR";"M" **\dir{-};
    "M";"B" **\dir{-};
(-16,14)*{}="a";
    "a";"B" **\dir{-};
   "b";"B" **\dir{-};
   "b'";"tr" **\dir{-} ;
  "TR";"tr" **\crv{(8,18) & (16,12)};
  "b";"b'" **\crv{ (-14,-16)&(-6,-22) };
 \endxy}
 }

\newcommand{\YYYQ}{
\psset{unit=0.5cm}
\begin{pspicture}[.6](4.5,5)
 \psline[linestyle=dashed](0,3)(4,3.5)
 \psline(0,3)(3,2)
 \psline(4,3.5)(3,2)
 \psline(0,3)(2,5)
 \psline(2,5)(3,2)
 \psline(2,5)(4,3.5)
\end{pspicture}
\quad \xy {\ar@{<->} (0,0);(6,0)}; \endxy \quad
\begin{pspicture}[.6](4.5,5)
\pscustom[fillstyle=gradient,
    gradbegin=white, gradend=lightgray,
    gradmidpoint=.5,gradangle=140]{
  \psbezier(0,3)(1,3.8)(3,4.3)(4,3.5)
   \psline(3,2)
   \psline(0,3)
  }
  \pscustom[fillstyle=gradient,
    gradbegin=white, gradend=gray,
    gradmidpoint=.5,gradangle=140]{
  \psbezier(0,3)(1.3,.8)(3.7,.8)(4,3.5)
   \psline(3,2)
   \psline(0,3)
  }
 \psline[linestyle=dashed](0,3)(4,3.5)
 \psline(0,3)(2,5)
 \psline(2,5)(3,2)
 \psline(2,5)(4,3.5)
 \psline[linestyle=dashed](2.25,2.75)(0,3)
 \psline[linestyle=dashed](2.25,2.75)(4,3.5)
 \psline[linestyle=dashed](2.25,2.75)(3,2)
 \psdots(2.25,2.75)
\end{pspicture}
\quad \xy {\ar@{<->} (0,0);(6,0)}; \endxy \quad
\begin{pspicture}[.6](4.5,5)
 \psline[linestyle=dashed](0,3)(4,3.5)
 \psline(0,3)(3,2)
 \psline(4,3.5)(3,2)
 \psline(0,3)(2,5)
 \psline(2,5)(3,2)
 \psline(2,5)(4,3.5)
  \psline(1.9,3.6)(0,3)
  \psline(1.9,3.6)(2,5)
 \psline(1.9,3.6)(4,3.5)
 \psline(1.9,3.6)(3,2)
 \psdots(1.9,3.6)
\end{pspicture}
}

\newcommand{\DiscCircle}{ \psset{xunit=1.2cm,yunit=1.2cm}
 \xy
  (-15,0)*{\begin{pspicture}(.6,.6)
            \pscircle(.3,.3){.3}
            \rput(0,.8){$S^{1}$}        
            \end{pspicture}}="l";
  (15,0)*{\begin{pspicture}(.6,.6)
        \pscircle(.3,.3){.3}
        \rput(.7,.8){$S^{1}$}         
        \end{pspicture}}="r";
  (0,15)*{\begin{pspicture}(.6,.6)
            \pscircle[linewidth=0.3pt,fillcolor=lightgray,fillstyle=solid](.3,.3){.3}
            \rput(0,.8){$D^{2}$}       
            \end{pspicture}}="t";
  {\ar^{i} "l";"t"};
   {\ar^{r} "t";"r"};
    {\ar_{1_{S^{1}}} "l";"r"};
  (0,22)*{};  
 \endxy
 }

\newcommand{\DDDI}{
\psset{linewidth=0.3pt,dimen=middle,
linearc=.05pt,cornersize=absolute}
\begin{pspicture}(3.5,3.5)
     \psline[linestyle=dashed,dash=3pt 2.5pt,linewidth=0.4pt]{c-c}(0,.5)(4,2)
     \pspolygon[linestyle=solid,linecolor=black, fillstyle=solid,fillcolor=lightgray]
        (1.9,1.9)(2.8,2.55)(2.9,1.9)(2,1.6)(1.9,1.9)
     \pspolygon[linestyle=solid,linecolor=black, fillstyle=solid,fillcolor=gray]
        (2,.8)(2,1.5)(2.9,1.9)(3.5,1)(2,.8)
     \pspolygon[linestyle=solid,linecolor=black, fillstyle=solid,fillcolor=lightgray]
        (1.6,1.4)(1,1.9)(1.9,1.9)(2,1.6)(1.6,1.4)
     \pspolygon[linestyle=solid,linecolor=black, fillstyle=solid,fillcolor=gray]
        (1.6,1.4)(1.4,.26)(2,.8)(2,1.5)(1.6,1.4)
     \pspolygon[linestyle=solid,linecolor=black, fillstyle=solid,fillcolor=darkgray]
        (2.2,1.9)(1.6,1.4)(2,1.5)(2.9,1.9)(2.25,1.9)
     \psline[linestyle=dashed,dash=3pt 2pt, linewidth=0.4pt]{c-c}(0,.5)(3,0)
     \psline[linestyle=dashed, dash=3pt 2pt,linewidth=0.4pt]{c-c}(3,0)(4,2)
     \psline[linestyle=dashed,dash=3pt 2pt,linewidth=0.4pt]{c-c}(4,2)(1.75,3)
     \psline[linestyle=dashed,dash=3pt 2pt,linewidth=0.4pt]{c-c}(3,0)(1.75,3)
     \psline[linestyle=dashed,dash=3pt 2pt,linewidth=0.4pt]{c-c}(0,.5)(1.75,3)
\end{pspicture}
}

\newcommand{\DDDII}{
\psset{linewidth=0.3pt,dimen=middle,
linearc=.05pt,cornersize=absolute}
\begin{pspicture}(3.5,3.5)
     \psline[linestyle=dashed,dash=3pt 2.5pt,linewidth=0.4pt]{c-c}(0,.5)(4,2)
     \pspolygon[linestyle=solid,linecolor=black, fillstyle=solid,fillcolor=lightgray]
        (1.9,1.9)(2.8,2.55)(2.9,1.9)(2,1.6)(1.9,1.9)
     \pspolygon[linestyle=solid,linecolor=black, fillstyle=solid,fillcolor=gray]
        (2,.8)(2,1.5)(2.9,1.9)(3.5,1)(2,.8)
     \pspolygon[linestyle=solid,linecolor=black, fillstyle=solid,fillcolor=lightgray]
        (1.6,1.4)(1,1.9)(1.9,1.9)(2,1.6)(1.6,1.4)
     \pspolygon[linestyle=solid,linecolor=black, fillstyle=solid,fillcolor=gray]
        (1.6,1.4)(1.4,.26)(2,.8)(2,1.5)(1.6,1.4)
     \psline[linestyle=solid, linecolor=red, linewidth=1pt]{c-c}(1.9,1.9)(2,1.6)
     \pspolygon[linestyle=solid,linecolor=black, fillstyle=solid,fillcolor=darkgray]
        (2.2,1.9)(1.6,1.4)(2,1.5)(2.9,1.9)(2.2,1.9)
     \psline[linestyle=dashed,dash=3pt 2pt, linewidth=0.4pt]{c-c}(0,.5)(3,0)
     \psline[linestyle=dashed, dash=3pt 2pt,linewidth=0.4pt]{c-c}(3,0)(4,2)
     \psline[linestyle=dashed,dash=3pt 2pt,linewidth=0.4pt]{c-c}(4,2)(1.75,3)
     \psline[linestyle=dashed,dash=3pt 2pt,linewidth=0.4pt]{c-c}(3,0)(1.75,3)
     \psline[linestyle=dashed,dash=3pt 2pt,linewidth=0.4pt]{c-c}(0,.5)(1.75,3)
     \psline[linestyle=solid,linecolor=red, linewidth=1pt]{c-c}(1,1.9)(1.9,1.9)
     \psline[linestyle=solid,linecolor=red, linewidth=1pt]{c-c}(1.9,1.9)(2.8,2.55)
     \psline[linestyle=solid,linecolor=red, linewidth=1pt]{c-c}(2,.8)(2,1.47)
     \psline[linestyle=solid,linecolor=red, linewidth=1pt]{c-c}(3.5,1)(2,.8)
     \psline[linestyle=solid,linecolor=red, linewidth=1pt]{c-c}(1.4,.26)(2,.8)
\end{pspicture}
\qquad \qquad
\psset{linewidth=0.3pt,dimen=middle,
linearc=.05pt,cornersize=absolute}
\begin{pspicture}(3.5,3.5)
     \psline[linestyle=dashed,dash=3pt 2.5pt,linewidth=0.4pt]{c-c}(0,.5)(4,2)
     \pspolygon[linestyle=solid,linecolor=black, fillstyle=solid,fillcolor=lightgray]
        (1.9,1.9)(2.8,2.55)(2.9,1.9)(2,1.6)(1.9,1.9)
     \pspolygon[linestyle=solid,linecolor=black, fillstyle=solid,fillcolor=gray]
        (2,.8)(2,1.5)(2.9,1.9)(3.5,1)(2,.8)
     \pspolygon[linestyle=solid,linecolor=black, fillstyle=solid,fillcolor=lightgray]
        (1.6,1.4)(1,1.9)(1.9,1.9)(2,1.6)(1.6,1.4)
     \pspolygon[linestyle=solid,linecolor=black, fillstyle=solid,fillcolor=gray]
        (1.6,1.4)(1.4,.26)(2,.8)(2,1.5)(1.6,1.4)
     \pspolygon[linestyle=solid,linecolor=black, fillstyle=solid,fillcolor=darkgray]
        (2.2,1.9)(1.6,1.4)(2,1.5)(2.9,1.9)(2.2,1.9)
     \psline[linestyle=dashed,dash=3pt 2pt, linewidth=0.4pt]{c-c}(0,.5)(3,0)
     \psline[linestyle=dashed, dash=3pt 2pt,linewidth=0.4pt]{c-c}(3,0)(4,2)
     \psline[linestyle=dashed,dash=3pt 2pt,linewidth=0.4pt]{c-c}(4,2)(1.75,3)
     \psline[linestyle=dashed,dash=3pt 2pt,linewidth=0.4pt]{c-c}(3,0)(1.75,3)
     \psline[linestyle=dashed,dash=3pt 2pt,linewidth=0.4pt]{c-c}(0,.5)(1.75,3)
     \psline[linestyle=solid,linecolor=red, linewidth=1pt]{c-c}(1.4,.26)(1.6,1.4)
     \psline[linestyle=solid,linecolor=red, linewidth=1pt]{c-c}(1.6,1.4)(1,1.9)
     \psline[linestyle=solid,linecolor=red, linewidth=1pt]{c-c}(2.2,1.9)(1.6,1.4)
     \psline[linestyle=solid,linecolor=red, linewidth=1pt]{c-c}(2.9,1.9)(2.2,1.9)
     \psline[linestyle=solid,linecolor=red, linewidth=1pt]{c-c}(2.8,2.55)(2.9,1.9)
     \psline[linestyle=solid,linecolor=red, linewidth=1pt]{c-c}(2.9,1.9)(3.5,1)
\end{pspicture}
}

\newcommand{\multl}{
  \pscustom[fillcolor=lightgray, fillstyle=solid]{
        \psbezier(1.5,2.5)(1.5,1.1)(.4,1.6)(.5,0)
        \psline(-0.5,0)
        \psbezier(-0.5,0)(-.4,1.6)(-1.5,1.1)(-1.5,2.5)
        \psline(-.5,2.5)
        \psbezier(-.5,2.5)(-.6,1.5)(0.6,1.5)(.5,2.5)
        \psline(1.5,2.5)
    }
}

\newcommand{\comultl}{
  \pscustom[fillcolor=lightgray, fillstyle=solid]{
        \psbezier(1.5,0)(1.5,1.4)(.4,.9)(.5,2.5)
        \psline(-0.5,2.5)
        \psbezier(-0.5,2.5)(-.4,.9)(-1.5,1.4)(-1.5,0)
        \psline(-.5,0)
        \psbezier(-.5,0)(-.6,1)(0.6,1)(.5,0)
        \psline(1.5,0)
    }
}

\newcommand{\ctl}{
  \begin{psclip}{
    \pscustom{
        \psline(-.58,2)(-.58,0)
        \psline(-.58,0)(.42,0)
        \psline(.42,0)(.42,2)
        \psellipse(-.08,2)(.5,.2)
    }
  }
    \pspolygon[fillcolor=lightgray,fillstyle=gradient,
    gradbegin=lightgray,gradend=gray,gradmidpoint=1,gradangle=110](-.58,0)(-.58,2.4)(.42,2.4)(.42,0)(-.58,0)
 \end{psclip}
 \pscustom[fillcolor=lightgray,fillstyle=gradient,
        gradbegin=white, gradend=gray,gradmidpoint=0,gradangle=88]{
    \psline(-.58,2)(-.58,0)
    \psbezier(-.58,0)(-.48,.5)(-.48,.7)(-.08,1)
    \psbezier(-.08,1)(.32,.7)(.32,.5)(.42,0)
    \psellipse(-.08,2)(.5,.2)
 }
 \psellipse[fillcolor=lightgray,fillstyle=gradient,
        gradbegin=lightgray, gradend=gray,gradmidpoint=1,gradangle=110](-.08,2)(.5,.2)
}

\newcommand{\ltc}{
    \pspolygon[fillcolor=lightgray,fillstyle=gradient,
    gradbegin=lightgray,gradend=gray,gradmidpoint=1,gradangle=60](.58,2)(.58,.4)(-.42,.4)(-.42,2)(.58,2)
 \pscustom[fillcolor=lightgray,fillstyle=gradient,
        gradbegin=white, gradend=gray,gradmidpoint=0,gradangle=88]{
    \psline(.58,0)(.58,2)
    \psbezier(.58,2)(.48,1.5)(.48,1.3)(.08,1)
    \psbezier(.08,1)(-.32,1.3)(-.32,1.5)(-.42,2)
    \psline(-.42,0)
    \psbezier(-.42,0)(-.32,-.25)(.48,-.25)(.58,0)
 }
 \begin{psclip}{
 \pspolygon[linestyle=none](.58,0)(.58,.3)(-.42,.3)(-.42,0)(.58,0)
 }
 \psellipse[linestyle=dotted](.08,0)(.5,0.2)
 \end{psclip}
}

 \newcommand{\birthl}{
 \pscustom[fillcolor=lightgray, fillstyle=solid]{
        \psbezier(-.5,0)(-.5,.9)(0.5,.9)(.5,0)
        \psline(-.5,0)
    }
 }

  \newcommand{\deathl}{
 \pscustom[fillcolor=lightgray, fillstyle=solid]{
        \psbezier(-.5,0)(-.5,-.9)(0.5,-.9)(.5,0)
        \psline(-.5,0)    }
 }

\newcommand{\multc}{
      \pscustom[fillstyle=gradient,
    gradbegin=white, gradend=gray,gradmidpoint=0,gradangle=70]{
        \psbezier(1.5,2.5)(1.5,1.1)(.4,1.6)(.5,0)
        \psbezier(.5,0)(.4,-.25)(-.4,-.25)(-.5,0)
        \psbezier(-0.5,0)(-.4,1.6)(-1.5,1.1)(-1.5,2.5)
        \psline(-.5,2.5)
        \psbezier(-.5,2.5)(-.6,1.5)(0.6,1.5)(.5,2.5)
        \psline(1.5,2.5)
    }
    \psellipse[fillcolor=lightgray,fillstyle=gradient,
        gradbegin=lightgray, gradend=gray,gradmidpoint=1,gradangle=110](-1,2.5)(.5,.2)
    \psellipse[fillcolor=lightgray,fillstyle=gradient,
        gradbegin=lightgray, gradend=gray,gradmidpoint=1,gradangle=110](1,2.5)(.5,.2)
     \begin{psclip}{
 \pspolygon[linestyle=none](.5,0)(.5,.3)(-.5,.3)(-.5,0)(.5,0)
 }
 \psellipse[linestyle=dotted](0,0)(.5,0.2)
 \end{psclip}
 }

\newcommand{\comultc}{
  \pscustom[fillstyle=gradient,
    gradbegin=white, gradend=gray,gradmidpoint=0,gradangle=110]{
        \psbezier(1.5,0)(1.5,1.4)(.4,.9)(.5,2.5)
        \psline(-0.5,2.5)
        \psbezier(-0.5,2.5)(-.4,.9)(-1.5,1.4)(-1.5,0)
        \psbezier(-1.5,0)(-1.4,-.25)(-.6,-.25)(-.5,0)
        \psbezier(-.5,0)(-.6,1)(0.6,1)(.5,0)
        \psbezier(.5,0)(.6,-.25)(1.4,-.25)(1.5,0)
    }
  \psellipse[fillcolor=lightgray,fillstyle=gradient,
        gradbegin=lightgray, gradend=gray,gradmidpoint=1,gradangle=110](0,2.5)(.5,.2)
\begin{psclip}{
 \pspolygon[linestyle=none](1.5,0)(1.5,.3)(-1.5,.3)(-1.5,0)(1.5,0)
 }
 \psellipse[linestyle=dotted](1,0)(.5,0.2)
 \psellipse[linestyle=dotted](-1,0)(.5,0.2)
 \end{psclip}
 }

\newcommand{\birthc}{
 \pscustom[fillstyle=gradient,
    gradbegin=white, gradend=gray,gradmidpoint=0,gradangle=110]{
        \psbezier(-.5,0)(-.5,.9)(0.5,.9)(.5,0)
        \psbezier(.5,0)(.4,-.25)(-.4,-.25)(-.5,0)
    }
 \begin{psclip}{
 \pspolygon[linestyle=none](.5,0)(.5,.3)(-.5,.3)(-.5,0)(.5,0)
 }
 \psellipse[linestyle=dotted](0,0)(.5,0.2)
 \end{psclip}
 }

\newcommand{\deathc}{
 \pscustom[fillstyle=gradient,
    gradbegin=white, gradend=gray,gradmidpoint=0,gradangle=70]{
        \psbezier(-.5,1)(-.5,.1)(0.5,.1)(.5,1)
        \psline(-.5,1)
 }
  \psellipse[fillcolor=lightgray,fillstyle=gradient,
        gradbegin=lightgray, gradend=gray,gradmidpoint=1,gradangle=110](0,1)(.5,.2)
 }

\newcommand{\identc}{
 \pscustom[fillcolor=lightgray,fillstyle=gradient,
        gradbegin=white, gradend=gray,gradmidpoint=0,gradangle=88]{
 \psline(.5,0)(.5,2.5)
 \psline(-.5,2.5)
 \psline(-.5,0)
 \psbezier(-.5,0)(-.4,-.25)(.4,-.25)(.5,0)
 }
\psellipse[fillcolor=lightgray,fillstyle=gradient,
        gradbegin=lightgray, gradend=gray,gradmidpoint=1,gradangle=110](0,2.5)(.5,.2)
 \begin{psclip}{
 \pspolygon[linestyle=none](.5,0)(.5,.3)(-.5,.3)(-.5,0)(.5,0)
 }
 \psellipse[linestyle=dotted](0,0)(.5,0.2)
 \end{psclip}
 }

 \newcommand{\smallidentc}{
     \pscustom[fillcolor=lightgray,fillstyle=gradient,
        gradbegin=white, gradend=gray,gradmidpoint=0,gradangle=88]{
        \psline(-.5,1)(-.5,0)
        \psbezier(-.5,0)(-.4,-.25)(.4,-.25)(.5,0)
        \psline(.5,1)
        \psline(-.5,1)
    }
\psellipse[fillcolor=lightgray,fillstyle=gradient,
        gradbegin=lightgray, gradend=gray,gradmidpoint=1,gradangle=110](0,1)(.5,.2)
 \begin{psclip}{
 \pspolygon[linestyle=none](.5,0)(.5,.3)(-.5,.3)(-.5,0)(.5,0)
 }
 \psellipse[linestyle=dotted](0,0)(.5,0.2)
 \end{psclip}
}

\newcommand{\ucrossc}{
\pscustom[fillcolor=lightgray,fillstyle=gradient,
        gradbegin=white, gradend=gray,gradmidpoint=0,gradangle=125]{
 \psline(.5,0)(-1.5,2.5)
 \psline(-.5,2.5)
 \psline(1.5,0)
 \psbezier(1.5,0)(1.4,-.25)(.6,-.25)(.5,0)
 }
  \pscustom[fillcolor=lightgray,fillstyle=gradient,
        gradbegin=white, gradend=gray,gradmidpoint=0,gradangle=70]{
 \psline(-.5,0)(1.5,2.5)
 \psline(.5,2.5)
 \psline(-1.5,0)
 \psbezier(-1.5,0)(-1.4,-.25)(-.6,-.25)(-.5,0)
 }
\psellipse[fillcolor=lightgray,fillstyle=gradient,
        gradbegin=lightgray, gradend=gray,gradmidpoint=1,gradangle=110](-1,2.5)(.5,.2)
\psellipse[fillcolor=lightgray,fillstyle=gradient,
        gradbegin=lightgray, gradend=gray,gradmidpoint=1,gradangle=110](1,2.5)(.5,.2)
 \psline[linestyle=dotted](-1.5,2.5)(.5,0)
 \psline[linestyle=dotted](-.5,2.5)(1.5,0)
 \begin{psclip}{
 \pspolygon[linestyle=none](1.5,0)(1.5,.3)(-1.5,.3)(-1.5,0)(1.5,0)
 }
 \psellipse[linestyle=dotted](1,0)(.5,0.2)
 \psellipse[linestyle=dotted](-1,0)(.5,0.2)
 \end{psclip}
}

\newcommand{\curverightc}{
  \pscustom[fillstyle=gradient,
    gradbegin=white, gradend=gray,gradmidpoint=0,gradangle=65]{
        \psbezier(1.5,2.5)(1.5,1.5)(.4,1.3)(.5,0)
        \psbezier(.5,0)(.4,-.25)(-.4,-.25)(-.5,0)
        \psbezier(-.5,0)(-.6,1.3)(.5,1.5)(.5,2.5)
        \psline(1.5,2.5)
    }
     \psellipse[fillcolor=lightgray,fillstyle=gradient,
        gradbegin=lightgray, gradend=gray,gradmidpoint=1,gradangle=110](1,2.5)(.5,.2) \begin{psclip}{
 \pspolygon[linestyle=none](.5,0)(.5,.3)(-.5,.3)(-.5,0)(.5,0)
 }
 \psellipse[linestyle=dotted](0,0)(.5,0.2)
 \end{psclip}
}

\newcommand{\curveleftc}{
  \pscustom[fillstyle=gradient,
    gradbegin=white, gradend=gray,gradmidpoint=0,gradangle=115]{
        \psbezier(-1.5,2.5)(-1.5,1.5)(-.4,1.3)(-.5,0)
        \psbezier(-.5,0)(-.4,-.25)(.4,-.25)(.5,0)
        \psbezier(.5,0)(.6,1.3)(-.5,1.5)(-.5,2.5)
        \psline(-1.5,2.5)
    }
    \psellipse[fillcolor=lightgray,fillstyle=gradient,
        gradbegin=lightgray, gradend=gray,gradmidpoint=1,gradangle=110](-1,2.5)(.5,.2)
 \begin{psclip}{
 \pspolygon[linestyle=none](.5,0)(.5,.3)(-.5,.3)(-.5,0)(.5,0)
 }
 \psellipse[linestyle=dotted](0,0)(.5,0.2)
 \end{psclip}
}

\newcommand{\stringZIGZAGi}{
\xy
    (-10,-12)*{}="1E";
    (-10,0)*{}="1";
    (0,0)*{}="2";
    (10,0)*{}="3";
    (10,12)*{}="3B";
 "2";"1" **\crv{(0,10)& (-10,10)}
     ?(.03)*\dir{>}  ?(1)*\dir{>};
 "3";"2" **\crv{(10,-10)& (0,-10)}
     ?(.03)*\dir{>}  ;
 "1";"1E" **\dir{-};
 "3B";"3" **\dir{-};
     (-5,8.5)*{\scriptstyle };
     (5,-9)*{\scriptstyle };
\endxy
\qquad = \xy (-6,10)*{}; (0,12)*{}; (0,-12)*{}; **\dir{-}
?(.47)*\dir{<}; (6,-10)*{}; (4,0)*{\scriptstyle }
\endxy
}

\newcommand{\stringZIGZAGii}{
  \xy
    (-10,12)*{}="1E";
    (-10,0)*{}="1";
    (0,0)*{}="2";
    (10,0)*{}="3";
    (10,-12)*{}="3B";
 "2";"1" **\crv{(0,-10)& (-10,-10)}
     ?(.03)*\dir{>}  ?(1)*\dir{>};
 "3";"2" **\crv{(10,10)& (0,10)}
     ?(.03)*\dir{>}  ;
 "1";"1E" **\dir{-};
 "3B";"3" **\dir{-};
      (5,10)*{\scriptstyle  };
     (-5,-10.5)*{\scriptstyle };
\endxy
\qquad = \xy (-6,10)*{}; (0,12)*{}; (0,-12)*{}; **\dir{-}
?(.53)*\dir{>}; (6,-10)*{}; (4,0)*{\scriptstyle };
\endxy
}

\newcommand{\stringDIM}{
 \xy
    (-5,8)*{}="x1";
    (5,8)*{}="x2";
    (-5,5)*{}="m1";
    (5,5)*{}="m2";
    (-5,-5)*{}="k1";
    (5,-5)*{}="k2";
    (-5,-8)*{}="y1";
    (5,-8)*{}="y2";
 \vtwist~{"m1"}{"m2"}{"k1"}{"k2"};
 "x1";"m1" **\dir{-} ?(.25)*\dir{>};
 "x2";"m2" **\dir{-} ?(0)*\dir{<};
 "k1";"y1" **\dir{-} ?(0)*\dir{<};
 "k2";"y2" **\dir{-} ?(.27)*\dir{>};
    "x1";"x2" **\crv{(-5,16) & (5,16)};
    "y1";"y2" **\crv{(-5,-14) & (5,-14)};
        (-7.5,8)*{H};
\endxy
}

\newcommand{\DIMofH}{
 \vcenter{\xy
    (-5,8)*{}="x1";
    (5,8)*{}="x2";
    (-5,5)*{}="m1";
    (5,5)*{}="m2";
    (-5,-5)*{}="k1";
    (5,-5)*{}="k2";
    (-5,-8)*{}="y1";
    (5,-8)*{}="y2";
 \vtwist~{"m1"}{"m2"}{"k1"}{"k2"};
 "x1";"m1" **\dir{-} ?(.25)*\dir{>};
 "x2";"m2" **\dir{-} ?(0)*\dir{<};
 "k1";"y1" **\dir{-} ?(0)*\dir{<};
 "k2";"y2" **\dir{-} ?(.27)*\dir{>};
    "x1";"x2" **\crv{(-5,16) & (5,16)};
    "y1";"y2" **\crv{(-5,-14) & (5,-14)};
        (-7.5,8)*{H};
\endxy}
 \qquad \qquad
 \vcenter{ \xy
    (0,14)*+{1}="1";
    (0,5)*+{e^i \tensor e_i}="2";
    (0,-5)*+{e_i \tensor e^i}="3";
    (0,-14)*+{\delta^i_i = \dim(H)}="4";
        {\ar@{|->} "1";"2"};
        {\ar@{|->} "2";"3"};
        {\ar@{|->} "3";"4"};
\endxy }
}

\newcommand{\DIMofx}{
 \vcenter{\xy
    (-5,8)*{}="x1";
    (5,8)*{}="x2";
    (-5,5)*{}="m1";
    (5,5)*{}="m2";
    (-5,-5)*{}="k1";
    (5,-5)*{}="k2";
    (-5,-8)*{}="y1";
    (5,-8)*{}="y2";
 \vtwist~{"m1"}{"m2"}{"k1"}{"k2"};
 "x1";"m1" **\dir{-} ?(.25)*\dir{>};
 "x2";"m2" **\dir{-} ?(0)*\dir{<};
 "k1";"y1" **\dir{-} ?(0)*\dir{<};
 "k2";"y2" **\dir{-} ?(.27)*\dir{>};
    "x1";"x2" **\crv{(-5,16) & (5,16)};
    "y1";"y2" **\crv{(-5,-14) & (5,-14)};
        (-7.5,8)*{x};
\endxy}
 \qquad \qquad
 \vcenter{ \xy
    (0,14)*+{1}="1";
    (0,5)*+{e^i \tensor e_i}="2";
    (0,-5)*+{e_i \tensor e^i}="3";
    (0,-14)*+{\delta^i_i = \dim(H)}="4";
        {\ar@{|->} "1";"2"};
        {\ar@{|->} "2";"3"};
        {\ar@{|->} "3";"4"};
\endxy }
}

\newcommand{\feynmanI}{
\xy
 (0,10)*{}="1";
 (0,4)*{}="2";
 (0,-4)*{}="3";
 (0,-10)*{}="4";
    "1";"2"+(0,-1) **\dir{~};
    "3"+(0,+.9);"4" **\dir{~};
    "2";"3" **\crv{(-3,4) & (-3,-4)}?(.6)*\dir{>};
    "2";"3" **\crv{(3,4) & (3,-4)}?(.45)*\dir{<};
\endxy
}

\newcommand{\feynmanII}{
\xy
 (0,10)*{}="1";
 (6,4)*{}="2";
 (6,-4)*{}="3";
 (0,-10)*{}="4";
    "1";"4" **\dir{-}?(.55)*\dir{>};
    "2";"3" **\crv{(3,4) & (3,-4)}?(.4)*\dir{<};
    "2";"3" **\crv{(9,4) & (9,-4)}?(.55)*\dir{>};
\endxy
}

\newcommand{\binorID}{
 \xy
 (-4,8)*{}="TL"; (4,8)*{}="TR";
 (-4,-8)*{}="BL"; (4,-8)*{}="BR";
    \vtwist~{"TL"}{"TR"}{"BL"}{"BR"};
    (-6,6.5)*{\scriptstyle \frac{1}{2}};
    (6,6.5)*{\scriptstyle \frac{1}{2}};
 \endxy
\quad = \quad
 \xy
 (-4,8)*{}="TL"; (4,8)*{}="TR";
 (-4,-8)*{}="BL"; (4,-8)*{}="BR";
    "TL";"TR" **\crv{(-3,0) & (3,0)};
    "BL";"BR" **\crv{(-3,0) & (3,0)};
    (-6,6.5)*{\scriptstyle \frac{1}{2}};
    (6,6.5)*{\scriptstyle \frac{1}{2}};
    (-6,-6.5)*{\scriptstyle \frac{1}{2}};
    (6,-6.5)*{\scriptstyle \frac{1}{2}};
 \endxy
\quad + \quad
  \xy
 (-4,8)*{}="TL"; (4,8)*{}="TR";
 (-4,-8)*{}="BL"; (4,-8)*{}="BR";
    "TL";"BL" **\dir{-};
    "TR";"BR" **\dir{-};
    (-6,6.5)*{\scriptstyle \frac{1}{2}};
    (6,6.5)*{\scriptstyle \frac{1}{2}};
 \endxy
 }

\newcommand{\useBINOR}{
\xy
  (0,15)*{}="T";
  (0,-15)*{}="B";
  (0,7.5)*{}="T'";
  (0,-7.5)*{}="B'";
    "T";"T'" **\dir{-};
    "B";"B'" **\dir{-};
    (-4.5,0)*{}="MB";
    (-10.5,0)*{}="LB";
    "T'";"LB" **\crv{(-1.5,-6) & (-10.5,-6)}; \POS?(.25)*{\hole}="2z";
    "LB"; "2z" **\crv{(-12,9) & (-3,9)};
    "2z";"B'" **\crv{(0,-4.5)};
    (2,13)*{\scs \frac{1}{2}};
    \endxy
    \quad = \quad
    \xy
  (0,15)*{}="T";
  (0,-15)*{}="B";
  (0,4)*{}="T'";
  (0,-4)*{}="B'";
    "T";"T'" **\dir{-};
    "B";"B'" **\dir{-};
    (-4,4)*{}="t1";
    (-10,4)*{}="t2";
    (-4,-4)*{}="b1";
    (-10,-4)*{}="b2";
    "T'"; "t1" **\crv{(0,0) & (-4,0)};
    "t1"; "t2" **\crv{(-4,7) & (-10,7)};
    "t2";"b2" **\dir{-};
    "B'"; "b1" **\crv{(0,0) & (-4,0)};
    "b1"; "b2" **\crv{(-4,-7) & (-10,-7)};
    (2,13)*{\scs \frac{1}{2}};
    \endxy
    \quad + \quad
    \xy
  (0,15)*{}="T";
  (0,-15)*{}="B";
  (0,4)*{}="T'";
  (0,-4)*{}="B'";
    "T";"B" **\dir{-};
    (-4,4)*{}="t1";
    (-10,4)*{}="t2";
    (-4,-4)*{}="b1";
    (-10,-4)*{}="b2";
    "t1"; "t2" **\crv{(-4,7) & (-10,7)};
    "t2";"b2" **\dir{-};
    "t1";"b1" **\dir{-};
    "b1"; "b2" **\crv{(-4,-7) & (-10,-7)};
    (2,13)*{\scs \frac{1}{2}};
    \endxy
        \quad =\quad -\;\;
    \xy
  (0,15)*{}="T";
  (0,-15)*{}="B";
    "T";"B" **\dir{-};
    (2,13)*{\scs \frac{1}{2}};
    \endxy
    }

\newcommand{\spinNET}{
\xy
    (-4,0)*{}="b";
    (4,8)*{}="t";
    (9,-6)*{}="v";
    (14,6)*{}="u";
    (-5,8)*{}="e";
    (-20,1)*{}="q";
    (-10,-10)*{}="c";
    (10,-15)*{}="d";
    (22,4)*{}="z";
    (24,-9)*{}="w";
        {\ar@/^1pc/@{-} "q";"e"};
        {\ar@/_.1pc/@{-} "q";"b"};
        {\ar@{-} "b";"e"};
        {\ar@/^.2pc/@{-} "e";"t"};
        {\ar@/_.5pc/@{-} "u";"t"};
        {\ar@/^.5pc/@{-} "u";"t"};
        {\ar@/_.2pc/@{-} "b";"v"};
        {\ar@/_.2pc/@{-} "z";"u"};
        {\ar@/_1pc/@{-} "q";"c"};
        {\ar@/_.8pc/@{-} "c";"d"};
        {\ar@/^.6pc/@{-} "c";"d"};
        {\ar@/^.3pc/@{-} "z";"v"};
        {\ar@/^.2pc/@{-} "z";"w"};
        {\ar@/^.4pc/@{-} "w";"d"};
        {\ar@{-} "v";"w"};
 \endxy
 }

\newcommand{\spinNETII}{
\def\objectstyle{\scriptscriptstyle}
\xy
    (-4,0)*{\bullet}="b";
    (4,8)*{\bullet}="t";
    (9,-6)*{\bullet}="v";
    (14,6)*{\bullet}="u";
    (-5,8)*{\bullet}="e";
    (-20,1)*{\bullet}="q";
    (-10,-10)*{\bullet}="c";
    (10,-15)*{\bullet}="d";
    (22,4)*{\bullet}="z";
    (24,-9)*{\bullet}="w";
        {\ar@/^1pc/@{-}^1 "q";"e"};
        {\ar@/_.1pc/@{-}^1 "q";"b"};
        {\ar@{-}^1 "b";"e"};
        {\ar@/^.2pc/@{-}_1 "e";"t"};
        {\ar@/_.5pc/@{-}_1 "u";"t"};
        {\ar@/^.5pc/@{-}^1 "u";"t"};
        {\ar@/_.2pc/@{-}^1 "b";"v"};
        {\ar@/_.2pc/@{-}^1 "z";"u"};
        {\ar@/_1pc/@{-}_1 "q";"c"};
        {\ar@/_.8pc/@{-}_1 "c";"d"};
        {\ar@/^.6pc/@{-}^1 "c";"d"};
        {\ar@/^.3pc/@{-}^1 "z";"v"};
        {\ar@/^.2pc/@{-}^1 "z";"w"};
        {\ar@/^.4pc/@{-}^1 "w";"d"};
        {\ar@{-}_1 "v";"w"};
 \endxy
 }

\newcommand{\PLUGzigzag}{ \xy
    (-10,-12)*{}="1E";
    (-10,0)*{}="1";
    (0,0)*{}="2";
    (10,0)*{}="3"+(3,0)*{H};
    (10,12)*{}="3B";
 "2";"1" **\crv{(0,10)& (-10,10)}
     ?(.03)*\dir{>}  ?(1)*\dir{>};
 "3";"2" **\crv{(10,-10)& (0,-10)}
     ?(.03)*\dir{>}  ;
 "1";"1E" **\dir{-};
 "3B";"3" **\dir{-};
     (-5,8.5)*{\scriptstyle e_H};
     (5,-9)*{\scriptstyle i_H};
\endxy
 \qquad \qquad
\xy
    (0,11)*+{\psi}="1";
    (0,0)*+{e_i \tensor e^i \tensor \psi}="2";
    (0,-11)*+{e_i \tensor \psi^i = \psi}="3";
        {\ar@{|->} "1";"2"};
        {\ar@{|->} "2";"3"};
\endxy
}

\newcommand{\SIXjI}{
\psset{xunit=1.2cm,yunit=1.2cm}
\psset{linewidth=0.3pt,dimen=middle,linearc=.05pt,cornersize=absolute}
\begin{pspicture}[.3](3.5,3.5)
     \pspolygon[linewidth=0.5pt,linestyle=solid,linecolor=black, fillstyle=gradient,gradbegin=white,
        gradend=lightgray,gradmidpoint=1,gradangle=130](0,.5)(3,0)(1.75,3)(0,.5)
     \pspolygon[linewidth=0.5pt,linestyle=solid,linecolor=black, fillstyle=gradient,gradbegin=lightgray,
        gradend=gray,gradmidpoint=0,gradangle=310](3,0)(1.75,3)(4,2)(3,0)
     \psline[linestyle=dotted,linewidth=0.5pt]{c-c}(0,.5)(4,2)
     \rput(.7,2){$i$}
     \rput(1.2,.1){$p$}
     \rput(1.2,1.2){$l$}
     \rput(3,2.7){$q$}
     \rput(2.8,1){$j$}
     \rput(3.8,.8){$k$}
\end{pspicture}
}

\newcommand{\SIXjII}{
\psset{xunit=1.2cm,yunit=1.2cm}
\psset{linewidth=0.3pt,dimen=middle,linearc=.05pt,cornersize=absolute}
\begin{pspicture}[.3](3.5,3.5)
     \pspolygon[linewidth=0.5pt,linestyle=solid,linecolor=black, fillstyle=gradient,gradbegin=white,
        gradend=lightgray,gradmidpoint=1,gradangle=130](0,.5)(3,0)(1.75,3)(0,.5)
     \pspolygon[linewidth=0.5pt,linestyle=solid,linecolor=black, fillstyle=gradient,gradbegin=lightgray,
        gradend=gray,gradmidpoint=0,gradangle=310](3,0)(1.75,3)(4,2)(3,0)
     \pspolygon(1.6,1.4)(2.9,1.9)(2.4,.8)(1.6,1.4)
     \pspolygon(1.6,1.4)(1.9,1.9)(2.9,1.9)(1.6,1.4)
    \psline(1.9,1.9)(2,1.7)
    \psline(2.08,1.5)(2.4,.8)
     \psline[linestyle=dotted,linewidth=0.5pt]{c-c}(0,.5)(1.85,1.2)
     \psline[linestyle=dotted,linewidth=0.5pt]{c-c}(2.75,1.52)(4,2)
     \rput(.7,2){$i$}
     \rput(1.2,.1){$p$}
     \rput(1.2,1.2){$l$}
     \rput(3,2.7){$q$}
     \rput(2.8,1){$j$}
     \rput(3.8,.8){$k$}
\end{pspicture}
}

\newcommand{\SIXjIII}{
\psset{xunit=1.8cm,yunit=1.8cm}
\psset{linewidth=0.3pt,dimen=middle,linearc=.05pt,cornersize=absolute}
\begin{pspicture}(3.5,1.3)
     \pspolygon(1.6,.6)(2.9,1.1)(2.4,0)(1.6,.6)
     \pspolygon(1.6,.6)(1.9,1.1)(2.9,1.1)(1.6,.6)
    \psline(1.9,1.1)(2,.9)
    \psline(2.08,.7)(2.4,0)
    \psdots[dotscale=.7 .7](1.6,.6)(1.9,1.1)(2.9,1.1)(2.4,0)
     \rput(1.6,1){$i$}
     \rput(1.8,.2){$p$}
     \rput(2.4,1.23){$q$}
     \rput(2.8,.6){$k$}
      \rput(2.5,.8){$j$}
     \rput(2.05,.55){$l$}
\end{pspicture}
}

\newcommand{\opetopicCHART}{
\begin{center}\makebox[0pt]{
\begin{tabular}{|c|c|c|c|c|}
  \hline
   \textbf{objects} & \textbf{morphisms} & \textbf{2-morphisms} & \textbf{3-morphisms} & $\cdots$ \\
  \hline \hline
   $\bullet$ & $\xy
  (-6,0)*+{\bullet}="1";
  (0,8)*+{}; 
  (0,-8)*+{}; 
  (6,0)*+{\bullet}="2";
  {\ar "1";"2"};
 \endxy$
    & $
  \def\objectstyle{\scriptstyle}
\xy 
 (-10,-5)*+{\bullet}="x";
 (10,-5)*+{\bullet}="y";
 (-10,1)*{\bullet}="1";
 (-7,6)*{\bullet}="2";
 (0,6)*+{\dots}="3";
 (7,6)*{\bullet}="4";
 (10,1)*{\bullet}="5";
  {\ar "x"; "y"};
  {\ar "x"; "1"};
  {\ar "1"; "2"};
  {\ar "2"; "3"};
  {\ar "3"; "4"};
  {\ar "4"; "5"};
  {\ar "5"; "y"};
  {\ar@{=>} (0,3); (0,-3)};
\endxy$
 & $\def\objectstyle{\scriptstyle} \xy 0;/r.14pc/:
 (-10,-5)*+{\bullet}="1";
 (10,-5)*+{\bullet}="2";
 (-6,6)*+{\bullet}="3";
 (6,6)*+{\bullet}="3'";
  {\ar "1"; "2"};
  {\ar "3'"; "2"};
  {\ar "1"; "3"};
  {\ar "3"; "3'"};
  {\ar "1"; "3'"};
\endxy
\;\; \xy {\ar@3{->}^{} (-2,0);(2,0)}; \endxy \; \xy 0;/r.14pc/:
 (-10,-5)*+{\bullet}="1";
 (10,-5)*+{\bullet}="2";
 (-6,6)*+{\bullet}="3";
 (6,6)*+{\bullet}="3'";
  {\ar "1"; "2"};
  {\ar "3'"; "2"};
  {\ar "1"; "3"};
  {\ar "3"; "3'"};
\endxy$ & Opetopes  \\
  \hline
\end{tabular}
}\end{center} }

\newcommand{\projector}{
 \xy 0;/r.16pc/:
   (0,-8)*\xycircle(1.5,1.5){-}="m";
   (0,-14)*{}="b";
   (-3,-5)*{}="l";
   (3,-5)*{}="r";
   (-3,5)*{}="l1";
   (3,5)*{}="r1";
   (0,8)*\xycircle(1.5,1.5){-}="m1";
   (0,14)*{}="b1";
         "b";"m" **\dir{-};
         "m";"l" **\crv{(-3,-8)};
         "m";"r" **\crv{(3,-8)};
         "m1";"l1" **\crv{(-3,8)};
         "m1";"r1" **\crv{(3,8)};
         "b1";"m1" **\dir{-};
         \vtwist~{"l1"}{"r1"}{"l"}{"r"}
 \endxy
}

\newcommand{\PROJcenterI}{
  \xy 0;/r.16pc/:
   (0,-8)*\xycircle(1.5,1.5){-}="m";
   (0,-14)*{}="b";
   (-3,-5)*{}="l";
   (3,-5)*{}="r";
   (-3,5)*{}="l1";
   (3,5)*{}="r1";
   (0,8)*\xycircle(1.5,1.5){-}="m1";
   (0,14)*\xycircle(2.7,2.7){-}="b1";
    (0,14)*{a};
         "b";"m" **\dir{-};
         "m";"l" **\crv{(-3,-8)};
         "m";"r" **\crv{(3,-8)};
         "m1";"l1" **\crv{(-3,8)};
         "m1";"r1" **\crv{(3,8)};
         "b1";"m1" **\dir{-};
         \vtwist~{"l1"}{"r1"}{"l"}{"r"}
 \endxy
 \quad = \quad
 \xy 0;/r.16pc/:
   (0,-8)*\xycircle(1.5,1.5){-}="m";
   (0,-14)*{}="b";
   (-3,-4)*\xycircle(1.5,1.5){-}="l";
   (3,-2)*{}="r";
   (-3,7)*{}="l1";
   (3,7)*{}="r1";
   (8,12)*\xycircle(2.7,2.7){-}="z";
    (8,12)*{a};
   (0,10)*{}="m1";
         "b";"m" **\dir{-};
         "m";"l" **\crv{(-3,-8)};
         "m";(3,-4) **\crv{(3,-8)};
         "m1";"l1" **\crv{(-3,10)};
         "m1";"r1" **\crv{(3,10)};
         "l";"z" **\crv{(10,-4)};
         \vtwist~{"l1"}{"r1"}{"l"}{"r"}
 \endxy
  \quad = \quad
 \xy 0;/r.16pc/:
   (0,-8)*\xycircle(1.5,1.5){-}="m";
   (0,-14)*{}="b";
   (-3,-4)*\xycircle(1.5,1.5){-}="l";
   (-8,12)*\xycircle(2.7,2.7){-}="z";
    (-8,12)*{a};
   (3,-2)*{}="r";
   (-3,7)*{}="l1";
   (3,7)*{}="r1";
   (0,10)*{}="m1";
         "b";"m" **\dir{-};
         "m";"l" **\crv{(-3,-8)};
         "m";"r" **\crv{(3,-8)};
         "m1";"l1" **\crv{(-3,10)};
         "m1";"r1" **\crv{(3,10)};
         "l";"z" **\crv{(-10,-4)};
         \vtwist~{"l1"}{"r1"}{"l"}{"r"}
 \endxy
   \quad = \quad
 \xy 0;/r.16pc/:
   (0,-8)*\xycircle(1.5,1.5){-}="m";
   (0,-14)*{}="b";
   (-3,-4)*\xycircle(1.5,1.5){-}="l";
   (3,-2)*{}="r";
   (0,10)*{}="m1";
   (-8,12)*\xycircle(2.7,2.7){-}="z";
    (-8,12)*{a};
         "b";"m" **\dir{-};
         "m";"l" **\crv{(-3,-8)};
         "m";"r" **\crv{(5,-8)};
         "l";"z" **\crv{(-10,-4)};
         "l";"r" **\crv{(-1,2) & (2,0) };
 \endxy
 \qquad = \qquad
 \xy 0;/r.16pc/:
   (0,12)*\xycircle(2.7,2.7){-}="b";
   (0,-14)*{}="m1";
   (0,12)*{a};
          "b";"m1" **\dir{-};
 \endxy
}

\newcommand{\PROJcenterII}{
 \xy 0;/r.12pc/:
   (-5,-8)*\xycircle(1.5,1.5){-}="m";
   (-5,14)*\xycircle(2.7,2.7){-}="b1"="b";
    (-5,14)*{a};
   (-8,-5)*{}="l";
   (-2,-5)*{}="r";
   (-8,5)*{}="l1";
   (-2,5)*{}="r1";
   (-5,8)*\xycircle(1.5,1.5){-}="m1";
   (0,-15)*\xycircle(1.5,1.5){-}="b1";
   (5,-8)*{}="m'";
          "b";"m1" **\dir{-};
          "m'";(5,15) **\dir{-};
          "b1";(0,-23) **\dir{-};
         "m";"l" **\crv{(-8,-8)};
         "m";"r" **\crv{(-2,-8)};
         "m1";"l1" **\crv{(-8,8)};
         "m1";"r1" **\crv{(-2,8)};
         "b1";"m" **\crv{(-5,-15)};
         "b1";"m'" **\crv{(5,-15)};
         \vtwist~{"l1"}{"r1"}{"l"}{"r"};
 \endxy
\qquad = \qquad
 \xy 0;/r.12pc/:
   (5,-8)*\xycircle(1.5,1.5){-}="m";
   (5,14)*\xycircle(2.7,2.7){-}="b1"="b";
    (5,14)*{a};
   (8,-5)*{}="l";
   (2,-5)*{}="r";
   (8,5)*{}="l1";
   (2,5)*{}="r1";
   (5,8)*\xycircle(1.5,1.5){-}="m1";
   (0,-15)*\xycircle(1.5,1.5){-}="b1";
   (-5,-8)*{}="m'";
          "b";"m1" **\dir{-};
          "m'";(-5,15) **\dir{-};
          "b1";(0,-23) **\dir{-};
         "m";"l" **\crv{(8,-8)};
         "m";"r" **\crv{(2,-8)};
         "m1";"l1" **\crv{(8,8)};
         "m1";"r1" **\crv{(2,8)};
         "b1";"m" **\crv{(5,-15)};
         "b1";"m'" **\crv{(-5,-15)};
         \vtwist~{"l1"}{"r1"}{"l"}{"r"};
 \endxy
 }

 \newcommand{\PROJcenterIII}{
 \xy 0;/r.12pc/:
   (-5,-8)*\xycircle(1.5,1.5){-}="m";
   (-5,14)*\xycircle(2.7,2.7){-}="b1"="b";
    (-5,14)*{a};
   (-8,-5)*{}="l";
   (-2,-5)*{}="r";
   (-8,5)*{}="l1";
   (-2,5)*{}="r1";
   (-5,8)*\xycircle(1.5,1.5){-}="m1";
   (0,-15)*\xycircle(1.5,1.5){-}="b1";
   (5,-8)*{}="m'";
          "b";"m1" **\dir{-};
          "m'";(5,15) **\dir{-};
          "b1";(0,-23) **\dir{-};
         "m";"l" **\crv{(-8,-8)};
         "m";"r" **\crv{(-2,-8)};
         "m1";"l1" **\crv{(-8,8)};
         "m1";"r1" **\crv{(-2,8)};
         "b1";"m" **\crv{(-5,-15)};
         "b1";"m'" **\crv{(5,-15)};
         \vtwist~{"l1"}{"r1"}{"l"}{"r"};
 \endxy
 \quad = \quad
 \xy 0;/r.12pc/:
   (5,-8)*\xycircle(1.5,1.5){-}="m";
   (-5,14)*\xycircle(2.7,2.7){-}="b";
    (-5,14)*{a};
   (8,-5)*{}="l";
   (2,-5)*{}="r";
   (-8,5)*{}="l1";
   (-2,5)*{}="r1";
   (-5,8)*\xycircle(1.5,1.5){-}="m1";
   (0,-15)*\xycircle(1.5,1.5){-}="b1";
   (-5,-8)*{}="m'";
          "b";"m1" **\dir{-};
          "l";(8,15) **\dir{-};
          "b1";(0,-23) **\dir{-};
         "m";"l" **\crv{(8,-8)};
         "m";"r" **\crv{(2,-8)};
         "m1";"l1" **\crv{(-8,8)};
         "m1";"r1" **\crv{(-2,8)};
         "b1";"m" **\crv{(5,-15)};
         "b1";"m'" **\crv{(-5,-15)};
         \vtwist~{"l1"}{"r1"}{"m'"}{"r"};
 \endxy
\quad = \quad
 \xy 0;/r.12pc/:
   (-5,14)*\xycircle(2.7,2.7){-}="b1"="b";
    (-5,14)*{a};
   (-5,8)*\xycircle(1.5,1.5){-}="m1";
   (-8,-2)*\xycircle(1.5,1.5){-}="l";
   (0,-15)*\xycircle(1.5,1.5){-}="b1";
   (3,-10)*{}="m'";
          "b";"m1" **\dir{-};
          "b1";(0,-23) **\dir{-};
         "b1";"m'" **\crv{(5,-15)};
         "m1";"l" **\crv{(-10,6)};
         "l";(-4,2) **\crv{(-5,1)};
         (-2,4);(3,15) **\crv{(3,8)};
         \vtwist~{"l"}{"m1"}{"b1"}{"m'"};
 \endxy
\quad = \quad
 \xy  0;/r.12pc/:
   (-5,14)*\xycircle(2.7,2.7){-}="b";
    (-5,14)*{a};
   (-5,8)*\xycircle(1.5,1.5){-}="m1";
   (-8,-2)*{}="l";
   (0,0)*\xycircle(1.5,1.5){-}="x";
   (0,-15)*\xycircle(1.5,1.5){-}="b1";
   (3,-10)*{}="m'";
          "b";"m1" **\dir{-};
          "b1";(0,-23) **\dir{-};
         "b1";"m'" **\crv{(5,-15)};
         "m1";"l" **\crv{(-10,6)};
         "x";"m1" **\crv{(4,8)};
        "x";(-1,-3) **\crv{(-2,7) & (-8,-5)};
        (1,-3);(6,15) **\crv{(8,-1) & (5,5)};
         \vtwist~{"l"}{"x"}{"b1"}{"m'"};
 \endxy
\quad = \quad
 \xy 0;/r.12pc/:
   (0,-15)*\xycircle(1.5,1.5){-}="m";
   (0,-23)*{}="b";
   (-5,-10)*\xycircle(1.5,1.5){-}="l";
   (3,-10)*{}="r";
   (-3,-6)*{}="l'";
   (3,-6)*{}="r'";
   (-3,5)*{}="l1";
   (3,5)*{}="r1";
   (0,8)*\xycircle(1.5,1.5){-}="m1";
   (0,14)*\xycircle(2.7,2.7){-}="b1";
    (0,14)*{a};
         "b";"m" **\dir{-};
         "m";"l" **\crv{(-3,-14)};
         "m";"r" **\crv{(3,-14)};
         "m1";"l1" **\crv{(-3,8)};
         "m1";"r1" **\crv{(3,8)};
         "b1";"m1" **\dir{-};
         "r'";"r" **\dir{-};
         "l";"l'" **\crv{(-3,-8)};
         "l";(-4,-5) **\crv{(-10,-5)};
         (-2,-5);(2,-5) **\crv{(0,-5)};
         (4,-5);(8,16) **\crv{(7,-5) & (8,1)};
         \vtwist~{"l1"}{"r1"}{"l'"}{"r'"}
 \endxy
\quad = \quad
 \xy 0;/r.12pc/:
   (0,-8)*\xycircle(1.5,1.5){-}="m";
   (0,-15)*\xycircle(1.5,1.5){-}="z";
   (0,-23)*{}="z'";
   (0,-12)*{}="b";
   (-3,-5)*{}="l";
   (3,-5)*{}="r";
   (-3,5)*{}="l1";
   (3,5)*{}="r1";
   (0,8)*\xycircle(1.5,1.5){-}="m1";
   (0,14)*\xycircle(2.7,2.7){-}="b1";
    (0,14)*{a};
         "b";"m" **\dir{-};
         "m";"l" **\crv{(-3,-8)};
         "m";"r" **\crv{(3,-8)};
         "m1";"l1" **\crv{(-3,8)};
         "m1";"r1" **\crv{(3,8)};
         "b1";"m1" **\dir{-};
         "z";"z'" **\dir{-};
         "z";"m" **\dir{-};
         "z";(-1,-12) **\crv{(-8,-12)};
         (1,-12);(10,15) **\crv{(12,-12)};
         \vtwist~{"l1"}{"r1"}{"l"}{"r"}
 \endxy
 \quad = \quad
  \xy 0;/r.12pc/:
   (5,-8)*\xycircle(1.5,1.5){-}="m";
   (5,14)*\xycircle(2.7,2.7){-}="b";
    (5,14)*{a};
   (8,-5)*{}="l";
   (2,-5)*{}="r";
   (8,5)*{}="l1";
   (2,5)*{}="r1";
   (5,8)*\xycircle(1.5,1.5){-}="m1";
   (0,-15)*\xycircle(1.5,1.5){-}="b1";
   (-5,-8)*{}="m'";
          "b";"m1" **\dir{-};
          "m'";(-5,14) **\dir{-};
          "b1";(0,-23) **\dir{-};
         "m";"l" **\crv{(8,-8)};
         "m";"r" **\crv{(2,-8)};
         "m1";"l1" **\crv{(8,8)};
         "m1";"r1" **\crv{(2,8)};
         "b1";"m" **\crv{(5,-15)};
         "b1";"m'" **\crv{(-5,-15)};
         \vtwist~{"l1"}{"r1"}{"l"}{"r"};
 \endxy
 }

 \newcommand{\BUBimpliesTOi}{
  \def\objectstyle{\scriptstyle}
\xy
 (-10,-8)*{\bullet}="L";
 (10,-8)*{\bullet}="R";
 (0,8)*{\bullet}="T";
    "L";"T" **\dir{-};
    "R";"T" **\dir{-};
    "L";"R" **\dir{-};
 \endxy
\quad \xy {\ar^{\txt\bf{2-2}} (-5,0); (5,0)}; \endxy \quad
 \xy
 (-10,-8)*{\bullet}="L";
 (10,-8)*{\bullet}="R";
 (0,8)*{\bullet}="T";
 (0,-8)*{\bullet}="M";
    "L";"T" **\dir{-};
    "R";"T" **\dir{-};
    {\ar@/_1.5pc/@{-}"L";"R"};
    {\ar@{-}@/^1.5pc/ "L";"R"};
    "R";"M" **\dir{-};
    "L";"M" **\dir{-};
 \endxy
\quad \xy {\ar^{\txt\bf{Bubble}} (-8,0); (8,0)}; \endxy \;\;
  \xy
 (-10,-8)*{\bullet}="L";
 (10,-8)*{\bullet}="R";
 (0,8)*{\bullet}="T";
 (0,-8)*{\bullet}="M";
    "L";"T" **\dir{-};
    "R";"T" **\dir{-};
    {\ar@/_1.5pc/@{-}"L";"R"};
    "T";"M" **\dir{-};
    "R";"M" **\dir{-};
    "L";"M" **\dir{-};
 \endxy
}

 \newcommand{\TOimpliesBUBi}{
 \def\objectstyle{\scriptstyle}
  \xy
 (-10,-8)*{\bullet}="L";
 (10,-8)*{\bullet}="R";
 (0,8)*{\bullet}="T";
    "L";"T" **\dir{-};
    "R";"T" **\dir{-};
    "L";"R" **\dir{-};
 \endxy
\quad \xy {\ar^{\txt\bf{3-1}} (-8,0); (8,0)}; \endxy \;\;
  \xy
 (-10,-8)*{\bullet}="L";
 (10,-8)*{\bullet}="R";
 (0,8)*{\bullet}="T";
 (0,-8)*{\bullet}="M";
    "L";"T" **\dir{-};
    "R";"T" **\dir{-};
    {\ar@/_1.5pc/@{-}"L";"R"};
    "T";"M" **\dir{-};
    "R";"M" **\dir{-};
    "L";"M" **\dir{-};
 \endxy
 \quad \xy {\ar^{\txt\bf{2-2}} (-5,0); (5,0)}; \endxy \quad
 \xy
 (-10,-8)*{\bullet}="L";
 (10,-8)*{\bullet}="R";
 (0,8)*{\bullet}="T";
 (0,-8)*{\bullet}="M";
    "L";"T" **\dir{-};
    "R";"T" **\dir{-};
    {\ar@/_1.5pc/@{-}"L";"R"};
    {\ar@{-}@/^1.5pc/ "L";"R"};
    "R";"M" **\dir{-};
    "L";"M" **\dir{-};
 \endxy
}

\newcommand{\TWOTHREEi}{
 \xy
 (0,30)*+{
    \xy 0;/r.10pc/:
    (6.18,19)*{}="t1"; 
    (-16.18,11.74)*{}="t2";
    (-16.18,-11.74)*{}="t3";
    (6.18,-19)*{}="t4";
    (20,0)*{}="t5";
        {\ar@{-}"t1";"t2"};
        {\ar@{-} "t2";"t3"};
        {\ar@{-} "t3";"t4"};
        {\ar@{-} "t4";"t5"};
        {\ar@{-} "t1";"t5"};
        {\ar@{-} "t1";"t3"}; {\ar@{-} "t3";"t5"};
   \endxy}="t";
 (-30,15)*+{
    \xy 0;/r.10pc/:
    (6.18,19)*{}="t1"; 
    (-16.18,11.74)*{}="t2";
    (-16.18,-11.74)*{}="t3";
    (6.18,-19)*{}="t4";
    (20,0)*{}="t5";
        {\ar@{-}"t1";"t2"};
        {\ar@{-} "t2";"t3"};
        {\ar@{-} "t3";"t4"};
        {\ar@{-} "t4";"t5"};
        {\ar@{-} "t1";"t5"};
        {\ar@{-} "t1";"t3"}; {\ar@{-} "t3";"t5"};
        {\ar@{-}|<<<<<<<{ \hole } "t2";"t5"};
   \endxy}="l1";
 (-30,-15)*+{
    \xy 0;/r.10pc/:
    (6.18,19)*{}="t1"; 
    (-16.18,11.74)*{}="t2";
    (-16.18,-11.74)*{}="t3";
    (6.18,-19)*{}="t4";
    (20,0)*{}="t5";
        {\ar@{-}"t1";"t2"};
        {\ar@{-} "t2";"t3"};
        {\ar@{-} "t3";"t4"};
        {\ar@{-} "t4";"t5"};
        {\ar@{-} "t1";"t5"};
        {\ar@{-} "t1";"t3"}; {\ar@{-} "t3";"t5"};
        {\ar@{-}|<<<<<<<{ \hole } "t2";"t5"};
        {\ar@{-}|<<<<<<<{  \hole}|>>>>>>{  \hole} "t2";"t4"};
   \endxy}="l2";
 (0,-30)*+{
    \xy 0;/r.10pc/:
    (6.18,19)*{}="t1"; 
    (-16.18,11.74)*{}="t2";
    (-16.18,-11.74)*{}="t3";
    (6.18,-19)*{}="t4";
    (20,0)*{}="t5";
        {\ar@{-}"t1";"t2"};
        {\ar@{-} "t2";"t3"};
        {\ar@{-} "t3";"t4"};
        {\ar@{-} "t4";"t5"};
        {\ar@{-} "t1";"t5"};
        {\ar@{-} "t1";"t3"}; {\ar@{-} "t3";"t5"};
        {\ar@{-}|<<<<<<<{ \hole \; \hole}|>>>>>>>{ \hole \; \hole} "t1";"t4"};
        {\ar@{-}|<<<<<<<{ \hole } "t2";"t5"};
        {\ar@{-}|<<<<<<<{  \hole}|>>>>>>{  \hole} "t2";"t4"};
   \endxy}="b";
 (30,15)*+{
    \xy 0;/r.10pc/:
    (6.18,19)*{}="t1"; 
    (-16.18,11.74)*{}="t2";
    (-16.18,-11.74)*{}="t3";
    (6.18,-19)*{}="t4";
    (20,0)*{}="t5";
        {\ar@{-}"t1";"t2"};
        {\ar@{-} "t2";"t3"};
        {\ar@{-} "t3";"t4"};
        {\ar@{-} "t4";"t5"};
        {\ar@{-} "t1";"t5"};
        {\ar@{-} "t1";"t3"}; {\ar@{-} "t3";"t5"};
        {\ar@{-}|<<<<<<<{ \hole } "t2";"t5"};
   \endxy}="r1";
 (30,-15)*+{
    \xy 0;/r.10pc/:
    (6.18,19)*{}="t1"; 
    (-16.18,11.74)*{}="t2";
    (-16.18,-11.74)*{}="t3";
    (6.18,-19)*{}="t4";
    (20,0)*{}="t5";
        {\ar@{-}"t1";"t2"};
        {\ar@{-} "t2";"t3"};
        {\ar@{-} "t3";"t4"};
        {\ar@{-} "t4";"t5"};
        {\ar@{-} "t1";"t5"};
        {\ar@{-} "t1";"t3"}; {\ar@{-} "t3";"t5"};
        {\ar@{-}|<<<<<<<{ \hole } "t2";"t5"};
        {\ar@{-}|<<<<<<<{  \hole}|>>>>>>{  \hole} "t2";"t4"};
   \endxy}="r2";
    {\ar@3{->} "t";"r1" };
    {\ar@3{->} "t";"l1" };
    {\ar@3{->} "l1";"l2" };
    {\ar@3{->} "r1";"r2" };
    {\ar@3{->} "r2";"b" };
    {\ar@3{->} "l2";"b" };
 \endxy
 }

 \newcommand{\CATchartI}{
 \begin{center}  \makebox[0pt]{
\begin{tabular}{|p{2.2in}|p{2.4in}|}
  \hline
  \textbf{Set-based mathematics} & \textbf{Category-based mathematics} \\
  \hline \hline
  \textbf{Sets}
 $ \xy
 (0,0)*{\includegraphics{blob30.eps}};
 (0,0)*{\scs \bullet};
 (2,4)*{\scs \bullet};
 (-5,-4)*{\scs \bullet};
 (6,-1)*{\scs \bullet};
 \endxy $ &  \textbf{Categories}
 $ \xy
 (0,0)*{\includegraphics{blob30.eps}};
 (2,0)*{\scs \bullet}="1";
 (-3,6)*{\scs \bullet}="2";
 (-5,-4)*{\scs \bullet}="3";
 (2,-7)*{\scs \bullet}="4";
  "1";"2" **\crv{(-2,6)}?(.25)*\dir{<};
  "1";"3" **\crv{(-2,-4)}?(.25)*\dir{<};
  "3";"2" **\crv{(-6,0)}?(.5)*\dir{<};
  "1";"4" **\crv{}?(.25)*\dir{<};
 \endxy $\\
 \textbf{Functions} \quad$ \xy
 (0,0)*{\includegraphics{blob30.eps}};
 (30,0)*{\includegraphics{blobII30.eps}};
 (0,-5)*{\scs \bullet}="1";
 (2,4)*{\scs \bullet}="2";
 (6,-1)*{\scs \bullet}="3";
 (33,-2)*{\scs \bullet}="1'";
 (27,-4)*{\scs \bullet}="2'";
 (28,1)*{\scs \bullet}="3'";
 (31,5)*{\scs \bullet}="4'";
 {\ar@{>} "1";"1'"};
 {\ar@{>} "2";"4'"};
 {\ar@{>} "3";"3'"};
 \endxy$  & \textbf{Functors} \quad $ \xy
 (0,0)*{\includegraphics{functor30.eps}};
 (-14,2)*{\scs \bullet}="2";
 (-8,-3)*{\scs \bullet}="3";
 (21,-2)*{\scs \bullet}="1'";
 (17,-4)*{\scs \bullet}="3'";
 (18,5)*{\scs \bullet}="4'";
  "2";"3" **\crv{}?(.5)*\dir{>};
  "3'";"4'" **\crv{}?(.5)*\dir{>};
  "1'";"4'" **\crv{(22,2)}?(.55)*\dir{>};
 {\ar@{>} "2";"4'"};
 {\ar@{>} "3";"3'"};
 \endxy$ \\
   &  \textbf{Natural Transformations}
 $ \xy
 (0,0)*{\includegraphics{natural40.eps}};
 (-17,1)*{\scs \bullet}="1";
 (-17,-7)*{\scs \bullet}="2";
 (12,4)*{\scs \bullet}="tl";
 (20,4)*{\scs \bullet}="tr";
 (12,-5)*{\scs \bullet}="bl";
 (20,-5)*{\scs \bullet}="br";
  "1";"2" **\crv{}?(.58)*\dir{>};
  "tl";"tr" **\crv{}?(.5)*\dir{>};
  "bl";"br" **\crv{}?(.55)*\dir{>};
  "tl";"bl" **\crv{}?(.55)*\dir{>}+(-3,0)*{\scs \alpha_y};
  "tr";"br" **\crv{}?(.55)*\dir{>}+(3,0)*{\scs \alpha_x};
 \endxy $\\
  \hline
\end{tabular} }
\end{center}
}

\newcommand{\BB}{
\xybox{%
  (-6,0)*{};
  (6,0)*{};
 (4,8)*{}="tr";(-4,8)*{}="tl";
 (4,-8)*{}="br";(-4,-8)*{}="bl";
 (0,0)*\xycircle(2.2,2.2){-}="m";
 (0,0)*{B};
 "tl";"m" **\dir{-};?(.3)*\dir{>};
 "tr";"m" **\dir{-};?(.3)*\dir{>};
 "bl";"m" **\dir{-};?(.205)*\dir{<};
 "br";"m" **\dir{-}; ?(.205)*\dir{<};
 }}

 \newcommand{\BByang}{
 \vcenter{\xy  0;/r.19pc/:
 (-5,-10)*\xycircle(3,3){-}="b";
 (-5,10)*\xycircle(3,3){-}="t";
 (5,0)*\xycircle(3,3){-}="m";
 "t";"m" **\dir{-} ?(.5)*\dir{>};
 "m";"b" **\dir{-} ?(.5)*\dir{>};
 "t";"b" **\dir{-} ?(.5)*\dir{>};
 (-5,20);"t" **\dir{-} ?(.5)*\dir{>};
 (-15,20);"t" **\dir{-} ?(.5)*\dir{>};
 (20,15);"m" **\dir{-} ?(.5)*\dir{>};
 (-15,20);"t" **\dir{-} ?(.5)*\dir{>};
  "b";(-5,-20)**\dir{-} ?(.5)*\dir{>};
  "m";(20,-15)**\dir{-} ?(.5)*\dir{>};
  "b";(-15,-20)**\dir{-} ?(.5)*\dir{>};
  (-5,10)*{B};
  (-5,-10)*{B};
  (5,0)*{B};
 \endxy}
  \quad = \quad
   \vcenter{\xy  0;/r.19pc/:
 (5,-10)*\xycircle(3,3){-}="b";
 (5,10)*\xycircle(3,3){-}="t";
 (-5,0)*\xycircle(3,3){-}="m";
 "t";"m" **\dir{-} ?(.5)*\dir{>};
 "m";"b" **\dir{-} ?(.5)*\dir{>};
 "t";"b" **\dir{-} ?(.5)*\dir{>};
 (5,20);"t" **\dir{-} ?(.5)*\dir{>};
 (15,20);"t" **\dir{-} ?(.5)*\dir{>};
 (-20,15);"m" **\dir{-} ?(.5)*\dir{>};
  "b";(5,-20)**\dir{-} ?(.5)*\dir{>};
  "m";(-20,-15)**\dir{-} ?(.5)*\dir{>};
  "b";(15,-20)**\dir{-} ?(.5)*\dir{>};
  (5,10)*{B};
  (5,-10)*{B};
  (-5,0)*{B};
 \endxy}
 }

\newcommand{\twothreebraid}{
\xy 0;/r.16pc/:
 (-18,-20)*{ }="b1";
 (-10,-20)*{ }="b2";
 (-2,-20)*{ }="b3";
 (14,-20)*{ }="b4";
 (22,-20)*{ }="b5";
 (-18,20)*{ }="T1";
 (-10,20)*{ }="T2";
 (6,20)*{ }="T3";
 (14,20)*{ }="T4";
 (22,20)*{ }="T5";
    "T1"; "b4" **\crv{ (-18,10) & (14,-10)} ?(.15)*\dir{>} ?(.85)*\dir{>};
    "T2"; "b5" **\crv{ (-10,10) & (22,-10)} ?(.15)*\dir{>} ?(.85)*\dir{>};
  (-1,5)*{
  \begin{pspicture}(1,1)
     \pspolygon[fillcolor=white,fillstyle=solid,linewidth=0pt,linecolor=white](0,0)(.3,0)(.3,.4)(0,.4)(0,0)
  \end{pspicture}};
  (2,9)*{
  \begin{pspicture}(1,1)
     \pspolygon[fillcolor=white,fillstyle=solid,linewidth=0pt,linecolor=white](0,0)(.3,0)(.3,.4)(0,.4)(0,0)
  \end{pspicture}};
  (7,5)*{
  \begin{pspicture}(1,1)
     \pspolygon[fillcolor=white,fillstyle=solid,linewidth=0pt,linecolor=white](0,0)(.3,0)(.3,.4)(0,.4)(0,0)
  \end{pspicture}};
  (4,1)*{
  \begin{pspicture}(1,1)
     \pspolygon[fillcolor=white,fillstyle=solid,linewidth=0pt,linecolor=white](0,0)(.3,0)(.3,.4)(0,.4)(0,0)
  \end{pspicture}};
  (12,1)*{
  \begin{pspicture}(1,1)
     \pspolygon[fillcolor=white,fillstyle=solid,linewidth=0pt,linecolor=white](0,0)(.3,0)(.3,.4)(0,.4)(0,0)
  \end{pspicture}};
  (9,-3)*{
  \begin{pspicture}(1,1)
     \pspolygon[fillcolor=white,fillstyle=solid,linewidth=0pt,linecolor=white](0,0)(.3,0)(.3,.4)(0,.4)(0,0)
  \end{pspicture}};
    "T5"; "b3" **\crv{ (22,10) & (-2,-10)} ?(.15)*\dir{>} ?(.85)*\dir{>};
    "T4"; "b2" **\crv{ (14,10) & (-10,-10)} ?(.15)*\dir{>} ?(.85)*\dir{>};
    "T3"; "b1" **\crv{ (6,10) & (-18,-10)} ?(.15)*\dir{>} ?(.85)*\dir{>};
\endxy }

\newcommand{\fancytangle}{
 \xy
  (5,10)*{}; (-10,-10)*{} **\crv{(6,-2)&(-12,4)}
 \POS?(.25)*{\hole}="x" \POS?(.45)*{\hole}="y" \POS?(.6)*{\hole}="z";
 "y"+(0,-1); (2,-10)*{} **\crv{}\POS?(.2)*{\hole}="M";
  (-10,10)*{}; "z" **\crv{(-9,0)};
 "z";"M" **\crv{};
 "M";"x" **\crv{(5,0)};
 "x";"y" **\crv{(0,7) & (-5,6)};
   (2.5,-2.5)*{\begin{pspicture}(1,1)
     \pspolygon[fillcolor=white,fillstyle=solid,linewidth=0pt,linecolor=white](0,0)(.3,0)(.3,.3)(0,.3)(0,0)
  \end{pspicture}};
  (-5,-10);(10,-10) **\crv{(-5,-3) & (10,-3)};
  (9,-3)*{\begin{pspicture}(1,1)
     \pspolygon[fillcolor=white,fillstyle=solid,linewidth=0pt,linecolor=white](0,0)(.3,0)(.3,.3)(0,.3)(0,0)
  \end{pspicture}};
  (10,10);(5,-10) **\crv{(11,5) & (6,-5)};
  (15,3)*\xycircle(2.2,3.2){-};
\endxy
}

\begin{document}

\title{A Prehistory of $n$-Categorical Physics}

\author{John C.\ Baez\thanks{Department of Mathematics,
University of California,
Riverside, CA 92521, USA.
Email: \texttt{baez@math.ucr.edu}}
\and
Aaron Lauda\thanks{Department of Mathematics,
Columbia University,
New York, NY 10027, USA.
Email: \texttt{lauda@math.columbia.edu}}
}

\date{August 18, 2009}

\maketitle

\begin{abstract}
This paper traces the growing role of categories and $n$-categories in
physics, starting with groups and their role in relativity, and
leading up to more sophisticated concepts which manifest themselves in
Feynman diagrams, spin networks, string theory, loop quantum gravity, and
topological quantum field theory.  Our chronology ends around 2000,
with just a taste of later developments such as open-closed
topological string theory, the categorification of quantum groups,
Khovanov homology, and Lurie's work on the classification of
topological quantum field theories.
\end{abstract}



%
\section{Introduction}
%

This paper is a highly subjective chronology describing how physicists
have begun to use ideas from $n$-category theory in their work, often
without making this explicit.  Somewhat arbitrarily, we start around
the discovery of relativity and quantum mechanics, and lead up to
conformal field theory and topological field theory.  In parallel, we
trace a bit of the history of $n$-categories, from Eilenberg and Mac
Lane's introduction of categories, to later work on monoidal and
braided monoidal categories, to Grothendieck's dreams involving
$\infty$-categories, and subsequent attempts to realize this dream.
Our chronology ends at the dawn of the 21st century; after then,
developments have been coming so thick and fast that we have not had
time to put them in proper perspective.

We call this paper a `prehistory' because $n$-categories and
their applications to physics are still in their infancy.  We call it
`a' prehistory because it represents just one view of a
multi-faceted subject: many other such stories can and should be told.
Ross Street's {\sl Conspectus of Australian Category Theory}
\cite{Street3} is a good example: it overlaps with ours, but only
slightly.  There are many aspects of $n$-categorical physics that our
chronology fails to mention, or touches on very briefly; other stories
could redress these deficiencies.  It would also be good to have a
story of $n$-categories that focused on algebraic topology, one that
focused on algebraic geometry, and one that focused on logic.  For
$n$-categories in computer science, we have John Power's {\sl Why
Tricategories?}  \cite{Power}, which while not focused on history at
least explains some of the issues at stake.

What is the goal of \textit{this} prehistory?  We are scientists
rather than historians of science, so we are trying to make a specific
scientific point, rather than accurately describe every twist and turn
in a complex sequence of events.  We want to show how categories and
even $n$-categories have slowly come to be seen as a good way to
formalize physical theories in which `processes' can be drawn as
diagrams---for example Feynman diagrams---but interpreted
algebraically---for example as linear operators.  To minimize the
prerequisites, we include a gentle introduction to $n$-categories (in
fact, mainly just categories and bicategories).  We also include a
review of some relevant aspects of 20th-century physics.

The most obvious roads to $n$-category theory start from issues
internal to pure mathematics.  Applications to physics only became
visible much later, starting around the 1980s.  So far, these
applications mainly arise around theories of quantum gravity,
especially string theory and `spin foam models' of loop quantum
gravity.  These theories are speculative and still under development,
not ready for experimental tests.  They may or may not succeed.  So,
it is too early to write a real history of $n$-categorical physics, or
even to know if this subject will become important.  We believe it
will---but so far, all we have is a `prehistory'.

%
\section{Road Map}
%

Before we begin our chronology, to help the reader keep from getting lost
in a cloud of details, it will be helpful to sketch the road ahead.
Why did categories turn out to be useful in physics?  The reason
is ultimately very simple.  A category consists of `objects' $x,y,z,
\dots$ and `morphisms' which go between objects, for example
\[                f \maps x \to y  .\]
A good example is the category of Hilbert spaces, where the objects
are Hilbert spaces and the morphisms are bounded operators.
In physics we can think of an object as a `state space' for some
physical system, and a morphism as a `process' taking states of
one system to states of another (perhaps the same one).   In
short, we use objects to describe \textit{kinematics}, and morphisms
to describe \textit{dynamics}.

Why $n$-categories?  For this we need to understand a bit about categories
and their limitations.  In a category, the only thing we can do with
morphisms is `compose' them: given a morphism $f \maps x \to y$
and a morphism $g \maps y \to z$, we can compose them and obtain
a morphism $g f \maps x \to z$.  This corresponds to our basic
intuition about processes, namely that one can occur after
another.  While this intuition is temporal in nature, it lends
itself to a nice spatial metaphor.  We can draw a morphism $f \maps x
\to y$ as a `black box' with an input of type $x$ and an output of
type $y$:
\[
 \xy (0,0)*{f};
(0,0)*\xycircle(2.65,2.65){-}="f"; (0,10)**\dir{-}
?(.5)*\dir{<}+(-3,0)*{ x}; "f";(0,-10)**\dir{-}
?(.75)*\dir{>}+(-3,0)*{ y};
\endxy
\]
Composing morphisms then corresponds to feeding the output of
one black box into another:
\[
 \xy (0,0)*{f};
(0,0)*\xycircle(2.65,2.65){-}="f"; (0,10)**\dir{-}
?(.5)*\dir{<}+(-3,0)*{ x}; "f";(0,-10)**\dir{-}
?(.75)*\dir{>}+(-3,0)*{ y};
  (0,-13)*{g};
(0,-13)*\xycircle(2.65,2.65){-}="g";
"g";(0,-23)**\dir{-} ?(.75)*\dir{>}+(-3,0)*{ z};
\endxy
\]

This sort of diagram might be sufficient to represent physical
processes if the universe were 1-dimensional: no dimensions of space,
just one dimension of time.  But in reality, processes can
occur not just in \textit{series} but also in \textit{parallel}---`side
by side', as it were:
\[
 \xy
(-5,0)*{f};
(-5,0)*\xycircle(2.65,2.65){-}="f"; (-5,10)**\dir{-}
?(.5)*\dir{<}+(-3,0)*{ x}; "f";(-5,-10)**\dir{-}
?(.75)*\dir{>}+(-3,0)*{ y};
(5,0)*{f'};
(5,0)*\xycircle(2.65,2.65){-}="f'"; (5,10)**\dir{-}
?(.5)*\dir{<}+(3,0.5)*{ x'}; "f'";(5,-10)**\dir{-}
?(.75)*\dir{>}+(3,0.5)*{ y'};
\endxy
\]
To formalize this algebraically, we need something more than a category:
at the very least a `monoidal category', which is a special
sort of `bicategory'.   The term `bicategory' hints at the two
ways of combining processes: in series and in parallel.

Similarly, the mathematics of bicategories might be sufficient for
physics if the universe were only 2-dimensional: one dimension of
space, one dimension of time.  But in our universe, is also possible
for physical systems to undergo a special sort of process where
they `switch places':
\[ \xy
 \vtwist~{(-5,8)}{(5,8)}{(-5,-8)}{(5,-8)}<>|>>><;
 (-7,5)*{ x};
 (7,5)*{ y};
\endxy \]
To depict this geometrically requires a third dimension, hinted
at here by the crossing lines.  To formalize it algebraically, we
need something more than a monoidal category: at the very least a
`braided monoidal category', which is a special sort of `tricategory'.

This escalation of dimensions can continue.  In the diagrams Feynman
used to describe interacting particles, we can continuously interpolate
between this way of switching two particles:
\[ \xy
 \vtwist~{(-5,8)}{(5,8)}{(-5,-8)}{(5,-8)}<>|>>><;
 (-7,5)*{ x};
 (7,5)*{ y};
\endxy \]
and this:
\[ \xy
 \vtwistneg~{(-5,8)}{(5,8)}{(-5,-8)}{(5,-8)}<>|>>><;
 (-7,5)*{ x};
 (7,5)*{ y};
\endxy \]
This requires four dimensions: one of time and three of space.
To formalize this algebraically we need a `symmetric monoidal
category', which is a special sort of `tetracategory'.

More general $n$-categories, including those for higher values of $n$,
may also be useful in physics.  This is especially true in string
theory and spin foam models of quantum gravity.  These theories
describe strings, graphs, and their higher-dimensional generalizations
propagating in spacetimes which may themselves have more than 4 dimensions.

So, in abstract the idea is simple: we can use $n$-categories to
\textit{algebraically} formalize physical theories in which processes
can be depicted \textit{geometrically} using $n$-dimensional diagrams.
But the development of this idea has been long and convoluted.  It is
also far from finished.  In our chronology we describe its development
up to the year 2000.  To keep the tale from becoming unwieldy,
we have been ruthlessly selective in our choice of topics.

In particular, we can roughly distinguish two lines of thought
leading towards $n$-categorical physics: one beginning with quantum
mechanics, the other with general relativity.  Since a major challenge
in physics is reconciling quantum mechanics and general relativity,
it is natural to hope that these lines of thought will eventually merge.
We are not sure yet how this will happen, but the two lines
have already been interacting throughout the 20th century.
Our chronology will focus on the first.  But before we start,
let us give a quick sketch of both.

The first line of thought starts with quantum mechanics and the
realization that in this subject, \textit{symmetries} are all-important.
Taken abstractly, the symmetries of any system form a group $G$.
But to describe how these symmetries act on states of a quantum system, we
need a `unitary representation' $\rho$ of this group on some Hilbert space
$H$.  This sends any group element $g \in G$ to a unitary operator
$\rho(g) \maps H \to H$.

The theory of $n$-categories allows for drastic generalizations of
this idea.  We can see any group $G$ as a category with one object where
all the morphisms are invertible: the morphisms of this category are
just the elements of the group, while composition is multiplication.
There is also a category $\Hilb$ where objects are Hilbert spaces and
morphisms are linear operators.  A representation of $G$ can be
seen as a map from the first category to the second:
\[                \rho \maps G \to \Hilb  .  \]
Such a map between categories is called a `functor'.  The functor $\rho$
sends the one object of $G$ to the Hilbert space $H$, and it sends each
morphism $g$ of $G$ to a unitary operator $\rho(g) \maps H \to H$.
In short, it realizes elements of the abstract group $G$ as actual
transformations of a specific physical system.

The advantage of this viewpoint is that now the group $G$ can be
replaced by a more general category.  Topological quantum field
theory provides the most famous example of such a generalization,
but in retrospect the theory of Feynman diagrams provides another,
and so does Penrose's theory of `spin networks'.

More dramatically, both $G$ and $\Hilb$ may be replaced by a more
general sort of $n$-category.  This allows for a rigorous treatment
of physical theories where physical processes are described by
$n$-dimensional diagrams.  The basic idea, however, is always the same:
\textit{a physical theory is a map sending `abstract'
processes to actual transformations of a specific physical system.}

The second line of thought starts with Einstein's theory of general
relativity, which explains gravity as the curvature of spacetime.
Abstractly, the presence of `curvature' means that as a particle
moves through spacetime from one point to another, its internal
state transforms in a manner that depends nontrivially on the path
it takes.  Einstein's great insight was that this notion of
curvature completely subsumes the older idea of gravity as
a `force'.  This insight was later generalized to electromagnetism
and the other forces of nature: we now treat them all as various kinds
of curvature.

In the language of physics, theories where forces are explained in
terms of curvature are called `gauge theories'.  Mathematically, the key
concept in a gauge theory is that of a `connection' on a `bundle'.
The idea here is to start with a manifold $M$ describing
spacetime.  For each point $x$ of spacetime, a bundle gives a set $E_x$
of allowed internal states for a particle at this point.  A connection
then assigns to each path $\gamma$ from $x \in M$ to $y \in M$ a
map $\rho(\gamma) \maps E_x \to E_y$.  This map, called `parallel
transport', says how a particle starting at $x$ changes state if
it moves to $y$ along the path $\gamma$.

Category theory lets us see that a connection is also a kind
of functor.  There is a category called the `path groupoid' of $M$,
denoted $\Path_1(M)$, whose objects are points of $M$: the morphisms are
paths, and composition amounts to concatenating paths.  Similarly, any bundle
$E$ gives a 'transport category', denoted $\Trans(E)$, where the
objects are the sets $E_x$ and the morphisms are maps between these.
A connection gives a functor
\[             \rho \maps \Path_1(M) \to \Trans(P) . \]
This functor sends each object $x$ of $P_1(M)$ to the set $E_x$, and
sends each path $\gamma$ to the map $\rho(\gamma)$.

So, the `second line of thought', starting from general relativity,
leads to a picture strikingly similar to the first!  Just as a
unitary group representation is a functor sending abstract symmetries
to transformations of a specific physical system, a connection is a
functor sending paths in spacetime to transformations of a specific
physical system: a particle.  And just as unitary group
representations are a special case of physical theories described as
maps between $n$-categories, when we go from point particles to
higher-dimensional objects we meet `higher gauge theories', which
use maps between $n$-categories to describe how such objects change
state as they move through spacetime \cite{BS}.  In short: the first
and second lines of thought are evolving in parallel---and
intimately linked, in ways that still need to be understood.

Sadly, we will not have much room for general relativity, gauge
theories, or higher gauge theories in our chronology.   We will be
fully occupied with group representations as applied to quantum
mechanics, Feynman diagrams as applied to quantum field theory, how
these diagrams became better understood with the rise of $n$-categories,
and how higher-dimensional generalizations of Feynman diagrams
arise in string theory, loop quantum gravity, topological quantum
field theory, and the like.

%
\section{Chronology}
%

\subsection*{Maxwell (1876)}

In his book {\it Matter and Motion}, Maxwell \cite{maxwell} wrote:

\begin{quote}
Our whole progress up to this point may be described as a gradual
development of the doctrine of relativity of all physical phenomena.
Position we must evidently acknowledge to be relative, for we cannot
describe the position of a body in any terms which do not express
relation.  The ordinary language about motion and rest does not so
completely exclude the notion of their being measured absolutely, but
the reason of this is, that in our ordinary language we tacitly assume
that the earth is at rest.... There are no landmarks in space; one
portion of space is exactly like every other portion, so that we cannot
tell where we are.  We are, as it were, on an unruffled sea, without
stars, compass, sounding, wind or tide, and we cannot tell in what
direction we are going.  We have no log which we can case out to take a
dead reckoning by; we may compute our rate of motion with respect to
the neighboring bodies, but we do not know how these bodies may be
moving in space.
\end{quote}

Readers less familiar with the history of physics may be surprised to
see these words, written 3 years before Einstein was born.  In fact,
the relative nature of velocity was already known to Galileo, who also
used a boat analogy to illustrate this.  However, Maxwell's equations
describing light made relativity into a hot topic.  First, it was
thought that light waves needed a medium to propagate in, the
`luminiferous aether', which would then define a rest frame.  Second,
Maxwell's equations predicted that waves of light move at a fixed
speed in vacuum regardless of the velocity of the source!  This seemed
to contradict the relativity principle.  It took the genius of
Lorentz, Poincar\'e, Einstein and Minkowski to realize that this
behavior of light is compatible with relativity of motion if we assume
space and time are united in a geometrical structure we now call {\it
Minkowski spacetime}.  But when this realization came, the importance
of the relativity principle was highlighted, and with it the
importance of {\it symmetry groups} in physics.

\subsection*{Poincar\'e (1894)}

In 1894, Poincar\'e invented the {\bf fundamental group}: for any
space $X$ with a basepoint $\ast$, homotopy classes of loops based at
$\ast$ form a group $\pi_1(X)$.  This hints at the unification of {\it
space} and {\it symmetry}, which was later to become one of the main
themes of $n$-category theory.  In 1945, Eilenberg and Mac Lane
described a kind of `inverse' to the process taking a space to its
fundamental group.  Since the work of Grothendieck in the 1960s, many
have come to believe that homotopy theory is secretly just the study
of certain vast generalizations of groups, called `$n$-groupoids'.
From this point of view, the fundamental group is just the tip of an
iceberg.

\subsection*{Lorentz (1904)}

Already in 1895 Lorentz had invented the notion of `local time'
to explain the results of the Michelson--Morley experiment, but
in 1904 he extended this work and gave formulas for what are
now called `Lorentz transformations' \cite{lorentz}.

\subsection*{Poincar\'e (1905)}

In his opening address to the Paris Congress in 1900, Poincar\'e asked
`Does the aether really exist?'  In 1904 he gave a talk at the
International Congress of Arts and Science in St.\ Louis, in which he
noted that ``\dots as demanded by the relativity principle the
observer cannot know whether he is at rest or in absolute motion''.

On the 5th of June, 1905, he wrote a paper `Sur la dynamique de
l'electron' \cite{Poincare} in which he stated: ``It seems that this
impossibility of demonstrating absolute motion is a general law of
nature''.  He named the Lorentz transformations after Lorentz, and
showed that these transformations, together with the rotations, form a
group.  This is now called the `Lorentz group'.

\subsection*{Einstein (1905)}

Einstein's first paper on relativity, `On the electrodynamics of
moving bodies' \cite{Einstein} was received on June 30th, 1905.  In
the first paragraph he points out problems that arise from applying
the concept of absolute rest to electrodynamics.  In the second, he
continues:

\begin{quote}
Examples of this sort, together with the unsuccessful attempts to
discover any motion of the earth relative to the `light medium,'
suggest that the phenomena of electrodynamics as well as of mechanics
possess no properties corresponding to the idea of absolute rest. They
suggest rather that, as already been shown to the first order of small
quantities, the same laws of electrodynamics and optics hold for all
frames of reference for which the equations of mechanics hold good.
We will raise this conjecture (the purport of which will hereafter be
called the `Principle of Relativity') to the status of a postulate,
and also introduce another postulate, which is only apparently
irreconcilable with the former, namely, that light is always
propagated in empty space with a definite velocity $c$ which is
independent of the state of motion of the emitting body.
\end{quote}

 From these postulates he derives formulas for the transformation
of coordinates from one frame of reference to another in uniform
motion relative to the first, and shows these transformations form
a group.

\subsection*{Minkowski (1908)}

In a famous address delivered at the 80th Assembly of German Natural
Scientists and Physicians on September 21, 1908, Hermann Minkowski
declared:

\begin{quote}
The views of space and time which I wish to lay before you have sprung
from the soil of experimental physics, and therein lies their
strength.  They are radical.  Henceforth space by itself, and time by
itself, are doomed to fade away into mere shadows, and only a kind of
union of the two will preserve an independent reality.
\end{quote}

He formalized special relativity by treating space and time as two
aspects of a single entity: {\it spacetime}.  In simple terms we may
think of this as $\R^4$, where a point ${\bf x} = (t,x,y,z)$ describes
the time and position of an event.  Crucially, this $\R^4$ is equipped
with a bilinear form, the {\bf Minkowski metric}:
\[         {\bf x} \cdot {\bf x'} = tt' - xx' - yy' - zz' \]
which we use as a replacement for the usual dot product
when calculating times and distances.
With this extra structure, $\R^4$ is now called {\bf Minkowski
spacetime}.  The group of all linear transformations
\[           T \maps \R^4 \to \R^4 \]
preserving the Minkowski metric is called the {\bf Lorentz group},
and denoted $\O(3,1)$.

\subsection*{Heisenberg (1925)}

In 1925, Werner Heisenberg came up with a radical new approach to
physics in which processes were described using matrices \cite{MehraMatrix}.
What makes this especially remarkable is that Heisenberg, like
most physicists of his day, had not heard of matrices!  His
idea was that given a system with some set of states, say
$\{1,\dots,n\}$, a process $U$ would be described by a bunch
of complex numbers $U^i_j$ specifying the `amplitude' for
any state $i$ to turn into any state $j$.  He composed processes
by summing over all possible intermediate states:
\[       (VU)^i_k = \sum_j V^j_k U^i_j . \]
Later he discussed his theory with his thesis advisor,
Max Born, who informed him that he had reinvented matrix multiplication.

Heisenberg never liked the term `matrix mechanics' for his work,
because he thought it sounded too abstract.  However, it is an apt
indication of the {\it algebraic} flavor of quantum physics.

\subsection*{Born (1928)}

In 1928, Max Born figured out what Heisenberg's mysterious
`amplitudes' actually meant: the absolute value squared $|U^i_j|^2$
gives the {\it probability} for the initial state $i$ to become the
final state $j$ via the process $U$.  This spelled the end of the
deterministic worldview built into Newtonian mechanics
\cite{Greenspan}.  More shockingly still, since amplitudes are
complex, a sum of amplitudes can have a smaller absolute value than
those of its terms.  Thus, quantum mechanics exhibits destructive
interference: allowing more ways for something to happen may reduce
the chance that it does!

\subsection*{Von Neumann (1932)}

In 1932, John von Neumann published a book on the foundations of
quantum mechanics \cite{vN}, which helped crystallize the now-standard
approach to this theory.  We hope that the experts will forgive us for
omitting many important subtleties and caveats in the following
sketch.

Every quantum system has a Hilbert space of states, $H$.  A {\bf
state} of the system is described by a unit vector $\psi \in H$.
Quantum theory is inherently probabilistic: if we put the system in
some state $\psi$ and immediately check to see if it is in the state
$\phi$, we get the answer `yes' with probability equal to $|\langle
\phi,\psi\rangle|^2$.

A reversible process that our system can undergo is called a
{\bf symmetry}.  Mathematically, any symmetry is described by a unitary
operator $U \maps H \to H$.  If we put the system in some state
$\psi$ and apply the symmetry $U$ it will then be in the state $U\psi$.
If we then check to see if it is in some state $\phi$, we get
the answer `yes' with probability $|\langle \phi,U\psi\rangle|^2$.
The underlying complex number $\langle \phi, U \psi \rangle$ is
called a {\bf transition amplitude}.  In particular, if we have
an orthonormal basis $e^i$ of $H$, the numbers
\[            U^i_j = \langle e^j , U e^i \rangle \]
are Heisenberg's matrices!

Thus, Heisenberg's matrix mechanics is revealed to be part of a
framework in which unitary operators describe physical processes.
But, operators also play another role in quantum theory.
A real-valued quantity that we can measure by doing experiments
on our system is called an {\bf observable}.  Examples include
energy, momentum, angular momentum and the like.
Mathematically, any observable is described by a
self-adjoint operator $A$ on the Hilbert space $H$ for the system
in question.  Thanks to the probabilistic nature of quantum mechanics,
we can obtain various different values when we measure the observable
$A$ in the state $\psi$, but the average or `expected' value will be
$\langle \psi, A \psi \rangle$.

If a group $G$ acts as symmetries of some quantum system, we
obtain a {\bf unitary representation} of $G$, meaning a Hilbert
space $H$ equipped with unitary operators
\[       \rho(g) \maps H \to H , \]
one for each $g \in G$, such that
\[           \rho(1) = 1_H  \]
and
\[           \rho(gh) = \rho(g) \rho(h) . \]
Often the group $G$ will be equipped with a topology.  Then we want
symmetry transformation close to the identity to affect the system
only slightly, so we demand that if $g_i \to 1$ in $G$, then
$\rho(g_i)\psi \to \psi$ for all $\psi \in H$.  Professionals use the
term \textbf{strongly continuous} for representations with this
property, but we shall simply call them \textbf{continuous}, since
we never discuss any other sort of continuity.

Continuity turns out to have powerful consequences, such as the Stone--von
Neumann theorem: if $\rho$ is a
continuous representation of $\R$ on $H$, then
\[         \rho(s) = \exp(-isA)   \]
for a unique self-adjoint operator $A$ on $H$.  Conversely, any
self-adjoint operator gives a continuous representation
of $\R$ this way.  In short, there is a correspondence between
observables and one-parameter groups of symmetries.  This links
the two roles of operators in quantum mechanics: self-adjoint
operators for observables, and unitary operators for symmetries.

\subsection*{Wigner (1939)}

We have already discussed how the Lorentz group $\O(3,1)$ acts
as symmetries of spacetime in special relativity: it is the group
of all linear transformations
\[           T \maps \R^4 \to \R^4 \]
preserving the Minkowski metric.  However, the full symmetry group
of Minkowski spacetime is larger: it includes translations as well.
So, the really important group in special relativity is the so-called
`Poincar\'e group':
\[          \P = \O(3,1) \ltimes \R^4  \]
generated by Lorentz transformations and translations.

Some subtleties appear when we take some findings from particle
physics into account.
Though time reversal
\[   (t,x,y,z) \mapsto (-t,x,y,z)   \]
and parity
\[     (t,x,y,z) \mapsto (t,-x,-y,-z) \]
are elements of $\P$, not every physical system has them as symmetries.
So it is better to exclude such elements of the Poincar\'e group by
working with the connected component of the identity, $\P_0$.  Furthermore,
when we rotate an electron a full turn, its
state vector does not come back to where it stated: it gets multiplied
by -1.  If we rotate it two full turns, it gets back to where it
started.  To deal with this, we should replace $\P_0$ by its universal
cover, $\tilde \P_0$.  For lack of a snappy name, in what follows
we call {\it this} group the {\bf Poincar\'e group}.

We have seen that in quantum mechanics, physical systems
are described by continuous unitary representations of
the relevant symmetry group.  In relativistic quantum mechanics,
this symmetry group is $\tilde \P_0$.  The Stone-von Neumann theorem
then associates observables to one-parameter subgroups of this group.
The most important observables in physics---energy, momentum,
and angular momentum---all arise this way!

For example, time translation
\[       g_s \maps (t,x,y,z) \mapsto (t+s,x,y,z)  \]
gives rise to an observable $A$ with
\[           \rho(g_s) = \exp(-isA).   \]
and this observable is the {\it energy} of the system, also
known as the {\bf Hamiltonian}.  If the system is in
a state described by the unit vector $\psi \in H$, the expected
value of its energy is $\langle \psi, A \psi \rangle$.
In the context of special relativity, the energy of a system
is always greater than or equal to that of the vacuum (the
empty system, as it were).  The energy of the vacuum is zero,
so it makes sense to focus attention on continuous
unitary representations of the Poincar\'e group with
\[         \langle \psi, A \psi \rangle \ge 0  . \]
These are usually called {\bf positive-energy representations}.

In a famous 1939 paper, Eugene Wigner \cite{Wigner} classified the
positive-energy representations of the Poincar\'e group.
All these representations can be built as direct sums of
irreducible ones, which serve as candidates for describing
`elementary particles': the building blocks of matter.
To specify one of these representations, we need to give a
number $m \ge 0$ called the `mass' of the particle, a number
$j = 0, \frac{1}{2}, 1, \dots$ called its `spin', and sometimes
a little extra data.

For example, the photon has spin $1$ and
mass $0$, while the electron has spin $\frac{1}{2}$ and mass equal
to about $9 \cdot 10^{-31}$ kilograms.  Nobody knows why particles
have the masses they do---this is one of the main unsolved problems
in physics---but they all fit nicely into Wigner's
classification scheme.

\subsection*{Eilenberg--Mac Lane (1945)}

Eilenberg and Mac Lane \cite{EM} invented the notion of a
`category' while working on algebraic topology.   The idea is that
whenever we study mathematical gadgets of any sort---sets,
or groups, or topological spaces, or positive-energy
representations of the Poincar\'e group, or whatever---we should
also study the structure-preserving maps between these gadgets.
We call the gadgets `objects' and the maps `morphisms'. The identity
map is always a morphism, and we can compose morphisms in an associative
way.

Eilenberg and Mac Lane thus defined a {\bf category} $C$ to consist of:
\begin{itemize}
\item a collection of {\bf objects},
\item for any pair of objects $x,y$, a set of $\hom(x,y)$ of
{\bf morphisms} from $x$ to $y$, written $f \maps x \to y$,
\end{itemize}
equipped with:
\begin{itemize}
\item for any object $x$, an {\bf identity morphism} $1_x \maps x
\to x$,
\item for any pair of morphisms $f \maps x \to y$ and $g
\maps y \to z$, a morphism $gf \maps x \to z$ called the {\bf
composite} of $f$ and $g$,
\end{itemize}
such that:
\begin{itemize}
\item for any morphism $f \maps x \to y$, the {\bf left and right
unit laws} hold: $1_y f = f = f 1_x$.
\item for any triple of
morphisms $f \maps w \to x$, $g \maps x \to y$, $h \maps y \to z$,
the {\bf associative law} holds: $(hg)f = h(gf)$.
\end{itemize}
Given a morphism $f \maps x \to y$, we call $x$ the {\bf source}
of $f$ and $y$ the {\bf target} of $y$.

Eilenberg and Mac Lane did much more than just define the concept
of category.  They also defined maps between categories, which they called
`functors'.  These send objects to objects, morphisms to morphisms, and
preserve all the structure in sight.  More precisely, given categories
$C$ and $D$, a {\bf functor} $F \maps C \to D$ consists of:
\begin{itemize}
\item a function $F$ sending objects in $C$ to objects in $D$, and
\item for any pair of objects $x,y \in \Ob(C)$, a function also
called $F$ sending morphisms in $\hom(x,y)$ to morphisms
in $\hom(F(x),F(y))$
\end{itemize}
such that:
\begin{itemize}
\item $F$ preserves identities: for any object $x \in C$,
$F(1_x) = 1_{F(x)}$;
\item $F$ preserves composition: for
any pair of morphisms $f \maps x \to y$, $g \maps y \to z$ in $C$,
$F(gf) = F(g) F(f)$.
\end{itemize}

Many of the famous invariants in algebraic topology are
actually functors, and this is part of how we convert topology problems
into algebra problems and solve them.    For example, the fundamental group
is a functor
\[            \pi_1 \maps \Top_* \to \Grp  . \]
from the category of topological spaces equipped with a basepoint
to the category of groups.
In other words, not only does any topological space with basepoint $X$ have a
fundamental group $\pi_1(X)$, but also any continuous map $f \maps
X \to Y$ preserving the basepoint
gives a homomorphism $\pi_1(f) \maps \pi_1(X) \to \pi_1(Y)$,
in a way that gets along with composition.
So, to show that the inclusion of the circle in the disc
\[ \psset{xunit=1.2cm,yunit=1.2cm} \xy
  (-12,0)*+{\begin{pspicture}(.6,.6)
            \pscircle(.3,.3){.3}
            \rput(0,.8){$S^{1}$}        
            \end{pspicture}}="l";
  (12,0)*+{\begin{pspicture}(.6,.6)
            \pscircle[linewidth=0.3pt,fillcolor=lightgray,fillstyle=solid](.3,.3){.3}
            \rput(.7,.8){$D^{2}$}       
            \end{pspicture}}="r";
    {\ar^{i} "l";"r"};
 \endxy \]
does not admit a retraction---that is, a map
\[    \psset{xunit=1.2cm,yunit=1.2cm}  \xy
  (12,0)*+{\begin{pspicture}(.6,.6)
            \pscircle(.3,.3){.3}
            \rput(.7,.8){$S^{1}$}        
            \end{pspicture}}="r";
  (-12,0)*+{\begin{pspicture}(.6,.6)
            \pscircle[linewidth=0.3pt,fillcolor=lightgray,fillstyle=solid](.3,.3){.3}
            \rput(0,.8){$D^{2}$}       
            \end{pspicture}}="l";
    {\ar^{r} "l";"r"};
 \endxy  \]
such that this diagram commutes:
\[
\DiscCircle
\]
we simply hit this question with the functor $\pi_1$ and note
that the homomorphism
\[       \pi_1(i) \maps \pi_1(S^1) \to \pi_1(D^2)  \]
cannot have a homomorphism
\[       \pi_1(r) \maps \pi_1(D^2) \to \pi_1(S^1)  \]
for which $\pi_1(r)\pi_1(i)$ is the identity, because
$\pi_1(S^1) = \Z$ and $\pi_1(D^2) = 0$.

However, Mac Lane later wrote that the real point of this paper
was not to define categories, nor to define functors between
categories, but to define `natural transformations' between functors!
These can be drawn as follows:
\[
\xy (-10,0)*+{\bullet}="4"; (10,0)*+{\bullet}="6";
{\ar@/^1.65pc/^{F} "4";"6"}; {\ar@/_1.65pc/_{G} "4";"6"};
{\ar@{=>}^<<<{\scriptstyle \alpha} (0,3)*{};(0,-3)*{}} ;
(-13,0)*{C};(13,0)*{D};
\endxy
\]
Given functors $F,G \maps C \to D$, a {\bf natural transformation}
$\alpha \maps F \To G$ consists of:
\begin{itemize}
\item a function $\alpha$ mapping each object $x \in C$ to a
morphism $\alpha_x \maps F(x) \to G(x)$
\end{itemize}
such that:
\begin{itemize}
\item for any morphism $f \maps x \to y$ in $C$, this diagram
commutes:
\[
\xymatrix{
 F(x)
   \ar[r]^{F(f)}
   \ar[d]_{\alpha_x}
&  F(y)
  \ar[d]^{\alpha_y}
\\
 G(x)
   \ar[r]_{G(f)}
&  G(y) }
\]
\end{itemize}
The commuting square here conveys the ideas that $\alpha$ not only gives
a morphism $\alpha_x \maps F(x) \to G(x)$ for each object $x \in C$, but
does so `naturally'---that is, in a way that is compatible with
all the morphisms in $C$.

The most immediately interesting natural transformations are the
natural isomorphisms.  When Eilenberg and Mac Lane were writing their
paper, there were many different recipes for computing the homology
groups of a space, and they wanted to formalize the notion that these
different recipes give groups that are not only isomorphic, but
`naturally' so.  In general, we say a morphism $g \maps y \to x$ is an
{\bf isomorphism} if it has an inverse: that is, a morphism $f \maps x
\to y$ for which $fg$ and $gf$ are identity morphisms.  A {\bf natural
isomorphism} between functors $F,G \maps C \to D$ is then a natural
transformation $\alpha \maps F \To G$ such that $\alpha_x$ is an
isomorphism for all $x \in C$.  Alternatively, we can define how to
compose natural transformations, and say a natural isomorphism is a
natural transformation with an inverse.

Invertible functors are also important---but here an important theme
known as `weakening' intervenes for the first time.  Suppose we have
functors $F \maps C \to D$ and $G \maps D \to C$.  It is unreasonable
to demand that if we apply first $F$ and then $G$, we get back exactly
the object we started with.  In practice all we really need, and all
we typically get, is a naturally isomorphic object.  So, we say a
functor $F \maps C \to D$ is an {\bf equivalence} if it has a {\bf
weak inverse}, that is, a functor $G \maps D \to C$ such that there
exist natural isomorphisms $\alpha \maps GF \To 1_C$, $\beta \maps FG
\To 1_D$.

In the first applications to topology, the categories involved were
mainly quite large: for example, the category of all topological
spaces, or all groups.  In fact, these categories are even `large' in
the technical sense, meaning that their collection of objects is not a
set but a proper class.  But later applications of category theory to
physics often involved small categories.

For example, any group $G$ can be thought of as a category with
one object and only invertible morphisms: the morphisms are the
elements of $G$, and composition is multiplication in the group.
A representation of $G$ on a Hilbert space is then the same as a
functor
\[   \rho \maps G \to \Hilb, \]
where $\Hilb$ is the category with Hilbert spaces as objects and
bounded linear operators as morphisms.  While this viewpoint may seem
like overkill, it is a prototype for the idea of describing theories
of physics as functors, in which `abstract' physical processes (e.g.\
symmetries) get represented in a `concrete' way (e.g.\ as operators).
However, this idea came long after the work of Eilenberg and Mac Lane:
it was born sometime around Lawvere's 1963 thesis, and came to
maturity in Atiyah's 1988 definition of `topological quantum field
theory'.

\subsection*{Feynman (1947)}

After World War II, many physicists who had been working in
the Manhattan project to develop the atomic bomb returned to
work on particle physics.  In 1947, a small conference on this
subject was held at Shelter Island, attended by luminaries such
as Bohr, Oppenheimer, von Neumann, Weisskopf, and Wheeler.  Feynman
presented his work on quantum field theory, but it seems nobody
understood it except Schwinger, who was later to share the Nobel
prize with him and Tomonaga.  Apparently it was a bit too far-out
for most of the audience.

Feynman described a formalism in which time evolution for quantum systems
was described using an integral over the space of all classical
histories:  a `Feynman path integral'.  These are notoriously hard to
make rigorous.  But, he also described a way to compute these
perturbatively as a sum over diagrams: `Feynman diagrams'.   For example,
in QED, the amplitude for an electron to absorb a photon is given by:
\[
 \feynmandiagram
\]
All these diagrams describe ways for an electron and photon to
come in and an electron to go out.
Lines with arrows pointing downwards stand for electrons.
Lines with arrows pointing upwards stand for positrons:
the positron is the `antiparticle' of an electron, and Feynman
realized that this could thought of as an electron going backwards in
time.  The wiggly lines stand for photons.   The photon is its own
antiparticle, so we do not need arrows on these wiggly lines.

Mathematically, each of the diagrams shown above
is shorthand for a linear operator
\[      f\maps H_e \tensor H_\gamma \to H_e  \]
where $H_e$ is the Hilbert space for an electron, and $H_\gamma$ is
a Hilbert space for a photon. We take the tensor product of group
representations when combining two systems, so $H_e \tensor
H_\gamma$ is the Hilbert space for an photon together with an
electron.

As already mentioned, elementary particles are described by
certain special representations of the Poincar\'e group---the
irreducible positive-energy ones.  So, $H_e$ and $H_\gamma$
are representations of this sort.  We can tensor these to obtain
positive-energy representations describing collections of elementary
particles.  Moreover, each Feynman diagram describes an {\bf
intertwining operator}: an operator that commutes with
the action of the Poincar\'e group.  This expresses the fact that if
we, say, rotate our laboratory before doing an experiment, we just
get a rotated version of the result we would otherwise get.

So, Feynman diagrams are {\it a notation for intertwining
operators between positive-energy representations of the Poincar\'e
group.}  However, they are so powerfully evocative that they are
much more than a mere trick!  As Feynman recalled later \cite{MehraFeynman}:

\begin{quote}
The diagrams were intended to represent physical processes and
the mathematical expressions used to describe them.  Each diagram
signified a mathematical expression.  In these diagrams I was
seeing things that happened in space and time.  Mathematical quantities
were being associated with points in space and time.  I would
see electrons going along, being scattered at one point, then going
over to another point and getting scattered there, emitting a
photon and the photon goes there.  I would make little pictures of all
that was going on; these were physical pictures involving the
mathematical terms.
\end{quote}

Feynman first published papers containing such diagrams in 1949
\cite{Feynman2,Feynman1}.  However, his work reached many physicists
through expository articles published even earlier by one of the
few people who understood what he was up to: Freeman Dyson
\cite{Dyson1,Dyson2}.  For more on the history of Feynman diagrams,
see the book by Kaiser \cite{Kaiser}.

The general context for such diagrammatic reasoning
came much later, from category theory.  The idea is that we
can draw a morphism $f \maps x \to y$ as an arrow going down:
\[ 
\xy {\ar_f (0,7)*+{x};(0,-7)*+{y}}
\endxy
\]
but then we can switch to a style of drawing
in which the objects are depicted not as dots but
as `wires', while the morphisms are drawn not as
arrows but as `black boxes' with one input wire and
one output wire:
 \[ 
\xy (-3,0)*{f}; (0,0)*{\bullet}="f"; (0,10)**\dir{-}
?(.5)*\dir{<}+(-3,2)*{ x}; "f";(0,-10)**\dir{-}
?(.75)*\dir{>}+(-3,0)*{ y};
\endxy
\qquad {\rm or} \qquad
 \xy (0,0)*{f};
(0,0)*\xycircle(2.65,2.65){-}="f"; (0,10)**\dir{-}
?(.5)*\dir{<}+(-3,2)*{ x}; "f";(0,-10)**\dir{-}
?(.75)*\dir{>}+(-3,0)*{ y};
\endxy
\]
This is starting to look a bit like a Feynman diagram!  However, to
get really interesting Feynman diagrams we need black boxes with many
wires going in and many wires going out.  The mathematics necessary
for this was formalized later, in Mac Lane's 1963 paper on monoidal
categories (see below) and Joyal and Street's 1980s work on `string
diagrams' \cite{JS0}.

\subsection*{Yang--Mills (1953)}

In modern physics the electromagnetic force is described by a $\U(1)$
gauge field.  Most mathematicians prefer to call this a `connection on
a principal $\U(1)$ bundle'.  Jargon aside, this means that if we
carry a charged particle around a loop in spacetime, its state will be
multiplied by some element of $\U(1)$---that is, a phase---thanks to
the presence of the electromagnetic field.  Moreover, everything about
electromagnetism can be understood in these terms!

In 1953, Chen Ning Yang and Robert Mills \cite{YM} formulated a
generalization of Maxwell's equations in which forces other
than electromagnetism can be described by connections on $G$-bundles
for groups other than $\U(1)$.  With a vast amount of work by many
great physicists, this ultimately led to the `Standard Model', a theory
in which {\it all forces other than gravity} are described using a
connection on a principal $G$-bundle where
\[     G = \U(1) \times \SU(2) \times \SU(3)  .\]
Though everyone would like to more deeply understand this curious
choice of $G$, at present it is purely a matter of fitting
the experimental data.

In the Standard Model, elementary particles are described as irreducible
positive-energy representations of $\tilde \P_0 \times G$.
Perturbative calculations in this theory can be done using
souped-up Feynman diagrams, which are a notation for intertwining
operators between positive-energy representations of $\tilde \P_0 \times G$.

While efficient, the mathematical jargon in the previous paragraphs does
little justice to how physicists actually think about these things.
For example, Yang and Mills {\it did not know about bundles and
connections} when formulating their theory.  Yang later wrote \cite{Yang}:

\begin{quote}
What Mills and I were doing in 1954 was generalizing Maxwell's
theory.  We knew of no geometrical meaning of Maxwell's theory, and we
were not looking in that direction. To a physicist, gauge potential is a
concept {\rm rooted} in our description of the electromagnetic
field.  Connection is a geometrical concept which I only learned around
1970.
\end{quote}

\subsection*{Mac Lane (1963)}

In 1963 Mac Lane published a paper describing the notion of a
`monoidal category' \cite{MacLane:1963}.  The idea was that in many
categories there is a way to take the `tensor product' of two objects,
or of two morphisms.  A famous example is the category $\Vect$, where
the objects are vector spaces and the morphisms are linear operators.
This becomes a monoidal category with the usual tensor product of
vector spaces and linear maps.  Other examples include the category
$\Set$ with the cartesian product of sets, and the category $\Hilb$
with the usual tensor product of Hilbert spaces.  We will also be
interested in $\Fin\Vect$ and $\Fin\Hilb$, where the objects are
\textit{finite-dimensional} vector spaces (resp.\ Hilbert spaces) and
the morphisms are linear maps.  We will also get many examples from
categories of representations of groups.  The theory of Feynman
diagrams, for example, turns out to be based on the symmetric monoidal
category of positive-energy representations of the Poincar\'e group!

In a monoidal category, given morphisms $f \maps x \to y$ and
$g \maps x' \to y'$ there is a morphism
\[         f \tensor g \maps x \tensor x' \to y \tensor y'  .\]
We can also draw this as follows:
\[ \xy   
 (-8,0)*{f};
  (-5,12)*{}; (-5,0)*{\bullet}="1_x"; **\dir{-}
?(.5)*\dir{<}+(3,0)*{\scriptstyle x}; "1_x";(-5,-12)*{}; **\dir{-}
?(.4)*\dir{<}+(3,0)*{\scriptstyle y}; (5,12)*{};
(5,0)*{\bullet}="1_x"; **\dir{-} ?(.5)*\dir{<}+(3,0)*{\scriptstyle
x'}; "1_x";(5,-12)*{}; **\dir{-} ?(.4)*\dir{<}+(3,0)*{\scriptstyle
y'}; (8,0)*{g};
\endxy \]
This sort of diagram is sometimes called a `string diagram'; the
mathematics of these was formalized later \cite{JS0}, but we can't
resist using them now, since they are so intuitive.  Notice that the
diagrams we could draw in a mere category were intrinsically
1-dimensional, because the only thing we could do is compose
morphisms, which we draw by sticking one on top of another.  In a
monoidal category the string diagrams become 2-dimensional, because
now we can also tensor morphisms, which we draw by placing them side
by side.

This idea continues to work in higher dimensions as well.  The kind of
category suitable for 3-dimensional diagrams is called a `braided
monoidal category'.  In such a category, every pair of objects $x,y$
is equipped with an isomorphism called the `braiding', which switches
the order of factors in their tensor product:
\[    B_{x,y} \maps x \tensor y \to y \tensor x  .\]
We can draw this process of switching as a diagram in 3 dimensions:
\[ \xy
 \vtwist~{(-5,8)}{(5,8)}{(-5,-8)}{(5,-8)}<>|>>><;
 (-7,5)*{\scriptstyle x};
 (7,5)*{\scriptstyle y};
\endxy \]
and the braiding $B_{x,y}$ satisfies axioms that are related to
the topology of 3-dimensional space.

All the examples of monoidal categories given above are also
braided monoidal categories.  Indeed, many mathematicians would
shamelessly say that given vector spaces $V$ and $W$, the tensor
product $V \tensor W$ is `equal to' the tensor product $W \tensor V$.
But this is not really true; if you examine the fine print you will
see that they are just isomorphic, via this braiding:
\[
\begin{array}{rcl}
         B_{V,W} \maps v \tensor w &\mapsto& w \tensor v .
\end{array}
\]

Actually, all the examples above are not just braided
but also `symmetric' monoidal categories.  This means
that if you switch two things and then switch them again, you
get back where you started:
\[        B_{x,y} B_{y,x} = 1_{x \tensor y}  .\]
Because all the braided monoidal categories Mac Lane knew satisfied
this extra axiom, he only considered symmetric monoidal
categories. In diagrams, this extra axiom says that:
\[ \vcenter{\xy
 \vtwist~{(-5,15)}{(5,15)}{(-5,5)}{(5,5)};
 \vtwist~{(-5,5)}{(5,5)}{(-5,-5)}{(5,-5)};
 (-5,18.5)*{};(-5,-9.5)*{} **\dir{}?(.5)*\dir{>}+(-3,1)*{\scriptstyle x};
 (5,18.5)*{};(5,-9.5)*{} **\dir{}?(.5)*\dir{>}+(3,1)*{\scriptstyle y};
\endxy} \qquad = \qquad
  \xy
 (-4,10)*{}="TL";
 (4,10)*{}="TR";
 (-4,-10)*{}="BL";
 (4,-10)*{}="BR";
    "TL";"BL" **\dir{-}?(.5)*\dir{>}+(-3,1)*{\scriptstyle x};
    "TR";"BR" **\dir{-}?(.5)*\dir{>}+(3,1)*{\scriptstyle y};
 \endxy\]
In 4 or more dimensions, any knot can be untied by just this
sort of process.  Thus, the string diagrams for symmetric monoidal
categories should really be drawn in 4 or more dimensions!
But, we can cheat and draw them in the plane, as we have above.

It is worth taking a look at Mac Lane's precise definitions, since
they are a bit subtler than our summary suggests, and these subtleties
are actually very interesting.

First, he demanded that a monoidal category have a unit for the tensor
product, which he call the `unit object', or `$1$'.  For example, the
unit for tensor product in $\Vect$ is the ground field, while the unit
for the Cartesian product in $\Set$ is the one-element set.  ({\it
Which} one-element set?  Choose your favorite one!)

Second, Mac Lane did not demand that the tensor product be
associative `on the nose':
\[         (x \tensor y) \tensor z = x \tensor (y \tensor z) ,\]
but only up a specified isomorphism called the `associator':
\[     a_{x,y,z} \maps (x \tensor y) \tensor z \to
                        x \tensor (y \tensor z)  .\]
Similarly, he didn't demand that $1$ act as the unit for
the tensor product `on the nose', but only up to specified isomorphisms
called the `left and right unitors':
\[       \ell_x \maps 1 \tensor x \to x  , \qquad
            r_x \maps x \tensor 1 \to x  .\]
The reason is that in real life, it is usually too much to
expect equations between objects in a category: usually we just
have isomorphisms, and this is good enough!  Indeed this is a
basic moral of category theory: equations between objects are
bad; we should instead specify isomorphisms.

Third, and most subtly of all, Mac Lane demanded that the associator
and left and right unitors satisfy certain `coherence laws', which
let us work with them as smoothly as if they {\it were} equations.
These laws are called the pentagon and triangle identities.

Here is the actual definition.  A {\bf monoidal category} consists of:
\begin{itemize}
\item a category $M$.
\item a functor called the {\bf tensor product} $
\tensor \maps M \times M \to M$, where we
write $\tensor(x,y)=x \tensor y$ and $\tensor(f,g)=f \tensor g$
for objects $x, y \in M$ and morphisms $f, g$ in $M$.
\item an object called the {\bf identity object} $1 \in M$.
\item natural isomorphisms called the {\bf associator}:
  \[ a_{x,y,z} \maps (x \tensor y) \tensor z \to x \tensor (y \tensor z), \]
the {\bf left unit law}:
  \[ \ell_x \maps 1 \ten x \to x , \]
and the {\bf right unit law}:
  \[ r_x \maps x \ten 1 \to x.  \]
\end{itemize}
such that the following diagrams commute for all objects $w,x,y,z
\in M$:
\begin{itemize}
\item the {\bf pentagon identity}:
\[
\xy 0;/r.35pc/:
    (-24.73,8.03)*+{ (w \otimes (x \otimes y)) \otimes z}="l";
    (0,26)*+{ ((w \otimes x) \otimes y) \otimes z}="t";
    (24.73,8.03)*+{ (w \otimes x) \otimes (y \otimes z)}="r";
    (15.28,-21.03)*+{ w \otimes (x \otimes (y \otimes z))}="br";
    (-15.28,-21.03)*+{ w \otimes ((x \otimes y) \otimes z)}="bl";
     {\ar_{a_{w, x, y} \otimes 1_z} "t";"l"};
     {\ar_{a_{w, x \otimes y, z}} "l";"bl"};
     {\ar_{1_w \otimes a_{x,y,z}} "bl";"br"};
     {\ar^{a_{w, x, y \otimes z}} "r";"br"};
     {\ar^{a_{w \otimes x, y, z}} "t";"r"};
\endxy
\]
governing the associator.
\vskip 1em
\item the {\bf triangle
identity}:
\[
\xymatrix{
 (x \ten 1) \ten y
     \ar[rr]^{a_{x,1,y}}
     \ar[dr]_{r_x \ten 1_y}
  &&  x \ten (1 \ten y)
     \ar[dl]^{1_x \ten \ell_y } \\
 & x \ten y   }  \\
\]
\end{itemize}
governing the left and right unitors.

The pentagon and triangle identities are the least obvious part of
this definition---but also the truly brilliant part.  The point of the
pentagon identity is that when we have a tensor product of four
objects, there are five ways to parenthesize it, and at first glance
the associator gives two different isomorphisms from $w \ten (x \ten
(y \ten z))$ to $((w \ten x) \ten y) \ten z$.  The pentagon identity
says these are in fact the same!  Of course when we have tensor
products of even more objects there are even more ways to parenthesize
them, and even more isomorphisms between them built from the
associator.  However, Mac Lane showed that the pentagon identity
implies these isomorphisms are all the same.  If we also assume the
triangle identity, all isomorphisms with the same source and target
built from the associator, left and right unit laws are equal.

In fact, the pentagon was also introduced in 1963 by James Stasheff
\cite{Stasheff1}, as part of an infinite sequence of polytopes called
`associahedra'.  Stasheff defined a concept of `$A_\infty$-space',
which is roughly a topological space having a product that is
associative up to homotopy, where this homotopy satisfies the pentagon
identity up homotopy, that homotopy satisfies yet another identity up
to homotopy, and so on, \textit{ad infinitum}.  The $n$th of these
identities is described by the $n$-dimensional associahedron.  The
first identity is just the associative law, which plays a crucial role
in the definition of \textbf{monoid}: a set with associative product
and identity element.  Mac Lane realized that the second, the pentagon
identity, should play a similar role in the definition of monoidal
category.  The higher ones show up in the theory of monoidal
bicategories, monoidal tricategories and so on.

With the concept of monoidal category in hand, one can define
a {\bf braided monoidal category} to consist of:
\begin{itemize}
\item a monoidal category $M$, and
\item a natural isomorphism called
the {\bf braiding}:
\[   B_{x,y} \maps x \tensor y \to y \tensor x. \]
\end{itemize}
such that these two diagrams commute, called the {\bf hexagon
identities}:

\[
\xy
   (-20,12)*+{(x \ten y) \ten z}="tl";
   (20,12)*+{(y \ten x) \ten z}="tr";
   (-40,0)*+{x \ten (y \ten z)}="ml";
   (40,0)*+{y \ten (x \ten z)}="mr";
   (-20,-12)*+{(y \ten z) \ten x}="bl";
   (20,-12)*+{y \ten (z \ten x)}="br";
        {\ar^{ B _{x,y} \ten z} "tl";"tr"};
        {\ar^{ a_{y,x,z}} "tr";"mr"};
        {\ar^{ a ^{-1}_{x,y,z}} "ml";"tl"};
        {\ar_{ B _{x,y \ten z}} "ml";"bl"};
        {\ar_{ a_{y,z,x}} "bl";"br"};
        {\ar^{ y \ten  B _{x,z} } "mr";"br"};
\endxy
\]
\vskip 1em
\[
\xy
   (-20,12)*+{x \ten (y \ten z)}="tl";
   (20,12)*+{x \ten (z \ten y)}="tr";
   (-40,0)*+{(x \ten y) \ten z}="ml";
   (40,0)*+{(x \ten z) \ten y}="mr";
   (-20,-12)*+{z \ten (x  \ten y)}="bl";
   (20,-12)*+{(z \ten x) \ten y}="br";
        {\ar^{ x \ten B _{y,z}} "tl";"tr"};
        {\ar^{ a _{x,z,y}^{-1}} "tr";"mr"};
        {\ar^{ a _{x,y,z}} "ml";"tl"};
        {\ar_{ B _{x\ten y,z}} "ml";"bl"};
        {\ar_{ a _{z,x,y}^{-1}} "bl";"br"};
        {\ar^{B _{x,z}\ten y} "mr";"br"};
\endxy
\]
The first hexagon equation says that switching the object $x$ past
$y \tensor z$ all at once is the same as switching it past $y$ and then
past $z$ (with some associators thrown in to move the parentheses).
The second one is similar: it says switching $x \tensor y$ past $z$
all at once is the same as doing it in two steps.

We define a {\bf symmetric monoidal category}
to be a braided monoidal category $M$ for which the braiding
satisfies $B_{x,y} = B_{y,x}^{-1}$ for all objects $x$ and $y$.
A monoidal, braided monoidal, or symmetric monoidal category is called
{\bf strict} if $a_{x,y,z},\ell_x,$ and $r_x$ are always identity
morphisms.  In this case we have
\[ (x \tensor y) \tensor z = x \tensor (y \tensor z) , \]
\[        1 \tensor x = x, \qquad x \tensor 1 = x .\]
Mac Lane showed in a certain precise sense, every monoidal or
symmetric monoidal category is equivalent to a strict one.  The same
is true for braided monoidal categories. However, the examples that
turn up in nature, like $\Vect$, are rarely strict.

\subsection*{Lawvere (1963)}

The famous category theorist F.\ William Lawvere began his graduate
work under Clifford Truesdell, an expert on `continuum mechanics',
that very practical branch of classical field theory which deals with
fluids, elastic bodies and the like.  In the process, Lawvere got very
interested in the foundations of physics, particularly the notion of
`physical theory', and his research took a very abstract turn.  Since
Truesdell had worked with Eilenberg and Mac Lane during World War II,
he sent Lawvere to visit Eilenberg at Columbia University, and that is
where Lawvere wrote his thesis.

In 1963, Lawvere finished a thesis on `functorial semantics'
\cite{Lawvere}.  This is a general framework for theories of
mathematical or physical objects in which a `theory' is described by a
category $C$, and a `model' of this theory is described by a functor
$Z \maps C \to D$.  Typically $C$ and $D$ are equipped with extra
structure, and $Z$ is required to preserve this structure.  The
category $D$ plays the role of an `environment' in which the models
live; often we take $D = \Set$.

Variants of this idea soon became important in topology, especially
`PROPs' and `operads'.  In the late 1960's and early 70's, Mac Lane
\cite{MacLane:1965}, Boardmann and Vogt \cite{BV}, May \cite{May} and
others used these variants to study `homotopy-coherent' algebraic
structures: that is, structures with operations satisfying laws only
up to homotopy, with the homotopies themselves obeying certain laws,
but only up to homotopy, \textsl{ad infinitum}.  The easiest examples
are Stasheff's $A_\infty$-spaces, which we mentioned in the previous
section.  The laws governing $A_\infty$-spaces are encoded in
`associahedra' such as the pentagon.  In later work, it was seen that
the associahedra form an operad.  By the 90's, operads had become
very important both in mathematical physics \cite{Stasheff2,Stasheff3} and
the theory of $n$-categories \cite{Leinster2}.  Unfortunately,
explaining this line of work would take us far afield.

Other outgrowths of Lawvere's vision of functorial semantics include
the definitions of `conformal field theory' and `topological quantum
field theory', propounded by Segal and Atiyah in the late 1980s.  We
will have much more to say about these.  In keeping with physics
terminology, these later authors use the word `theory' for what
Lawvere called a `model': namely, a structure-preserving functor $Z
\maps C \to D$.  There is, however, a much more important difference.
Lawvere focused on classical physics, and took $C$ and $D$ to be
categories with cartesian products.  Segal and Atiyah focused on
quantum physics, and took $C$ and $D$ to be symmetric monoidal
categories, not necessarily cartesian.

\subsection*{B\'{e}nabou (1967)}

In 1967 B\'{e}nabou \cite{Benabou} introduced the notion of a
`bicategory', or as it is sometimes now called, a `weak 2-category'.
The idea is that besides objects and morphisms, a bicategory has
2-morphisms going between morphisms, like this:
\[
 \xy (-35,0)*{\bullet}+(-2,2)*{x}; {\ar^f (-8,0)*+{\bullet};(8,0)*+{\bullet}};
 (40,0)*{
 \xy (-8,0)*+{\bullet}="4"; (8,0)*+{\bullet}="6";
 {\ar@/^1.65pc/^{f} "4";"6"}; {\ar@/_1.65pc/_{g} "4";"6"};
 {\ar@{=>}^<<<{\scriptstyle \alpha} (0,3)*{};(0,-3)*{}} ;
 \endxy};
  (-11,0)*{x};(11,0)*{y}; (11,20)*{};
  (-35,15)*{\txt\bf{objects}};
  (0,15)*{\txt\bf{morphisms}};
  (40,15)*{\txt\bf{2-morphisms}};\endxy
\]
In a bicategory we can compose morphisms as in an ordinary
category, but also we can compose 2-morphisms in two ways:
vertically and horizontally:
\[
\xy (-8,0)*+{\bullet}="4"; (8,0)*+{\bullet}="6"; {\ar "4";"6"};
{\ar@/^1.75pc/^{f} "4";"6"}; {\ar@/_1.75pc/_{g} "4";"6"};
{\ar@{=>}^<<{\scriptstyle \alpha} (0,6)*{};(0,1)*{}} ;
{\ar@{=>}^<<{\scriptstyle \beta} (0,-1)*{};(0,-6)*{}} ;
\endxy
\qquad \qquad  \xy (-16,0)*+{\bullet}="4"; (0,0)*+{\bullet}="6";
{\ar@/^1.65pc/^{f} "4";"6"}; {\ar@/_1.65pc/_{g} "4";"6"};
{\ar@{=>}^<<<{\scriptstyle \alpha} (-8,3)*{};(-8,-3)*{}} ;
(0,0)*+{\bullet}="4"; (16,0)*+{\bullet}="6"; {\ar@/^1.65pc/^{f'}
"4";"6"}; {\ar@/_1.65pc/_{g'} "4";"6"}; {\ar@{=>}^<<<{\scriptstyle
\beta} (8,3)*{};(8,-3)*{}} ;
\endxy
\]
There are also identity morphisms and identity 2-morphisms, and
various axioms governing their behavior.  Most importantly, the usual
laws for composition of morphisms---the left and right unit laws and
associativity---hold only {\it up to specified 2-isomorphisms}.  (A
\textbf{2-isomorphism} is a 2-morphism that is invertible with respect
to vertical composition.)  For example, given morphisms $h \maps w \to
x$, $g \maps x \to y$ and $f \maps y \to z$, we have a 2-isomorphism
called the `associator':
\[     a_{f,g,h} \maps (fg)h \to f(gh) .\]
As in a monoidal category, this should satisfy the pentagon identity.

Bicategories are everywhere once you know how to look.
For example, there is a bicategory $\Cat$ in which:
\begin{itemize}
\item the objects are categories,
\item the morphisms are functors,
\item the 2-morphisms are natural transformations.
\end{itemize}
This example is unusual, because composition of morphisms
happens to satisfy the left and right unit laws and associativity
on the nose, as equations.  A more typical example is $\Bimod$,
in which:
\begin{itemize}
\item the objects are rings,
\item the morphisms from $R$ to $S$ are $R-S$-bimodules,
\item the 2-morphisms are bimodule homomorphisms.
\end{itemize}
Here composition of morphisms is defined by tensoring:
given an $R-S$-bimodule $M$ and an $S-T$-bimodule, we can tensor them
over $S$ to get an $R-T$-bimodule.   In this example the laws
for composition hold only up to specified 2-isomorphisms.

Another class of examples comes
from the fact that a monoidal category is secretly a bicategory
with one object!  The correspondence involves a kind of `reindexing'
as shown in the following table:
\[
\begin{tabular}{|c|c|}
  \hline
   \textbf{Monoidal Category} & \textbf{Bicategory}  \\
  \hline \hline
   ---& objects  \\
  objects &  morphisms\\
  morphisms & 2-morphisms  \\
  tensor product of objects &composite of morphisms  \\
  composite of morphisms & vertical composite of 2-morphisms  \\
  tensor product of morphisms &  horizontal composite of 2-morphisms \\
  \hline
\end{tabular}
\]
In other words, to see a monoidal category as a bicategory with only one
object, we should call the objects of the monoidal category `morphisms',
and call its morphisms `2-morphisms'.

A good example of this trick involves the monoidal category $\Vect$.
Start with $\Bimod$ and pick out your favorite object, say the ring of
complex numbers.  Then take all those bimodules of this ring that are
complex vector spaces, and all the bimodule homomorphisms between
these.  You now have a sub-bicategory with just one object---or in
other words, a monoidal category!  This is $\Vect$.

The fact that a monoidal category is secretly just a degenerate
bicategory eventually stimulated a lot of interest in higher
categories: people began to wonder what kinds of degenerate higher
categories give rise to braided and symmetric monoidal categories.
The impatient reader can jump ahead to 1995, when the pattern
underlying all these monoidal structures and their higher-dimensional
analogs became more clear.

\subsection*{Penrose (1971)}

In general relativity people had been using index-ridden
expressions for a long time.  For example, suppose we have a
binary product on a vector space $V$:
\[
    m \maps V\otimes V\to V .
\]
A normal person would abbreviate
$m(v \tensor w)$ as $v \cdot w$ and write the associative law
as
\[    (u \cdot v) \cdot w  = u \cdot (v \cdot w)  .\]
A mathematician might show off by writing
\[          m(m \tensor 1) = m(1 \tensor m)  \]
instead.  But physicists would pick a basis $e^i$ of $V$ and set
\[
    m(e^i \otimes e^j)= \sum_k m^{ij}_{k} e^k
\]
or
\[
    m(e^i \otimes e^j)= m^{ij}_k e^k
\]
for short, using the `Einstein summation convention' to sum over any
repeated index that appears once as a superscript and once as
a subscript.  Then, they would write the associative law as follows:
\[
    m^{ij}_{p} m^{pk}_{l}= m^{iq}_{l} m^{jk}_{q}.
\]
Mathematicians would mock them for this, but until Penrose
came along there was really no better completely general
way to manipulate tensors.  Indeed, before Einstein introduced
his summation convention in 1916, things were even worse.
He later joked to a friend \cite{Pais}:

\begin{quote}
I have made a great discovery in mathematics; I have suppressed
the summation sign every time that the summation must be made over
an index which occurs twice....
\end{quote}

In 1971, Penrose \cite{Penrose1}
introduced a new notation where tensors are
drawn as `black boxes', with superscripts corresponding to wires
coming in from above, and subscripts corresponding to wires
going out from below.
For example, he might draw $m \maps V \tensor V \to V$ as:
\[
 \xy
  (0,0)*\xycircle(3,3){-}="c";
  (0,0)*{m};
  (-6,10)*{}="tl"+(-1,-3)*{\scriptstyle i};
  (6,10)*{}="tr"+(1,-3)*{\scriptstyle j};
  (0,-10)*{}="b"+(-1,1)*{\scriptstyle k};
     "tr"; "c" **\dir{-};
     "tl"; "c" **\dir{-};
     "c"; "b" **\dir{-};
 \endxy
\]
and the associative law as:
\[
 \xy
  (0,0)*\xycircle(3,3){-}="c1";
  (0,0)*{m};
  (-6,10)*{}="tl"+(-1,-3)*{\scriptstyle i};
  (6,10)*{}="tr"+(1,-3)*{\scriptstyle j};
  (0,-10)*{}="b"+(0,4)*{\scriptstyle p};
     "tr"; "c1" **\dir{-};
     "tl"; "c1" **\dir{-};
  (5,-10)*\xycircle(3,3){-}="c";
  (5,-10)*{m};
  (2,0)*{}="tl";
  (18,10)*{}="tr"+(1,-3)*{\scriptstyle k};
  (5,-25)*{}="b"+(-1,5)*{\scriptstyle l};
     "tr"; "c" **\dir{-};
     "c"; "c1" **\dir{-};
     "c"; "b" **\dir{-};
 \endxy
\qquad = \qquad
 \xy
  (0,0)*\xycircle(3,3){-}="c1";
  (0,0)*{m};
  (6,10)*{}="tl"+(1,-3)*{\scriptstyle k};
  (-6,10)*{}="tr"+(-1,-3)*{\scriptstyle j};
  (0,-10)*{}="b"+(0,4)*{\scriptstyle q};
     "tr"; "c1" **\dir{-};
     "tl"; "c1" **\dir{-};
  (-5,-10)*\xycircle(3,3){-}="c";
  (-5,-10)*{m};
  (-2,0)*{}="tl";
  (-18,10)*{}="tr"+(-1,-3)*{\scriptstyle i};
  (-5,-25)*{}="b"+(-1,5)*{\scriptstyle l};
     "tr"; "c" **\dir{-};
     "c"; "c1" **\dir{-};
     "c"; "b" **\dir{-};
 \endxy
\]
In this notation we sum over the indices labelling `internal wires'---by
which we mean wires that are the output of one box and an input
of another.  This is just the Einstein summation convention in disguise:
so the above picture is merely an artistic way of drawing this:
\[
    m^{ij}_{p} m^{pk}_{l}= m^{iq}_{l} m^{jk}_{q}.
\]
But it has an enormous advantage: {\it no ambiguity is introduced if
we leave out the indices}, since the wires tell us how the
tensors are hooked together:
\[
 \xy
  (0,0)*\xycircle(3,3){-}="c1";
  (0,0)*{m};
  (-6,10)*{}="tl";
  (6,10)*{}="tr";
  (0,-10)*{}="b";
     "tr"; "c1" **\dir{-};
     "tl"; "c1" **\dir{-};
  (5,-10)*\xycircle(3,3){-}="c";
  (5,-10)*{m};
  (2,0)*{}="tl";
  (18,10)*{}="tr";
  (5,-25)*{}="b";
     "tr"; "c" **\dir{-};
     "c"; "c1" **\dir{-};
     "c"; "b" **\dir{-};
 \endxy
\qquad = \qquad
 \xy
  (0,0)*\xycircle(3,3){-}="c1";
  (0,0)*{m};
  (6,10)*{}="tl";
  (-6,10)*{}="tr";
  (0,-10)*{}="b";
     "tr"; "c1" **\dir{-};
     "tl"; "c1" **\dir{-};
  (-5,-10)*\xycircle(3,3){-}="c";
  (-5,-10)*{m};
  (-2,0)*{}="tl";
  (-18,10)*{}="tr";
  (-5,-25)*{}="b";
     "tr"; "c" **\dir{-};
     "c"; "c1" **\dir{-};
     "c"; "b" **\dir{-};
 \endxy
\]
This is a more vivid way of writing the mathematician's equation
\[
m(m \tensor 1_V) = m(1_V \tensor m)
\]
because tensor products are written horizontally and composition
vertically, instead of trying to compress them into a single line of text.

In modern language, what Penrose had noticed here was that
$\Fin\Vect$, the category of finite-dimensional vector spaces and
linear maps, is a symmetric monoidal category, so we can draw
morphisms in it using string diagrams.  But he probably wasn't
thinking about categories: he was probably more influenced by the
analogy to Feynman diagrams.

Indeed, Penrose's pictures are very much like Feynman diagrams, but
simpler.  Feynman diagrams are pictures of morphisms in the symmetric
monoidal category of positive-energy representations of the Poincar\'e
group!  It is amusing that this complicated example was considered
long before $\Vect$.  But that is how it often works: simple ideas
rise to consciousness only when difficult problems make them
necessary.

Penrose also considered some examples more complicated than $\Fin\Vect$
but simpler than full-fledged Feynman diagrams.  For any compact Lie
group $K$, there is a symmetric monoidal category $\Rep(K)$.  Here the
objects are finite-dimensional continuous unitary representations of
$K$---that's a bit of a mouthful, so we will just call them
`representations'.  The morphisms are {\bf intertwining operators}
between representations: that is, operators $f \maps H \to H'$ with
\[          f(\rho(g)\psi) = \rho'(g) f(\psi)  \]
for all $g \in K$ and $\psi \in H$, where $\rho(g)$ is the
unitary operator by which $g$ acts on $H$, and $\rho'(g)$ is
the one by which $g$ acts on $H'$.  The category $\Rep(K)$ becomes
symmetric monoidal with the usual tensor product of
group representations:
\[        (\rho \tensor \rho')(g) = \rho(g) \tensor \rho(g')  \]
and the obvious braiding.

As a category, $\Rep(K)$ is easy to describe.
Every object is a direct sum of finitely many {\bf
irreducible} representations: that is, representations that are not
themselves a direct sum in a nontrivial way.  So, if we pick a
collection $E_i$ of irreducible representations, one from each
isomorphism class, we can write any object $H$ as
\[        H \cong \bigoplus_i H^i \otimes E_i     \]
where the $H^i$ is the finite-dimensional Hilbert space describing
the multiplicity with which the irreducible $E_i$ appears in $H$:
\[         H^i = \hom(E_i, H)  \]
Then, we use Schur's Lemma, which describes the morphisms between
irreducible representations:
\begin{itemize}
\item
When $i = j$, the space $\hom(E_i,E_j)$ is
1-dimensional: all morphisms from $E_i$ to $E_j$
are multiples of the identity.
\item
When $i \ne j$, the space $\hom(E_i,E_j)$ is 0-dimensional:
all morphisms from $E$ to $E'$ are zero.
\end{itemize}
\noindent
So, every representation is a direct sum of irreducibles,
and every morphism between irreducibles is a multiple of the
identity (possibly zero).  Since composition is linear in each argument,
this means there's only one way composition of morphisms can
possibly work.  So, the category is completely pinned down as
soon as we know the set of irreducible representations.

One nice thing about $\Rep(K)$ is that every object has a dual.
If $H$ is some representation, the dual vector space $H^\ast$
also becomes a representation, with
\[            (\rho^\ast(g)f)(\psi) = f(\rho(g)\psi)  \]
for all $f \in H^\ast$, $\psi \in H$.  In our string diagrams, we can
use little arrows to distinguish between $H$ and $H^\ast$: a
downwards-pointing arrow labelled by $H$ stands for the object $H$,
while an upwards-pointing one stands for $H^\ast$.  For example, this:
\[
 \xy
  (0,10)*{}="1";
  (0,-10)*{}="2";
  "1";"2" **\dir{-}?(.45)*\dir{<}+(3.5,0)*{H};
 \endxy
\]
is the string diagram for the identity morphism $1_{H^\ast}$. This
notation is meant to remind us of Feynman's idea of antiparticles
as particles going backwards in time.

The dual pairing
\[
   \begin{array}{rcl}
             e_H \maps H^\ast \tensor H &\to& \C  \\
                      f \tensor v &\mapsto& f(v)
\end{array}
\]
is an intertwining operator, as is the operator
\[   \begin{array}{rcl}
            i_H \maps \C &\to& H \tensor H^\ast  \\
                      c &\mapsto& c\; 1_H
\end{array}
\]
where we think of $1_H \in \hom(H,H)$ as 
an element of $H \tensor H^\ast$.
We can draw these operators as a `cup':
\[
\xy
 (-6,6)*{};(6,6)*{};
   **\crv{(6,-8) & (-6,-8)}
    ?(.20)*\dir{>}
    ?(.89)*\dir{>};
   (1,-6)*{      };
   (-9,1)*{H}; (9,1)*{H};
\endxy
\qquad {\rm stands \; for}\qquad
 \xy
 (0,6)*+{H^{*} \ten H}="1";
 (0,-6)*+{\C}="2";
   {\ar_{\scriptstyle e_H} "1";"2"};
\endxy
\]
and a `cap':
\[
\xy
 (-6,-6)*{};
 (6,-6)*{};
   **\crv{(6,8) & (-6,8)} ?(.16)*\dir{>} ?(.9)*\dir{>};
 (1,7)*{    };
 (-9,-1)*{H}; (9,-1)*{H};
\endxy
\qquad {\rm stands \; for} \qquad
 \xy
 (0,6)*+{\C}="1";
 (0,-6)*+{H \ten H^*}="2";
   {\ar_{\scriptstyle i_H} "1";"2"};
\endxy
\]
Note that if no edges reach the bottom (or top) of a diagram,
it describes a morphism to (or from) the trivial representation of
$G$ on $\C$---since this is the tensor product of {\it no}
representations.

The cup and cap satisfy the {\bf zig-zag identities}:
\[ \stringZIGZAGi
\]
\[
\stringZIGZAGii
\]

\noindent
These identities are easy to check.  For example, the first
zig-zag gives a morphism from $H$ to $H$ which we can compute by
feeding in a vector $\psi \in H$:
\[
\PLUGzigzag
\]
So indeed, this is the identity morphism. But, the beauty of these
identities is that they let us straighten out a portion of a
string diagram as if it were actually a piece of string!  Algebra
is becoming topology.

Furthermore, we have:
\[
 \stringDIM \; \; = \dim(H)
\]
This requires a little explanation.
A `closed' diagram---one with no edges coming in and no edges
coming out---denotes an intertwining operator from the trivial
representation to itself.  Such a thing is just multiplication by
some number.  The equation above says the operator on the left is
multiplication by $\dim(H)$.  We can check this as follows:
\[
\DIMofH
\]
So, {\it a loop gives a dimension}.  This explains a big
problem that plagues Feynman diagrams in quantum field theory---namely,
the `divergences' or `infinities' that show up in diagrams containing
loops, like this:
\[
\feynmanII
\]
or more subtly, like this:
\[
\feynmanI
\]
These infinities come from the fact that most
positive-energy representations of the Poincar\'e group are
infinite-dimensional.  The reason is that this group is
noncompact.  For a compact Lie group, all the irreducible
continuous representations are finite-dimensional.

So far we have been discussing representations of compact Lie groups
quite generally.  In his theory of `spin networks' \cite{Penrose2,
Penrose3}, Penrose worked out all the details for $\SU(2)$: the group
of $2 \times 2$ unitary complex matrices with determinant 1.  This
group is important because it is the universal cover of the 3d
rotation group.  This lets us handle particles like the electron,
which doesn't come back to its original state after one full
turn---but does after two!

The group $\SU(2)$ is the subgroup of the Poincar\'e group
whose corresponding observables are the components of angular momentum.
Unlike the Poincar\'e group, it is compact.  As already mentioned,
we can specify an irreducible positive-energy representation of
the Poincar\'e group by choosing a mass $m \ge 0$, a spin
$j = 0, \frac{1}{2}, 1, \frac{3}{2}, \dots$
and sometimes a little extra data.  Irreducible unitary representations
of $\SU(2)$ are simpler: for these, we just need to choose a spin.
The group $\SU(2)$ has one irreducible unitary representation of each
dimension.  Physicists call the representation of dimension $2j+1$
the `spin-$j$' representation, or simply `$j$' for short.

Every representation of $\SU(2)$ is isomorphic to its dual, so
we can pick an isomorphism
\[           \sharp \maps j \to j^\ast  \]
for each $j$.  Using this, we can stop writing little arrows on
our string diagrams.  For example, we get a new `cup'
\[
 \vcenter{\xy
 (-6,8)*{};(6,8)*{};
   **\crv{(6,-10) & (-6,-10)}
    ?(.20)*\dir{}
    ?(.89)*\dir{};
   (1,-6)*{      };
   (-9,4)*{j}; (9,4)*{j};
\endxy }
\qquad \qquad
 \vcenter{ \xy
 (0,16)*+{j \ten j}="1";
 (0,6)*+{j^{*} \ten j}="2";
 (0,-3)*+{\C}="3";
    {\ar_{\sharp \ten 1} "1";"2"};
    {\ar_{\scriptstyle e_j} "2";"3"};
\endxy }
\]
and similarly a new cap.  These satisfy an interesting relation:
\[ \xy
  (0,15)*{}="T";
  (0,-15)*{}="B";
  (0,7.5)*{}="T'";
  (0,-7.5)*{}="B'";
    "T";"T'" **\dir{-};
    "B";"B'" **\dir{-};
    (-4.5,0)*{}="MB";
    (-10.5,0)*{}="LB";
    "T'";"LB" **\crv{(-1.5,-6) & (-10.5,-6)}; \POS?(.25)*{\hole}="2z";
    "LB"; "2z" **\crv{(-12,9) & (-3,9)};
    "2z";"B'" **\crv{(0,-4.5)};
    (2,13)*{j};
    \endxy
    \quad = \quad
  (-1)^{2j} \;\;
  \xy
  (0,15)*{}="T";
  (0,-15)*{}="B";
  (0,7.5)*{}="T'";
  (0,-7.5)*{}="B'";
    "T";"B" **\dir{-};
    (2,13)*{j};
    \endxy
 \]
\noindent
Physically, this means that when we give a spin-$j$ particle
a full turn, its state transforms trivially when $j$ is an
integer:
\[         \psi \mapsto \psi  \]
but it picks up a sign when $j$ is an integer plus $\frac{1}{2}$:
\[         \psi \mapsto -\psi . \]
Particles of the former sort are called {\bf bosons}; those of
the latter sort are called {\bf fermions}.

The funny minus sign for fermions also shows up when we build a loop
with our new cup and cap:
\[ \xy
    (-5,4)*{}="x1";
    (5,4)*{}="x2";
    (-5,-4)*{}="y1";
    (5,-4)*{}="y2";
 "x1";"y1" **\dir{-}; 
 "x2";"y2" **\dir{-};
    "x1";"x2" **\crv{(-5,10) & (5,10)};
    "y1";"y2" **\crv{(-5,-10) & (5,-10)};
\endxy
    \quad = \quad
(-1)^{2j} \, (2j+1)
 \]
\noindent
We get, not the usual dimension of the spin-$j$ representation, but the
dimension times a sign depending on whether this representation is
bosonic or fermionic!  This is sometimes called the {\bf
superdimension}, since its full explanation involves what physicists
call `supersymmetry'.  Alas, we have no time to discuss this here: we
must hasten on to Penrose's theory of spin networks!

Spin networks are a nice notation for morphisms between tensor
products of irreducible representations of $\SU(2)$.
The key underlying fact is that:
\[     j \tensor k \; \iso \; |j - k| \; \oplus \; |j - k| + 1 \; \oplus
\cdots \; \oplus \; j + k   \]
Thus, the space of intertwining
operators $\hom(j\tensor k,l)$ has dimension 1 or 0 depending
on whether or not $l$ appears in this direct sum.
We say the triple $(j,k,l)$ is {\bf admissible} when this
space has dimension 1.  This happens
when the triangle inequalities are satisfied:
\[      |j - k| \le l \le j + k \]
and also $j + k + l \in \Z$.

For any admissible triple $(j,k,l)$
we can choose a nonzero intertwining operator from $j \tensor k$ to $l$,
which we draw as follows:
\[
\\
 \xy
  (0,0)*{\scriptscriptstyle \bullet}="c";
  (-8,8)*{}="tl"+(-2,-3)*{j};
  (8,8)*{}="tr"+(2,-3)*{k};
  (0,-10)*{}="b"+(-2,02)*{l};
     "tr"; "c" **\dir{-};
     "tl"; "c" **\dir{-};
     "c"; "b" **\dir{-};
\endxy
\\
\]
Using the fact that a closed diagram gives a number, we can
normalize these intertwining operators so that the `theta
network' takes a convenient value, say:
 \[
 \xy
  (-8,0)*{\scriptscriptstyle \bullet}="l";
  (8,0)*{\scriptscriptstyle \bullet}="r";
     "l"; "r" **\crv{(-8,14) & (8,14)};
     "l"; "r" **\crv{(-8,-14) & (8,-14)};
     "l"; "r" **\dir{-};
  (0,12)*{j}; (0,2)*{k}; (0,-13)*{l};
 \endxy
\qquad = \qquad 1
 \]
When the triple $(j,k,l)$ is not admissible, we define
\[
\\
 \xy
  (0,0)*{\scriptscriptstyle \bullet}="c";
  (-8,8)*{}="tl"+(-2,-3)*{j};
  (8,8)*{}="tr"+(2,-3)*{k};
  (0,-10)*{}="b"+(-2,02)*{l};
     "tr"; "c" **\dir{-};
     "tl"; "c" **\dir{-};
     "c"; "b" **\dir{-};
\endxy
\\
\]
to be the zero operator, so that
 \[
 \xy
  (-8,0)*{\scriptscriptstyle \bullet}="l";
  (8,0)*{\scriptscriptstyle \bullet}="r";
     "l"; "r" **\crv{(-8,14) & (8,14)};
     "l"; "r" **\crv{(-8,-14) & (8,-14)};
     "l"; "r" **\dir{-};
  (0,12)*{j}; (0,2)*{k}; (0,-13)*{\ell};
 \endxy
\qquad = \qquad 0
 \]

We can then build more complicated intertwining operators by composing
and tensoring the ones we have described so far.  For example, this
diagram shows an intertwining operator from the representation
$2 \tensor \frac{3}{2} \tensor 1$ to the representation
$\frac{5}{2} \tensor 2$:
\[
\\
 \xy
  (-4,8)*{\scriptscriptstyle \bullet}="x"+(-2,-4)*{\frac{5}{2}};
  (-2,0)*{\scriptscriptstyle \bullet}="y";
  (0,-8)*{\scriptscriptstyle \bullet}="z"+(1,4)*{\frac{7}{2}};
  (-8,15)*{}="tl"+(-1.5,-3)*{2};
  (0,15)*{}="tm"+(1.5,-3)*{\frac{3}{2}};
  (8,15)*{}="tr"+(1.5,-3)*{1};
  (-6,-15)*{}="bl"+(-2,2)*{\frac{5}{2}};
  (6,-15)*{}="br"+(2,2)*{2};
     "tl"; "x" **\dir{-};
     "tm"; "x" **\dir{-};
     "tr"; "y" **\dir{-};
     "x"; "y" **\dir{-};
     "y"; "z" **\dir{-};
     "z"; "bl" **\dir{-};
     "z"; "br" **\dir{-};
\endxy
\\
\]
A diagram of this sort is called a `spin network'.  The resemblance to
a Feynman diagram is evident.  There is a category where the morphisms
are spin networks, and a functor from this category to $\Rep(\SU(2))$.
A spin network with no edges coming in from the top and no edges
coming out at the bottom is called {\bf closed}.  A closed spin
network determines an intertwining operator from the trivial
representation of $\SU(2)$ to itself, and thus a complex number.  For
more details, see the paper by Major \cite{Major}.

Penrose noted that spin networks satisfy a bunch of interesting rules.
For example, we can deform a spin network in various ways without
changing the operator it describes.  We have already seen the zig-zag
identity, which is an example of this.  Other rules involve changing
the topology of the spin network.  The most important of these is the
{\bf binor identity} for the spin-$\frac{1}{2}$ representation:
\[
\binorID
\]
We can use this to prove something we have already
seen:
\[
\useBINOR
\]
\noindent
Physically, this says that turning a spin-$\frac{1}{2}$
particle around 360 degrees multiplies its state by $-1$.

There are also interesting rules involving the spin-1
representation, which imply some highly nonobvious results. For
example, every trivalent planar graph with no edge-loops and all
edges labelled by the spin-1 representation:
\[
 \spinNETII
\]

\noindent
evaluates to a nonzero number \cite{Penrose4}.  But, Penrose
showed this fact is equivalent to the four-color theorem!

By now, Penrose's diagrammatic approach to the finite-dimensional
representations of $\SU(2)$ has been generalized to many compact
simple Lie groups.  A good treatment of this material is the free
online book by Cvitanovi\'c \cite{Cvitanovic}.  His book includes a
brief history of diagrammatic methods that makes a nice companion to
the present paper.  Much of the work in his book was done in the
1970's.  However, the huge burst of work on diagrammatic methods for
algebra came later, in the 1980's, with the advent of `quantum
groups'.

\subsection*{Ponzano--Regge (1968)}

Sometimes history turns around and goes back in time, like
an antiparticle.  This seems like the only sensible explanation
of the revolutionary work of Ponzano and Regge \cite{PR}, who
applied Penrose's theory of spin networks {\it before it was invented}
to relate tetrahedron-shaped spin networks to gravity in 3 dimensional
spacetime.   Their work eventually led to a theory called the
Ponzano--Regge model, which allows for an exact solution of many
problems in 3d quantum gravity \cite{Carlip}.

In fact, Ponzano and Regge's paper on this topic appeared in the
proceedings of a conference on spectroscopy, because the $6j$
symbol is important in chemistry.  But for our purposes, the
$6j$ symbol is just the number obtained by evaluating this spin network:
\[
\SIXjIII
\]
depending on six spins $i,j,k,l,p,q$.

In the Ponzano--Regge model of $3d$ quantum gravity, spacetime is made
of tetrahedra, and we label the edges of tetrahedra with spins to
specify their {\it lengths}.  To compute the amplitude for spacetime
to have a particular shape, we multiply a bunch of amplitudes (that
is, complex numbers): one for each tetrahedron, one for each triangle,
and one for each edge.  The most interesting ingredient in this recipe
is the amplitude for a tetrahedron.  This is given by the $6j$ symbol.

But, we have to be a bit careful!  Starting from a tetrahedron
whose edge lengths are given by spins:
\[
\SIXjI
\]
\vskip 1em \noindent
we compute its amplitude using the `Poincar\'e dual' spin
network, which has:
\begin{itemize}
\item
one vertex at the center of each face of the original tetrahedron;
\item
one edge crossing each edge of the original tetrahedron.
\end{itemize}
It looks like this:
\[
 \SIXjII
\]
\vskip 1em \noindent
Its edges inherit spin labels from the edges of the
original tetrahedron:
\[
\SIXjIII
\]
\noindent
{\em Voil\`a!}  The $6j$ symbol!

It is easy to get confused, since the Poincar\'e dual of a tetrahedron
just happens to be another tetrahedron.  But, there are good reasons
for this dualization process.  For example, the $6j$ symbol vanishes if the
spins labelling three edges meeting at a vertex violate the triangle
inequalities, because then these spins will be `inadmissible'.  For
example, we need
\[   |i - j| \le p \le i+j    \]
or the intertwining operator
\[
 \xy
  (0,0)*{\scriptscriptstyle \bullet}="c";
  (-8,8)*{}="tl"+(-2,-3)*{i};
  (8,8)*{}="tr"+(2,-3)*{j};
  (0,-10)*{}="b"+(-2,02)*{p};
     "tr"; "c" **\dir{-};
     "tl"; "c" **\dir{-};
     "c"; "b" **\dir{-};
 \endxy
 \]
will vanish, forcing the $6j$ symbols to vanish as well.
But in the original tetrahedron, these spins label the three
sides of a triangle:
\[
 \xy
  (-8,5)*{i};
  (8,5)*{j};
  (0,-8)*{p};
  (0,10)*{}="t";
 (-10,-6)*{}="bl";
 (10,-6)*{}="br";
   {\ar@{-} "bl";"br" };
   {\ar@{-} "bl";"t" };
   {\ar@{-} "t";"br" };
 \endxy
\]

\noindent
So, {\it the amplitude for a tetrahedron vanishes
if it contains a triangle that violates the triangle inequalities!}

This is exciting because it suggests that the representations of
$\SU(2)$ somehow know about the geometry of tetrahedra.   Indeed,
there are other ways for a tetrahedron to be `impossible' besides having
edge lengths that violate the triangle inequalities.  The $6j$ symbol
does not vanish for all these tetrahedra, but it is exponentially
damped---very much as a particle in quantum mechanics can tunnel through
barriers that would be impenetrable classically, but with an amplitude that
decays exponentially with the width of the barrier.

In fact the relation between $\Rep(\SU(2))$ and 3-dimensional geometry
goes much deeper.  Regge and Ponzano found an excellent asymptotic
formula for the $6j$ symbol that depends entirely on geometrically
interesting aspects of the corresponding tetrahedron: its volume, the
dihedral angles of its edges, and so on.  But, what is truly amazing
is that this asymptotic formula also matches what one would want from
a theory of quantum gravity in 3 dimensional spacetime!

More precisely, the Ponzano--Regge model is a theory of `Riemannian'
quantum gravity in 3 dimensions.  Gravity in our universe is described
with a Lorentzian metric on 4-dimensional spacetime, where each
tangent space has the Lorentz group acting on it.  But, we can imagine
gravity in a universe where spacetime is 3-dimensional and the metric
is Riemannian, so each tangent space has the rotation group $\SO(3)$
acting on it.  The quantum description of gravity in this universe
should involve the double cover of this group, $\SU(2)$ ---
essentially because it should describe not just how particles of
integer spin transform as they move along paths, but also particles of
half-integer spin.  And it seems the Ponzano--Regge model is the right
theory to do this.

A rigorous proof of Ponzano and Regge's asymptotic formula was
given only in 1999, by Justin Roberts \cite{JustinRoberts}.
Physicists are still finding wonderful surprises in the
Ponzano--Regge model.  For example, if we study it on a 3-manifold
with a Feynman diagram removed, with edges labelled by suitable
representations, it describes not only `pure' quantum gravity but
also \textit{matter!}  The series of papers by Freidel and Louapre
explain this in detail \cite{FL}.

Besides its meaning for geometry and physics, the $6j$ symbol
also has a purely category-theoretic significance: it is a concrete
description of the associator in $\Rep(\SU(2))$.  The associator
gives a linear operator
\[  a_{i,j,k} \maps (i \tensor j) \tensor k \to i \tensor (j \tensor k). \]
The $6j$ symbol is a way of expressing this operator as a bunch of numbers.
The idea is to use our basic intertwining operators to construct
operators
\[     S \maps (i \tensor j) \tensor k \to l , \qquad \qquad
       T \maps l \to i \tensor (j \tensor k)  ,\]
namely:
\[
S \qquad = \qquad
 \xy
  (0,7.5)*{\scriptscriptstyle \bullet}="c1";
  (-6,17.5)*{}="tl"+(-1,-3)*{ i};
  (6,17.5)*{}="tr"+(1,-3)*{ j};
  (0,7.5)*{}="b"+(0,-4)*{ p};
     "tr"; "c1" **\dir{-};
     "tl"; "c1" **\dir{-};
  (5,-2.5)*{\scriptscriptstyle \bullet}="c";
  (2,7.5)*{}="tl";
  (18,17.5)*{}="tr"+(1,-3)*{ k};
  (5,-17.5)*{}="b"+(-1,5)*{ l};
     "tr"; "c" **\dir{-};
     "c"; "c1" **\dir{-};
     "c"; "b" **\dir{-};
 \endxy
\qquad \qquad
T \qquad = \qquad
 \xy
  (0,-7.5)*{\scriptscriptstyle \bullet}="c1";
  (6,-17.5)*{}="tl"+(1,3)*{ k};
  (-6,-17.5)*{}="tr"+(-1,3)*{ j};
  (0,0)*{}="b"+(0,-4)*{ q};
     "tr"; "c1" **\dir{-};
     "tl"; "c1" **\dir{-};
  (-5,2.5)*{\scriptscriptstyle \bullet}="c";
  (-2,-7.5)*{}="tl";
  (-18,-17.5)*{}="tr"+(-1,3)*{ i};
  (-5,17.5)*{}="b"+(-1,-5)*{ l};
     "tr"; "c" **\dir{-};
     "c"; "c1" **\dir{-};
     "c"; "b" **\dir{-};
 \endxy
\\
\]
Using the associator to bridge the gap between
$(i \tensor j) \tensor k$ and $i \tensor (j \tensor k)$,
we can compose $S$ and $T$ and take the trace of the resulting
operator, obtaining a number.  These numbers encode
everything there is to know about the associator in the monoidal
category $\Rep(\SU(2))$.  Moreover, these numbers are just the $6j$
symbols:
\vskip 1em
\[
\tr(T a_{i,j,k} S) \qquad = \qquad
\begin{pspicture}[.5](2,2)
\rput(1,1.5){\SIXjIII}
\end{pspicture}
\]
\vskip 1em \noindent
This can be proved by gluing the pictures for $S$ and $T$
together and warping the resulting spin network until it looks
like a tetrahedron!  We leave this as an exercise for the reader.

The upshot is a remarkable and mysterious fact: the associator in the
monoidal category of representations of $\SU(2)$ encodes information
about 3-dimensional quantum gravity!  This fact will become less
mysterious when we see that 3-dimensional quantum gravity is almost a
topological quantum field theory, or TQFT.  In our discussion of
Barrett and Westbury's 1992 paper on TQFTs, we will see that a large
class of 3d TQFTs can be built from monoidal categories.

\subsection*{Grothendieck (1983)}

In his 600--page letter entitled \textit{Pursuing Stacks},
Grothendieck fantasized about $n$-categories for higher $n$---even $n
= \infty$---and their relation to homotopy theory \cite{Gro}.  The
rough idea of an $\infty$-category is that it should be a
generalization of a category which has objects, morphisms, 2-morphisms
and so on forever.  In the fully general, `weak' $\infty$-categories, all the
laws governing composition of $j$-morphisms should hold only up to a
specified $(j+1)$-morphisms, which in turn satisfy laws of their own,
but only up to specified $(j+2)$-morphisms, and so on.  Furthermore,
all these higher morphisms which play the role of `laws' should be
equivalences---where a $k$-morphism is an `equivalence' if it is
invertible {\it up to equivalence}.  The circularity here is not
necessarily vicious, but it hints at how tricky $\infty$-categories
can be.

Grothendieck believed that among the weak $\infty$-categories there should
be a special class, the `weak $\infty$-groupoids', in which all
$j$-morphisms ($j \ge 1$) are equivalences.  He also believed that
every space $X$ should have a weak $\infty$-groupoid $\Pi_\infty(X)$
called its `fundamental $\infty$-groupoid', in which:
\begin{itemize}
\item the objects are points of $X$,
\item the morphisms are paths in $X$,
\item the 2-morphisms are paths of paths in $X$,
\item the 3-morphisms are paths of paths of paths in $X$,
\item {\it etc.}
\end{itemize}
Moreover, $\Pi_\infty(X)$ should be a complete invariant of the
homotopy type of $X$, at least for nice spaces like CW complexes.  In
other words, two nice spaces should have equivalent fundamental
$\infty$-groupoids if and only if they are homotopy equivalent.

The above sketch of Grothendieck's dream is phrased in terms of a
`globular' approach to $n$-categories, where the $n$-morphisms are
modeled after $n$-dimensional discs:
\vskip 0.2em
\[
\begin{tabular}{|c|c|c|c|c|}
  \hline
   \textbf{objects} & \textbf{morphisms} & \textbf{2-morphisms} & \textbf{3-morphisms} &
   $\cdots$ \\
  \hline \hline
   $\bullet$ & $\xy
  (-8,0)*+{\bullet}="1";
  (0,10)*+{}; 
  (0,-10)*+{}; 
  (8,0)*+{\bullet}="2";
  {\ar "1";"2"};
 \endxy$ & $\xy (-8,0)*+{\bullet}="4"; (8,0)*+{\bullet}="6";
  {\ar@/^1.65pc/^{}"4";"6"};
  {\ar@/_1.65pc/_{} "4";"6"};
{\ar@{=>}(0,3)*{};(0,-3)*{}} ;
\endxy$ & $\xy 0;/r.22pc/:
  (-20,0)*+{\bullet}="1";
  (0,0)*+{\bullet}="2";
  {\ar@/^2pc/ "1";"2"};
  {\ar@/_2pc/ "1";"2"};
   (-10,6)*+{}="A";
   (-10,-6)*+{}="B";
  {\ar@{=>}@/_.7pc/ "A"+(-2,0) ; "B"+(-1,-.8)};
  {\ar@{=}@/_.7pc/ "A"+(-2,0)  ; "B"+(-2,0)};
  {\ar@{=>}@/^.7pc/ "A"+(2,0)  ; "B"+(1,-.8)};
  {\ar@{=}@/^.7pc/ "A"+(2,0)  ; "B"+(2,0)};
  {\ar@3{->} (-12,0)*{}; (-8,0)*{}};
 \endxy$ & Globes  \\
  \hline
\end{tabular}
\]

\noindent
However, he also imagined other approaches based on $j$-morphisms
with different shapes, such as simplices:
\vskip 0.2em
\[
\begin{tabular}{|c|c|c|c|c|}
  \hline
   \textbf{objects} & \textbf{morphisms} & \textbf{2-morphisms} & \textbf{3-morphisms} &
   $\cdots$ \\
  \hline \hline
   $\bullet$ & $\xy
  (-8,0)*+{\bullet}="1";
  (0,10)*+{}; 
  (0,-10)*+{}; 
  (8,0)*+{\bullet}="2";
  {\ar "1";"2"};
 \endxy$ & $\vcenter{\xy 0;/r.17pc/:
   (-10,-5)*+{ \scriptstyle \bullet}="1";
   (10,-5)*+{ \scriptstyle \bullet}="2";
   (0,12)*+{ \scriptstyle \bullet}="3";
    {\ar "1"; "2"};
    {\ar "3"; "2"};
    {\ar "1"; "3"};
    \endxy}$ & $\xy 0;/r.17pc/:
    (-10,-5 )*+{ \scriptstyle \bullet}="1";
    (8,-10)*+{ \scriptstyle \bullet}="2";
    (15,0)*+{ \scriptstyle \bullet}="3";
   (1,12)*+{ \scriptstyle \bullet}="4";
       {\ar "1";"2" };
       {\ar "2";"3" };
       {\ar "4";"3" };
    {\ar "1";"4" };
    {\ar "4";"2" };
       {\ar @{.>}|>>>>>>{\hole \hole} "1";"3"};
       \endxy$ & Simplices  \\
  \hline
\end{tabular}
\]

\noindent
In fact, simplicial weak $\infty$-groupoids had already been developed in a
1957 paper by Kan \cite{Kan}; these are now called `Kan complexes'.
In this framework $\Pi_\infty(X)$ is indeed a complete invariant of
the homotopy type of any nice space $X$.  So, the real challenge is to
define {\it weak $\infty$-categories} in the simplicial and other
approaches, and then define weak $\infty$-groupoids as special cases of
these, and prove their relation to homotopy theory.

Great progress towards fulfilling Grothendieck's dream has been made
in recent years.  We cannot possibly do justice to the enormous body
of work involved, so we simply offer a quick thumbnail sketch.
Starting around 1977, Street began developing a simplicial approach to
$\infty$-categories \cite{Street1,Street2} based on ideas from the
physicist Roberts \cite{JohnRoberts}.  Thanks in large part to the
recently published work of Verity, this approach has begun to
really take off \cite{Verity1,Verity2,Verity3}.

In 1995, Baez and Dolan initiated another approach to weak
$n$-categories, the `opetopic' approach \cite{hda3}:

\vskip 1em
\opetopicCHART
\vskip 1em

\noindent
The idea here is that an $(n+1)$-dimensional opetope describes a way
of gluing together $n$-dimensional opetopes.  The opetopic approach
was corrected and clarified by various authors \cite{Leinster1, HMP,
Ch1, Ch2, Ch3, Ch4}, and by now it has been developed by Makkai
\cite{Makkai} into a full-fledged foundation for mathematics.  We have
already mentioned how in category theory it is considered a mistake to
assert equations between objects: instead, one should specify an
isomorphism between them.  Similarly, in $n$-category theory it is a
mistake to assert an equation between $j$-morphisms for any $j < n$:
one should instead specify an equivalence.  In Makkai's approach to
the foundations of mathematics based on weak $\infty$-categories,
\textit{equality plays no role, so this mistake is impossible to
make}.  Instead of stating equations one must always specify
equivalences.

Also starting around 1995, Tamsamani \cite{Tam1} and Simpson
\cite{Simpson1,Simpson2,Simpson3,Simpson4} developed a
`multisimplicial' approach to weak $n$-categories.  In a 1998 paper,
Batanin \cite{Batanin,Street4} initiated a globular approach to weak
$\infty$-categories.  Penon \cite{Penon} gave a related, very compact
definition of $\infty$-category, which was later improved by Batanin,
Cheng and Makkai \cite{Batanin4,ChMakkai}.  There is also a
topologically motivated approach using operads due to Trimble
\cite{Trimble}, which was studied and generalized by Cheng and Gurski
\cite{Ch5,ChGur}.  Yet another theory is due to Joyal, with
contributions by Berger \cite{Joyal,Berger1}.

This great diversity of approaches raises the question of when two
definitions of $n$-category count as `equivalent'.  In
\textsl{Pursuing Stacks}, Grothendieck proposed the following answer.
Suppose that for all $n$ we have two different definitions of weak
$n$-category, say `$n$-${\rm category}_1$' and `$n$-${\rm
category}_2$'.  Then we should try to construct the $(n+1)$-${\rm
category}_1$ of all $n$-${\rm categories}_1$ and the $(n+1)$-${\rm
category}_1$ of all $n$-${\rm categories}_2$ and see if these are
equivalent as objects of the $(n+2)$-${\rm category}_1$ of all
$(n+1)$-${\rm categories}_1$.  If so, we may say the two definitions
are equivalent as seen from the viewpoint of the first definition.

Of course, this strategy for comparing definitions of weak $n$-category
requires a lot of work.  Nobody has carried it out for any pair of
significantly different definitions.  There is also some freedom of
choice involved in constructing the two $(n+1)$-${\rm categories}_1$
in question.  One should do it in a `reasonable' way, but what does
that mean?  And what if we get a different answer when we reverse the
roles of the two definitions?

A somewhat less strenuous strategy for comparing definitions is
suggested by homotopy theory.  Many different approaches to homotopy
theory are in use, and though superficially very different, there is
by now a well-understood sense in which they are fundamentally the
same.  Different approaches use objects from different categories to
represent topological spaces, or more precisely, the
homotopy-invariant information in topological spaces, called their
`homotopy types'.  These categories are not equivalent, but each one
is equipped with a class of morphisms called `weak equivalences',
which play the role of homotopy equivalences.  Given a category $C$
equipped with a specified class of weak equivalences, under mild
assumptions one can throw in inverses for these morphisms and obtain a
category called the `homotopy category' $\mathrm{Ho}(C)$.  Two
categories with specified equivalences may be considered the same for
the purposes of homotopy theory if their homotopy categories are
equivalent in the usual sense of category theory.  The same strategy
---or more sophisticated variants---can be applied to comparing
definitions of $n$-category, so long as one can construct a
\textsl{category} of $n$-categories.

Starting around 2000, work began on comparing different approaches to
$n$-category theory \cite{Ch3,Ch5,Berger1,Makkai2,Simpson5}.  There
has also been significant progress towards achieving Grothendieck's
dream of relating weak $n$-groupoids to homotopy theory \cite{Tam2,
Batanin2, Batanin3, Berger2, Cisinski, Pa1, Pa2}.  But $n$-category
theory is still far from mature.  This is \textsl{one} reason the
present paper is just a `prehistory'.

Luckily, Leinster has written a survey of definitions of $n$-category
\cite{Leinster.survey}, and also a textbook on the role of operads and
their generalizations in higher category theory \cite{Leinster.book}.
Cheng and Lauda have prepared an `illustrated guidebook' of higher
categories, for those who like to visualize things \cite{ChengLauda}.
The forthcoming book by Baez and May \cite{BaezMay} provides more
background for readers who want to learn the subject.  And for
applications to algebra, geometry and physics, try the conference
proceedings edited by Getzler and Kapranov \cite{GK} and by Davydov
\textit{et al} \cite{Davydov}.

\subsection*{String theory (1980's)}

In the 1980's there was a huge outburst of work on string theory.
There is no way to summarize it all here, so we shall content
ourselves with a few remarks about its relation to $n$-categorical
physics.  For a general overview the reader can start with the
introductory text by Zweibach \cite{Zweibach}, and then turn
to the book by Green, Schwarz and Witten \cite{GSW}, which was
written in the 1980s, or the book by Polchinski \cite{Polchinski},
which covers more recent developments.

String theory goes beyond ordinary quantum field theory by replacing
0-dimensional point particles by 1-dimensional objects: either
circles, called `closed strings', or intervals, called `open strings'.
So, in string theory, the essentially 1-dimensional Feynman diagrams
depicting worldlines of particles are replaced by 2-dimensional
diagrams depicting `string worldsheets':
\vskip 1em
\[
 {\xy 0;/r.20pc/:
 (-10,10)*{}="l";
 (10,10)*{}="r";
 (0,0)*{}="m";
 (0,-13)*{}="b";
  "l";"m" **\dir{-}?(.5)*\dir{>};
  "m";"r" **\dir{-}?(.5)*\dir{>};
  {\ar@{~} "m";"b"};
 \endxy}
 \qquad \mapsto \qquad
 \begin{pspicture}[.5](4,2.5)
  \rput(2,0){\multc}
\end{pspicture}
\]
\vskip 1em
\noindent
This is a hint that as we pass from ordinary quantum field theory
to string theory, the mathematics of \textit{categories} is replaced
by the mathematics of \textit{bicategories}.  However, this hint took
a while to be recognized.

To compute an operator from a Feynman diagram, only the topology of
the diagram matters, including the specification of which edges are
inputs and which are outputs.  In string theory we need to equip the
string worldsheet with a conformal structure, which is a recipe for
measuring angles.  More precisely: a \textbf{conformal structure} on a
surface is an orientation together with an equivalence class of
Riemannian metrics, where two metrics counts as equivalent if they
give the same answers whenever we use them to compute angles between
tangent vectors.

A conformal structure is also precisely what we need to do
\textit{complex analysis} on a surface.  The power of complex analysis
is what makes string theory so much more tractable than theories of
higher-dimensional membranes.

\subsection*{Joyal--Street (1985)}

Around 1985, Joyal and Street introduced braided monoidal categories
\cite{JS1}.   The story is nicely told in Street's \textsl{Conspectus}
\cite{Street3}, so here we focus on the mathematics.

As we have seen, braided monoidal categories are just like Mac Lane's
symmetric monoidal categories, but without the law
\[     B_{x,y} = B_{y,x}^{-1}  .\]
The point of dropping this law becomes clear if we draw
the isomorphism $B_{x,y} \maps x \tensor y \to y \tensor x$ as a little
braid:
\[ \xy
 \vtwist~{(-5,8)}{(5,8)}{(-5,-8)}{(5,-8)}<>|>>><;
 (-7,5)*{\scriptstyle x};
 (7,5)*{\scriptstyle y};
\endxy \]
Then its inverse is naturally drawn as
\[ \xy
 \vtwistneg~{(-5,8)}{(5,8)}{(-5,-8)}{(5,-8)}<>|>>><;
 (-7,5)*{\scriptstyle y};
 (7,5)*{\scriptstyle x};
\endxy \]
since then the equation $B_{x,y} B_{x,y}^{-1} = 1$ makes topological
sense:
\[
\vcenter{\begin{xy}
 \vtwist~{(-5,5)}{(5,5)}{(-5,-5)}{(5,-5)}|>>><;
 \vtwistneg~{(-5,15)}{(5,15)}{(-5,5)}{(5,5)}<>|>>><;
    (-6,16.4)*{y};
    (6,16.4)*{x};
    (-6,-7)*{y};
    (6,-7)*{x};
\end{xy}}
\quad = \quad \vcenter{\begin{xy}
 (-5,-5);(-5,15) **\dir{-}?(.5)*\dir{<};
 (5,-5);(5,15)  **\dir{-}?(.5)*\dir{<};
    (-6,16.4)*{y};
    (6,16.4)*{x};
    (-6,-7)*{y};
    (6,-7)*{x};
\end{xy}}
\]
and similarly for $B_{x,y}^{-1} B_{x,y} = 1$:
\[
\vcenter{\begin{xy}
\vtwistneg~{(-5,5)}{(5,5)}{(-5,-5)}{(5,-5)}|>>><;
\vtwist~{(-5,15)}{(5,15)}{(-5,5)}{(5,5)}<>|>>><;
    (-6,16.4)*{x};
    (6,16.4)*{y};
    (-6,-7)*{x};
    (6,-7)*{y};
\end{xy}}
\quad = \quad \vcenter{\begin{xy}
 (-5,-5);(-5,15) **\dir{-}?(.5)*\dir{<};
 (5,-5);(5,15) **\dir{-}?(.5)*\dir{<};
    (-6,16.4)*{x};
    (6,16.4)*{y};
    (-6,-7)*{x};
    (6,-7)*{y};
\end{xy}}
\]
In fact, these equations are familiar in knot theory, where they
describe ways of changing a 2-dimensional picture of a knot (or braid,
or tangle) without changing it as a 3-dimensional topological entity.
Both these equations are called the {\bf second Reidemeister move}.

On the other hand, the law $B_{x,y} = B_{y,x}^{-1}$
would be drawn as
\[
\vcenter{ \xy
 \vtwist~{(-5,8)}{(5,8)}{(-5,-8)}{(5,-8)}<>|>>><;
 (-7,5)*{\scriptstyle x};
 (7,5)*{\scriptstyle y};
\endxy  }
\quad = \quad
\vcenter{ \xy
 \vtwistneg~{(-5,8)}{(5,8)}{(-5,-8)}{(5,-8)}<>|>>><;
 (-7,5)*{\scriptstyle x};
 (7,5)*{\scriptstyle y};
\endxy }
\]
and this is {\it not} a valid move in knot theory: in fact, using this
move all knots become trivial.  So, it make some sense to drop it, and
this is just what the definition of braided monoidal category does.

Joyal and Street constructed a very important braided monoidal
category called $\Braid$.  Every object in this category is a
tensor product of copies of a special object $x$, which we draw as
a point.  So, we draw the object $x^{\tensor n}$ as a row of
$n$ points.  The unit for the tensor product, $I = x^{\tensor 0}$,
is drawn as a blank space.  All the morphisms in $\Braid$ are
{\bf endomorphisms}: they go from an object to itself.  In particular,
a morphism $f \maps x^{\tensor n} \to x^{\tensor n}$ is an $n$-strand
braid:
\[ \xy 0;/r.16pc/:
 (-15,-20)*{}="b1";
 (-5,-20)*{}="b2";
 (5,-20)*{}="b3";
(14,-20)*{}="b4";
 (-14,20)*{}="T1";
 (-5,20)*{}="T2";
 (5,20)*{}="T3";
(15,20)*{}="T4";
 "b1"; "T4" **\crv{(-15,-7) & (15,-5)} \POS?(.25)*{\hole}="2x" \POS?(.47)*{\hole}="2y"
\POS?(.65)*{\hole}="2z" ?(.9)*\dir{<} ?(.1)*\dir{<};
 "b2";"2x" **\crv{(-5,-15)} ?(.5)*\dir{<};
 "b3";"2y" **\crv{(5,-10)} ?(.35)*\dir{<};
 "b4";"2z" **\crv{(14,-9)} ?(.3)*\dir{<};
 (-15,-5)*{}="3x";
 "2x"; "3x" **\crv{(-15,-10)};
 "3x"; "T3" **\crv{(-15,15) & (5,10)} \POS?(.38)*{\hole}="4y"
     \POS?(.65)*{\hole}="4z" ?(.1)*\dir{<};
     "T1";"4y" **\crv{(-14,16)} ?(.5)*\dir{>};
     "T2";"4z" **\crv{(-5,16)} ?(.5)*\dir{>};
     "2y";"4z" **\crv{(-10,3) & (10,2)} \POS?(.6)*{\hole}="5z";
     "4y";"5z" **\crv{(-5,5)}; "5z";"2z" **\crv{(5,4)};
\endxy \]
and composition is defined by stacking one braid on top of another.
We tensor morphisms in $\Braid$ by setting braids side by side.  The
braiding is defined in an obvious way: for example, the braiding
\[
B_{2,3} \colon x^{\tensor 2} \tensor x^{\tensor 3} \to x^{\tensor 3}
\tensor x^{\tensor 2}
\]
looks like this:
\[
 \twothreebraid
\]

Joyal and Street showed that $\Braid$ is the `free braided
monoidal category on one object'.  This and other results of
theirs justify the use of string diagrams as a technique
for doing calculations in braided monoidal categories.  They
published a paper on this in 1991, aptly titled \textsl{The Geometry
of Tensor Calculus} \cite{JS0}.

Let us explain more precisely what it means that $\Braid$ is the free
braided monoidal category on one object.  For starters, $\Braid$ is a
braided monoidal category containing a special object $x$: the point.
But when we say $\Braid$ is the \textit{free} braided monoidal
category on this object, we are saying much more.  Intuitively, this
means two things.  First, every object and morphism in $\Braid$ can be
built from $1$ using operations that are part of the definition of
`braided monoidal category'.  Second, every equation that holds in
$\Braid$ follows from the definition of `braided monoidal category'.

To make this precise, consider a simpler but related example.  The
group of integers $\Z$ is the free group on one element, namely the
number $1$.  Intuitively speaking this means that every integer can be
built from the integer $1$ using operations built into the definition
of `group', and every equation that holds in $\Z$ follows from the
definition of `group'.  For example, $(1 + 1) + 1 = 1 + (1 + 1)$
follows from the associative law.

To make these intuitions precise it is good to use the idea of a
`universal property'.  Namely: for any group $G$ containing an element
$g$ there exists a unique homomorphism
\[            \rho \maps \Z \to G  \]
such that
\[              \rho(1) = g .\]
The uniqueness clause here says that every integer is built from $1$
using the group operations: that is why knowing what $\rho$ does to $1$
determines $\rho$ uniquely.  The existence clause says that every equation
between integers follows from the definition of a group: if there
were extra equations, these would block the existence of homomorphisms
to groups where these equations failed to hold.

So, when we say that $\Braid$ is the `free' braided monoidal category
on the object $1$, we mean something \textit{roughly} like this: for
any braided monoidal category $C$, and any object $c \in C$, there
is a unique map of braided monoidal categories
\[           Z \maps \Braid \to C  \]
such that
\[           Z(x) = c  .\]

This will not be not precise until we define a map of braided monoidal
categories.  The correct concept is that of a `braided monoidal
functor'.  But we also need to weaken the universal property.  To say
that $Z$ is `unique' means that any two candidates sharing the desired
property are \textit{equal}.  But this is too strong: it is bad to
demand equality between functors.  Instead, we should say that any two
candidates are \textit{isomorphic}.  For this we need the concept of
`braided monoidal natural isomorphism'.

Once we have these concepts in hand, the correct theorem is as
follows.  For any braided monoidal category $C$, and any object $x
\in C$, there exists a braided monoidal functor
\[           Z \maps \Braid \to C  \]
such that
\[           Z(x) = c  .\]
Moreover, given two such braided monoidal functors, there is a braided
monoidal natural isomorphism between them.

Now we just need to define the necessary concepts.  The definitions
are a bit scary at first sight, but they illustrate the idea of
\textbf{weakening}: that is, replacing equations by isomorphisms which
satisfy equations of their own.  They will also be needed for the
definition of `topological quantum field theories', which we will
present in our discussion of Atiyah's 1988 paper.

To begin with, a functor $F \maps C \to D$ between monoidal categories
is {\bf monoidal} if it is equipped with:
\begin{itemize}
\item a natural isomorphism $\Phi_{x,y} \maps F(x) \tensor F(y)
\to F(x \tensor y)$, and
\item an isomorphism $\phi \maps 1_D \to F(1_C)$
\end{itemize}
such that:
\begin{itemize}
\item the following diagram commutes for any objects $x,y,z \in
C$:
\[
\begin{diagram} [F(x) \tensor (F(y) \tensor F(z))]
\node{(F(x) \tensor F(y)) \tensor F(z)} \arrow{e,t}{\Phi_{x,y}\,
\tensor 1_{F(z)}}  \arrow{s,l}{a_{F(x),F(y),F(z)}} \node{F(x
\tensor y) \tensor F(z)} \arrow{e,t}{\Phi_{x\tensor y,z}}
\node{F((x \tensor y)\tensor z)}
\arrow{s,r}{F(a_{x,y,z})}  \\
\node{F(x) \tensor (F(y) \tensor F(z))} \arrow{e,b}{1_{F(x)}
\tensor \Phi_{y,z}} \node{F(x) \tensor F(y \tensor z)}
\arrow{e,b}{\Phi_{x,y\tensor z}} \node{F(x \tensor (y \tensor z))}
\end{diagram}
\]
\item the following diagrams commute for any object $x \in C$:
\[
\begin{diagram}[F(1) \tensor F(x)]
\node{1 \tensor F(x)} \arrow{e,t}{\ell_{F(x)}} \arrow{s,l}{\phi
\tensor 1_{F(x)}}
\node{F(x)} \\
\node{F(1) \tensor F(x)} \arrow{e,b}{\Phi_{1,x}} \node{F(1 \tensor
x)} \arrow{n,r}{F(\ell_x)}
\end{diagram}
\]
\[
\begin{diagram}[F(1) \tensor F(x)]
\node{F(x) \tensor 1} \arrow{e,t}{r_{F(x)}}
\arrow{s,l}{1_{F(x)}\tensor \phi}
\node{F(x)} \\
\node{F(x) \tensor F(1)} \arrow{e,b}{\Phi_{x,1}} \node{F(x \tensor
1)} \arrow{n,r}{F(r_x)}
\end{diagram}
\]
\end{itemize}
Note that we do not require $F$ to preserve the tensor product
or unit `on the nose'.  Instead, it is enough that it preserve them
\textit{up to specified isomorphisms}, which must in turn satisfy some
plausible equations called `coherence laws'.  This is typical of weakening.

A functor $F \maps C \to D$ between braided monoidal categories
is \textbf{braided monoidal} if it is monoidal and it makes the
following diagram commute for all $x,y \in C$:
\[
\begin{diagram}[F(x) \tensor F(x) \tensor F(y)]
\node{F(x) \tensor F(y)\quad } \arrow{e,t}{B_{F(x),F(y)}}
\arrow{s,l}{\Phi_{x,y}} \node{\quad F(y) \tensor F(x)}
\arrow{s,r}{\Phi_{y,x}} \\
\node{F(x \tensor y)} \arrow{e,b}{F(B_{x,y})} \node{F(y \tensor
x)}
\end{diagram}
\]
This condition says that $F$ preserves the braiding as best it can,
given the fact that it only preserves tensor products up to a
specified isomorphism.  A {\bf symmetric monoidal functor} is just a
braided monoidal functor that happens to go between symmetric monoidal
categories.  No extra condition is involved here.

Having defined monoidal, braided monoidal and symmetric monoidal
functors, let us next do the same for natural transformations.  Recall
that a monoidal functor $F \maps C \to D$ is really a triple
$(F,\Phi,\phi)$ consisting of a functor $F$, a natural isomorphism
$\Phi_{x,y} \maps F(x) \tensor F(y) \to F(x \tensor y),$ and an
isomorphism $\phi \maps 1_{D} \to F(1_{C})$.  Suppose that
$(F,\Phi,\phi)$ and $(G,\Gamma,\gamma)$ are monoidal functors from the
monoidal category $C$ to the monoidal category $D$.  Then a natural
transformation $\alpha \maps F \To G$ is {\bf monoidal} if the
diagrams
\[
\begin{diagram}[F(x) \tensor F(y)]
\node{F(x) \tensor F(y)} \arrow{e,t}{\alpha_x \tensor \alpha_y}
\arrow{s,l}{\Phi_{x,y}}
\node{G(x) \tensor G(y)} \arrow{s,r}{\Gamma_{x,y}}
\\
\node{F(x \tensor y)} \arrow{e,b}{\alpha_{x \tensor y}}
\node{G(x \tensor y)}
\end{diagram}
\]
and
\[
\begin{diagram}[F(x) \tensor F(y)]
\node{1_D}
\arrow{s,l}{\phi}
\arrow{se,t}{\gamma}
\\
\node{F(1_C)} \arrow{e,b}{\alpha_{1_C}} \node{G(1_C)}
\end{diagram}
\]
commute.  There are no extra condition required of {\bf braided monoidal}
or {\bf symmetric monoidal} natural transformations.

The reader, having suffered through these definitions, is entitled to
see an application besides Joyal and Street's algebraic description of
the category of braids.  At the end of our discussion of Mac Lane's
1963 paper on monoidal categories, we said that in a certain sense
every monoidal category is equivalent to a strict one.  Now we can
make this precise!  Suppose $C$ is a monoidal category.  Then there is
a strict monoidal category $D$ that is {\bf monoidally equivalent} to
$C$.  That is: there are monoidal functors $F \maps C \to D$, $G \maps
D \to C$ and monoidal natural isomorphisms $\alpha \maps F G \To 1_D$,
$\beta \maps G F \To 1_C$.

This result allows us to work with strict monoidal categories, even
though most monoidal categories found in nature are not strict: we can
take the monoidal category we are studying and replace it by a
monoidally equivalent strict one.  The same sort of result is true for
braided monoidal and symmetric monoidal categories.

A very similar result holds for bicategories: they are all equivalent
to \textbf{strict 2-categories}: that is, bicategories where all the
associators and unitors are identity morphisms.  However, the pattern
breaks down when we get to tricategories: not every tricategory is
equivalent to a strict 3-category!  At this point the necessity for
weakening becomes clear.

\subsection*{Jones (1985)}

A \textbf{knot} is a circle smoothly embedded in $\R^3$:
\[
\xy 0;/r.15pc/:
(6,9)*{}="1";
(-8.5,-1)*{}="2";
"1";"2" **\crv{~*=<.5pt>{.} (0,30)}?(.75)*\dir{-}+(-2,2)*{};
(-6.5,8)*{}="1";
(-.5,-9)*{}="2";
"1";"2" **\crv{~*=<.5pt>{.} (-28.5,9.3)}?(.7)*\dir{-}+(-2,-2)*{};
(-9.5,-3.35)*{}="1";
(8.5,-3)*{}="2";
"1";"2" **\crv{~*=<.5pt>{.} (-17.67,-24.19)}?(.7)*\dir{-}+(-1,-3)*{};
(1,-10)*{}="1";
(6.5,7.13)*{}="2";
"1";"2" **\crv{~*=<.5pt>{.} (17.67,-24.19)}?(.7)*\dir{-}+(3,-1)*{};
(11,-1)*{}="1";
(-4,8)*{}="2";
"1";"2" **\crv{~*=<.5pt>{.} (28.5,9.3)}?(.93)*\dir{-}+(1,2)*{};
\endxy
\]
More generally, a \textbf{link} is a collection of disjoint knots.
In topology, we consider two links to be the same, or `isotopic',
if we can deform one smoothly without its strands crossing until
it looks like the other.  Classifying links up to isotopy is a
challenging task that has spawned many interesting theorems and
conjectures.  To prove these, topologists are always looking for
link invariants: that is, quantities they can compute from a link,
which are equal on isotopic links.

In 1985, Jones \cite{Jones} discovered a new link invariant,
now called the `Jones polynomial'.  To everyone's surprise he
defined this using some mathematics with no previously known connection to
knot theory: the operator algebras developed in the 1930s by Murray and
von Neumann \cite{vN} as part of a general formalism for quantum theory.
Shortly thereafter, the Jones polynomial was generalized by many authors
obtaining a large family of so-called `quantum invariants' of links.

Of all these new link invariants, the easiest to explain is the
`Kauffman bracket' \cite{Kauffman1987}.  The Kauffman bracket can
be thought of as a simplified version of the Jones polynomial.
It is also a natural development of Penrose's 1971 work on spin networks
\cite{Kauffman1998}.

As we have seen, Penrose gave a recipe for computing a number from any
spin network.  The case relevant here is a spin network with vertices
at all, with every edge labelled by the spin $\frac{1}{2}$.
For spin networks like this we can compute the number by repeatedly
using the binor identity:
\[
 \xy
 (-4,8)*{}="TL"; (4,8)*{}="TR";
 (-4,-8)*{}="BL"; (4,-8)*{}="BR";
    \vtwist~{"TL"}{"TR"}{"BL"}{"BR"};
 \endxy
\quad = \quad
 \xy
 (-4,8)*{}="TL"; (4,8)*{}="TR";
 (-4,-8)*{}="BL"; (4,-8)*{}="BR";
    "TL";"TR" **\crv{(-3,0) & (3,0)};
    "BL";"BR" **\crv{(-3,0) & (3,0)};
 \endxy
\quad + \quad
  \xy
 (-4,8)*{}="TL"; (4,8)*{}="TR";
 (-4,-8)*{}="BL"; (4,-8)*{}="BR";
    "TL";"BL" **\dir{-};
    "TR";"BR" **\dir{-};
 \endxy
\]
and this formula for the `unknot':
\[ \xy
    (-5,4)*{}="x1";
    (5,4)*{}="x2";
    (-5,-4)*{}="y1";
    (5,-4)*{}="y2";
 "x1";"y1" **\dir{-}; 
 "x2";"y2" **\dir{-};
    "x1";"x2" **\crv{(-5,10) & (5,10)};
    "y1";"y2" **\crv{(-5,-10) & (5,-10)};
\endxy
    \quad = \quad
-2
 \]

The Kauffman bracket obeys modified versions of the above identities.
These involve a parameter that we will call $q$:
\[
 \xy
 (-4,8)*{}="TL"; (4,8)*{}="TR";
 (-4,-8)*{}="BL"; (4,-8)*{}="BR";
    \vtwist~{"TL"}{"TR"}{"BL"}{"BR"};
 \endxy
\quad =
\quad
 \xy
 (-8,0)*{q};
 (-4,8)*{}="TL"; (4,8)*{}="TR";
 (-4,-8)*{}="BL"; (4,-8)*{}="BR";
    "TL";"TR" **\crv{(-3,0) & (3,0)};
    "BL";"BR" **\crv{(-3,0) & (3,0)};
 \endxy
\quad + \quad
  \xy
 (-9,0)*{q^{-1}};
 (-4,8)*{}="TL"; (4,8)*{}="TR";
 (-4,-8)*{}="BL"; (4,-8)*{}="BR";
    "TL";"BL" **\dir{-};
    "TR";"BR" **\dir{-};
 \endxy
\]
and
\[ \xy
    (-5,4)*{}="x1";
    (5,4)*{}="x2";
    (-5,-4)*{}="y1";
    (5,-4)*{}="y2";
 "x1";"y1" **\dir{-}; 
 "x2";"y2" **\dir{-};
    "x1";"x2" **\crv{(-5,10) & (5,10)};
    "y1";"y2" **\crv{(-5,-10) & (5,-10)};
\endxy
    \quad = \quad
-(q^2 + q^{-2})
 \]
Among knot theorists, identities of this sort are called `skein
relations'.

Penrose's original recipe is unable to detecting linking or knotting,
since it also satisfies this identity:
\[
\vcenter{ \xy
 \vtwist~{(-5,8)}{(5,8)}{(-5,-8)}{(5,-8)};
\endxy  }
\quad = \quad
\vcenter{ \xy
 \vtwistneg~{(-5,8)}{(5,8)}{(-5,-8)}{(5,-8)};
\endxy }
\]
coming from the fact that $\Rep(\SU(2))$ is a \textsl{symmetric}
monoidal category.  The Kauffman bracket arises from a more
interesting braided monoidal category: the category of representations
of the `quantum group' associated to $\SU(2)$.  This quantum group depends
on a parameter $q$, which conventionally is related to quantity we
are calling $q$ above by a mildly annoying formula.  To keep our
story simple, we identify these two parameters.

When $q = 1$, the category of representations of the quantum group
associated to $\SU(2)$ reduces to $\Rep(\SU(2))$, and the Kauffman
bracket reduces to Penrose's original recipe.  At other values of $q$,
this category is not symmetric, and the Kauffman bracket detects
linking and knotting.

In fact, all the quantum invariants of links discovered around this
time turned out to come from braided monoidal categories: namely,
categories of representations of quantum groups.  When $q = 1$, these
quantum groups reduce to ordinary groups, their categories of
representations become symmetric, and the quantum invariants of links
become boring.

A basic result in knot theory says that given diagrams of two isotopic
links, we can get from one to the other by warping the page on which
they are drawn, together with a finite sequence of steps where we
change a small portion of the diagram.  There are three such steps,
called the \textbf{first Reidemeister move}:
\[ \xy
  (0,15)*{}="T";
  (0,-15)*{}="B";
  (0,7.5)*{}="T'";
  (0,-7.5)*{}="B'";
    "T";"T'" **\dir{-};
    "B";"B'" **\dir{-};
    (-4.5,0)*{}="MB";
    (-10.5,0)*{}="LB";
    "T'";"LB" **\crv{(-1.5,-6) & (-10.5,-6)}; \POS?(.25)*{\hole}="2z";
    "LB"; "2z" **\crv{(-12,9) & (-3,9)};
    "2z";"B'" **\crv{(0,-4.5)};
    (2,13)*{};
    \endxy
    \quad = \quad
    \xy
  (0,15)*{}="T";
  (0,-15)*{}="B";
  (0,7.5)*{}="T'";
  (0,-7.5)*{}="B'";
    "T";"B" **\dir{-};
    (2,13)*{};
    \endxy
 \]
the \textbf{second Reidemeister move}:
\[
\vcenter{\begin{xy}0;/r.30pc/:
\vtwistneg~{(-5,5)}{(5,5)}{(-5,-5)}{(5,-5)};
\vtwist~{(-5,15)}{(5,15)}{(-5,5)}{(5,5)};
\end{xy}}
\quad = \quad \vcenter{\begin{xy}
 (-5,-5);(-5,15) **\dir{-};
 (5,-5);(5,15) **\dir{-};
    (-6,16.4)*{};
    (6,16.4)*{};
    (-6,-7)*{};
    (6,-7)*{};
\end{xy}}
\]
and the \textbf{third Reidemeister move}:
\vskip 1em
\noindent
\[
\def\objectstyle{\scriptstyle}
\def\labelstyle{\scriptstyle}
  \xy
   (12,15)*{}="C";
   (4,15)*{}="B";
   (-7,15)*{}="A";
   (12,-15)*{}="3";
   (-3,-15)*{}="2";
   (-12,-15)*{}="1";
       "C";"1" **\crv{(15,0)& (-15,0)};
       (-5,-5)*{}="2'";
       (7,2)*{}="3'";
     "2'";"2" **\crv{};
     "3'";"3" **\crv{(7,-8)};
       \vtwist~{"A"}{"B"}{(-6,-1)}{(6,5.7)};
\endxy
 \qquad = \qquad
   \xy
   (-12,-15)*{}="C";
   (-4,-15)*{}="B";
   (7,-15)*{}="A";
   (-12,15)*{}="3";
   (3,15)*{}="2";
   (12,15)*{}="1";
       "C";"1" **\crv{(-15,0)& (15,0)};
       (5,5)*{}="2'";
       (-7,-2)*{}="3'";
     "2'";"2" **\crv{(4,6)};
     "3'";"3" **\crv{(-7,8)};
       \vtwist~{"A"}{"B"}{(6,1)}{(-6,-5.7)};
\endxy
\]
\vskip 0.5 em
\noindent
Kauffman gave a beautiful purely diagrammatic argument that his
bracket was invariant under the second and third Reidemeister moves.
We leave it as a challenge to the reader to find this argument, which
looks very simple \textit{after one has seen it}.  On the other hand,
the bracket is not invariant under the first Reidemeister move.
But, it transforms in a simple way, as this calculation shows:
\[
\xy
  (0,15)*{}="T";
  (0,-15)*{}="B";
  (0,7.5)*{}="T'";
  (0,-7.5)*{}="B'";
    "T";"T'" **\dir{-};
    "B";"B'" **\dir{-};
    (-4.5,0)*{}="MB";
    (-10.5,0)*{}="LB";
    "T'";"LB" **\crv{(-1.5,-6) & (-10.5,-6)}; \POS?(.25)*{\hole}="2z";
    "LB"; "2z" **\crv{(-12,9) & (-3,9)};
    "2z";"B'" **\crv{(0,-4.5)};
    (2,13)*{};
    \endxy
    \quad = \quad q \; \;
    \xy
  (0,15)*{}="T";
  (0,-15)*{}="B";
  (0,4)*{}="T'";
  (0,-4)*{}="B'";
    "T";"T'" **\dir{-};
    "B";"B'" **\dir{-};
    (-4,4)*{}="t1";
    (-10,4)*{}="t2";
    (-4,-4)*{}="b1";
    (-10,-4)*{}="b2";
    "T'"; "t1" **\crv{(0,0) & (-4,0)};
    "t1"; "t2" **\crv{(-4,7) & (-10,7)};
    "t2";"b2" **\dir{-};
    "B'"; "b1" **\crv{(0,0) & (-4,0)};
    "b1"; "b2" **\crv{(-4,-7) & (-10,-7)};
    (2,13)*{};
    \endxy
    \; \; + \quad q^{-1} \;
    \xy
  (0,15)*{}="T";
  (0,-15)*{}="B";
  (0,4)*{}="T'";
  (0,-4)*{}="B'";
    "T";"B" **\dir{-};
    (-4,4)*{}="t1";
    (-10,4)*{}="t2";
    (-4,-4)*{}="b1";
    (-10,-4)*{}="b2";
    "t1"; "t2" **\crv{(-4,7) & (-10,7)};
    "t2";"b2" **\dir{-};
    "t1";"b1" **\dir{-};
    "b1"; "b2" **\crv{(-4,-7) & (-10,-7)};
    (2,13)*{};
    \endxy
    \quad = \quad
  -q^{-3} \;
    \xy
  (0,15)*{}="T";
  (0,-15)*{}="B";
    "T";"B" **\dir{-};
    (2,13)*{};
    \endxy
\]
where we used the skein relations and did a little algebra.  So, while
the Kauffman bracket is not an isotopy invariant of links, it comes
close: we shall later see that it is an invariant of `framed' links,
made from ribbons.  And with a bit of tweaking, it gives the Jones
polynomial, which \textit{is} an isotopy invariant.

This and other work by Kauffman helped elevate string diagram
techniques from a curiosity to a mainstay of modern mathematics.  His
book \textit{Knots and Physics} was especially influential in this
respect \cite{Kauffman2}.  Meanwhile, the work of Jones led researchers
towards a wealth of fascinating connections between von Neumann algebras,
higher categories, and quantum field theory in 2- and 3-dimensional
spacetime.

\subsection*{Freyd--Yetter (1986)}

Among the many quantum invariants of links that appeared after Jones
polynomial, one of the most interesting is the `HOMFLY-PT' polynomial,
which, it later became clear, arises from the category of
representations of the quantum group associated to $\SU(n)$.  This
polynomial got its curious name because it was independently
discovered by many mathematicians, some of whom teamed up to write a
paper about it for the \textsl{Bulletin of the American Mathematical
Society} in 1985: Hoste, Ocneanu, Millet, Freyd, Lickorish and Yetter
\cite{HOMFLY}.  The `PT' refers to Przytycki and Traczyk, who
published separately \cite{PT}.

Different authors of this paper took different approaches.  Freyd and
Yetter's approach is particularly germane to our story because they
used a category where morphisms are tangles.  A `tangle' is a
generalization of a braid that allows strands to double back, and also
allows closed loops:
\[
 \fancytangle
\]
So, a link is just a tangle with no strands coming in on top, and
none leaving at the bottom.  The advantage of tangles is that we can
take a complicated link and chop it into simple pieces, which are tangles.

Shortly after Freyd heard Street give a talk on braided monoidal
categories and the category of braids, Freyd and Yetter found a
similar purely algebraic description of the category of oriented tangles
\cite{FY}.  A tangle is `oriented' if each strand is equipped with a
smooth nowhere vanishing field of tangent vectors, which we can
draw as little arrows.  We have already seen what an orientation
is good for: it lets us distinguish between representations and
their duals---or in physics, particles and antiparticles.

There is a precisely defined but also intuitive notion of when two
oriented tangles count as the same: roughly speaking, whenever we can
go from the first to the second by smoothly moving the strands without
moving their ends or letting the strands cross.  In this case we say
these oriented tangles are `isotopic'.

The category of oriented tangles has isotopy classes of oriented
tangles as morphisms.  We compose tangles by sticking one on top of
the other.  Just like Joyal and Street's category of braids, $\Tang$
is a braided monoidal category, where we tensor tangles by placing
them side by side, and the braiding is defined using the fact that a
braid is a special sort of tangle.

In fact, Freyd and Yetter gave a purely algebraic description of the
category of oriented tangles as a `compact' braided monoidal category.
Here a monoidal category $C$ is {\bf compact} if every object $x \in
C$ has a {\bf dual}: that is, an object $x^\ast$ together with
morphisms called the {\bf unit}:
\[
\xy
 (-6,-6)*{};
 (6,-6)*{};
   **\crv{(6,8) & (-6,8)} ?(.16)*\dir{>} ?(.9)*\dir{>};
 (1,7)*{    };
 (-9,-1)*{ }; (9,-1)*{ };
\endxy
\qquad {=} \qquad
 \xy
 (0,6)*+{1}="1";
 (0,-6)*+{x \ten x^*}="2";
   {\ar_{\scriptstyle i_x} "1";"2"};
\endxy
\]
and the {\bf counit}:
\[
\xy
 (-6,6)*{};(6,6)*{};
   **\crv{(6,-8) & (-6,-8)}
    ?(.20)*\dir{>}
    ?(.89)*\dir{>};
   (1,-6)*{      };
   (-9,1)*{ }; (9,1)*{ };
\endxy
\qquad {=}\qquad
 \xy
 (0,6)*+{x^{*} \ten x}="1";
 (0,-6)*+{1}="2";
   {\ar_{\scriptstyle e_x} "1";"2"};
\endxy
\]
satisfying the {\bf zig-zag identities}:
\[ \stringZIGZAGi
\]
\[
\stringZIGZAGii
\]
We have already seen these identities in our discussion of Penrose's
work.  Indeed, some classic examples of compact {\it symmetric}
monoidal categories include $\Fin\Vect$, where $x^\ast$ is the usual
dual of the vector space $x$, and $\Rep(K)$ for any compact Lie group
$K$, where $x^\ast$ is the dual of the representation $x$.  But the
zig-zag identities clearly hold in the category of oriented tangles,
too, and this example is braided but not symmetric.

There are some important subtleties that our sketch has overlooked so
far.  For example, for any object $x$ in a compact braided monoidal
category, this string diagram describes an isomorphism $d_x \maps
x \to x^{**}$:
\[ \xy
  (0,15)*{}="T";
  (0,-15)*{}="B";
  (0,7.5)*{}="T'";
  (0,-7.5)*{}="B'";
  "T";"T'" **\dir{-}?(.5)*\dir{>};
    "B";"B'" **\dir{-}?(.5)*\dir{<};
    (-4.5,0)*{}="MB";
    (-10.5,0)*{}="LB";
    "T'";"LB" **\crv{(-1.5,-6) & (-10.5,-6)}; \POS?(.25)*{\hole}="2z";
    "LB"; "2z" **\crv{(-12,9) & (-3,9)};
    "2z";"B'" **\crv{(0,-4.5)};
    (2,13)*{x};
    (3,-13)*{x^{**}}
    \endxy
\]
But if we think of this diagram as an oriented tangle, it is
isotopic to a straight line.  This suggests that $d_x$ should be
an identity morphism.  To implement this idea, Freyd and Yetter used
braided monoidal categories where each object has a \textsl{chosen}
dual, and this equation holds: $x^{**} = x$.  Then they imposed
the equation $d_x = 1_x$, which says that
\[ \xy
  (0,15)*{}="T";
  (0,-15)*{}="B";
  (0,7.5)*{}="T'";
  (0,-7.5)*{}="B'";
  "T";"T'" **\dir{-}?(.5)*\dir{>};
    "B";"B'" **\dir{-};
    (-4.5,0)*{}="MB";
    (-10.5,0)*{}="LB";
    "T'";"LB" **\crv{(-1.5,-6) & (-10.5,-6)}; \POS?(.25)*{\hole}="2z";
    "LB"; "2z" **\crv{(-12,9) & (-3,9)};
    "2z";"B'" **\crv{(0,-4.5)};
    (2,13)*{};
    \endxy
    \qquad = \qquad
  \xy
  (0,15)*{}="T";
  (0,-15)*{}="B";
  (0,7.5)*{}="T'";
  (0,-7.5)*{}="B'";
  "T";"B" **\dir{-}?(.5)*\dir{>};
    (2,13)*{};
    \endxy
 \]

This seems sensible, but it in category theory it is always dangerous
to impose equations between objects, like $x^{**} = x$.  And indeed,
the danger becomes clear when we remember that Penrose's spin networks
\textsl{violate} the above rule: instead, they satisfy
\[ \xy
  (0,15)*{}="T";
  (0,-15)*{}="B";
  (0,7.5)*{}="T'";
  (0,-7.5)*{}="B'";
    "T";"T'" **\dir{-};
    "B";"B'" **\dir{-};
    (-4.5,0)*{}="MB";
    (-10.5,0)*{}="LB";
    "T'";"LB" **\crv{(-1.5,-6) & (-10.5,-6)}; \POS?(.25)*{\hole}="2z";
    "LB"; "2z" **\crv{(-12,9) & (-3,9)};
    "2z";"B'" **\crv{(0,-4.5)};
    (2,13)*{j};
    \endxy
    \quad = \quad
  (-1)^{2j + 1} \;\;
  \xy
  (0,15)*{}="T";
  (0,-15)*{}="B";
  (0,7.5)*{}="T'";
  (0,-7.5)*{}="B'";
    "T";"B" **\dir{-};
    (2,13)*{j};
    \endxy
 \]
The Kauffman bracket violates the rule in an even more complicated way.
As mentioned in our discussion of Jones' 1985 paper, the Kauffman
bracket satisfies
\[ \xy
  (0,15)*{}="T";
  (0,-15)*{}="B";
  (0,7.5)*{}="T'";
  (0,-7.5)*{}="B'";
    "T";"T'" **\dir{-};
    "B";"B'" **\dir{-};
    (-4.5,0)*{}="MB";
    (-10.5,0)*{}="LB";
    "T'";"LB" **\crv{(-1.5,-6) & (-10.5,-6)}; \POS?(.25)*{\hole}="2z";
    "LB"; "2z" **\crv{(-12,9) & (-3,9)};
    "2z";"B'" **\crv{(0,-4.5)};
    (2,13)*{};
    \endxy
    \quad = \quad
  -q^{-3} \;\;
  \xy
  (0,15)*{}="T";
  (0,-15)*{}="B";
  (0,7.5)*{}="T'";
  (0,-7.5)*{}="B'";
    "T";"B" **\dir{-};
    (2,13)*{};
    \endxy
 \]
So, while Freyd and Yetter's theorem is correct, it needs some
fine-tuning to cover all the interesting examples.

For this reason, Street's student Shum \cite{Shum} considered tangles
where each strand is equipped with both an orientation and a
\textbf{framing} --- a nowhere vanishing smooth field of unit normal
vectors.  We can draw a framed tangle as made of ribbons, where one
edge of each ribbon is black, while the other is red.  The black edge
is the actual tangle, while the normal vector field points from the
black edge to the red edge.  But in string diagrams, we usually avoid
drawing the framing by using a standard choice: the \textbf{blackboard
framing}, where the unit normal vector points at right angles to the
page, towards the reader.

There is an evident notion of when two framed oriented tangles count
as the same, or `isotopic'.  Any such tangle is isotopic to one where
we use the blackboard framing, so we lose nothing by making this choice.
And with this choice, the following framed tangles are not isotopic:
\[ \xy
  (0,15)*{}="T";
  (0,-15)*{}="B";
  (0,7.5)*{}="T'";
  (0,-7.5)*{}="B'";
  "T";"T'" **\dir{-}?(.5)*\dir{>};
    "B";"B'" **\dir{-};
    (-4.5,0)*{}="MB";
    (-10.5,0)*{}="LB";
    "T'";"LB" **\crv{(-1.5,-6) & (-10.5,-6)}; \POS?(.25)*{\hole}="2z";
    "LB"; "2z" **\crv{(-12,9) & (-3,9)};
    "2z";"B'" **\crv{(0,-4.5)};
    (2,13)*{};
    \endxy
    \qquad \ne \qquad
  \xy
  (0,15)*{}="T";
  (0,-15)*{}="B";
  (0,7.5)*{}="T'";
  (0,-7.5)*{}="B'";
  "T";"B" **\dir{-}?(.5)*\dir{>};
    (2,13)*{};
    \endxy
 \]
\noindent
The problem is that if we think of these tangles as ribbons, 
and pull the left one tight, it has a 360 degree twist in it.

What is the framing good for in physics?  The picture above is the
answer.  We can think of each tangle as a physical
process involving particles.   The presence of the framing means
that the left-hand process is topologically different than the
right-hand process, in which a particle just sits there unchanged.

This is worth pondering in more detail.  Consider the left-hand picture:
\[ \xy
  (0,15)*{}="T";
  (0,-15)*{}="B";
  (0,7.5)*{}="T'";
  (0,-7.5)*{}="B'";
  "T";"T'" **\dir{-}?(.5)*\dir{>};
    "B";"B'" **\dir{-};
    (-4.5,0)*{}="MB";
    (-10.5,0)*{}="LB";
    "T'";"LB" **\crv{(-1.5,-6) & (-10.5,-6)}; \POS?(.25)*{\hole}="2z";
    "LB"; "2z" **\crv{(-12,9) & (-3,9)};
    "2z";"B'" **\crv{(0,-4.5)};
    (2,13)*{};
    \endxy
\]
Reading this from top to bottom, it starts with a single particle.  Then
a virtual particle-antiparticle pair is created on the left.  Then the 
new virtual particle and the original particle switch places by moving
around each other clockwise.  Finally, the original particle and its
antiparticle annihilate each other.  So, this is all about \textit{a
particle that switches places with a copy of itself}.

But we can also think of this picture as a ribbon.  If we pull it
tight, we get a ribbon that is topologically equivalent---that is,
isotopic.  It has a $360^\circ$ clockwise twist in it.  This describes
\textit{a particle that rotates a full turn}:
\[
 \xy
 (0,0)*{\includegraphics[angle=180,scale=0.5]{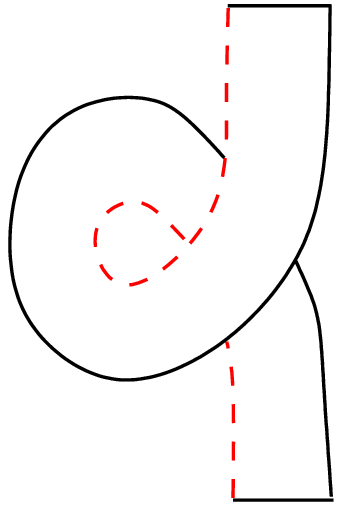}};
 \endxy
\qquad = \qquad
 \xy
 (0,0)*{\includegraphics[angle=180,scale=0.5]{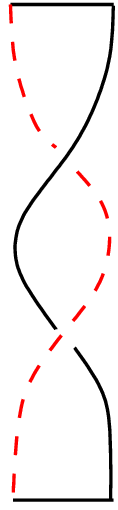}};
 \endxy
\]
So, as far as topology is concerned, we can express the concept of
rotating a single particle a full turn in terms of
switching two identical particles---at least in situations where
creation and annihilation of particle-antiparticle pairs is possible.  This
fact is quite remarkable.  As emphasized by Feynman \cite{Feynman3}, it
lies at the heart of the famous `spin-statistics theorem' in quantum
field theory.  We have already seen that in theories of physics where
spacetime is 4-dimensional, the phase of a particle is multiplied by
either $1$ or $-1$ when we rotate it a full turn: $1$ for bosons, and
$-1$ for fermions.  The spin-statistics theorem says that switching
two identical copies of this particle has the same effect on their
phase: $1$ for bosons, $-1$ for fermions.

The story becomes even more interesting in theories of physics where
spacetime is 3-dimensional.  In this situation space is 2-dimensional,
so we can distinguish between clockwise and counterclockwise rotations.
Now the spin-statistics theorem says that rotating a single particle a
full turn clockwise gives the same phase as switching two identical particles
of this type by moving them around each other clockwise.  Rotating a
particle a full turn clockwise need not have the same effect as rotating
it counterclockwise, so this phase need not be its own inverse.  In fact,
it can be \textit{any} unit complex number.  This allows for
`exotic' particles that are neither bosons nor fermions.  In 1982, such
particles were dubbed \textbf{anyons} by Frank Wilczek \cite{Wilczek}.

Anyons are not just mathematical curiosities.  Superconducting thin
films appear to be well described by theories in which the dimension
of spacetime is 3: two dimensions for the film, and one for time.  In
such films, particle-like excitations arise, which act like anyons
to a good approximation.  The presence of these `quasiparticles'
causes the film to respond in a surprising way to magnetic fields when
current is running through it.  This is called the `fractional quantum
Hall effect' \cite{Ezawa}.

In 1983, Robert Laughlin \cite{Laughlin} published an explanation of
the fractional quantum Hall effect in terms of anyonic quasiparticles.
He won the Nobel prize for this work in 1998, along with Horst
St\"ormer and Daniel Tsui, who observed this effect in the lab
\cite{GST}.  By now we have an increasingly good understanding of
anyons in terms of a quantum field theory called Chern--Simons theory,
which also explains knot invariants such as the Kauffman bracket.  For
a bit more on this, see our discussion of Witten's 1989 paper on
Chern--Simons theory.

But we are getting ahead of ourselves!  Let us return to the work of
Shum.  She constructed a category where the objects are finite
collections of oriented points in the unit square.  By `oriented' we
mean that each point is labelled either $x$ or $x^*$.  We call a point
labelled by $x$ \textbf{positively oriented}, and one labelled by
$x^*$ \textbf{negatively oriented}.  The morphisms in Shum's category
are isotopy classes of framed oriented tangles.  As usual, composition
is defined by gluing the top of one tangle to the bottom of the other.
We shall call this category $1\Tang_2$.  The reason for this curious
notation is that the tangles themselves have dimension 1, but they
live in a space --- or spacetime, if you prefer --- of dimension $1+2
= 3$.  The number 2 is called the `codimension'.  It turns out that
varying these numbers leads to some very interesting patterns.

Shum's theorem gives a purely algebraic description of $1\Tang_2$ in
terms of `ribbon categories'.  We have already seen that in a compact
braided monoidal category $C$, every object $x \in C$ comes equipped
with an isomorphism to its double dual, which we denoted $d_x
\maps x \to x^{**}$.  A ribbon category is a compact braided monoidal
category where each object $x$ is also equipped with another
isomorphism, $c_x \maps x^{**} \to x$, which must satisfy a short
list of axioms.  We call this a `ribbon structure'.  Composing this
ribbon structure with $d_x$, we get an isomorphism
\[          b_x = c_x d_x \maps x \to x  .\]
Now the point is that we can draw a string diagram for $b_x$
which is very much like the diagram for $d_x$, but with
$x$ as the output instead of $x^{**}$:
\[ \xy
  (0,15)*{}="T";
  (0,-15)*{}="B";
  (0,7.5)*{}="T'";
  (0,-7.5)*{}="B'";
  "T";"T'" **\dir{-}?(.5)*\dir{>};
    "B";"B'" **\dir{-}?(.5)*\dir{<};
    (-4.5,0)*{}="MB";
    (-10.5,0)*{}="LB";
    "T'";"LB" **\crv{(-1.5,-6) & (-10.5,-6)}; \POS?(.25)*{\hole}="2z";
    "LB"; "2z" **\crv{(-12,9) & (-3,9)};
    "2z";"B'" **\crv{(0,-4.5)};
    (2,13)*{x};
    (2,-13)*{x}
    \endxy
\]
This is the composite of $d_x$, which we know how to draw, and $c_x$,
which we leave invisible, since we do not know how to draw it.

In modern language, Shum's theorem says that $1\Tang_2$ is the `free
ribbon category on one object', namely the positively oriented point,
$x$.  The definition of ribbon category is designed to make it obvious
that $1\Tang_2$ is a ribbon category.  But in what sense is it `free
on one object'?  For this we define a `ribbon functor' to be a braided
monoidal functor between ribbon categories that preserves the ribbon
structure.  Then the statement is this.  First, given any ribbon
category $C$ and any object $c \in C$, there is a ribbon functor
\[            Z \maps 1\Tang_2 \to C  \]
such that
\[             Z(x) = c   .\]
Second, $Z$ is unique up to a braided monoidal natural isomorphism.

For a thorough account of Shum's theorem and related results, see
Yetter's book \cite{Yetter}.  We emphasized some technical aspects of
this theorem because they are rather strange.  As
we shall see, the theme of $n$-categories `with duals' becomes increasingly
important as our history winds to its conclusion, but duals remain a bit
mysterious.  Shum's theorem is the first hint of this: to avoid the equation
between objects $x^{**} = x$, it seems we are forced to introduce an
isomorphism $c_x \maps x^{**} \to x$ with no clear interpretation
as a string diagram.  We will see similar mysteries later.

Shum's theorem should remind the reader of Joyal and Street's theorem
saying that $\Braid$ is the free braided monoidal category on one
object.  They are the first in a long line of results that describe
interesting topological structures as free structures on one object,
which often corresponds to a point.  This
idea has been dubbed ``the primacy of the point''.

\subsection*{Drinfel'd (1986)}

In 1986, Vladimir Drinfel'd won the Fields medal for his work on
quantum groups \cite{Drinfeld}.  This was the culmination of a long
line of work on exactly solvable problems in low-dimensional physics,
which we can only briefly sketch.

Back in 1926, Heisenberg \cite{Heisenberg} considered a simplified
model of a ferromagnet like iron, consisting of spin-$\frac{1}{2}$
particles---electrons in the outermost shell of the iron
atoms---sitting in a cubical lattice and interacting only with their
nearest neighbors.  In 1931, Bethe \cite{Bethe} proposed an ansatz
which let him exactly solve for the eigenvalues of the Hamiltonian in
Heisenberg's model, at least in the even simpler case of a {\it
1-dimensional} crystal.  This was subsequently generalized by Onsager
\cite{Onsager}, C.\ N.\ and C.\ P.\ Yang \cite{YangYang}, Baxter
\cite{Baxter} and many others.

The key turns out to be something called the `Yang--Baxter
equation'. It's easiest to understand this in the context of
2-dimensional quantum field theory.  Consider a Feynman diagram
where two particles come in and two go out:
\[
\BB
\]
This corresponds to some operator
\[  B \maps H \tensor H \to H \tensor H  \]
where $H$ is the Hilbert space of states of the particle.  It turns out
that the physics simplifies immensely, leading to exactly solvable
problems, if:
\[
\BByang
\]
This says we can slide the lines around in a certain way without
changing the operator described by the Feynman diagram.   In
terms of algebra:
\[          (B \otimes 1)(1 \otimes B)(B \otimes 1) =
            (1 \otimes B)(B \otimes 1)(1 \otimes B).
\]
This is the {\bf Yang--Baxter equation}; it makes sense in any
monoidal category.

In their 1985 paper, Joyal and Street noted that given any
object $x$ in a braided monoidal category, the braiding
\[     B_{x,x} \maps x \tensor x \to x \tensor x \]
is a solution of the Yang--Baxter equation.  If we draw this equation
using string diagrams, it looks like the third Reidemeister move in
knot theory:
\vskip 0.5em
\noindent
\[
\def\objectstyle{\scriptstyle}
\def\labelstyle{\scriptstyle}
  \xy
   (12,15)*{}="C";
   (4,15)*{}="B";
   (-7,15)*{}="A";
   (12,-15)*{}="3";
   (-3,-15)*{}="2";
   (-12,-15)*{}="1";
       "C";"1" **\crv{(15,0)& (-15,0)};
       (-5,-5)*{}="2'";
       (7,2)*{}="3'";
     "2'";"2" **\crv{};
     "3'";"3" **\crv{(7,-8)};
       \vtwist~{"A"}{"B"}{(-6,-1)}{(6,5.7)};
\endxy
 \qquad = \qquad
   \xy
   (-12,-15)*{}="C";
   (-4,-15)*{}="B";
   (7,-15)*{}="A";
   (-12,15)*{}="3";
   (3,15)*{}="2";
   (12,15)*{}="1";
       "C";"1" **\crv{(-15,0)& (15,0)};
       (5,5)*{}="2'";
       (-7,-2)*{}="3'";
     "2'";"2" **\crv{(4,6)};
     "3'";"3" **\crv{(-7,8)};
       \vtwist~{"A"}{"B"}{(6,1)}{(-6,-5.7)};
\endxy
\]
\vskip 0.5 em
\noindent
Joyal and Street also showed that given any solution of the
Yang--Baxter equation in any monoidal category, we can build a
braided monoidal category.

Mathematical physicists enjoy exactly solvable problems, so after the
work of Yang and Baxter a kind of industry developed, devoted to
finding solutions of the Yang--Baxter equation.  The Russian school,
led by Faddeev, Sklyanin, Takhtajan and others, were especially
successful \cite{FST}.  Eventually Drinfel'd discovered how to get
solutions of the Yang--Baxter equation from any simple Lie algebra.
The Japanese mathematician Jimbo did this as well, at about the
same time \cite{Jimbo}.

What they discovered was that the universal enveloping algebra $\U\g$
of any simple Lie algebra $\g$ can be `deformed' in a manner depending
on a parameter $q$, giving a one-parameter family of `Hopf algebras'
$\U_q \g$.  Since Hopf algebras are mathematically analogous to groups
and in some physics problems the parameter $q$ is related to Planck's
constant $\hbar$ by $q = e^\hbar$, the Hopf algebras $\U_q \g$ are
called `quantum groups'.  There is by now an extensive theory of these
\cite{CP,Kassel,Majid}.

Moreover, these Hopf algebras have a special property which implies
that any representation of $\U_q \g$ on a
vector space $V$ comes equipped with an operator
\[  B \colon V \tensor V \to V \tensor V \]
satisfying the Yang--Baxter equation.  We shall say a bit more about
this in our discussion of a 1989 paper by Reshetikhin and Turaev.

This work led to a far more thorough understanding of exactly
solvable problems in 2d quantum field theory \cite{EL}.  It was also
the first big \textit{explicit} intrusion of category theory into
physics.  As we shall see, Drinfel'd's constructions can be nicely
explained in the language of braided monoidal categories.  This
led to the widespread adoption of this language, which was then
applied to other problems in physics.  Everything beforehand only
looks category-theoretic in retrospect.

\subsection*{Segal (1988)}

In an attempt to formalize some of the key mathematical structures
underlying string theory, Graeme Segal \cite{Segal} proposed axioms
describing a `conformal field theory'.  \textit{Roughly}, these say
that it is a symmetric monoidal functor
\[
    Z \maps 2\Cob_\C \to \Hilb
\]
with some nice extra properties.  Here $2\Cob_\C$ is the
category whose morphisms are `string worldsheets', like this:
\vskip 1em
\noindent
\[
 \begin{pspicture}[.5](4,2.5)
  \rput(2,0){\multc}
\end{pspicture}
\qquad
 \xy
 {\ar_M (0,12)*+{S}; (0,-12)*+{S'}};
 \endxy
\]
We compose these morphisms by gluing them end to end, like this:
\[
\begin{pspicture}[.5](4,5)
  \rput(1,0){\comultc}
  \rput(1,2.5){\multc}
\end{pspicture}
\qquad
 \xy
 {\ar_M (0,22)*+{S}; (0,0)*+{S'}};
 {\ar_{M'} (0,-2)*+{}; (0,-22)*+{S''}};
 \endxy
\]

A bit more precisely, an object $2\Cob_\C$ as a union of parametrized
circles, while a morphism $M \maps S \to S'$ is a 2-dimensional
`cobordism' equipped with some extra structure.  Here an
$n$-dimensional `cobordism' is roughly an $n$-dimensional compact
oriented manifold with boundary, $M$, whose boundary has been written as
the disjoint union of two $(n-1)$-dimensional manifolds $S$ and $S'$,
called the `source' and `target'.

In the case of $2\Cob_\C$, we need these cobordisms to be equipped
with a conformal structure and a parametrization of each boundary
circle.  The parametrization lets us give the composite of two
cobordisms a conformal structure built from the conformal structures
on the two parts.

In fact we are glossing over many subtleties here; we hope the
above sketch gets the idea across.   In any event,
$2\Cob_\C$ is a symmetric monoidal category, where we tensor objects
or morphisms by setting them side by side:
\[
\begin{pspicture}[.5](7,2.5)
  \rput(1,0){\comultc}
  \rput(5,0){\identc}
\end{pspicture}
\qquad
 \xy
 {\ar_{M \tensor M'} (0,12)*+{S_1 \tensor S_2}; (0,-12)*+{S'_1 \otimes S'_2}};
 \endxy
\]
Similarly, $\Hilb$ is a symmetric monoidal category with the usual
tensor product of Hilbert spaces.  A basic rule of
quantum physics is that the Hilbert space for a disjoint union of two
physical systems should be the tensor product of their Hilbert spaces.
This suggests that a conformal field theory, viewed as a functor $Z
\maps 2\Cob_\C \to \Hilb$, should preserve tensor products---at least
up to a specified isomorphism.  So, we should demand that $Z$ be a
monoidal functor.  A bit more reflection along these lines leads us to
demand that $Z$ be a symmetric monoidal functor.

There is more to the full definition of a conformal field theory than
merely a symmetric monoidal functor $Z \maps 2\Cob_\C \to \Hilb$.  For
example, we also need a `positive energy' condition reminiscent of the
condition we already met for representations of the Poincar\'e group.
Indeed there is a profusion of different ways to make the idea of
conformal field theory precise, starting with Segal's original
definition.  But the different approaches are nicely related, and the
subject of conformal field theory is full of deep results, interesting
classification theorems, and applications to physics and mathematics.
A good introduction is the book by Di Francesco, Mathieu and Senechal
\cite{FMS}.

\subsection*{Atiyah (1988)}

Shortly after Segal proposed his definition of `conformal field
theory', Atiyah \cite{Atiyah} modified it by dropping the conformal
structure and allowing cobordisms of an arbitrary fixed dimension.  He
called the resulting structure a `topological quantum field theory',
or `TQFT' for short.  One of his goals was to formalize some work
by Witten \cite{Witten2} on invariants of 4-dimensional manifolds
coming from a quantum field theory sometimes called `Donaldson theory',
which is related to Yang--Mills theory.  These invariants have led
to a revolution in our understanding of 4-dimensional topology---but
ironically, Donaldson theory has never been successfully dealt with
using Atiyah's axiomatic approach.  We will say more about this in
our discussion of Crane and Frenkel's 1994 paper.  For now, let
us simply explain Atiyah's definition of a TQFT.

In modern language, an \textbf{\textit{n}-dimensional TQFT} is
a symmetric monoidal functor
\[
    Z\maps n\Cob \to \Fin\Vect.
\]
Here $\Fin\Vect$ stands for the category of finite-dimensional
complex vector spaces and linear operators between them, while
$n\Cob$ is the category with:
\begin{itemize}
\item
compact oriented $(n-1)$-dimensional manifolds as objects;
\item
oriented $n$-dimensional cobordisms as morphisms.
\end{itemize}
Taking the disjoint union of manifolds makes $n\Cob$ into a monoidal
category.  The braiding in $n\Cob$ can be drawn like this:
\[
\begin{pspicture}[.5](4,2.5)
  \rput(2,0){\ucrossc}
\end{pspicture}
\qquad
 \xy
 {\ar_{B_{S,S'}} (0,12)*+{S \otimes S'}; (0,-14)*+{S' \otimes S}};
 \endxy
\]
but because we are interested in `abstract' cobordisms, not
embedded in any ambient space, this braiding will be
symmetric.

Physically, idea of a TQFT is that it describes a featureless
universe that looks \textit{locally} the same in every state.
In such an imaginary universe, the only way to distinguish
different states is by doing `global' observations, for example
by carrying a particle around a noncontractible loop in space.
Thus, TQFTs appear to be very simple toy models of physics, which
ignore most of the interesting features of what we see around us.
It is precisely for this reason that TQFTs are more tractable than
full-fledged quantum field theories.  In what follows we shall spend
quite a bit of time explaining how TQFTs are related to $n$-categories.
If $n$-categorical physics is ever to blossom, we must someday go
further.  There are some signs that this may be starting \cite{Schreiber}.
But attempting to discuss this would lead us out of our `prehistory'.

Mathematically, the study of topological quantum field theories quickly
leads to questions involving duals.  In our explanation of the work of Freyd
and Yetter we mentioned `compact' monoidal categories, where every
object has a dual.  One can show that $n\Cob$ is compact, with the
dual $x^\ast$ of an object $x$ being the same manifold equipped with
the opposite orientation.  Similarly, $\Fin\Vect$ is compact with the
usual notion of dual for vector spaces.  The categories $\Vect$ and
$\Hilb$ are not compact, since we can always define a `dimension' of
an object in a compact braided monoidal category by
\[
\DIMofx
\]
but this diverges for an infinite-dimensional vector space, or
Hilbert space.  As we have seen, the infinities that plague
ordinary quantum field theory arise from his fact.

As a category, $\Fin\Vect$ is equivalent to $\Fin\Hilb$, the
category of finite-dimensional complex Hilbert spaces and linear
operators.  However, $\Fin\Hilb$ and also $\Hilb$ have something
in common with $n\Cob$ that $\Vect$ lacks: they have `duals for
morphisms'.  In $n\Cob$, given a morphism
\[
 \begin{pspicture}[.5](4,2.5)
  \rput(2,0){\multc}
\end{pspicture}
\qquad
 \xy
 {\ar_M (0,12)*+{S}; (0,-14)*+{S'}};
 \endxy
\]
we can reverse its orientation and switch its source
and target to obtain a morphism going `backwards in time':
\[
 \begin{pspicture}[.5](4,2.5)
  \rput(2,0){\comultc}
\end{pspicture}
\qquad
 \xy
 {\ar_{M^\dagger} (0,12)*+{S'}; (0,-14)*+{S}};
 \endxy
\]
Similarly, given a linear operator $T \maps H \to H'$
between Hilbert spaces, we can define
an operator $T^\dagger \maps H' \to H$ by demanding that
\[         \langle T^\dagger \phi, \psi \rangle =
           \langle \phi, T \psi \rangle \]
for all vectors $\psi \in H, \phi \in H'$.

Isolating the common properties of these constructions, we say
a category {\bf has duals for morphisms} if for any morphism
$f \maps x \to y$ there is a morphism $f^\dagger \maps y \to x$
such that
\[ (f^\dagger)^\dagger = f, \quad (fg)^\dagger = g^\dagger f^\dagger,
\quad 1_x^\dagger = 1_x.\]
We then say morphism $f$ is {\bf unitary} if $f^\dagger$
is the inverse of $f$.   In the case of $\Hilb$ this is just
a unitary operator in the usual sense.

As we have seen, symmetries in quantum physics are described not just
by group representations on Hilbert spaces, but by \textit{unitary}
representations.  This is a hint of the importance of `duals for
morphisms' in physics.  We can always think of a group $G$ as a
category with one object and with all morphisms invertible.  This
becomes a category with duals for morphisms by setting $g^\dagger =
g^{-1}$ for all $g \in G$.  A representation of $G$ on a Hilbert space
is the same as a functor $\rho \maps G \to \Hilb$, and this
representation is unitary precisely when
\[             \rho(g^\dagger) = \rho(g)^\dagger  .\]
The same sort of condition shows up in many other contexts
in physics.  So, quite generally, given any functor $F \maps C \to D$ between
categories with duals for morphisms, we say $F$ is \textbf{unitary}
if $F(f^\dagger) = F(f)^\dagger$ for every morphism in $C$.
It turns out that the physically most interesting TQFTs are
the \textbf{unitary} ones, which are \textit{unitary} symmetric
monoidal functors
\[              Z \maps n\Cob \to \Fin\Hilb  .\]

While categories with duals for morphisms play a crucial role in this
definition, and also 1989 paper by Doplicher and Roberts, and also the
1995 paper by Baez and Dolan, they seem to have been a bit neglected
by category theorists until 2005, when Selinger \cite{Selinger}
introduced them under the name of `dagger categories' as part of his
work on the foundations of quantum computation.  Perhaps one reason
for this neglect is that their definition implicitly involves an
equation between objects---something normally shunned in category
theory.

To see this equation between objects explicitly, note that a category
with duals for morphisms, or \textbf{dagger category}, may be defined
as a category $C$ equipped with a contravariant functor $\dagger \maps
C \to C$ such that
\[              \dagger^2 = 1_C  \]
and $x^\dagger = x$ for every object $x \in C$.  Here by a
\textbf{contravariant} functor we mean one that reverses the order of
composition: this is a just way of saying that $(fg)^\dagger = g^\dagger
f^\dagger$.

Contravariant functors are well-accepted in category theory, but it
raises eyebrows to impose equations between objects, like $x^\dagger =
x$.  This is not just a matter of fashion.  Such equations cause real
trouble: if $C$ is a dagger category, and $F \maps C \to D$ is an
equivalence of categories, we cannot use $F$ to give $D$ the structure
of a dagger category, precisely because of this equation.
Nonetheless, the concept of dagger category seems crucial in quantum
physics.  So, there is a tension that remains to be resolved here.

The reader may note that this is not the first time an equation
between objects has obtruded in the study of duals.  We have already
seen one in our discussion of Freyd and Yetter's 1986 paper.  In that
case the problem involved duals for objects, rather than morphisms.
And in that case, Shum found a way around the problem.  When it comes
to duals for morphisms, no comparable fix is known.  However, in our
discussion of Doplicher and Roberts 1989 paper, we will see that the
two problems are closely connected.

\subsection*{Dijkgraaf (1989)}

Shortly after Atiyah defined TQFTs, Dijkgraaf gave a purely algebraic
characterization of 2d TQFTs in terms of commutative Frobenius
algebras \cite{Dij1}.

Recall that a 2d TQFT is a symmetric monoidal functor $Z\maps 2\Cob
\to \Vect$.  An object of $2\Cob$ is a compact oriented 1-dimensional
manifold---a disjoint union of copies of the circle $S^1$.  A morphism
of $2\Cob$ is a 2d cobordism between such manifolds.  Using `Morse
theory', we can chop any 2d cobordism $M$ into elementary building
blocks that contain only a single critical point.  These are called
the \textbf{birth of a circle}, the {\bf upside-down pair of pants},
the \textbf{death of a circle} and the {\bf pair of pants}:
\[ \psset{xunit=.4cm,yunit=.4cm}
\begin{pspicture}(4,4)
  \rput(2,1.6){\birthc}
\end{pspicture}
\qquad\quad
\begin{pspicture}(4,4)
  \rput(2,1){\multc}
\end{pspicture}
\qquad\quad
\begin{pspicture}(4,4)
  \rput(2,1.6){\deathc}
\end{pspicture}
\qquad\quad
\begin{pspicture}(4,4)
  \rput(2,1){\comultc}
\end{pspicture}
\]
Every 2d cobordism is built from these by composition, tensoring,
and the other operations present in any symmetric monoidal
category.  So, we say that $2\Cob$ is `generated' as a symmetric
monoidal category by the object $S^1$ and these morphisms.
Moreover, we can list a complete set of relations that these
generators satisfy:
\begin{equation}\label{algebra}
\psset{xunit=.3cm,yunit=.3cm}
\begin{pspicture}[.5](4,5)
  \rput(2,0){\multc}
  \rput(3,2.5){\multc}
  \rput(1,2.5){\curveleftc}
\end{pspicture}
\quad = \quad
\begin{pspicture}[.5](4,5)
  \rput(2,0){\multc}
  \rput(1,2.5){\multc}
  \rput(3,2.5){\curverightc}
\end{pspicture}
\qquad\qquad
\begin{pspicture}[.5](4,4)
  \rput(2,0){\multc}
  \rput(3,2.5){\smallidentc}
  \rput(1,2.5){\birthc}
\end{pspicture}
\quad = \quad
\begin{pspicture}[.5](2,4)
  \rput(1,0){\identc}
  \rput(1,2.5){\smallidentc}
\end{pspicture}
\quad = \quad
\begin{pspicture}[.5](4,4)
  \rput(2,0){\multc}
  \rput(1,2.5){\smallidentc}
  \rput(3,2.5){\birthc}
\end{pspicture}
\end{equation}
\begin{equation}\label{coalgebra}
\psset{xunit=.3cm,yunit=.3cm}
\begin{pspicture}[.5](4,5.8)
  \rput(1,0){\comultc}
  \rput(4,0){\curveleftc}
  \rput(2,2.5){\comultc}
\end{pspicture}
\quad = \quad
\begin{pspicture}[.5](4,5.8)
  \rput(0,0){\curverightc}
  \rput(3,0){\comultc}
  \rput(2,2.5){\comultc}
\end{pspicture}
\qquad\qquad
\begin{pspicture}[.5](4,4)
  \rput(.8,0){\deathc}
  \rput(3,0){\smallidentc}
  \rput(2,1){\comultc}
\end{pspicture}
\quad = \quad
\begin{pspicture}[.5](2,4)
  \rput(1,0){\identc}
  \rput(1,2.5){\smallidentc}
\end{pspicture}
\quad = \quad
\begin{pspicture}[.5](4,4)
  \rput(1,0){\smallidentc}
  \rput(2.8,0){\deathc}
  \rput(2,1){\comultc}
\end{pspicture}
\end{equation}

\begin{equation}\label{frobenius}
\psset{xunit=.3cm,yunit=.3cm}
\begin{pspicture}[.5](5,5.8)
  \rput(0,0){\identc}
  \rput(3,0){\multc}
  \rput(1,2.5){\comultc}
  \rput(4,2.5){\identc}
\end{pspicture}
\quad = \quad
\begin{pspicture}[.5](2,5.8)
  \rput(1,0){\comultc}
  \rput(1,2.5){\multc}
\end{pspicture}
\quad = \quad
\begin{pspicture}[.5](5,5.8)
  \rput(5,0){\identc}
  \rput(2,0){\multc}
  \rput(4,2.5){\comultc}
  \rput(1,2.5){\identc}
\end{pspicture}
\end{equation}

\begin{equation} \label{commutative}
\psset{xunit=.3cm,yunit=.3cm}
\begin{pspicture}[.5](2,5.8)
  \rput(1,0){\multc}
  \rput(1,2.5){\ucrossc}
\end{pspicture}
\quad = \quad
\begin{pspicture}[.5](2,5.8)
  \rput(1,0){\multc}
  \rput(0,2.5){\identc}
  \rput(2,2.5){\identc}
\end{pspicture}
\end{equation}
\vskip 0.5em
\noindent
$2\Cob$ is completely described as a symmetric monoidal category by
means of these generators and relations.

Applying the functor $Z$ to the circle gives a vector space
$F = Z(S^1)$, and applying it to the cobordisms shown below gives
certain linear maps:
\[ \psset{xunit=.4cm,yunit=.4cm}
\begin{pspicture}(4,4)
  \rput(2,1.6){\birthc}
  \rput(2,0){$i \maps \C \to F$}
\end{pspicture}
\qquad\quad
\begin{pspicture}(4,4)
  \rput(2,1){\multc}
  \rput(2,0){$m \maps F \otimes F\to F$}
\end{pspicture}
\qquad\quad
\begin{pspicture}(4,4)
  \rput(2,1.6){\deathc}
  \rput(2,0){$\varepsilon \maps F \to \C$}
\end{pspicture}
\qquad\quad
\begin{pspicture}(4,4)
  \rput(2,1){\comultc}
  \rput(2,0){$\Delta \maps F \to F \otimes F$}
\end{pspicture}
\]
This means that our
2-dimensional TQFT is completely determined by choosing a vector space
$F$ equipped with linear maps $i, m, \varepsilon, \Delta$ satisfying
the relations drawn as pictures above.  

Surprisingly, all this stuff
amounts to a well-known algebraic structure: `commutative Frobenius algebra'.
For starters, Equation \ref{algebra}:
\[
\psset{xunit=.3cm,yunit=.3cm}
 \xy
 (0,10)*+{F \otimes F \otimes F}="1";
 (0,0)*+{F \otimes F}="2";
 (0,-10)*+{F}="3";
  {\ar_{1_F \otimes m} "1";"2"};
  {\ar_{\mu} "2";"3"};
 \endxy
 \;\;
\begin{pspicture}[.5](4,5)
  \rput(2,0){\multc}
  \rput(3,2.5){\multc}
  \rput(1,2.5){\curveleftc}
\end{pspicture}
\qquad = \qquad
\begin{pspicture}[.5](4,5)
  \rput(2,0){\multc}
  \rput(1,2.5){\multc}
  \rput(3,2.5){\curverightc}
\end{pspicture}
\;\;\;
 \xy
 (0,10)*+{F \otimes F \otimes F}="1";
 (0,0)*+{F \otimes F}="2";
 (0,-10)*+{F}="3";
  {\ar_{m \otimes 1_F} "1";"2"};
  {\ar_{m} "2";"3"};
 \endxy
\]
says that the map $m$ defines an associative multiplication on
$F$. The second relation says that the map $i$ gives a unit for the
multiplication on $F$.  This makes $F$ into an {\bf algebra}.  The
upside-down versions of these relations appearing in \ref{coalgebra}
say that $F$ is also a {\bf coalgebra}.  An algebra that is also a
coalgebra where the multiplication and comultiplication are related by
Equation \ref{frobenius} is called a {\bf Frobenius algebra}.
Finally, Equation \ref{commutative} is the commutative law for
multiplication.

In 1996, Abrams \cite{Abrams1} was
able to construct a category of 2d TQFTs and prove it is equivalent to
the category of commutative Frobenius algebras.  This makes precise
the sense in which a 2-dimensional topological quantum field theory
`is' a commutative Frobenius algebra.  It implies that when one has a
commutative Frobenius algebra in the category $\Fin\Vect$, one immediately
gets a symmetric monoidal functor $Z \maps 2\Cob\to\Vect$, hence a
2-dimensional topological quantum field theory.  This perspective is
explained in great detail in the book by Kock~\cite{Kock}.

In modern language, the essence of Abrams' result is contained in the
following theorem: \emph{$2\Cob$ is the free symmetric monoidal
category on a commutative Frobenius algebra}.  To make this precise,
we first define a commutative Frobenius algebra in \textit{any} symmetric
monoidal category, using the same diagrams as above.  Next, suppose $C$
is any symmetric monoidal category and $c \in C$ is a commutative
Frobenius algebra in $C$. Then first, there exists a symmetric monoidal
functor
\[                Z \maps 2\Cob \to C   \]
with
\[                Z(S^1) = c  \]
and such that $Z$ sends the multiplication, unit, cocomultiplication
and counit for $S^1$ to those for $c$.  Second, $Z$ is unique up to
a symmetric monoidal natural isomorphism.

This result should remind the reader of Joyal and Street's algebraic
characterization of the category of braids, and Shum's
characterization of the category of framed oriented tangles.  It is a
bit more complicated, because the circle is a bit more complicated
than the point.  The idea of an `extended' TQFT, which we shall
describe later, strengthens the concept of a TQFT so as to restore
``the primacy of the point''.

\subsection*{Doplicher--Roberts (1989)}

In 1989, Sergio Doplicher and John Roberts published a paper
\cite{DR1} showing how to reconstruct a compact topological group
$K$---for example, a compact Lie group---from its category of
finite-dimensional continuous unitary representations, $\Rep(K)$.
They then used this to show one could start with a fairly general
quantum field theory and \textit{compute} its gauge group, instead of
putting the group in by hand \cite{DR2}.

To do this, they actually needed some extra structure on $\Rep(K)$.
For our purposes, the most interesting thing they needed was
its structure as a `symmetric monoidal category with duals'.
Let us define this concept.

In our discussion of Atiyah's 1988 paper on TQFTs, we explained what
it means for a category to be a `dagger category', or have `duals for
morphisms'.  When such a category is equipped with extra structure, it
makes sense to demand that this extra structure be compatible with this
duality.  For example, we can demand that an isomorphism $f \maps x \to y$
be \textbf{unitary}, meaning
\[                  f^\dagger f = 1_x , \qquad f f^\dagger = 1_y .\]
So, we say a monoidal category $C$ {\bf has duals for morphisms}
if its underlying category has duals for morphisms, the duality preserves
the tensor product:
\[            (f \otimes g)^\dagger = f^\dagger \otimes g^\dagger \]
and moreover all the relevant isomorphisms are unitary: the associators
$a_{x,y,z}$, and the left and right unitors $\ell_x$ and $r_x$.
We say a braided or symmetric monoidal category {\bf has
duals for morphisms} if all these conditions hold and in addition the
braiding $B_{x,y}$ is unitary.  There is an easy way to make $1\Tang_2$ into
a braided monoidal category with duals for morphisms.  Both $n\Cob$ and
$\Fin\Hilb$ are symmetric monoidal categories with duals for morphisms.

Besides duals for morphisms, we may consider duals for objects.  In
our discussion of Freyd and Yetter's 1986 paper, we said a
monoidal category has `duals for objects', or is `compact', if for
each object $x$ there is an object $x^\ast$ together with a unit
$i_x\maps 1 \to x \tensor x^\ast$ and counit $e_x \maps x^\ast \tensor
x \to 1$ satisfying the zig-zag identities.

Now suppose a braided monoidal category has both duals for morphisms
and duals for objects.  Then there is yet another compatibility
condition we can---and should---demand.  Any object has a counit,
shaped like a cup:
\[
\xy
 (-6,6)*{};(6,6)*{};
   **\crv{(6,-8) & (-6,-8)}
    ?(.20)*\dir{>}
    ?(.89)*\dir{>};
   (1,-6)*{      };
   (-9,1)*{ }; (9,1)*{ };
\endxy
\qquad\qquad
 \xy
 (0,6)*+{x^{*} \ten x}="1";
 (0,-6)*+{1}="2";
   {\ar_{\scriptstyle e_x} "1";"2"};
\endxy
\]
and taking the dual of this morphism we get a kind of cap:
\[
\xy
 (-6,-6)*{};
 (6,-6)*{};
   **\crv{(6,8) & (-6,8)} ?(.16)*\dir{>} ?(.9)*\dir{>};
 (1,7)*{    };
 (-9,-1)*{}; (9,-1)*{};
\endxy
\qquad\qquad
 \xy
 (0,6)*+{1}="1";
 (0,-6)*+{x^* \ten x}="2";
   {\ar_{\scriptstyle i_x} "1";"2"};
\endxy
\]
Combining these with the braiding we get a morphism like this:
\[ \xy
  (0,15)*{}="T";
  (0,-15)*{}="B";
  (0,7.5)*{}="T'";
  (0,-7.5)*{}="B'";
  "T";"T'" **\dir{-}?(.5)*\dir{>};
    "B";"B'" **\dir{-}?(.5)*\dir{<};
    (-4.5,0)*{}="MB";
    (-10.5,0)*{}="LB";
    "T'";"LB" **\crv{(-1.5,-6) & (-10.5,-6)}; \POS?(.25)*{\hole}="2z";
    "LB"; "2z" **\crv{(-12,9) & (-3,9)};
    "2z";"B'" **\crv{(0,-4.5)};
    (2,13)*{x};
    (2,-13)*{x}
    \endxy
\]
This looks just like the morphism $b_x \maps x \to x$ that we
introduced in our discussion of Freyd and Yetter's 1986 paper---only
now it is the result of combining duals for objects and duals for
morphisms!  Some string diagram calculations suggest that $b_x$ should
be unitary.  So, we say a braided monoidal category \textbf{has duals}
if it has duals for objects, duals for morphisms, and the twist
isomorphism $b_x \maps x \to x$, constructed as above, is unitary for
every object $x$.

In a symmetric monoidal category with duals on can show that $b_x^2 = 1_x$.
In physics this leads to the boson/fermion distinction mentioned earlier,
since a boson is any particle that remains unchanged when rotated a
full turn, while a fermion is any particle whose phase gets multiplied
by $-1$ when rotated a full turn.  Both $n\Cob$ and $\Hilb$ are
symmetric monoidal categories with duals, and both are `bosonic' in
the sense that $b_x = 1_x$ for every object.  The same is true for
$\Rep(K)$ for any compact group $K$.  This features prominently in
the paper by Doplicher and Roberts.

In recent years, interest has grown in understanding the foundations
of quantum physics with the help of category theory.  One reason is
that in theoretical work on quantum computation, there is a
fruitful overlap between the category theory used in quantum
physics and that used in computer science.  In a 2004 paper on this
subject, Abramsky and Coecke \cite{AC} introduced symmetric
monoidal categories with duals under the name of `strongly compact
closed categories'.  These entities were later dubbed `dagger compact
categories' by Selinger \cite{Selinger}, and this name seems to have
caught on.  What we are calling symmetric monoidal categories with
duals for morphisms, he calls `dagger symmetric monoidal categories'.

\subsection*{Reshetikhin--Turaev (1989)}

We have mentioned how Jones discovery in 1985 of a new invariant of
knots led to a burst of work on related invariants.  Eventually it was
found that all these so-called `quantum invariants' of knots can be
derived in a systematic way from quantum groups.  A particularly clean
treatment using braided monoidal categories can be found in a paper by
Nikolai Reshetikhin and Vladimir Turaev \cite{RT1}.  This is a good
point to summarize a bit of the theory of quantum groups in its modern
form.

The first thing to realize is that a quantum group is not a group:
it is a special sort of algebra.  What quantum groups
and groups have in common is that their categories of representations
have similar properties.   The category of finite-dimensional
representations of a group is a symmetric monoidal category with
duals for objects.  The category of finite-dimensional representations
of a quantum group is a \textit{braided} monoidal category with duals
for objects.

As we saw in our discussion of Freyd and Yetter's 1986 paper, the
category $1\Tang_2$ of tangles in 3 dimensions is the \textit{free}
braided monoidal category with duals on one object $x$.  So, if
$\Rep(A)$ is the category of finite-dimensional representations of a
quantum group $A$, any object $V \in \Rep(A)$ determines a braided
monoidal functor
\[               Z \maps 1\Tang_2 \to \Rep(A)  .\]
with
\[                  Z(x) =  V   .\]
This functor gives an invariant of tangles: a linear operator
for every tangle, and in particular a number for every knot or link.

So, what sort of algebra has representations that form a braided
monoidal category with duals for objects?  This turns out to be one of
a family of related questions with related answers.  The more extra
structure we put on an algebra, the nicer its category of
representations becomes:

\vskip 2em
\vbox{
\begin{center}
{\small
\begin{tabular}{|c|c|}                    \hline
algebra   &  category                    \\     \hline
bialgebra &  monoidal category           \\   \hline
quasitriangular bialgebra &  braided monoidal category  \\  \hline
triangular bialgebra  & symmetric monoidal category  \\   \hline
\hline
Hopf algebra          & monoidal category  \\
                      &  with duals for objects \\  \hline
quasitriangular & braided monoidal category \\
 Hopf algebra   &  with duals for objects \\  \hline
triangular      &  symmetric monoidal category  \\
 Hopf algebra   & with duals for objects \\ \hline
\end{tabular}} \vskip 1em
Algebras and their categories of representations
\end{center}
}
\vskip 1em

\noindent
For each sort of algebra $A$ in the left-hand column, its
category of representations $\Rep(A)$ becomes a category of the sort listed
in the right-hand column.  In particular, a quantum group is a
kind of `quasitriangular Hopf algebra'.

In fact, the correspondence between algebras and their categories
of representations works both ways.   Under some mild technical assumptions,
we can recover $A$ from $\Rep(A)$ together with the `forgetful
functor' $F \maps \Rep(A) \to \Vect$ sending each representation to its
underlying vector space.  The theorems guaranteeing this are
called `Tannaka--Krein reconstruction theorems' \cite{JS2}.
They are reminiscent of the Doplicher--Roberts reconstruction theorem,
which allows us to recover a compact topological group $G$ from its
category of representations.  However, they are easier to prove, and
they came earlier.

So, someone who strongly wishes to avoid learning about
quasitriangular Hopf algebras can get away with it, at least for a
while, if they know enough about braided monoidal categories with
duals for objects.  The latter subject is ultimately more fundamental.
Nonetheless, it is very interesting to see how the correspondence
between algebras and their categories of representations works.  So,
let us sketch how any bialgebra has a monoidal category of
representations, and then give some examples coming from groups and
quantum groups.

First, recall that an \textbf{algebra} is a vector space $A$
equipped with an associative multiplication
\[  \begin{array}{cccc}
m \maps & A \tensor A &\to& A  \\
        & a \tensor b &\mapsto& ab
\end{array}
\]
together with an element $1 \in A$ satisfying the left and right
unit laws: $1a = a = a1$ for all $a \in A$.  We can draw
the multiplication using a string diagram:
\[
 \xy
  (0,0)*\xycircle(3,3){-}="c";
  (0,0)*{m};
  (-6,10)*{}="tl"+(-1,-3)*{ };
  (6,10)*{}="tr"+(1,-3)*{ };
  (0,-10)*{}="b"+(-1,1)*{ };
     "tr"; "c" **\dir{~};
     "tl"; "c" **\dir{~};
     "c"; "b" **\dir{~};
 \endxy
\]
We can also describe the element $1 \in A$ using the
unique operator $i \maps \C \to A$ that sends the complex
number $1$ to $1 \in A$.  Then we can draw this operator
using a string diagram:
\[
 \xy
  (0,0)*\xycircle(3,3){-}="c";
  (0,0)*{i};
  (0,-10)*{}="b"+(-1,1)*{ };
     "c"; "b" **\dir{~};
 \endxy
\]
In this notation, the associative law looks like this:
\[
 \xy
  (0,10)*\xycircle(3,3){-}="c1";
  (0,10)*{m};
  (-6,20)*{}="tl";
  (6,20)*{}="tr";
  (0,0)*{}="b";
     "tr"; "c1" **\dir{~};
     "tl"; "c1" **\dir{~};
  (5,0)*\xycircle(3,3){-}="c";
  (5,0)*{m};
  (2,10)*{}="tl";
  (18,20)*{}="tr";
  (5,-15)*{}="b";
     "tr"; "c" **\dir{~};
     "c"; "c1" **\dir{~};
     "c"; "b" **\dir{~};
 \endxy
\qquad = \qquad
 \xy
  (0,10)*\xycircle(3,3){-}="c1";
  (0,10)*{m};
  (6,20)*{}="tl";
  (-6,20)*{}="tr";
  (0,0)*{}="b";
     "tr"; "c1" **\dir{~};
     "tl"; "c1" **\dir{~};
  (-5,0)*\xycircle(3,3){-}="c";
  (-5,0)*{m};
  (-2,10)*{}="tl";
  (-18,20)*{}="tr";
  (-5,-15)*{}="b";
     "tr"; "c" **\dir{~};
     "c"; "c1" **\dir{~};
     "c"; "b" **\dir{~};
 \endxy
\]
while the left and right unit laws look like this:
\[
 \xy
  (0,10)*\xycircle(3,3){-}="c1";
  (0,10)*{i};
  (-6,20)*{}="tl";
  (6,20)*{}="tr";
  (0,0)*{}="b";
  (5,0)*\xycircle(3,3){-}="c";
  (5,0)*{m};
  (2,10)*{}="tl";
  (18,20)*{}="tr";
  (5,-15)*{}="b";
      "tr"; "c" **\dir{~};
     "c"; "c1" **\dir{~};
     "c"; "b" **\dir{~};
 \endxy
\qquad = \qquad \qquad
\xy
  (0,20)*{}="t";
  (0,-15)*{}="b";
  "t"; "b" **\dir{~};
 \endxy
\qquad \qquad = \qquad
 \xy
  (0,10)*\xycircle(3,3){-}="c1";
  (0,10)*{i};
  (6,20)*{}="tl";
  (-6,20)*{}="tr";
  (0,0)*{}="b";
  (-5,0)*\xycircle(3,3){-}="c";
  (-5,0)*{m};
  (-2,10)*{}="tl";
  (-18,20)*{}="tr";
  (-5,-15)*{}="b";
     "tr"; "c" **\dir{~};
     "c"; "c1" **\dir{~};
     "c"; "b" **\dir{~};
 \endxy
\]

A representation of an algebra is a lot like a representation of
a group, except that instead of writing $\rho(g) v$ for the action
of a group element $g$ on a vector $v$, we write $\rho(a \otimes v)$
for the action of an algebra element $a$ on a vector $v$.  More
precisely, a {\bf representation} of an algebra $A$ is a
vector space $V$ equipped with an operator
\[       \rho \maps A \tensor V \to V  \]
satisfying these two laws:
\[           \rho(1 \tensor v) = v  , \qquad
           \rho(ab \tensor v) = \rho(a \tensor \rho(b \tensor v)) . \]
Using string diagrams can draw $\rho$ as follows:
\[
 \xy
  (0,0)*\xycircle(3,3){-}="c";
  (0,0)*{\rho};
  (-6,10)*{}="tl"+(-1,-3)*{ };
  (6,10)*{}="tr"+(1,-3)*{ };
  (0,-10)*{}="b"+(-1,1)*{ };
     "tr"; "c" **\dir{-};
     "tl"; "c" **\dir{~};
     "c"; "b" **\dir{-};
 \endxy
\]
Note that wiggly lines refer to the object $A$, while straight
ones refer to $V$.  Then the two laws obeyed by $\rho$
look very much like associativity and the left unit law:
\[
 \xy
  (0,10)*\xycircle(3,3){-}="c1";
  (0,10)*{m};
  (-6,20)*{}="tl";
  (6,20)*{}="tr";
  (0,0)*{}="b";
     "tr"; "c1" **\dir{~};
     "tl"; "c1" **\dir{~};
  (5,0)*\xycircle(3,3){-}="c";
  (5,0)*{\rho};
  (2,10)*{}="tl";
  (18,20)*{}="tr";
  (5,-15)*{}="b";
     "tr"; "c" **\dir{-};
     "c"; "c1" **\dir{~};
     "c"; "b" **\dir{-};
 \endxy
\qquad = \qquad
 \xy
  (0,10)*\xycircle(3,3){-}="c1";
  (0,10)*{\rho};
  (6,20)*{}="tl";
  (-6,20)*{}="tr";
  (0,0)*{}="b";
     "tr"; "c1" **\dir{~};
     "tl"; "c1" **\dir{-};
  (-5,0)*\xycircle(3,3){-}="c";
  (-5,0)*{\rho};
  (-2,10)*{}="tl";
  (-18,20)*{}="tr";
  (-5,-15)*{}="b";
     "tr"; "c" **\dir{~};
     "c"; "c1" **\dir{-};
     "c"; "b" **\dir{-};
 \endxy
\]
\[
 \xy
  (0,10)*\xycircle(3,3){-}="c1";
  (0,10)*{i};
  (-6,20)*{}="tl";
  (6,20)*{}="tr";
  (0,0)*{}="b";
  (5,0)*\xycircle(3,3){-}="c";
  (5,0)*{\rho};
  (2,10)*{}="tl";
  (18,20)*{}="tr";
  (5,-15)*{}="b";
      "tr"; "c" **\dir{-};
     "c"; "c1" **\dir{~};
     "c"; "b" **\dir{-};
 \endxy
\qquad = \qquad \qquad
 \xy
  (0,20)*{}="t";
  (0,-15)*{}="b";
     "t"; "b" **\dir{-};
 \endxy
\]

To make the representations of an algebra into the objects of
a category, we must define morphisms between them.  Given
two algebra representations, say $\rho \maps A \tensor
V \to V$ and $\rho' \maps A \tensor V' \to V'$,
we define an \textbf{intertwining operator} $f \maps V \to V'$
to be a linear operator such that
\[       f(\rho(a \otimes v)) = \rho'(a \otimes f(v)). \]
This closely resembles the definition of an
intertwining operator between group representations.  It says
that acting by $a \in A$ and then applying the intertwining operator
is the same as applying the intertwining operator and then acting
by $a$.

With these definitions, we obtain a category $\Rep(A)$ with
finite-dimensional representations of $A$ as objects and intertwining
operators as morphisms.  However, unlike group representations, there
is no way in general to define the tensor product of algebra
representations!  For this, we need $A$ to be a `bialgebra'.  To
understand what this means, first recall from our discussion of
Dijkgraaf's 1989 thesis that a {\bf coalgebra} is just like an algebra,
only upside-down.  More precisely, it is a vector space equipped with
a {\bf comultiplication}:
\[
 \xy
  (0,0)*\xycircle(3,3){-}="c";
  (0,0)*{\Delta};
  (-6,-10)*{}="tl"+(-1,-3)*{ };
  (6,-10)*{}="tr"+(1,-3)*{ };
  (0,10)*{}="b"+(-1,1)*{ };
     "tr"; "c" **\dir{~};
     "tl"; "c" **\dir{~};
     "c"; "b" **\dir{~};
 \endxy
\]
and {\bf counit}:
\[
 \xy
  (0,0)*\xycircle(3,3){-}="c";
  (0,0)*{\varepsilon};
  (0,10)*{}="b"+(-1,1)*{ };
     "c"; "b" **\dir{~};
 \endxy
\]
satisfying the {\bf coassociative law}:
\[
 \xy
  (0,-10)*\xycircle(3,3){-}="c1";
  (0,-10)*{\Delta};
  (-6,-20)*{}="tl";
  (6,-20)*{}="tr";
  (0,0)*{}="b";
     "tr"; "c1" **\dir{~};
     "tl"; "c1" **\dir{~};
  (5,0)*\xycircle(3,3){-}="c";
  (5,0)*{\Delta};
  (2,-10)*{}="tl";
  (18,-20)*{}="tr";
  (5,15)*{}="b";
     "tr"; "c" **\dir{~};
     "c"; "c1" **\dir{~};
     "c"; "b" **\dir{~};
 \endxy
\qquad = \qquad
 \xy
  (0,-10)*\xycircle(3,3){-}="c1";
  (0,-10)*{\Delta};
  (6,-20)*{}="tl";
  (-6,-20)*{}="tr";
  (0,0)*{}="b";
     "tr"; "c1" **\dir{~};
     "tl"; "c1" **\dir{~};
  (-5,0)*\xycircle(3,3){-}="c";
  (-5,0)*{\Delta};
  (-2,-10)*{}="tl";
  (-18,-20)*{}="tr";
  (-5,15)*{}="b";
     "tr"; "c" **\dir{~};
     "c"; "c1" **\dir{~};
     "c"; "b" **\dir{~};
 \endxy
\]
and left/right {\bf counit laws}:
\[
 \xy
  (0,-10)*\xycircle(3,3){-}="c1";
  (0,-10)*{\varepsilon};
  (-6,-20)*{}="tl";
  (6,-20)*{}="tr";
  (0,0)*{}="b";
  (5,0)*\xycircle(3,3){-}="c";
  (5,0)*{\Delta};
  (2,-10)*{}="tl";
  (18,-20)*{}="tr";
  (5,15)*{}="b";
      "tr"; "c" **\dir{~};
     "c"; "c1" **\dir{~};
     "c"; "b" **\dir{~};
 \endxy
\qquad = \qquad \qquad
\xy
  (0,15)*{}="t";
  (0,-20)*{}="b";
  "t"; "b" **\dir{~};
 \endxy
\qquad \qquad = \qquad
 \xy
  (0,-10)*\xycircle(3,3){-}="c1";
  (0,-10)*{\varepsilon};
  (6,-20)*{}="tl";
  (-6,-20)*{}="tr";
  (0,0)*{}="b";
  (-5,0)*\xycircle(3,3){-}="c";
  (-5,0)*{\Delta};
  (-2,-10)*{}="tl";
  (-18,-20)*{}="tr";
  (-5,15)*{}="b";
     "tr"; "c" **\dir{~};
     "c"; "c1" **\dir{~};
     "c"; "b" **\dir{~};
 \endxy
\]

A {\bf bialgebra} is a vector space equipped with an algebra and
coalgebra structure that are compatible in a certain way.  We have
already seen that a Frobenius algebra is both an algebra and a
coalgebra, with the multiplication and comultiplication obeying the
compatibility conditions in Equation \ref{frobenius}.  A bialgebra
obeys \textit{different} compatibility conditions.  These can be drawn
using string diagrams, but it is more enlightening to note that they
are precisely the conditions we need to make the category of
representations of an algebra $A$ into a {\it monoidal} category.  The
idea is that the comultiplication $\Delta \maps A \to A \otimes A$
lets us `duplicate' an element $A$ so it can act on both factors in a
tensor product of representations, say $\rho$ and $\rho'$:
\[
 \xy
  (-10,10)*\xycircle(3,3){-}="ct";  (-10,10)*{\Delta};
  (-5,-8)*\xycircle(3,3){-}="cbl";   (-5,-8)*{\rho};
  (10,-8)*\xycircle(3,3){-}="cbr";   (10,-8)*{\rho'};
  (-10,22)*{}="tl"; (5,22)*{}="tm";  (20,22)*{}="tr";
  (-5,-20)*{}="bl";
    (10,-20)*{}="br";
     "tl"; "ct" **\dir{~};
     "tm"; "cbl" **\crv{}; \POS?(.3)*{\hole}="break";
     {\ar@{~}@/^1pc/ "cbl"; "ct"};
     "break";"cbr" **\dir{~};
     "break";"ct" **\dir{~};
     "cbl"; "bl" **\dir{-};
     "cbr"; "br" **\dir{-};
     "tr"; "cbr" **\dir{-};
 \endxy
 \]
This gives $\Rep(A)$ a tensor product.  Similarly, we use the counit
to let $A$ act on $\C$ as follows:
\[
\xy
  (-5,0)*\xycircle(3,3){-}="c";
  (-5,0)*{\varepsilon};
  (-5,15)*{}="b"+(-1,1)*{ };
     "c"; "b" **\dir{~};
  (5,15)*{}="tr";
  (5,-15)*{}="br";
  "tr"; "br" **\dir{-};
  (8,10)*{\C};
 \endxy
\]
We can then write down equations saying that $\Rep(A)$ is a
monoidal category with the same associator and unitors as in $\Vect$,
and with $\C$ as its unit object.  These equations are then the
definition of `bialgebra'.

As we have seen, the category of representations of a compact Lie
group $K$ is also a monoidal category.  In this sense, bialgebras are
a generalization of such groups.  Indeed, there is a way to turn any
group of this sort into a bialgebra $A$, and when the group is
simply connected, this bialgebra has an equivalent category of
representations:
\[             \Rep(K) \simeq \Rep(A)   .\]
So, as far as its representations are concerned, there is really no
difference.  But a big advantage of bialgebras is that we can
often `deform' them to obtain new bialgebras that \textit{don't}
come from groups.

The most important case is when $K$ is not only simply-connected and
compact, but also \textbf{simple}, which for Lie groups means that all
its normal subgroups are finite.  We have already been discussing an example:
$\SU(2)$.  Groups of this sort were classified by \'Elie
Cartan in 1894, and by the mid-1900s their theory had grown to one of
the most enormous and beautiful edifices in mathematics.  The fact
that one can deform them to get interesting bialgebras called `quantum
groups' opened a brand new wing in this edifice, and the experts
rushed in.

A basic fact about groups of this sort is that they have `complex
forms'.  For example, $\SU(2)$ has the complex form $\SL(2)$,
consisting of $2 \times 2$ complex matrices with determinant 1.  This
group contains $\SU(2)$ as a subgroup.  The advantage of $\SU(2)$ is
that it is compact, which implies that its finite-dimensional
continuous representations can always be made unitary.  The advantage
of $\SL(2)$ is that it is a complex manifold, with all the group
operations being analytic functions; this allows us to define
`analytic' representations of this group.  For our purposes, another
advantage of $\SL(2)$ is that its Lie algebra is a complex vector space.
Luckily we do not have to choose one group over the other, since the
finite-dimensional continuous unitary representations of $\SU(2)$
correspond precisely to the finite-dimensional analytic representations
of $\SL(2)$.  And as emphasized by Hermann Weyl, \textit{every}
simply-connected compact simple Lie group $K$ has a complex Lie group
$G$ for which this relation holds!

These facts let us say a bit more about how to
get a bialgebra with the same representations as our group $K$.
First, we take the complex form $G$ of the group $K$, and consider its
Lie algebra, $\g$.  Then we let $\g$ freely generate an algebra in
which these relations hold:
\[                x y - y x = [x,y]   \]
for all $x,y \in \g$.  This algebra is called the \textbf{universal
enveloping algebra} of $\g$, and denoted $\U\g$.  It is in fact a
bialgebra, and we have an equivalence of monoidal categories:
\[             \Rep(K) \simeq \Rep(\U\g)  .\]

What Drinfel'd discovered is that we can `deform' $\U\g$ and get a
\textbf{quantum group} $\U_q \g$.  This is a family of bialgebras
depending on a complex parameter $q$, with the property that
$\U_q \g \cong U \g$ when $q = 1$.  Moreover, these bialgebras
are unique, up to changes of the parameter $q$ and other
inessential variations.

In fact, quantum groups are much better than mere bialgebras:
they are `quasitriangular Hopf algebras'.  This is just an
intimidating way of saying that $\Rep(\U_q \g)$ is not merely a
monoidal category, but in fact a braided monoidal category with duals
for objects.  And this, in turn, is just an intimidating way of saying
that any representation of $U_q \g$ gives an invariant of framed
oriented tangles!  Reshetikhin and Turaev's paper explained exactly
how this works.

If all this seems too abstract, take $K = \SU(2)$.  From what we have
already said, these categories are equivalent:
\[          \Rep(\SU(2)) \simeq  \Rep(\U \ssl(2))  \]
where $\ssl(2)$ is the Lie algebra of $\SL(2)$.  So, we get a braided
monoidal category with duals for objects, $\Rep(\U_q \ssl(2))$, which
reduces to $\Rep(\SU(2))$ when we set $q = 1$.  This is why $\U_q \ssl(2)$ is
often called `quantum $\SU(2)$', especially in the physics literature.

Even better, the quantum group $\U_q \ssl(2)$ has a 2-dimensional
representation which reduces to the usual spin-$\frac{1}{2}$
representation of $\SU(2)$ at $q = 1$.  Using this representation to
get a tangle invariant, we obtain the Kauffman bracket---at least up
to some minor normalization issues that we shall ignore here.  So,
Reshetikhin and Turaev's paper massively generalized the Kauffman
bracket and set it into its proper context: the representation theory
of quantum groups!

In our discussion of Kontsevich's 1993 paper we will sketch how
to actually get our hands on quantum groups.

\subsection*{Witten (1989)}

In the 1980s there was a lot of work on the Jones polynomial
\cite{Kohno}, leading up to the result we just sketched: a beautiful
description of this invariant in terms of representations of quantum
$\SU(2)$.  Most of this early work on the Jones polynomial used
2-dimensional pictures of knots and tangles---the string diagrams we
have been discussing here.  This was unsatisfying in one respect:
researchers wanted an intrinsically 3-dimensional description of the
Jones polynomial.

In his paper `Quantum field theory and the Jones polynomial'
\cite{Witten}, Witten gave such a description using a gauge field
theory in 3d spacetime, called Chern--Simons theory.  He also
described how the category of representations of $\SU(2)$ could be
deformed into the category of representations of quantum $\SU(2)$
using a conformal field theory called the Wess--Zumino--Witten model,
which is closely related to Chern--Simons theory.  We shall say a
little about this in our discussion of Kontsevich's 1993 paper.

\subsection*{Rovelli--Smolin (1990)}

Around 1986, Abhay Ashtekar discovered a new formulation of general
relativity, which made it more closely resemble gauge theories such as
Yang--Mills theory \cite{Ashtekar1}.  In 1990, Rovelli and Smolin
\cite{RS} published a paper that used this to develop a new approach
to the old and difficult problem of quantizing gravity --- that is,
treating it as a quantum rather than a classical field theory.  This
approach is usually called `loop quantum gravity', but in its later
development it came to rely heavily on Penrose's spin networks
\cite{RS2,Baez2}.  It reduces to the Ponzano--Regge model in the case
of 3-dimensional quantum gravity; the difficult and so far unsolved
challenge is finding a correct treatment of 4-dimensional quantum
gravity in this approach, if one exists.

As we have seen, spin networks are mathematically like Feynman
diagrams with the Poincar\'e group replaced by $\SU(2)$.  However,
Feynman diagrams describe \textit{processes} in ordinary quantum field
theory, while spin networks describe \textit{states} in loop quantum
gravity.  For this reason it seemed natural to explore the possibility
that some sort of 2-dimensional diagrams going between spin networks
are needed to describe processes in loop quantum gravity.  These were
introduced by Reisenberger and Rovelli in 1996 \cite{RR}, and further
formalized and dubbed `spin foams' in 1997 \cite{Baez2,Baez3}.  As we shall
see, just as Feynman diagrams can be used to do computations in
categories like the category of Hilbert spaces, spin foams can be used
to do computations in bicategories like the bicategory of `2-Hilbert
spaces'.

For a review of loop quantum gravity and spin foams with plenty of
references for further study, start with the article by Rovelli
\cite{Rovelli1}.  Then try his book \cite{Rovelli2} and the book
by Ashtekar \cite{Ashtekar2}.

\subsection*{Kashiwara and Lusztig~(1990)}

Every matrix can be written as a sum of a lower triangular matrix,
a diagonal matrix and an upper triangular matrix.  Similarly, for
every simple Lie algebra $\g$, the quantum group $U_q\mathfrak{g}$
has a `triangular decomposition'
\[
U_q\mathfrak{g} \cong
U_q^-\mathfrak{g} \tensor U_q^0\mathfrak{g} \tensor U_q^+\mathfrak{g}.
\]
If one is interested in the braided monoidal category of finite
dimensional representations of $U_q \mathfrak{g}$, then it
turns out that one only needs to understand the lower triangular part
$U_q^-\mathfrak{g}$ of the quantum group.  Using a sophisticated
geometric approach Lusztig~\cite{Lus1,Lus2} defined a
basis for $U_q^-\mathfrak{g}$ called the `canonical
basis', which has remarkable properties.  Using algebraic methods,
Kashiwara~\cite{Kas1, Kas2, Kas3} defined a `global crystal basis'
for $U_q^-(\mathfrak{g})$, which was later shown by Grojnowski and
Lusztig~\cite{GL2} to coincide with the canonical basis.

What makes the canonical basis so
interesting is that given two basis elements $e^i$ and $e^j$,
their product $e^i e^j$ can be expanded in terms of basis elements
\[
e^i e^j = \sum_{k} m^{ij}_k e^k
\]
where the constants $m^{ij}_k$ are \textit{polynomials} in $q$
and $q^{-1}$, and these polynomials have \textit{natural numbers} as
coefficients.   If we had chosen a basis at random, we would only
expect these constants to be rational functions of $q$, with rational
numbers as coefficients.

The appearance of natural numbers here hints that quantum groups are
just shadows of more interesting structures where the canonical basis
elements become objects of a category, multiplication becomes the
tensor product in this category, and addition becomes direct sum in
this category.  Such a structure could be called a
\textit{categorified} quantum group.  Its existence was explicitly
conjectured in a paper by Crane and Frenkel, which we will discuss
below.  Indeed, this was already visible in Lusztig's geometric
approach to studying quantum groups using so-called `perverse sheaves'
~\cite{Lus3}.

For a simpler example of this phenomenon, recall our discussion of
Penrose's 1971 paper.  We saw that if $K$ is a compact Lie group, the
category $\Rep(K)$ has a tensor product and direct sums.  If we pick
one irreducible representation $E^i$ from each isomorphism class, then
every object in $\Rep(K)$ is a direct sum of these objects $E^i$,
which thus act as a kind of `basis' for $\Rep(K)$.  As a result, we
have
\[
E^i \tensor E^j \iso \bigoplus_{k} M^{ij}_k \tensor E^k  \]
for certain finite-dimensional vector spaces $M^{ij}_k$.
The dimensions of these vector spaces, say
\[         m^{ij}_k = \dim(M^{ij}_k) , \]
are \textit{natural numbers}.  We can define an algebra with one
basis vector $e^i$ for each $E^i$, and with a multiplication defined
by
\[
e^i e^j = \sum_{k} m^{ij}_k e^k
\]
This algebra is called the \textbf{representation ring} of $K$, and
denoted $R(K)$.  It is associative because the tensor product in
$\Rep(K)$ is associative up to isomorphism.

In fact, representation rings were discovered before categories of
representations.   Suppose someone had handed us such  ring and
asked us to explain it.  Then the fact that it had a basis where the
constants $m^{ij}_k$ are natural numbers would be a clue that it came
from a monoidal category with direct sums!

The special properties of the canonical basis are a similar clue, but
here there is an extra complication: instead of natural numbers,
we are getting polynomials in $q$ and $q^{-1}$ with natural number
coefficients.   We shall give an explanation of this later, in our
discussion of Khovanov's 1999 paper.

\subsection*{Kapranov--Voevodsky (1991)}

Around 1991, Kapranov and Voevodsky made available a preprint in which
they initiated work on `2-vector spaces' and what we now call `braided
monoidal bicategories' \cite{KV}.  They also studied a
higher-dimensional analogue of the Yang--Baxter equation called the
`Zamolodchikov tetrahedron equation'.  Recall from our discussion of
Joyal and Street's 1985 paper that any solution of the Yang--Baxter
equation gives a braided monoidal category.  Kapranov and Voevodsky
argued that similarly, any solution of the Zamolodchikov tetrahedron
equation gives a braided monoidal bicategory.

The basic idea of a braided monoidal bicategory is straightforward: it
is like a braided monoidal category, but with a bicategory replacing
the underlying category.  This lets us `weaken' equational laws
involving 1-morphisms, replacing them by specified 2-isomorphisms.  To
obtain a useful structure we also need to impose equational laws on
these 2-isomorphisms---so-called `coherence laws'.  This is the tricky
part, which is why Kapranov and Voevodsky's original definition of
`semistrict braided monoidal 2-category' required a number of fixes
\cite{hda1,Crans,DS}, leading ultimately to the fully general concept
of braided monoidal bicategory introduced by McCrudden \cite{McCrudden}.

However, their key insight was striking and robust.  As we have seen,
any object in a braided monoidal category gives an isomorphism
\[          B = B_{x,x} \maps x \tensor x \to x \tensor x  \]
satisfying the Yang--Baxter equation
\[   (B \otimes 1)(1 \otimes B)(B \otimes 1) =
          (1 \otimes B)(B \otimes 1)(1 \otimes B)  \]
which in pictures corresponds to the third Reidemeister move.
In a braided monoidal bicategory, the Yang--Baxter equation holds only
up to a 2-isomorphism
\[  Y \maps (B \otimes 1)(1 \otimes B)(B \otimes 1) \To
          (1 \otimes B)(B \otimes 1)(1 \otimes B)  \]
which in turn satisfies the `Zamolodchikov tetrahedron equation'.

This equation is best understood using diagrams.  If we think of $Y$
as the surface in 4-space traced out by the process of performing the
third Reidemeister move:
\[   Y \maps
\def\objectstyle{\scriptstyle}
\def\labelstyle{\scriptstyle}
  \xy
   (12,15)*{}="C";
   (4,15)*{}="B";
   (-7,15)*{}="A";
   (12,-15)*{}="3";
   (-3,-15)*{}="2";
   (-12,-15)*{}="1";
       "C";"1" **\crv{(15,0)& (-15,0)};
       (-5,-5)*{}="2'";
       (7,2)*{}="3'";
     "2'";"2" **\crv{};
     "3'";"3" **\crv{(7,-8)};
       \vtwist~{"A"}{"B"}{(-6,-1)}{(6,5.7)};
\endxy
 \quad \To \quad \xy
   (-12,-15)*{}="C";
   (-4,-15)*{}="B";
   (7,-15)*{}="A";
   (-12,15)*{}="3";
   (3,15)*{}="2";
   (12,15)*{}="1";
       "C";"1" **\crv{(-15,0)& (15,0)};
       (5,5)*{}="2'";
       (-7,-2)*{}="3'";
     "2'";"2" **\crv{(4,6)};
     "3'";"3" **\crv{(-7,8)};
       \vtwist~{"A"}{"B"}{(6,1)}{(-6,-5.7)};
\endxy
\]
then the \textbf{Zamolodchikov tetrahedron equation} says the surface
traced out by first performing the third Reidemeister move on a
threefold crossing and then sliding the result under a fourth strand
is isotopic to that traced out by first sliding the threefold crossing
under the fourth strand and then performing the third Reidemeister
move.  So, this octagon commutes:
\[
  \xy 0;/r.13pc/:   
    (0,50)*+{       
 \xy 
 (-15,-20)*{}="T1";
 (-5,-20)*{}="T2";
 (5,-20)*{}="T3";
 (15,-20)*{}="T4";
 (-14,20)*{}="B1";
 (-5,20)*{}="B2";
 (5,20)*{}="B3";
 (15,20)*{}="B4";
    "T1"; "B4" **\crv{(-15,-7) & (15,-5)}
        \POS?(.25)*{\hole}="2x" \POS?(.47)*{\hole}="2y" \POS?(.6)*{\hole}="2z";
    "T2";"2x" **\crv{(-4,-12)};
    "T3";"2y" **\crv{(5,-10)};
    "T4";"2z" **\crv{(16,-9)};
 (-15,-5)*{}="3x";
    "2x"; "3x" **\crv{(-18,-10)};
    "3x"; "B3" **\crv{(-13,0) & (4,10)}
        \POS?(.3)*{\hole}="4x" \POS?(.53)*{\hole}="4y";
    "2y"; "4x" **\crv{};
    "2z"; "4y" **\crv{};
 (-15,10)*{}="5x";
    "4x";"5x" **\crv{(-17,6)};
    "5x";"B2" **\crv{(-14,12)}
        \POS?(.6)*{\hole}="6x";
    "6x";"B1" **\crv{(-14,18)};
    "4y";"6x" **\crv{(-8,10)};
 \endxy
    }="T";
    (-40,30)*+{
 \xy 
 (-15,-20)*{}="b1";
 (-5,-20)*{}="b2";
 (5,-20)*{}="b3";
 (14,-20)*{}="b4";
 (-14,20)*{}="T1";
 (-5,20)*{}="T2";
 (5,20)*{}="T3";
 (15,20)*{}="T4";
    "b1"; "T4" **\crv{(-15,-7) & (15,-5)}
        \POS?(.25)*{\hole}="2x" \POS?(.47)*{\hole}="2y" \POS?(.65)*{\hole}="2z";
    "b2";"2x" **\crv{(-5,-15)};
    "b3";"2y" **\crv{(5,-10)};
    "b4";"2z" **\crv{(14,-9)};
 (-15,-5)*{}="3x";
    "2x"; "3x" **\crv{(-15,-10)};
    "3x"; "T3" **\crv{(-15,15) & (5,10)}
        \POS?(.38)*{\hole}="4y" \POS?(.65)*{\hole}="4z";
    "T1";"4y" **\crv{(-14,16)};
    "T2";"4z" **\crv{(-5,16)};
    "2y";"4z" **\crv{(-10,3) & (10,2)} \POS?(.6)*{\hole}="5z";
    "4y";"5z" **\crv{(-5,5)};
    "5z";"2z" **\crv{(5,4)};
 \endxy
    }="TL";
    (-75,0)*+{
 \xy 
 (-14,20)*{}="T1";
 (-4,20)*{}="T2";
 (4,20)*{}="T3";
 (15,20)*{}="T4";
 (-15,-20)*{}="B1";
 (-5,-20)*{}="B2";
 (5,-20)*{}="B3";
 (15,-20)*{}="B4";
    "B1";"T4" **\crv{(-15,5) & (15,-5)}
        \POS?(.25)*{\hole}="2x" \POS?(.49)*{\hole}="2y" \POS?(.65)*{\hole}="2z";
    "2x";"T3" **\crv{(-20,10) & (5,10) }
        \POS?(.45)*{\hole}="3y" \POS?(.7)*{\hole}="3z";
    "2x";"B2" **\crv{(-5,-14)};
        "T1";"3y" **\crv{(-16,17)};
        "T2";"3z" **\crv{(-5,17)};
        "3z";"2z" **\crv{};
        "3y";"2y" **\crv{};
        "B3";"2z" **\crv{ (5,-5) &(20,-10)}
            \POS?(.4)*{\hole}="4z";
        "2y";"4z" **\crv{(6,-8)};
        "4z";"B4" **\crv{(15,-15)};
 \endxy
    }="ML";
    (-40,-30)*+{
 \xy 
 (-14,20)*{}="T1";
 (-4,20)*{}="T2";
 (4,20)*{}="T3";
 (15,20)*{}="T4";
 (-15,-20)*{}="B1";
 (-5,-20)*{}="B2";
 (5,-20)*{}="B3";
 (15,-20)*{}="B4";
    "B1";"T4" **\crv{(-15,-5) & (15,5)}
        \POS?(.38)*{\hole}="2x" \POS?(.53)*{\hole}="2y" \POS?(.7)*{\hole}="2z";
    "T1";"2x" **\crv{(-15,5)};
    "2y";"B2" **\crv{(10,-10) & (-6,-10)}
        \POS?(.45)*{\hole}="4x";
    "2z";"B3" **\crv{ (15,0)&(15,-10) & (6,-16)}
        \POS?(.7)*{\hole}="5x";
    "T3";"2y" **\crv{(5,10)& (-6,18) }
        \POS?(.5)*{\hole}="3x";
    "T2";"3x" **\crv{(-5,15)};
    "3x";"2z" **\crv{(7,11)};
    "2x";"4x" **\crv{(-3,-7)};
    "4x";"5x" **\crv{};
    "5x";"B4" **\crv{(15,-15)};
 \endxy
    }="BL";
    (40,30)*+{
 \xy 
 (14,-20)*{}="T1";
 (4,-20)*{}="T2";
 (-4,-20)*{}="T3";
 (-15,-20)*{}="T4";
 (15,20)*{}="B1";
 (5,20)*{}="B2";
 (-5,20)*{}="B3";
 (-15,20)*{}="B4";
    "B1";"T4" **\crv{(15,5) & (-15,-5)}
        \POS?(.38)*{\hole}="2x" \POS?(.53)*{\hole}="2y" \POS?(.7)*{\hole}="2z";
    "T1";"2x" **\crv{(15,-5)};
    "2y";"B2" **\crv{(-10,10) & (6,10)}
        \POS?(.45)*{\hole}="4x";
    "2z";"B3" **\crv{ (-15,0)&(-15,10) & (-6,16)}
        \POS?(.7)*{\hole}="5x";
    "T3";"2y" **\crv{(-5,-10)& (6,-18) }
        \POS?(.5)*{\hole}="3x";
    "T2";"3x" **\crv{(5,-15)};
    "3x";"2z" **\crv{(-7,-11)};
    "2x";"4x" **\crv{(3,7)};
    "4x";"5x" **\crv{};
    "5x";"B4" **\crv{(-15,15)};
 \endxy
    }="TR";
    (75,0)*+{
 \xy 
 (14,-20)*{}="T1";
 (4,-20)*{}="T2";
 (-4,-20)*{}="T3";
 (-15,-20)*{}="T4";
 (15,20)*{}="B1";
 (5,20)*{}="B2";
 (-5,20)*{}="B3";
 (-15,20)*{}="B4";
    "B1";"T4" **\crv{(15,-5) & (-15,5)}
        \POS?(.25)*{\hole}="2x" \POS?(.49)*{\hole}="2y" \POS?(.65)*{\hole}="2z";
    "2x";"T3" **\crv{(20,-10) & (-5,-10) }
        \POS?(.45)*{\hole}="3y" \POS?(.7)*{\hole}="3z";
    "2x";"B2" **\crv{(5,14)};
        "T1";"3y" **\crv{(16,-17)};
        "T2";"3z" **\crv{(5,-17)};
        "3z";"2z" **\crv{};
        "3y";"2y" **\crv{};
        "B3";"2z" **\crv{ (-5,5) &(-20,10)}
            \POS?(.4)*{\hole}="4z";
        "2y";"4z" **\crv{(-6,8)};
        "4z";"B4" **\crv{(-15,15)};
 \endxy
    }="MR";
    (40,-30)*+{
 \xy 
 (15,20)*{}="b1";
 (5,20)*{}="b2";
 (-5,20)*{}="b3";
 (-14,20)*{}="b4";
 (14,-20)*{}="T1";
 (5,-20)*{}="T2";
 (-5,-20)*{}="T3";
 (-15,-20)*{}="T4";
    "b1"; "T4" **\crv{(15,7) & (-15,5)}
        \POS?(.25)*{\hole}="2x" \POS?(.47)*{\hole}="2y" \POS?(.65)*{\hole}="2z";
    "b2";"2x" **\crv{(5,15)};
    "b3";"2y" **\crv{(-5,10)};
    "b4";"2z" **\crv{(-14,9)};
 (15,5)*{}="3x";
    "2x"; "3x" **\crv{(15,10)};
    "3x"; "T3" **\crv{(15,-15) & (-5,-10)}
        \POS?(.38)*{\hole}="4y" \POS?(.65)*{\hole}="4z";
    "T1";"4y" **\crv{(14,-16)};
    "T2";"4z" **\crv{(5,-16)};
    "2y";"4z" **\crv{(10,-3) & (-10,-2)} \POS?(.6)*{\hole}="5z";
    "4y";"5z" **\crv{(5,-5)};
    "5z";"2z" **\crv{(-5,-4)};
 \endxy
    }="BR";
    (0,-50)*+{
 \xy 
 (15,20)*{}="T1";
 (5,20)*{}="T2";
 (-5,20)*{}="T3";
 (-15,20)*{}="T4";
 (15,-20)*{}="B1";
 (5,-20)*{}="B2";
 (-5,-20)*{}="B3";
 (-15,-20)*{}="B4";
    "T1"; "B4" **\crv{(15,7) & (-15,5)}
        \POS?(.25)*{\hole}="2x" \POS?(.45)*{\hole}="2y" \POS?(.6)*{\hole}="2z";
    "T2";"2x" **\crv{(4,12)};
    "T3";"2y" **\crv{(-5,10)};
    "T4";"2z" **\crv{(-16,9)};
 (15,5)*{}="3x";
    "2x"; "3x" **\crv{(18,10)};
    "3x"; "B3" **\crv{(13,0) & (-4,-10)}
        \POS?(.3)*{\hole}="4x" \POS?(.53)*{\hole}="4y";
    "2y"; "4x" **\crv{};
    "2z"; "4y" **\crv{};
 (15,-10)*{}="5x";
    "4x";"5x" **\crv{(17,-6)};
    "5x";"B2" **\crv{(14,-12)}
        \POS?(.6)*{\hole}="6x";
    "6x";"B1" **\crv{};
    "4y";"6x" **\crv{};
 \endxy
    }="B";
            (-20,65)*{}="X1";
            (-35,55)*{}="X2";               
                {\ar@{=>} "X1";"X2"};       
            (20,65)*{}="X1";                
            (35,55)*{}="X2";                
                {\ar@{=>} "X1";"X2"};       
            (60,40)*{}="X1";                
            (75,25)*{}="X2";                
                {\ar@{=>} "X1";"X2"};       
            (-60,40)*{}="X1";               
            (-75,25)*{}="X2";               
                {\ar@{=>} "X1";"X2"};
            (-60,-40)*{}="X2";
            (-75,-25)*{}="X1";
                {\ar@{=>} "X1";"X2"};
            (60,-40)*{}="X2";
            (75,-25)*{}="X1";
                {\ar@{=>} "X1";"X2"};
            (-20,-65)*{}="X2";
            (-35,-55)*{}="X1";
                {\ar@{=>} "X1";"X2"};
            (20,-65)*{}="X2";
            (35,-55)*{}="X1";
                {\ar@{=>} "X1";"X2"};
  \endxy
\]

Just as the Yang--Baxter equation relates two different planar
projections of 3 lines in $\R^3$, the Zamolodchikov tetrahedron
relates two different projections onto $\R^3$ of 4 lines in $\R^4$.
This suggests that solutions of the Zamolodchikov equation can give
invariants of `2-dimensional tangles' in 4-dimensional space (roughly,
surfaces embedded in 4-space) just as solutions of the Yang--Baxter
equation can give invariants of tangles (roughly, curves embedded in
3-space).  Indeed, this was later confirmed \cite{CS,CRS,hda4}.

Drinfel'd's work on quantum groups naturally gives solutions of the
Yang--Baxter equation in the category of vector spaces.  This suggested
to Kapranov and Voevodsky the idea of looking for solutions of the
Zamolodchikov tetrahedron equation in some bicategory of `2-vector
spaces'.  They defined 2-vector spaces using the following analogy:

\vskip 1em
\begin{center}
{\small
\begin{tabular}{|c|c|}                               \hline
$\C$        &  $\Vect$                       \\
$+$         &  $\oplus$                      \\
$\times$    &  $\otimes$                     \\
$0$         &  $\{0\}$                       \\
$1$         &  $\C$                          \\    \hline
\end{tabular}} \vskip 1em
Analogy between ordinary linear algebra and higher linear algebra
\end{center}

So, just as a finite-dimensional vector space may be defined as a set
of the form $\C^n$, they defined a {\bf 2-vector space} to be a
category of the form $\Vect^n$.  And just as a linear operator $T
\maps \C^n \to \C^m$ may be described using an $m \times n$ matrix of
complex numbers, they defined a {\bf linear functor} between 2-vector
spaces to be an $m \times n$ matrix of vector spaces!  Such matrices
indeed act to give functors from $\Vect^n$ to $\Vect^m$.  We
can also add and multiply such matrices in the usual way, but with
$\oplus$ and $\otimes$ taking the place of $+$ and $\times$.

Finally, there is a new layer of structure: given two linear functors
$S,T \maps \Vect^n$ $\to \Vect^m$, Kapranov and Voevodsky
defined a {\bf linear natural transformation} $\alpha \maps S \To T$
to be an $m \times n$ matrix of linear operators
\[         \alpha_{ij} \maps S_{ij} \to T_{ij}  \]
going between the vector spaces that are the matrix entries for $S$
and $T$.  This new layer of structure winds up making 2-vector spaces
into the objects of a {\it bicategory}.

Kapranov and Voevodsky called this bicategory $2\Vect$.  They also
defined a tensor product for 2-vector spaces, which turns out to make
$2\Vect$ into a `monoidal bicategory'.  The Zamolodchikov tetrahedron
equation makes sense in any monoidal bicategory, and any solution
gives a \textsl{braided} monoidal bicategory.  Conversely, any object
in a braided monoidal bicategory gives a solution of the Zamolodchikov
tetrahedron equation.  These results hint that the relation between
quantum groups, solutions of the Yang--Baxter equation, braided
monoidal categories and 3d topology is not a freak accident: all these
concepts may have higher-dimensional analogues!  To reach these
higher-dimensional analogues, it seems we need to take concepts and
systematically `boost their dimension' by making the following
replacements:

\vskip 2em
\vbox{
\begin{center}
{\small
\begin{tabular}{|c|c|}                    \hline
elements    &  objects                       \\     \hline
equations   &  isomorphisms                  \\
between elements   &  between objects        \\     \hline
sets        &  categories                    \\     \hline
functions   &  functors                      \\     \hline
equations   &  natural isomorphisms    \\
between functions   & between functors       \\     \hline
\end{tabular}}
\vskip 1em
Analogy between set theory and category theory
\end{center}
}
\vskip 0.5em

\noindent
In their 1994 paper, Crane and Frenkel called this process of
dimension boosting \textbf{categorification}.  We have already
seen, for example, that the representation category $\Rep(K)$
of a compact Lie group is a categorification of its representation
ring $R(K)$.  The representation ring is a vector space; the
representation category is a 2-vector space.  In fact the representation
ring is an algebra, and as we shall in our discussion of Barrett
and Westbury's 1992 paper, the representation category is a
`2-algebra'.

\subsection*{Reshetikhin--Turaev (1991)}

In 1991, Reshetikhin and Turaev \cite{RT2} published a paper in which
they constructed invariants of 3-manifolds from quantum groups.  These
invariants were later seen to be part of a full-fledged 3d TQFT.
Their construction made rigorous ideas from Witten's 1989 paper on
Chern--Simons theory and the Jones polynomial, so this TQFT is now
usually called the Witten--Reshetikhin--Turaev theory.

Their construction uses representations of a quantum group $U_q \g$,
but not the whole category $\Rep(U_q \g)$.  Instead they use a special
subcategory, which can be constructed when $q$ is a suitable root of
unity.  This subcategory has many nice properties: for example, it is
a braided monoidal category with duals, and also a 2-vector space with
a \textit{finite} basis of simple object.  These and some extra
properties are summarized by saying that this subcategory is a
`modular tensor category'.  Such categories were later intensively
studied by Turaev \cite{Turaev} and many others
\cite{BakalovKirillov}.  In this work, the Witten--Reshetikhin--Turaev
construction was generalized to obtain a 3d TQFT from any modular
tensor category.  Moreover, it was shown that any quantum group $U_q
\g$ gives rise to a modular tensor category when $q$ is a suitable
root of unity.

However, it was later seen that in most cases there is a 4d TQFT of
which the Witten--Reshetikhin--Turaev TQFT in 3 dimensions is merely a
kind of side-effect.  So, for the purposes of understanding the
relation between $n$-categories and TQFTs in various dimensions, it is
better to postpone further treatment of the
Witten--Reshetikhin--Turaev theory until our discussion of Turaev's
1992 paper on the 4-dimensional aspect of this theory.

\subsection*{Turaev--Viro (1992)}

In 1992, the topologists Turaev and Viro \cite{TV} constructed another
invariant of 3-manifolds---which we now know is part of a full-fledged
3d TQFT---from the modular category arising from quantum $\SU(2)$.
Their construction was later generalized to all modular tensor
categories, and indeed beyond.  By now, any 3d TQFT arising via this
construction is called a Turaev--Viro model.

The relation between the Turaev--Viro model and the
Witten--Reshetikhin--Turaev theory is subtle and interesting, but for
our limited purposes a few words will suffice.  Briefly: it later
became clear that a sufficiently nice {\it braided monoidal} category
lets us construct a {\it 4-dimensional} TQFT, which has a
Witten--Reshetikhin--Turaev TQFT in 3 dimensions as a kind of shadow.
On the other hand, Barrett and Westbury discovered that we only need a
sufficiently nice {\it monoidal} category to construct a {\it
3-dimensional} TQFT---and the Turaev--Viro models are among these.
This outlook makes certain patterns clearer; we shall explain these
patterns further in sections to come.

When writing their original paper, Turaev and Viro did not know about
the Ponzano--Regge model of quantum gravity.  However, their
construction amounts to taking the Ponzano--Regge model and curing it
of its divergent sums by replacing $\SU(2)$ by the corresponding
quantum group.  Despite the many technicalities involved, the basic
idea is simple.  The Ponzano--Regge model is not a 3d TQFT, because it
assigns divergent values to the operator $Z(M)$ for many cobordisms
$M$.  The reason is that computing this operator involves
triangulating $M$, labelling the edges by spins $j = 0, \frac{1}{2},
1, \dots$, and summing over spins.  Since there are infinitely many
choices of the spins, the sum may diverge.  And since the spin
labelling an edge describes its length, this divergence arises
physically from the fact that we are summing over geometries that can
be \textit{arbitrarily large}.

Mathematically, spins correspond to irreducible representations of
$\SU(2)$.  There are, of course, infinitely many of these.  The same
is true for the quantum group $U_q \ssl(2)$.  But in the modular
tensor category, we keep only \textit{finitely many} of the
irreducible representations of $U_q \ssl(2)$ as objects, corresponding
to the spins $j = 0, \frac{1}{2}, 1, \dots, \frac{k}{2}$, where $k$
depends on the root of unity $q$.  This cures the Ponzano--Regge model
of its infinities.  Physically, introducing the parameter $q$
corresponds to introducing a nonzero `cosmological constant' in 3d
quantum gravity.  The cosmological constant endows the vacuum with a
constant energy density and forces spacetime to curl up instead of
remaining flat.  This puts an upper limit on the size of spacetime,
avoiding the divergent sum over arbitrarily large geometries.

We postpone a detailed description of the Turaev--Viro model until our
discussion of Barrett and Westbury's 1992 paper.  As mentioned, this
paper strips Turaev and Viro's construction down to its bare
essentials, building a 3d TQFT from any sufficiently nice monoidal
category: the braiding is inessential.  But the work of Barrett and
Westbury is a categorified version of Fukuma, Hosono and Kawai's work
on 2d TQFTs, so we should first discuss that.

\subsection*{Fukuma--Hosono--Kawai (1992)}

Fukuma, Hosono and Kawai found a way to construct two-dimensional
topological quantum field theories from semisimple algebras
\cite{FHK}.  Though they did not put it this way, they essentially
gave a recipe to turn any 2-dimensional cobordism
\[
 \includegraphics[width=5cm]{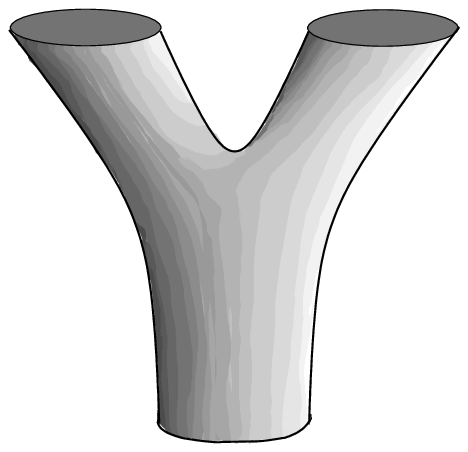}
 \qquad \qquad
\xy
    (-5,54)*+{S}="1";
    (-2,32)*+{M}="2";
    (-5,10)*+{S'}="3";
        {\ar@{->} "1";"3"};
\endxy
\]
into a string diagram, and use that diagram to define an operator
between vector spaces:
\[
\tilde{Z}(M) \maps \tilde{Z}(S) \to \tilde{Z}(S') \]
This gadget $\tilde{Z}$ is not quite a TQFT, but with a little
extra work it gives a TQFT which we will call $Z$.

The recipe begins as follows.   Triangulate the cobordism
$M$:
\[
 \includegraphics[width=5cm]{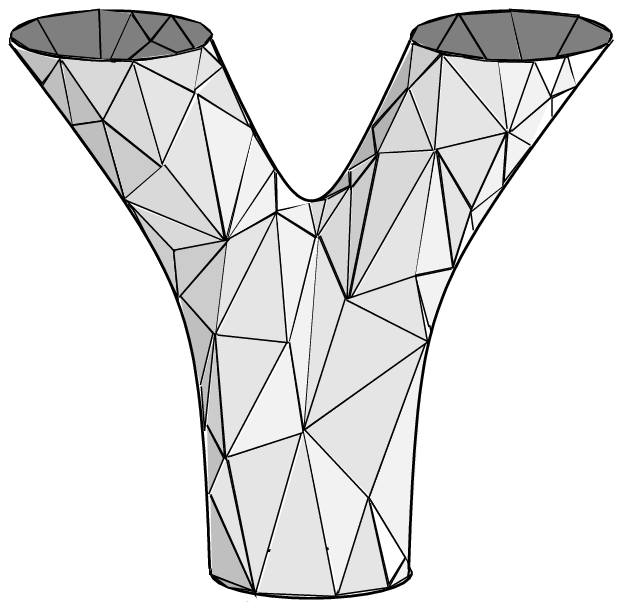}
\]
This picture already looks a bit like a string diagram, but never
mind that.  Instead, take the Poincar\'{e} dual of
the triangulation, drawing a string diagram with:
\begin{itemize}
\item
one vertex in the center of each triangle of the original triangulation;
\item
one edge crossing each edge of the original triangulation.
\end{itemize}
\[
 \includegraphics[width=5cm]{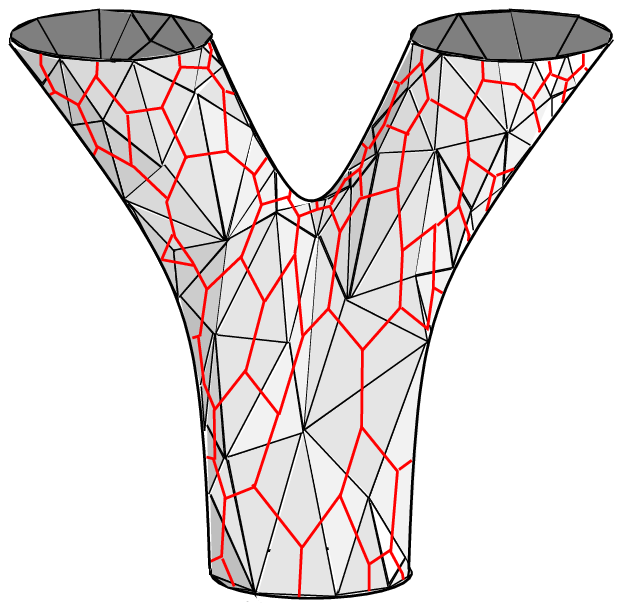}
\]
We then need a way to evaluate
this string diagram and get an operator.

For this, fix an associative algebra $A$.  Then using Poincar\'e
duality, each triangle in the triangulation can be reinterpreted
as a string diagram for multiplication in $A$:
\[
\xy
  (-8,5)*{};
  (8,5)*{};
  (0,-8)*{};
  (0,10)*{}="t";
 (-10,-6)*{}="bl";
 (10,-6)*{}="br";
   {\ar@{-} "bl";"br" };
   {\ar@{-} "bl";"t" };
   {\ar@{-} "t";"br" };
   {\ar@{=>} (0,2); (0,-2) };
 \endxy
\quad \rightsquigarrow \quad
 \xy
  (0,0)*\xycircle(3,3){-}="c";
  (0,0)*{m};
  (-8,8)*{}="tl"+(-2,-3)*{};
  (8,8)*{}="tr"+(2,-3)*{};
  (0,-10)*{}="b"+(-2,02)*{};
     "tr"; "c" **\dir{-};
     "tl"; "c" **\dir{-};
     "c"; "b"  **\dir{-};
  (0,10)*{}="t";
 (-10,-6)*{}="bl";
 (10,-6)*{}="br";
   {\ar@{--} "bl";"br" };
   {\ar@{--} "bl";"t" };
   {\ar@{--} "t";"br" };
 \endxy
\quad \rightsquigarrow \quad
\xy
  (0,0)*\xycircle(3,3){-}="c";
  (0,0)*{m};
  (-8,8)*{}="tl"+(-2,-3)*{};
  (8,8)*{}="tr"+(2,-3)*{};
  (0,-10)*{}="b"+(-2,02)*{};
     "tr"; "c" **\dir{-}?(.5)*\dir{>};
     "tl"; "c" **\dir{-}?(.5)*\dir{>};
     "c"; "b" **\dir{-}?(.7)*\dir{>};
 \endxy
\]
Actually there is a slight subtlety here.  The above string diagram
comes with some extra information: little arrows on the edges, which
tell us which edges are coming in and which are going out.  To avoid
the need for this extra information, let us equip $A$ with an
isomorphism to its dual vector space $A^*$.  Then we can take any
triangulation of $M$ and read it as a string diagram for an operator
$\tilde{Z}(M)$.  If our triangulation gives the manifold $S$ some
number of edges, say $n$, and gives $S'$ some other number of edges,
say $n'$, then we have
\[  \tilde{Z}(M) \maps \tilde{Z}(S) \to \tilde{Z}(S')  \]
where
\[     \tilde{Z}(S) = A^{\tensor n}, \qquad
       \tilde{Z}(S') = A^{\tensor n'}  .\]

We would like this operator $\tilde{Z}(M)$ to be well-defined and
independent of our choice of triangulation for $M$.  And now a miracle
occurs.  In terms of triangulations, the associative law:
\[
\vbox{
 \xy
  (0,10)*\xycircle(3,3){-}="c1";
  (0,10)*{m};
  (-6,20)*{}="tl"+(-1,-3)*{};
  (6,20)*{}="tr"+(1,-3)*{};
  (0,0)*{}="b"+(0,4)*{};
     "tr"; "c1" **\dir{-};
     "tl"; "c1" **\dir{-};
  (5,0)*\xycircle(3,3){-}="c";
  (5,0)*{m};
  (2,10)*{}="tl";
  (18,20)*{}="tr"+(1,-3)*{};
  (5,-15)*{}="b"+(-1,5)*{};
     "tr"; "c" **\dir{-};
     "c"; "c1" **\dir{-};
     "c"; "b" **\dir{-};
 \endxy
}
\qquad = \qquad
\vbox{
 \xy
  (0,10)*\xycircle(3,3){-}="c1";
  (0,10)*{m};
  (6,20)*{}="tl"+(1,-3)*{};
  (-6,20)*{}="tr"+(-1,-3)*{};
  (0,0)*{}="b"+(0,4)*{};
     "tr"; "c1" **\dir{-};
     "tl"; "c1" **\dir{-};
  (-5,0)*\xycircle(3,3){-}="c";
  (-5,0)*{m};
  (-2,0)*{}="tl";
  (-18,20)*{}="tr"+(-1,-3)*{};
  (-5,-15)*{}="b"+(-1,5)*{};
     "tr"; "c" **\dir{-};
     "c"; "c1" **\dir{-};
     "c"; "b" **\dir{-};
 \endxy
}
\]
can be redrawn as follows:
\[
 \vcenter{\xy
 (-10,-10)*{}="bl";
 (10,-10)*{}="br";
 (-10,10)*{}="tl";
 (10,10)*{}="tr";
    {\ar@{-} "bl";"br" };
    {\ar@{-} "br";"tr" };
    {\ar@{-} "tl";"tr" };
    {\ar@{-} "bl";"tl" };
    {\ar@{-} "bl";"tr" };
        (-12,0)*{};
        (0,12)*{};
        (0,2)*{};
        (12,0)*{};
        (0,-12)*{};
 \endxy}
 \qquad = \qquad
  \vcenter{\xy
 (10,-10)*{}="bl";
 (-10,-10)*{}="br";
 (10,10)*{}="tl";
 (-10,10)*{}="tr";
    {\ar@{-} "bl";"br" };
    {\ar@{-} "br";"tr" };
    {\ar@{-} "tl";"tr" };
    {\ar@{-} "bl";"tl" };
    {\ar@{-} "bl";"tr" };
        (-12,0)*{};
        (0,12)*{};
        (0,-4)*{};
        (12,0)*{};
        (0,-12)*{};
 \endxy}
\]
This equation is already famous in topology!  It is the {\bf 2-2
move}: one of two so-called \textbf{Pachner moves} for changing the
triangulation of a surface without changing the surface's topology.
The other is the {\bf 1-3 move}:
\[  \pachnerII \]
By repeatedly using these moves, we can go between any two
triangulations of $M$ that restrict to the same triangulation of its
boundary.

The associativity of the algebra $A$ guarantees that the operator
$\tilde{Z}(M)$ does not change when we apply the 2-2 move.  To ensure
that $\tilde{Z}(M)$ is also unchanged by 1-3 move, we require $A$
to be `semisimple'.  There are many equivalent ways of defining this
concept.  For example, given that we are working over the complex
numbers, we can define an algebra $A$ to be \textbf{semisimple} if it
is isomorphic to a finite direct sum of matrix algebras.  A more
conceptual definition uses the fact that any algebra $A$ comes
equipped with a bilinear form
\[          g(a,b) = \tr(L_a L_b)  \]
where $L_a$ stands for left multiplication by $a$:
\[
\begin{array}{rccc}
              L_a \maps &A &\to& A  \\
                        &x &\mapsto& a x
\end{array}
\]
and $\tr$ stands for the trace.  We can reinterpret $g$ as a linear
operator $g \maps A \otimes A \to \C$, which we can draw as a `cup':
\[
 \vcenter{\xy
 (-6,8)*{};(6,8)*{};
   **\crv{(6,-10) & (-6,-10)}
    ?(.20)*\dir{}
    ?(.89)*\dir{};
   (1,-6)*{      };
   (-9,4)*{}; (9,4)*{};
\endxy }
\qquad \qquad
 \vcenter{ \xy
 (0,16)*+{A \ten A}="1";
 (0,-3)*+{\C}="3";
    {\ar_{g} "1";"3"};
\endxy }
\]
We say $g$ is \textbf{nondegenerate} if we can find a
is a corresponding `cap' that satisfies the zig-zag equations.
Then we say the algebra $A$ is \textbf{semisimple} if $g$ is
nondegenerate.   In this case, the map $a \mapsto g(a,\cdot)$
gives an isomorphism $A \cong A^*$, which lets us avoid writing
little arrows on our string diagram.   Even better, with the chosen
cap and cup, we get the equation:
\[
 \xy 0;/r.16pc/:
   (0,-8)*\xycircle(1.5,1.5){-}="m";
   (0,-14)*{}="b";
   (-3,-4)*\xycircle(1.5,1.5){-}="l";
   (3,-2)*{}="r";
   (0,10)*{}="m1";
   (-8,12)*{}="z";
         "b";"m" **\dir{-};
         "m";"l" **\crv{(-3,-8)};
         "m";"r" **\crv{(5,-8)};
         "l";"z" **\crv{(-10,-4)};
         "l";"r" **\crv{(-1,2) & (2,0) };
 \endxy
 \qquad = \qquad
 \xy 0;/r.16pc/:
   (0,12)*{}="b";
   (0,-14)*{}="m1";
          "b";"m1" **\dir{-};
 \endxy
\]
where each circle denotes the multiplication $m \maps A \tensor A \to A$.
This equation then turns out to imply the 1-3 move!  Proving this
is a good workout in string diagrams and Poincar\'e duality.

So: starting from a semisimple algebra $A$, we obtain an operator
$\tilde{Z}(M)$ from any triangulated 2d cobordism $M$.  Moreover, this
operator is invariant under both Pachner moves.
But how does this construction give us a 2d TQFT?   It is easy to check
that
\[             \tilde{Z}(M M') = \tilde{Z}(M) \, \tilde{Z}(M') ,\]
which is a step in the right direction.  We have seen that $\tilde{Z}(M)$
is the same regardless of which triangulation we pick for $M$, as long
as we fix the triangulation of its boundary.  Unfortunately, it
depends on the triangulation of the boundary: after all, if
$S$ is the circle triangulated with $n$ edges then $\tilde{Z}(S) =
A^{\otimes n}$.   So, we need to deal with this problem.

Given two different
triangulations of the same 1-manifold, say $S$ and $S'$, we can always
find a triangulated cobordism $M \maps S \to S'$ which is a
\textbf{cylinder}, meaning it is homeomorphic to $S \times [0,1]$,
with $S$ and $S'$ as its two ends.  For example:
\[
 \xy
 (0,0)*{\includegraphics{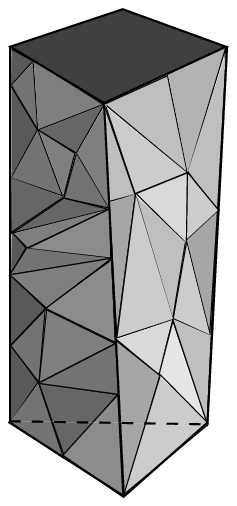}};
 {\ar^{M}(20,22)*+{S}; (20,-20)*+{S'}};
 \endxy
\]
This cobordism gives an operator $\tilde{Z}(M) \maps
\tilde{Z}(S) \to \tilde{Z}(S')$, and because this operator is
independent of the triangulation of the interior of $M$, we obtain a
canonical operator from $\tilde{Z}(S)$ to $\tilde{Z}(S')$.
In particular, when $S$ and $S'$ are equal as triangulated manifolds,
we get an operator
\[        p_S \maps \tilde{Z}(S) \to \tilde{Z}(S)  .\]
This operator is not the identity, but a simple calculation shows that it
is a \textbf{projection}, meaning
\[   p_S^2 = p_S  .\]
In physics jargon, this operator acts as a projection onto the space of
`physical states'.  And if we define $Z(S)$ to be the range of $p_S$,
and $Z(M)$ to be the restriction of $\tilde{Z}(M)$ to $Z(S)$, we
can check that $Z$ is a TQFT!

How does this construction relate to the construction of 2d TQFTs from
commutative Frobenius algebras explained our discussion of Dijkgraaf's
1989 thesis?  To answer this, we need to see how the commutative
Frobenius algebra $Z(S^1)$ is related to the semisimple algebra $A$.
In fact $Z(S^1)$ turns out to be the \textbf{center} of $A$: the set
of elements that commute with all other elements of $A$.

The proof is a nice illustration of the power of string diagrams.
Consider the simplest triangulated cylinder from $S^1$ to itself.
We get this by taking a square, dividing it into two triangles by
drawing a diagonal line, and then curling it up to form a cylinder:
\[
\includegraphics[scale=0.6]{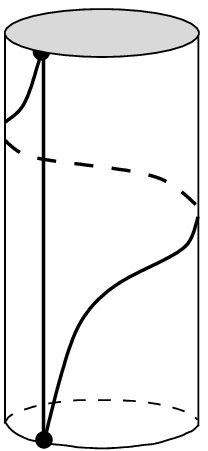}
\]
This gives a projection
\[      p = p_{S^1} \maps \tilde{Z}(S^1) \to \tilde{Z}(S^1)  \]
whose range is $Z(S^1)$.  Since we have triangulated $S^1$ with a
single edge in this picture, we have $\tilde{Z}(S^1) = A$.  So, the
commutative Frobenius algebra $Z(S^1)$ sits inside $A$ as the range of
the projection $p \maps A \to A$.

Let us show that the range of $p$ is precisely the center of
$A$.   First, take the triangulated cylinder above and draw the
Poincar\'e dual string diagram:
\[
\includegraphics[scale=0.6]{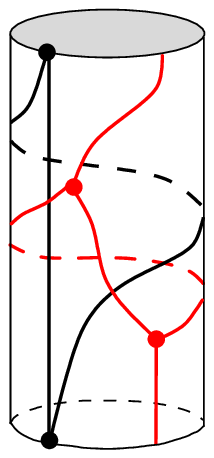}
\]
Erasing everything except this string diagram, we obtain a kind
of `formula' for $p$:
\[   p = \projector  \]
where the little circles stand for multiplication in $A$.  To see that
$p$ maps $A$ \textit{onto} into its center, it suffices to check that
if $a$ lies in the center of $A$ then $pa = a$.  This is a nice string
diagram calculation:
\[
\PROJcenterI
\]
In the second step we use the fact that $a$ is in the center of
$A$; in the last step we use semisimplicity.
Similarly, to see that $p$ maps $A$ \textit{into} its center,
it suffices to check that for any $a \in A$, the element $pa$
commutes with every other element of $A$.   In string diagram notation,
this says that:
\[
\PROJcenterII
\]
The proof is as follows:
\[
\PROJcenterIII
\]

\subsection*{Barrett--Westbury (1992)}

In 1992, Barrett and Westbury completed a paper that treated ideas
very similar to those of Turaev and Viro's paper from the same year
\cite{BW}.  Unfortunately, it only reached publication much later, so
everyone speaks of the Turaev--Viro model.  Barrett and Westbury
showed that to construct 3d TQFTs we only need a nice monoidal
category, not a braided monoidal category.  More technically: we do
not need a modular tensor category; a `spherical category' will
suffice \cite{BW2}.  Their construction can be seen as a categorified
version of the Fukuma--Hosono--Kawai construction, and we shall
present it from that viewpoint.

The key to the Fukuma--Hosono--Kawai construction was getting an
operator from a triangulated 2d cobordism and checking its
invariance under the 2-2 and 1-3 Pachner moves.  In both
these moves, the `before' and `after' pictures can be seen as the
front and back of a tetrahedron:
\[
 \vcenter{\xy 0;/r.12pc/:
 (-10,-10)*{}="bl";
 (10,-10)*{}="br";
 (-10,10)*{}="tl";
 (10,10)*{}="tr";
    {\ar@{-} "bl";"br" };
    {\ar@{-} "br";"tr" };
    {\ar@{-} "tl";"tr" };
    {\ar@{-} "bl";"tl" };
    {\ar@{-} "bl";"tr" };
        (-12,0)*{};
        (0,12)*{};
        (0,2)*{};
        (12,0)*{};
        (0,-12)*{};
 \endxy}
 \qquad \longleftrightarrow \qquad
  \vcenter{\xy 0;/r.12pc/:
 (10,-10)*{}="bl";
 (-10,-10)*{}="br";
 (10,10)*{}="tl";
 (-10,10)*{}="tr";
    {\ar@{-} "bl";"br" };
    {\ar@{-} "br";"tr" };
    {\ar@{-} "tl";"tr" };
    {\ar@{-} "bl";"tl" };
    {\ar@{-} "bl";"tr" };
        (-12,0)*{};
        (0,12)*{};
        (0,-4)*{};
        (12,0)*{};
        (0,-12)*{};
 \endxy}
\]
\[
 \vcenter{\xy 0;/r.15pc/:
 (-10,0)*{}="L";
 (10,0)*{}="R";
 (0,16)*{}="T";
 (0,6)*{}="M";
 (-10,12)*{}="TL";
 (10,12)*{}="TR";
    "T";"L" **\dir{-};
    "R";"T" **\dir{-};
    "L";"R" **\dir{-};
 \endxy}
\qquad \longleftrightarrow \qquad
 \vcenter{\xy 0;/r.15pc/:
 (-10,0)*{}="L";
 (10,0)*{}="R";
 (0,16)*{}="T";
 (0,6)*{}="M";
    "L";"T" **\dir{-};
    "R";"T" **\dir{-};
    "L";"R" **\dir{-};
    "T";"M" **\dir{-};
    "R";"M" **\dir{-};
    "L";"M" **\dir{-};
 \endxy}
\]

All this has an analogue one dimension up.  For starters, there are
also Pachner moves in 3 dimensions.  The \textbf{2-3 move} takes us
from two tetrahedra attached along a triangle to three sharing an edge,
or vice versa:
\[
    \xy 0;/r.16pc/:
     (-10.6,-14.63)*{}="t1";
     (-17.1,5.56)*{}="t2";
     (0,18)*{}="t3";
     (17.1,5.56)*{}="t4";
     (10.6,-14.63)*{}="t5";
   {\ar@{-}^{} "t1";"t2"};
   {\ar@{-}^{} "t2";"t3"};
   {\ar@{-}^{} "t3";"t4"};
   {\ar@{-}^{} "t4";"t5"};
   {\ar@{-} "t1";"t5"};
   {\ar@{.}|<<<<<<<<<{\hole \; \hole} "t1";"t3"};
   {\ar@{.}|<<<<<<<<<{\hole}|>>>>>>>>>{\hole} "t1";"t4"};
   {\ar@{-} "t2";"t5"};
   {\ar@{-} "t3";"t5"};
  \endxy
\qquad \longleftrightarrow \qquad
    \xy 0;/r.16pc/:
     (-10.6,-14.63)*{}="t1";
     (-17.1,5.56)*{}="t2";
     (0,18)*{}="t3";
     (17.1,5.56)*{}="t4";
     (10.6,-14.63)*{}="t5";
   {\ar@{-}^{} "t1";"t2"};
   {\ar@{-}^{} "t2";"t3"};
   {\ar@{-}^{} "t3";"t4"};
   {\ar@{-}^{} "t4";"t5"};
   {\ar@{-} "t1";"t5"};
   {\ar@{.}|<<<<<<<<<{\hole \; \hole}|>>>>>>>>>{\hole} "t1";"t3"};
   {\ar@{.}|<<<<<<<<<{\hole}|>>>>>>>>>{\hole} "t1";"t4"};
   {\ar@{-} "t2";"t5"};
   {\ar@{-} "t3";"t5"};
   {\ar@{--}|>>>>>>{\hole} "t2";"t4"};
  \endxy
\]
On the left side we see two tetrahedra sharing a triangle, the tall
isosceles triangle in the middle.  On the right we see three tetrahedra
sharing an edge, the dashed horizontal line.  The \textbf{1-4 move}
lets us split one tetrahedron into four, or merge four back into one:
\[
 \xy
 (-10,-5 )*{}="1";
 (8,-10)*{}="2";
 (15,0)*{}="3";
 (1,12)*{}="4";
    {\ar@{-} "1";"2" };
    {\ar@{-}"2";"3" };
    {\ar@{-} "4";"3" };
    {\ar@{-} "1";"4" };
    {\ar@{-} "4";"2" };
    {\ar@{.}|>>>>>>>>>>{\hole \hole} "1";"3"};
 \endxy
\qquad \longleftrightarrow \qquad
 \xy
 (-10,-5 )*{}="1";
 (8,-10)*{}="2";
 (15,0)*{}="3";
 (1,12)*{}="4";
 (0,3)*{}="m";
    {\ar@{--} "1";"m" };
    {\ar@{--} "2";"m" };
    {\ar@{--} "3";"m" };
    {\ar@{--} "4";"m" };
    {\ar@{-} "1";"2" };
    {\ar@{-} "2";"3" };
    {\ar@{-} "4";"3" };
    {\ar@{-} "1";"4" };
    {\ar@{-} "4";"2" };
    {\ar@{.}|<<<<<<<<<<<<<<{\hole \; \hole} "1";"3"};
 \endxy
\]
Given a 3d cobordism $M \maps S \to S'$, repeatedly applying these
moves lets us go between any two triangulations of $M$ that restrict
to the same triangulation of its boundary.  Moreover, for both these
moves, the `before' and `after' pictures can be seen as the front and
back of a \textit{4-simplex}: the 4-dimensional analogue of a tetrahedron.

Fukuma, Hosono and Kawai constructed 2d TQFTs from certain monoids:
namely, semisimple algebras.  As we have seen, the key ideas were these:
\begin{itemize}
\item
A triangulated 2d cobordism gives an operator by letting each
triangle correspond to multiplication in a semisimple algebra.
\item
Since the multiplication is associative, the resulting operator is
invariant under the 2-2 Pachner move.
\item
Since the algebra is semisimple, the operator is also invariant
under the 1-3 move.
\end{itemize}
In a very similar way, Barrett and
Westbury constructed 3d TQFTs from certain monoidal categories called
`spherical categories'.  We can think of a spherical category as a
categorified version of a semisimple algebra.  The key ideas are
these:
\begin{itemize}
\item
A triangulated compact 2d manifold gives a vector space by
letting each triangle correspond to tensor product in a spherical
category.
\item
A triangulated 3d cobordism gives an operator by letting each
tetrahedron correspond to the associator in the spherical category.
\item
Since the associator satisfies the pentagon identity, the resulting
operator is invariant under the 2-3 Pachner move.
\item
Since the spherical category is `semisimple', the operator is
also invariant under the 1-4 move.
\end{itemize}

The details are a bit elaborate, so let us just sketch some of the
simplest, most beautiful aspects.  Recall from our discussion of
Kapranov and Voevodsky's 1991 paper that categorifying the concept of
`vector space' gives the concept of `2-vector space'.  Just as there
is a category $\Vect$ of vector spaces, there is a bicategory $2\Vect$
of 2-vector spaces, with:
\begin{itemize}
\item 2-vector spaces as objects,
\item linear functors as morphisms,
\item linear natural transformations as 2-morphisms.
\end{itemize}
In fact $2\Vect$ is a monoidal bicategory, with a tensor product
satisfying
\[              \Vect^m \otimes \Vect^n \simeq \Vect^{m n} . \]
This lets us define a {\bf 2-algebra} to be a 2-vector space $A$
that is also a monoidal category for which the tensor product extends to
a linear functor
\[     m \maps A \otimes A \to A, \]
and for which the associator and unitors extend to linear natural
transformations.  We have already seen a nice example of a 2-algebra,
namely $\Rep(K)$ for a compact Lie group $K$.  Here the tensor product
is the usual tensor product of group representations.

Now let us fix a 2-algebra $A$.  Given a triangulated compact 2-dimensional
manifold $S$, we can use Poincar\'e duality to reinterpret
each triangle as a picture of the multiplication $m \maps A \otimes A
\to A$:
\[
\vbox{
\xy 0;/r.20pc/:
  (-8,5)*{};
  (8,5)*{};
  (0,-8)*{};
  (0,10)*{}="t";
 (-10,-6)*{}="bl";
 (10,-6)*{}="br";
   {\ar@{-} "bl";"br" };
   {\ar@{-} "bl";"t" };
   {\ar@{-} "t";"br" };
 \endxy
}
\quad \rightsquigarrow \quad
\vbox{
\xy 0;/r.20pc/:
  (0,0)*\xycircle(3,3){-}="c";
  (0,0)*{m};
  (-8,8)*{}="tl"+(-2,-3)*{};
  (8,8)*{}="tr"+(2,-3)*{};
  (0,-10)*{}="b"+(-2,02)*{};
     "tr"; "c" **\dir{-};
     "tl"; "c" **\dir{-};
     "c"; "b" **\dir{-};
 \endxy
}
\]
As in the Fukuma--Hosono--Kawai model, this lets us turn the
triangulated manifold into a string diagram.  And as before, if $A$ is
`semisimple'---or more precisely, if $A$ is a spherical category---we
do not need to write little arrows on the edges of this string diagram
for it to make sense.  But since everything is categorified, this
string diagram now describes linear \textit{functor}.  Since $S$
has no boundary, this string diagram starts and ends with no edges, so
it describes a linear functor from $A^{\tensor 0}$ to itself.  Just as
the tensor product of \textit{zero} copies of a vector space is defined
to be $\C$, the tensor product of no copies of 2-vector space is
defined to be $\Vect$.  But a linear functor from $\Vect$ to itself is
given by a $1 \times 1$ matrix of vector spaces---that is, a vector
space!  This recipe gives us a vector space $\tilde{Z}(S)$ for any
compact 2d manifold $S$.

Next, from a triangulated 3d cobordism $M \maps S \to S'$,
we wish to obtain a linear operator $\tilde{Z}(M) \maps \tilde{Z}(S)
\to \tilde{Z}(S')$.  For this, we can use Poincar\'e duality to reinterpret
each tetrahedron as a picture of the associator.  The `front' and `back' of
the tetrahedron correspond to the two functors that the associator
goes between:
\[
 \vcenter{\xy 0;/r.20pc/:
 (-10,-10)*{}="bl";
 (10,-10)*{}="br";
 (-10,10)*{}="tl";
 (10,10)*{}="tr";
    {\ar@{-} "bl";"br" };
    {\ar@{-} "br";"tr" };
    {\ar@{-} "tl";"tr" };
    {\ar@{-} "bl";"tl" };
    {\ar@{-} "bl";"tr" };
        (-12,0)*{};
        (0,12)*{};
        (0,2)*{};
        (12,0)*{};
        (0,-12)*{};
 \endxy}
 \;\; \To \;\;
  \vcenter{\xy 0;/r.20pc/:
 (10,-10)*{}="bl";
 (-10,-10)*{}="br";
 (10,10)*{}="tl";
 (-10,10)*{}="tr";
    {\ar@{-} "bl";"br" };
    {\ar@{-} "br";"tr" };
    {\ar@{-} "tl";"tr" };
    {\ar@{-} "bl";"tl" };
    {\ar@{-} "bl";"tr" };
        (-12,0)*{};
        (0,12)*{};
        (0,-4)*{};
        (12,0)*{};
        (0,-12)*{};
 \endxy}
\qquad \rightsquigarrow \qquad
\vbox{
 \xy 0;/r.20pc/:
  (-3,3)*\xycircle(3,3){-}="c1";
  (-3,3)*{m};
  (-10,3)*{}="tl"+(-1,-3)*{};
  (-3,10)*{}="tr"+(1,-3)*{};
  (2,-10)*{}="b"+(0,4)*{};
     "tr"; "c1" **\dir{-};
     "tl"; "c1" **\dir{-};
  (3,-3)*\xycircle(3,3){-}="c";
  (3,-3)*{m};
  (-10,3)*{}="tl";
  (10,-3)*{}="tr"+(1,-3)*{};
  (3,-10)*{}="b"+(-1,5)*{};
     "tr"; "c" **\dir{-};
     "c"; "c1" **\dir{-};
     "c"; "b" **\dir{-};
 \endxy
}
\; \To \;
\vbox{
 \xy 0;/r.20pc/:
  (3,3)*\xycircle(3,3){-}="c1";
  (3,3)*{m};
  (10,3)*{}="tl"+(-1,-3)*{};
  (3,10)*{}="tr"+(1,-3)*{};
  (-2,-10)*{}="b"+(0,4)*{};
     "tr"; "c1" **\dir{-};
     "tl"; "c1" **\dir{-};
  (-3,-3)*\xycircle(3,3){-}="c";
  (-3,-3)*{m};
  (10,3)*{}="tl";
  (-10,-3)*{}="tr"+(1,-3)*{};
  (-3,-10)*{}="b"+(-1,5)*{};
     "tr"; "c" **\dir{-};
     "c"; "c1" **\dir{-};
     "c"; "b" **\dir{-};
 \endxy
}
\]

A more 3-dimensional view is helpful here.  Starting from a
triangulated 3d cobordism $M \maps S \to S'$, we can use Poincar\'e
duality to build a piecewise-linear cell complex, or `2-complex' for
short.  This is a 2-dimensional generalization of a graph; just as a
graph has vertices and edges, a 2-complex has vertices, edges and
polygonal faces.  The 2-complex dual to the triangulation of a 3d
cobordism has:
\begin{itemize}
\item
one vertex in the center of each tetrahedron of the original
triangulation;
\item
one edge crossing each triangle of the original triangulation;
\item
one face crossing each edge of the original triangulation.
\end{itemize}
We can interpret this 2-complex as a higher-dimensional analogue of a
string diagram, and use this to compute an operator $\tilde{Z}(M)
\maps \tilde{Z}(S) \to \tilde{Z}(S')$.  This outlook is stressed in
`spin foam models' \cite{Baez2,Baez3}, of which the
Turaev--Viro--Barrett--Westbury model is the simplest and most
successful.

Each tetrahedron in $M$ gives a little piece of the 2-complex,
which looks like this:
\[
\DDDI
\]
If we look at the string diagrams on the front and back of this
picture, we see they describe the two linear functors that the
associator goes between:
\[
\DDDII
\]
This is just a deeper look at the something we already saw in our
discussion of Ponzano and Regge's 1968 paper.  There we saw a
connection between the tetrahedron, the $6j$ symbols, and the
associator in $\Rep(\SU(2))$.  Now we are seeing that for any
spherical category, a triangulated 3d cobordism gives a 2d cell
complex built out of pieces that we can interpret as associators.
So, just as triangulated 2-manifolds give us linear functors,
triangulated 3d cobordisms gives us linear natural transformations!

More precisely, recall that every compact triangulated 2-manifold $S$
gave a linear functor from $\Vect$ to $\Vect$, or $1 \times 1$ matrix
of vector spaces, which we reinterpreted as a vector space
$\tilde{Z}(S)$.  Similarly, every triangulated 3d cobordism $M \maps S
\to S'$ gives a linear natural transformation gives between such
linear functors.  This amounts to a $1 \times 1$ matrix of linear
operators, which we can reinterpret as a linear operator $\tilde{Z}(M)
\maps \tilde{Z}(S) \to \tilde{Z}(S')$.

The next step is to show that $\tilde{Z}(M)$ is invariant under the
2-3 and 1-4 Pachner moves.  If we can do this, the rest is easy: we
can follow the strategy we have already seen in the
Fukuma--Hosono--Kawai construction and obtain a 3d TQFT.

At this point another miracle comes to our rescue: the pentagon
identity gives invariance under the 2-3 move!  The 2-3 move goes from
two tetrahedra to three, but each tetrahedron corresponds to an
associator, so we can interpret this move as an equation between a
natural transformation built from two associators and one built from
three.  And this equation is just the pentagon identity.

\begin{figure}
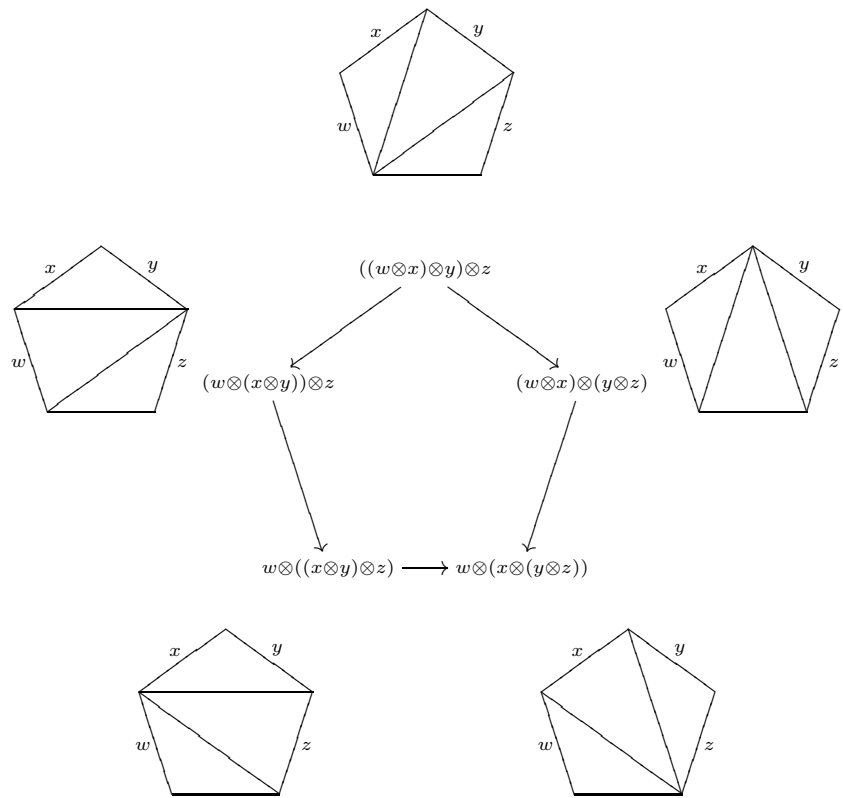

\begin{center}
\[
\bigpentagon
\]
\vskip 1em
\caption{Deriving the 2-3 Pachner move from the pentagon identity.}
\end{center}
\end{figure}

To see why, ponder the `pentagon of pentagons' in Figure 1.
This depicts five ways to parenthesize a tensor product of objects
$w,x,y,z$ in a monoidal category.  Each corresponds to a triangulation
of a pentagon.  (The repeated appearance of the number five here is
just a coincidence.)  We can go between these parenthesized tensor
products using the associator.  In terms of triangulations, each use
of the associator corresponds to a 2-2 move.  We can go from the top
of the picture to the lower right in two ways: one using two steps and
one using three.  The two-step method builds up this picture:
\[
    \xy 0;/r.16pc/:
     (-10.6,-14.63)*{}="t1";
     (-17.1,5.56)*{}="t2";
     (0,18)*{}="t3";
     (17.1,5.56)*{}="t4";
     (10.6,-14.63)*{}="t5";
   {\ar@{-}^{} "t1";"t2"};
   {\ar@{-}^{} "t2";"t3"};
   {\ar@{-}^{} "t3";"t4"};
   {\ar@{-}^{} "t4";"t5"};
   {\ar@{-} "t1";"t5"};
   {\ar@{.}|<<<<<<<<<{\hole \; \hole} "t1";"t3"};
   {\ar@{.}|<<<<<<<<<{\hole}|>>>>>>>>>{\hole} "t1";"t4"};
   {\ar@{-} "t2";"t5"};
   {\ar@{-} "t3";"t5"};
  \endxy
\]
which shows two tetrahedra attached along a triangle.  The three-step
method builds up this picture:
\[
    \xy 0;/r.16pc/:
     (-10.6,-14.63)*{}="t1";
     (-17.1,5.56)*{}="t2";
     (0,18)*{}="t3";
     (17.1,5.56)*{}="t4";
     (10.6,-14.63)*{}="t5";
   {\ar@{-}^{} "t1";"t2"};
   {\ar@{-}^{} "t2";"t3"};
   {\ar@{-}^{} "t3";"t4"};
   {\ar@{-}^{} "t4";"t5"};
   {\ar@{-} "t1";"t5"};
   {\ar@{.}|<<<<<<<<<{\hole \; \hole}|>>>>>>>>>{\hole} "t1";"t3"};
   {\ar@{.}|<<<<<<<<<{\hole}|>>>>>>>>>{\hole} "t1";"t4"};
   {\ar@{-} "t2";"t5"};
   {\ar@{-} "t3";"t5"};
   {\ar@{--}|>>>>>>{\hole} "t2";"t4"};
  \endxy
\]
which shows three tetrahedra sharing a common edge.
The pentagon identity thus yields the 2-3 move:
\[
    \xy 0;/r.16pc/:
     (-10.6,-14.63)*{}="t1";
     (-17.1,5.56)*{}="t2";
     (0,18)*{}="t3";
     (17.1,5.56)*{}="t4";
     (10.6,-14.63)*{}="t5";
   {\ar@{-}^{} "t1";"t2"};
   {\ar@{-}^{} "t2";"t3"};
   {\ar@{-}^{} "t3";"t4"};
   {\ar@{-}^{} "t4";"t5"};
   {\ar@{-} "t1";"t5"};
   {\ar@{.}|<<<<<<<<<{\hole \; \hole} "t1";"t3"};
   {\ar@{.}|<<<<<<<<<{\hole}|>>>>>>>>>{\hole} "t1";"t4"};
   {\ar@{-} "t2";"t5"};
   {\ar@{-} "t3";"t5"};
  \endxy
\qquad = \qquad
    \xy 0;/r.16pc/:
     (-10.6,-14.63)*{}="t1";
     (-17.1,5.56)*{}="t2";
     (0,18)*{}="t3";
     (17.1,5.56)*{}="t4";
     (10.6,-14.63)*{}="t5";
   {\ar@{-}^{} "t1";"t2"};
   {\ar@{-}^{} "t2";"t3"};
   {\ar@{-}^{} "t3";"t4"};
   {\ar@{-}^{} "t4";"t5"};
   {\ar@{-} "t1";"t5"};
   {\ar@{.}|<<<<<<<<<{\hole \; \hole}|>>>>>>>>>{\hole} "t1";"t3"};
   {\ar@{.}|<<<<<<<<<{\hole}|>>>>>>>>>{\hole} "t1";"t4"};
   {\ar@{-} "t2";"t5"};
   {\ar@{-} "t3";"t5"};
   {\ar@{--}|>>>>>>{\hole} "t2";"t4"};
  \endxy
\]
The other axioms in the definition of spherical category then yield
the 1-4 move, and so we get a TQFT.

At this point it is worth admitting that the link between the
associative law and 2-2 move and that between the pentagon identity
and 2-3 move are not really `miracles' in the sense of unexplained
surprises.  This is just the beginning of a pattern that relates the
$n$-dimensional simplex and the $(n-1)$-dimensional Stasheff
associahedron.  An elegant explanation of this can be found in
Street's 1987 paper `The algebra of oriented simplexes'
\cite{Street1}---the same one in which he proposed a simplicial
approach to weak $\infty$-categories.  Since there are also Pachner
moves in every dimension \cite{Pachner}, the Fukuma--Hosono--Kawai
model and the Turaev--Viro--Barrett--Westbury model should be just the
first of an infinite series of constructions building
$(n+1)$-dimensional TQFT from `semisimple $n$-algebras'.  But this is
largely open territory, apart from some important work in 4
dimensions, which we turn to next.

\subsection*{Turaev (1992)}

As we already mentioned, the Witten--Reshetikhin--Turaev construction
of 3-dimensional TQFTs from modular tensor categories is really just a
spinoff of a way to get \textit{4-dimensional} TQFTs from modular
tensor categories.  This began becoming visible in 1991, when Turaev
released a preprint \cite{Turaev2} on building 4d TQFTs from modular
tensor categories.  In 1992 he published a paper with more details
\cite{Turaev3}, and his book explains the ideas even more thoroughly
\cite{Turaev}.  His construction amounts to a 4-dimensional analogue
of the Turaev--Viro--Barrett--Westbury construction.  Namely, from a 4d
cobordism $M \maps S \to S'$, one can compute a linear operator
$\tilde{Z}(M) \maps \tilde{Z}(S) \to \tilde{Z}(S')$ with the help of a
2-dimensional CW complex sitting inside $M$.  As already mentioned, we
think of this complex as a higher-dimensional analogue of a string
diagram.

In 1993, following work by the physicist Ooguri \cite{Ooguri}, Crane
and Yetter \cite{CY} gave a different construction of 4d TQFTs from
the modular tensor category associated to quantum $\SU(2)$.  This
construction used a triangulation of $M$.  It was later generalized to
a large class of modular tensor categories \cite{CKY}, and thanks to
the work of Justin Roberts \cite{JustinRoberts2}, it is clear that
Turaev's construction is related to the Crane--Yetter construction by
Poincar\'e duality, following a pattern we have seen already.

At this point the reader, seeking simplicity amid these complex
historical developments, should feel a bit puzzled.  We have seen
that:
\begin{itemize}
\item
The Fukuma--Hosono--Kawai construction gives 2d TQFTs from
sufficiently nice monoids (semisimple algebras).
\item
The Turaev--Viro--Bartlett--Westbury construction gives 3d TQFTs
from sufficiently nice monoidal categories (spherical categories).
\end{itemize}
Given this, it would be natural to expect:
\begin{itemize}
\item
Some similar construction gives 4d TQFTs from sufficiently
nice monoidal bicategories.
\end{itemize}
Indeed, this is true!  Mackaay \cite{Mackaay} proved it in 1999.
But how does this square with the following fact?
\begin{itemize}
\item
The Turaev--Crane--Yetter construction gives 4d TQFTs from
sufficiently nice braided monoidal categories (modular tensor categories).
\end{itemize}
The answer is very nice: it turns out that braided monoidal categories
are a \textit{special case} of monoidal bicategories!

We should explain this, because it is part of a fundamental pattern
called the `periodic table of $n$-categories'.  As a warmup, let us
see why a commutative monoid is the same as a monoidal category with
only one object.  This argument goes back to work of Eckmann and
Hilton \cite{EH}, published in 1962.  A categorified version of their
argument shows that a braided monoidal category is the same as
monoidal bicategory with only one object.  This seems to have first
been noticed by Joyal and Tierney \cite{JT} around 1984.

Suppose first that $C$ is a category with one object $x$.  Then
composition of morphisms makes the set of morphisms from $x$ to
itself, denoted $\hom(x,x)$, into a \textbf{monoid}: a set with an
associative multiplication and an identity element.  Conversely, any
monoid gives a category with one object in this way.

But now suppose that $C$ is a monoidal category with one object $x$.
Then this object must be the unit for the tensor product.  As before,
$\hom(x,x)$ becomes a monoid using composition of morphisms.  But now
we can also tensor morphisms.  By Mac Lane's coherence theorem, we may
assume without loss of generality that $C$ is a strict monoidal
category.  Then the tensor product is associative, and we have $1_x
\tensor f = f = f \tensor 1_x $ for every $f \in \hom(x,x)$.  So,
$\hom(x,x)$ becomes a monoid in a second way, with the same identity
element.

However, the fact that tensor product is a functor implies
the \textbf{interchange law}:
\[  (f f') \tensor (g g') = (f \tensor g)(f' \tensor g'). \]
This lets us carry out the following remarkable argument, called
the \textbf{Eckmann--Hilton argument}:
\[
\begin{array}{ccl}
   f\tensor g  &=& (1 f) \tensor ( g 1) \\
     &=& (1 \tensor  g)(f \tensor 1)            \\
     &=&  gf                                    \\
     &=& (g \tensor 1)(1 \tensor f)            \\
     &=& (g 1) \tensor (1 f)                   \\
     &=& g \tensor f.
\end{array}
\]
In short: composition and tensor product are equal, and they are both
commutative!  So, $\hom(x,x)$ is a commutative monoid.  Conversely,
one can show that any commutative monoid can be thought of as the
morphisms in a monoidal category with just one object.

In fact, Eckmann and Hilton came up with their argument in work on
topology, and its essence is best revealed by a picture.  Let us draw
the composite of morphisms by putting one on top of the other, and
draw their tensor product by putting them side by side.  We have
often done this using string diagrams, but just for a change, let us
draw morphisms as squares.  Then the Eckmann--Hilton argument goes
as follows:
\[
 \xy
 (-6,6)*{}="TL";
 (-6,-6)*{}="BL";
 (6,6)*{}="TR";
 (6,-6)*{}="BR";
 (0,6)*{}="MT";
 (0,-6)*{}="MB";
 "TL";"TR" **\dir{-};
 "TL";"BL" **\dir{-};
 "TR";"BR" **\dir{-};
 "BL";"BR" **\dir{-};
 "MT";"MB" **\dir{-};
 (3,0)*{ g};
 (-3,0)*{f};
 (0,-9)*{\scriptstyle f \tensor  g};
 \endxy
\quad = \quad
 \xy
 (-6,6)*{}="TL";
 (-6,-6)*{}="BL";
 (6,6)*{}="TR";
 (6,-6)*{}="BR";
 (0,6)*{}="MT";
 (0,-6)*{}="MB";
 (-6,0)*{}="LM";
 (6,0)*{}="LR";
 "TL";"TR" **\dir{-};
 "TL";"BL" **\dir{-};
 "TR";"BR" **\dir{-};
 "BL";"BR" **\dir{-};
 "MT";"MB" **\dir{-};
  "LM";"LR" **\dir{-};
 (3,-3)*{ g};
 (-3,3)*{f};
 (-3,-3)*{1};
 (3,3)*{1};
 (0,-9)*{\scriptstyle (1 \tensor g)(f \tensor 1)};
 \endxy
 \quad = \quad
  \xy
 (-6,6)*{}="TL";
 (-6,-6)*{}="BL";
 (6,6)*{}="TR";
 (6,-6)*{}="BR";
 (-6,0)*{}="LM";
 (6,0)*{}="LR";
 "TL";"TR" **\dir{-};
 "TL";"BL" **\dir{-};
 "TR";"BR" **\dir{-};
 "BL";"BR" **\dir{-};
  "LM";"LR" **\dir{-};
 (0,-3)*{ g};
 (0,3)*{f};
 (0,-9)*{\scriptstyle  gf};
 \endxy
 \quad = \quad
  \xy
 (-6,6)*{}="TL";
 (-6,-6)*{}="BL";
 (6,6)*{}="TR";
 (6,-6)*{}="BR";
 (0,6)*{}="MT";
 (0,-6)*{}="MB";
 (-6,0)*{}="LM";
 (6,0)*{}="LR";
 "TL";"TR" **\dir{-};
 "TL";"BL" **\dir{-};
 "TR";"BR" **\dir{-};
 "BL";"BR" **\dir{-};
 "MT";"MB" **\dir{-};
  "LM";"LR" **\dir{-};
 (-3,-3)*{ g};
 (3,3)*{f};
 (3,-3)*{1};
 (-3,3)*{1};
 (0,-9)*{\scriptstyle ( g  \tensor 1)(1 \tensor f)};
 \endxy
 \quad = \quad
  \xy
 (-6,6)*{}="TL";
 (-6,-6)*{}="BL";
 (6,6)*{}="TR";
 (6,-6)*{}="BR";
 (0,6)*{}="MT";
 (0,-6)*{}="MB";
 "TL";"TR" **\dir{-};
 "TL";"BL" **\dir{-};
 "TR";"BR" **\dir{-};
 "BL";"BR" **\dir{-};
 "MT";"MB" **\dir{-};
 (-3,0)*{ g};
 (3,0)*{f};
 (0,-9)*{\scriptstyle   g \tensor f };
 \endxy
\]

We can categorify this whole discussion.  For starters, we noted in
our discussion of B\'enabou's 1967 paper that if $C$ is a
bicategory with one object $x$, then $\hom(x,x)$ is a monoidal
category---and conversely, any monoidal category arises in this way.
Then, the Eckmann--Hilton argument can be used to show that a
monoidal bicategory with one object is a braided monoidal category.
Since categorification amounts to replacing equations with
isomorphisms, each step in the argument now gives an isomorphism:
\[
\begin{array}{ccl}
   f\tensor g  &\iso & (1 f) \tensor ( g 1) \\
     &\iso & (1 \tensor  g)(f \tensor 1)            \\
     &\iso &  gf                                    \\
     &\iso& (g \tensor 1)(1 \tensor f)            \\
     &\iso& (g 1) \tensor (1 f)                   \\
     &\iso& g \tensor f.
\end{array}
\]
Composing these, we obtain an isomorphism from $f\tensor g$ to
$g\tensor f$, which we can think of as a braiding:
\[      B_{f,g} \maps f \tensor g \to g \tensor f .\]
We can even go further and check that this makes $\hom(x,x)$ into a
braided monoidal category.

A picture makes this plausible.  We can use the third dimension to
record the process of the Eckmann--Hilton argument.  If we compress
$f$ and $g$ to small discs for clarity, it looks like this:
\vskip 1em
\[
 \xy 0;/r.16pc/:
 (-14,10)*{}="TL";
 (14,10)*{}="TR";
 (14,-10)*{}="BR";
 (-14,-10)*{}="BL";
 (-6,20)*{}="xTL";
 (22,20)*{}="xTR";
 (22,0)*{}="xBR";
 (-6,0)*{}="xBL";
 (0,10)*{}="M1";
 (8,20)*{}="M2";
 (0,-10)*{}="MB1";
 (8,0)*{}="MB2";
 (5,0)*{}="LeftBackString";
 (2,0)*{}="RightFrontString";
 (-2.5,0)*{}="LeftFrontString";
 (6.75,-1.25)*{}="BottomBackString";
 (10,0)*{}="RightBackString";
     "TL";"TR" **\dir{-};
     "TR";"BR" **\dir{-};
     "BR";"BL" **\dir{-};
     "BL";"TL" **\dir{-};
     "xTL";"xTR" **\dir{-};
     "xTR";"xBR" **\dir{-};
     "xBR";"RightBackString" **\dir{.};
     "xBL";"LeftFrontString" **\dir{.};
     "LeftBackString";"RightFrontString" **\dir{.};
     "TL";"xTL" **\dir{-};
     "TR";"xTR" **\dir{-};
     "BL";"xBL" **\dir{.};
     "BR";"xBR" **\dir{-};
     "M1";"M2" **\dir{-};
     "MB1";"BottomBackString" **\dir{.};
     "xTL";"xBL" **\dir{.};
 (6,7.5)*\ellipse(2,1){-};
 (-1,-2.5)*\ellipse(2,1){-};
 (-2,7.5)*\ellipse(2,1){-};
 (5,-2.5)*\ellipse(2,1){-};
 (10,15)*{}="s1";
 (-4,-5)*{}="s2";
 "s1";"s2" **\crv{(10,5) & (-4,5)};
    \POS?(.67)*{\hole}="z2"; \POS?(.5)*{\hole}="z1";
 (14,15)*{}="s1";
 (0,-5)*{}="s2";
 "s1";"s2" **\crv{(14,5) & (0,5)}; \POS?(.71)*{}="y2"; \POS?(.61)*{}="y1";
 (-6,15)*{}="LL";
 (-2,15)*{}="LR";
 "LL";"z2" **\crv{(-6,4) & (0,5)};
 "LR";"z1" **\crv{(-2,4) & (4,5)};
 (8,-5)*{}="BL";
 (12,-5)*{}="BR";
 "y2";"BL" **\crv{(7.5,0)};
 "y1";"BR" **\crv{(10,4) & (12,-5)};
 (13,18)*{\scriptscriptstyle  g};
  (-5,-8)*{\scriptscriptstyle  g};
   (10,-8)*{\scriptscriptstyle f};
    (-3,18)*{\scriptscriptstyle f};
 \endxy
\]
\vskip 1em
\noindent
This clearly looks like a braiding!

In the above pictures we are moving $f$ around $g$ clockwise.  There is an
alternate version of the categorified Eckmann--Hilton argument that
amounts to moving $f$ around $g$ counterclockwise:
\[
\begin{array}{ccl}
f\tensor g  &\iso & (f 1)\tensor (1  g)\\
     &\iso & (f \tensor 1)(1 \tensor  g) \\
     &\iso & f g \\
     &\iso & (1 \tensor f)( g \tensor 1)\\
     &\iso & (1 g) \tensor (f 1) \\
     &\iso & g\tensor f.
\end{array}
\]
This gives the following picture:
\vskip 1em
\[
 \xy 0;/r.16pc/:
 (-22,10)*{}="TL";
 (6,10)*{}="TR";
 (6,-10)*{}="BR";
 (-22,-10)*{}="BL";
 (-14,20)*{}="xTL";
 (14,20)*{}="xTR";
 (14,0)*{}="xBR";
 (-14,0)*{}="xBL";
 (-8,10)*{}="M1";
 (0,20)*{}="M2";
 (-8,-10)*{}="MB1";
 (0,0)*{}="MB2";
 (-3,0)*{}="LeftBackString";
 (-6,0)*{}="RightFrontString";
 (-10.5,0)*{}="LeftFrontString";
 (-1.25,-1.25)*{}="BottomBackString";
 (2,0)*{}="RightBackString";
     "TL";"TR" **\dir{-};
     "TR";"BR" **\dir{-};
     "BR";"BL" **\dir{-};
     "BL";"TL" **\dir{-};
     "xTL";"xBL" **\dir{.};
     "xTL";"xTR" **\dir{-};
     "xTR";"xBR" **\dir{-};
     "xBR";"RightBackString" **\dir{.};
     "xBL";"LeftFrontString" **\dir{.};
     "LeftBackString";"RightFrontString" **\dir{.};
     "TL";"xTL" **\dir{-};
     "TR";"xTR" **\dir{-};
     "BL";"xBL" **\dir{.};
     "BR";"xBR" **\dir{-};
     "M1";"M2" **\dir{-};
     "MB1";"BottomBackString" **\dir{.};
 (-6,7.5)*\ellipse(2,1){-};
 (1,-2.5)*\ellipse(2,1){-};
 (2,7.5)*\ellipse(2,1){-};
 (-5,-2.5)*\ellipse(2,1){-};
 (-10,15)*{}="s1";
 (4,-5)*{}="s2";
 "s1";"s2" **\crv{(-10,5) & (4,5)};
    \POS?(.67)*{\hole}="z2"; \POS?(.5)*{\hole}="z1";
 (-14,15)*{}="s1";
 (0,-5)*{}="s2";
 "s1";"s2" **\crv{(-14,5) & (0,5)}; \POS?(.71)*{}="y2"; \POS?(.61)*{}="y1";
 (6,15)*{}="LL";
 (2,15)*{}="LR";
 "LL";"z2" **\crv{(6,4) & (0,5)};
 "LR";"z1" **\crv{(2,4) & (-4,5)};
 (-8,-5)*{}="BL";
 (-12,-5)*{}="BR";
 "y2";"BL" **\crv{(-7.5,0)};
 "y1";"BR" **\crv{(-10,4) & (-12,-5)};
 (-12,18)*{\scriptscriptstyle f};
  (2,-8)*{\scriptscriptstyle f};
   (-12,-8)*{\scriptscriptstyle  g};
    (5,18)*{\scriptscriptstyle  g};
 \endxy
\]
\vskip 1em
\noindent
This picture corresponds to a \textit{different} isomorphism from
$f \tensor g$ to $g \tensor f$, namely the reverse braiding
\[      B_{g,f}^{-1} \maps f \tensor g \to g \tensor f .\]
This is a great example of how different proofs of the
same equation may give different isomorphisms when we categorify
them.

The 4d TQFTs constructed from modular tensor categories were a
bit disappointing, in that they gave invariants of 4-dimensional
manifolds that were already known, and unable to shed light on
the deep questions of 4-dimensional topology.  The reason could be
that braided monoidal categories are rather degenerate examples of
monoidal bicategories.  In their 1994 paper, Crane and Frenkel
began the search for more interesting monoidal bicategories coming
from the representation theory of \textit{categorified} quantum groups.
As of now, it is still unknown if these give more interesting 4d
TQFTs.

\subsection*{Kontsevich (1993)}

In his famous paper of 1993, Kontsevich \cite{Kontsevich} arrived at a
deeper understanding of quantum groups, based on ideas of Witten, but
making less explicit use of the path integral approach to quantum
field theory.

In a nutshell, the idea is this.  Fix a compact simply-connected
simple Lie group $K$ and finite-dimensional representations $\rho_1, \dots,
\rho_n$.  Then there is a way to attach a vector space $Z(z_1, \dots z_n)$
to any choice of
distinct points $z_1, \dots, z_n$ in the plane, and a way
to attach a linear operator
\[ Z(f) \maps Z(z_1, \dots, z_n) \to Z(z'_1, \dots, z'_n) \]
to any $n$-strand braid going from the points $(z_1,\dots, z_n)$
to the points $z'_1, \dots, z'_n$.  The trick is to imagine each
strand of the braid as the worldline of a particle in 3d spacetime.
As the particles move, they interact with each other via a gauge
field satisfying the equations of Chern--Simons theory.  So, we
use parallel transport to describe how their internal states change.
As usual in quantum theory, this process is described by a linear
operator, and this operator is $Z(f)$.  Since Chern--Simons theory
describe a gauge field with zero curvature, this operator depends
only on the topology of the braid.  So, with some work we get a braided
monoidal category from this data.  With more work we can get operators
not just for braids but also tangles---and thus, a braided monoidal
category with duals for objects.    Finally, using a Tannaka--Krein
reconstruction theorem, we can show this category is the category of
finite-dimensional representations of a quasitriangular Hopf algebra:
the `quantum group' associated to $G$.

\subsection*{Lawrence (1993)}

In 1993, Lawrence wrote an influential paper on `extended topological
quantum field theories' \cite{Lawrence}, which she developed further
in later work \cite{Lawrence2}.  As we have seen, many TQFTs can be
constructed by first triangulating a cobordism, attaching a piece of
algebraic data to each simplex, and then using these to construct an
operator.  For the procedure to give a TQFT, the resulting operator
must remain the same when we change the triangulation by a Pachner move.
Lawrence tackled the question of precisely what is going on here.  Her
approach was to axiomatize a structure with operations corresponding
to ways of gluing together $n$-dimensional simplexes, satisfying
relations that guarantee invariance under the Pachner moves.

The use of simplexes is not ultimately the essential point here: the
essential point is that we can build any $n$-dimensional spacetime out
of a few standard building blocks, which can be glued together locally
in a few standard ways.  This lets us describe the topology of
spacetime purely combinatorially, by saying how the building blocks
have been assembled.  This reduces the problem of building TQFTs to an
essentially \textit{algebraic} problem, though one of a novel sort.

(Here we are glossing over the distinction between topological,
piecewise-linear and smooth manifolds.  Despite the term `TQFT', our
description is really suited to the case of \textsl{piecewise-linear}
manifolds, which can be chopped into simplexes or other polyhedra.
Luckily there is no serious difference between piecewise-linear and
smooth manifolds in dimensions below 7, and both these agree with
topological manifolds below dimension 4.)

Not every TQFT need arise from this sort of recipe: we loosely use the
term `extended TQFT' for those that do.  The idea is that while an
ordinary TQFT only gives operators for $n$-dimensional manifolds with
boundary (or more precisely, cobordisms), an `extended' one assigns
some sort of data to $n$-dimensional manifolds \textit{with
corners}---for example, simplexes and other polyhedra.  This is a
physically natural requirement, so it is believed that the most
interesting TQFT's are extended ones.

In ordinary algebra we depict multiplication by setting symbols
side-by-side on a line: multiplying $a$ and $b$ gives $ab$.  In
category theory we visualize morphisms as arrows, which we glue
together end to end in a one-dimensional way.  In studying TQFTs we
need `higher-dimensional algebra' to describe how to glue pieces of
spacetime together.

The idea of higher-dimensional algebra had been around for several
decades, but by this time it began to really catch on.  For example,
in 1992 Brown wrote a popular exposition of higher-dimensional
algebra, aptly titled `Out of line' \cite{Brown}.  It became clear
that $n$-categories should provide a very general approach to
higher-dimensional algebra, since they have ways of composing
$n$-morphisms that mimic ways of gluing together $n$-dimensional
simplexes, globes, or other shapes.  Unfortunately, the theory of
$n$-categories was still in its early stages of development, limiting
its potential as a tool for studying extended TQFT's.

For this reason, a partial implementation of the idea of extended TQFT
became of interest---see for example Crane's 1995 paper
\cite{Crane1995}.  Instead of working with the symmetric monoidal
category $n\Cob$, he began to grapple with the symmetric monoidal
bicategory $n\Cob_2$, where, roughly speaking:
\begin{itemize}
\item
objects are compact oriented $(n-2)$-dimensional manifolds;
\item
morphisms are $(n-1)$-dimensional cobordisms;
\item
2-morphisms are $n$-dimensional `cobordisms between cobordisms'.
\end{itemize}
His idea was that a `once extended TQFT' should be a symmetric
monoidal functor
\[        Z \maps n\Cob_2 \to 2\Vect . \]

In this approach, `cobordisms between cobordisms' are described using
manifolds with corners.  The details are still a bit tricky: it seems
the first precise general construction of $n\Cob_2$ as a bicategory
was given by Morton \cite{Morton} in 2006, and in 2009 Schommer-Pries
proved that $2\Cob_2$ was a symmetric monoidal bicategory
\cite{Schommer-Pries}.  Lurie's \cite{Lurie} more powerful approach
goes in a somewhat different direction, as we explain in our
discussion of Baez and Dolan's 1995 paper.

Since 2d TQFTs are completely classified by the result in Dijkgraaf's
1989 thesis, the concept of `once extended TQFT' may seem like
overkill in dimension 2.  But this would be a short-sighted attitude.
Around 2001, motivated in part by work on $D$-branes in string theory,
Moore and Segal \cite{MS,Segal2} introduced once extended 2d TQFTs under
the name of `open-closed topological string theories'.  However, they
did not describe these using the bicategory $2\Cob_2$.  Instead, they
considered a symmetric monoidal category $2\Cob^{{\rm ext}}$ whose
objects include not just compact 1-dimensional manifolds like the
circle (`closed strings') but also 1-dimensional manifolds with
boundary like the interval (`open strings').  Here are morphisms that
generate $2\Cob^{{\rm ext}}$ as a symmetric monoidal category:
\[
\psset{xunit=.4cm,yunit=.4cm}
\begin{pspicture}[.2](27,3.5)
  \rput(0,1){\multl}
  \rput(4,1){\comultl}
  \rput(7,2){\birthl}
  \rput(9,2.4){\deathl}
  \rput(12,1){\multc}
  \rput(16,1){\comultc}
  \rput(19,2){\birthc}
  \rput(21,1.6){\deathc}
  \rput(23,1){\ltc}
  \rput(26,1){\ctl}
\end{pspicture}
\]
Using these, Moore and Segal showed that a once extended 2d TQFT gives a
Frobenius algebra for the interval and a commutative Frobenius algebra
for the circle.  The operations in these Frobenius algebras account
for all but the last two morphisms shown above.  The last two give a
projection from the first Frobenius algebra to the second, and an
inclusion of the second into the center of the first.

Later, Lauda and Pfeiffer~\cite{LP1} gave a detailed proof that
$2\Cob^{{\rm ext}}$ is the free symmetric monoidal category on a
Frobenius algebra equipped with a projection into its center
satisfying certain relations.  Using this, they showed \cite{LP2} that
the Fukuma--Hosono--Kawai construction can be extended to obtain
symmetric monoidal functors $Z \maps 2\Cob^{{\rm ext}} \to \Vect$.
Fjelstad, Fuchs, Runkel and Schweigert have gone in a different direction,
describing full-fledged open-closed conformal field theories using
Frobenius algebras \cite{FS,RFFS,RFS}.

Once extended TQFTs should be even more interesting in dimension 3.
At least in a rough way, we can see how the
Turaev--Viro--Barrett--Westbury construction should generalize to give
examples of such theories.  Recall that this construction starts with
a 2-algebra $A \in 2\Vect$ satisfying some extra conditions.  Then:
\begin{itemize}
\item
A triangulated compact 1d manifold $S$ gives a 2-vector space
$\tilde{Z}(S)$ built by tensoring one copy of $A$ for each edge
in $S$.
\item
A triangulated 2d cobordism $M \maps S \to S'$ gives a linear functor
$\tilde{Z}(M) \maps$ $\tilde{Z}(S) \to \tilde{Z}(S')$ built out of
one multiplication functor $m \maps A \otimes A \to A$
for each triangle in $M$.
\item
A triangulated 3d cobordism between cobordisms $\alpha \maps
M \To M'$ gives a linear natural transformation
$\tilde{Z}(\alpha) \maps \tilde{Z}(M) \To \tilde{Z}(M')$ built out of
one associator for each tetrahedron in $\alpha$.
\end{itemize}
From $\tilde{Z}$ we should then be able to construct a once
extended 3d TQFT
\[  Z \maps 3\Cob_2 \to 2\Vect .\]
However, to the best of our knowledge, this construction has not been
carried out.  The work of Kerler and Lyubashenko constructs the
Witten--Reshetikhin--Turaev theory as a kind of extended 3d TQFT using
a somewhat different formalism: `double categories' instead of
bicategories \cite{KerlerLyubashenko}.

\subsection*{Crane--Frenkel (1994)}

In 1994, Louis Crane and Igor Frenkel wrote a paper entitled `Four
dimensional topological quantum field theory, Hopf categories, and the
canonical bases' \cite{CF}.  In this paper they discussed algebraic
structures that provide TQFTs in various low dimensions:
\[
\label{eq_ladder}
\xymatrix@!C=2.1pc@R=1pc{
  n=4&&\mbox{trialgebras}\ar@{-}[ddr]
     &&\mbox{Hopf categories}\ar@{-}[ddl]\ar@{-}[ddr]
     &&\mbox{monoidal bicategories}\ar@{-}[ddl]\\
  \\
  n=3&&&\mbox{Hopf algebras}\ar@{-}[ddr]
     &&\mbox{monoidal categories}\ar@{-}[ddl]\\
  \\
  n=2&&&&\mbox{algebras}
}
\]
This chart is a bit schematic, so let us expand on it a bit.  In our
discussion of Fukuma, Hosono and Kawai's 1992 paper, we have seen how
they constructed 2d TQFTs from certain algebras, namely semisimple
algebras.  In our discussion of Barrett and Westbury's paper from the
same year, we have seen how they constructed 3d TQFTs from certain
monoidal categories, namely spherical categories.  But any Hopf
algebra has a monoidal category of representations, and we can use
Tannaka--Krein reconstruction to recover a Hopf algebra from its
category of representations.  This suggests that we might be able
to construct 3d TQFTs directly from certain Hopf algebras.  Indeed,
this is the case, as was shown by Kuperberg \cite{Kuperberg} and
Chung--Fukuma--Shapere \cite{CFS}.  Indeed, there is a beautiful direct
relation between 3-dimensional topology and the Hopf algebra axioms.

Crane and Frenkel speculated on how this pattern continues in higher
dimensions.  To anyone who understands the `dimension-boosting' nature
of categorification, it is natural to guess that one can construct 4d
TQFTs from certain monoidal bicategories.  Indeed, as we have
mentioned, this was later shown by Mackaay \cite{Mackaay}, who was
greatly influenced by the Crane--Frenkel paper.  But this in turn
suggests that we could obtain monoidal bicategories by considering
`2-representations' of categorified Hopf algebras, or `Hopf
categories'---and that perhaps we could construct 4d TQFTs
\textit{directly} from certain Hopf categories.

This may be true.  In 1997, Neuchl \cite{Neuchl} gave a definition of
Hopf categories and showed that a Hopf category has a monoidal
bicategory of 2-representations on 2-vector spaces.  In 1998, Carter,
Kauffman and Saito \cite{CKS} found beautiful relations between
4-dimensional topology and the Hopf category axioms.

Crane and Frenkel also suggested that there should be some kind
of algebra whose category of representations was a Hopf category.
They called this a `trialgebra'.  They sketched the definition;
in 2004 Pfeiffer~\cite{Pfeiffer} gave a more precise treatment and
showed that any trialgebra has a Hopf category of representations.

However, \textit{defining} these structures is just the first step
toward constructing interesting 4d TQFTs.  As Crane and Frenkel put it:
\begin{quote}
To proceed any further we need a miracle, namely, the existence of
an interesting family of Hopf categories.
\end{quote}
Many of the combinatorial constructions of 3-dimensional TQFTs input a
Hopf algebra, or the representation category of a Hopf algebra, and
produce a TQFT.  However, the most interesting class of 3-dimensional
TQFTs come from Hopf algebras that are deformed universal enveloping
algebras $U_q\mathfrak{g}$.  The question is where can one find an
interesting class of Hopf categories that will give invariants that
are useful in 4d topology.

Topology in 4 dimensions is very different from lower dimensions: it
is the first dimension where homeomorphic manifolds can fail to
diffeomorphic.  In fact, there exist \textbf{exotic $\R^4$'s}:
manifolds homeomorphic to $\R^4$ but not diffeomorphic to it.  This is
the only dimension in which exotic $\R^n$'s exist!  The discovery of
exotic $\R^4$'s relied on invariants coming from quantum field theory
that can distinguish between homeomorphic 4-dimensional manifolds that
are not diffeomorphic.  Indeed this subject, known as `Donaldson
theory' \cite{DK}, is what motivated Witten to invent the term
`topological quantum field theory' in the first place \cite{Witten2}.
Later, Seiberg and Witten revolutionized this subject with a
streamlined approach \cite{SW1,SW2}, and Donaldson theory was
rebaptized `Seiberg--Witten theory'.  There are by now some good
introductory texts on these matters \cite{FU,MF,Morgan}.  The book by
Scorpan \cite{Scorpan} is especially inviting.

But this mystery remains: how---\textit{if at all!}---can the
4-manifold invariants coming from quantum field theory be computed
using Hopf categories, trialgebras or related structures?  While such
structures would give TQFTs suitable for piecewise-linear manifolds,
there is no essential difference between piecewise-linear and smooth
manifolds in dimension 4.  Unfortunately, interesting examples of Hopf
categories seem hard to construct.

Luckily Crane and Frenkel did more than sketch the definition
of a Hopf category.  They also conjectured where examples might arise:
\begin{quote}
The next important input is the existence of the canonical bases,
for a special family of Hopf algebras, namely, the quantum
groups. These bases are actually an indication of the existence of a
family of Hopf categories, with structures closely related to the
quantum groups.
\end{quote}
Crane and Frenkel suggested that the existence of the
Lusztig--Kashiwara canonical bases for upper triangular part of the
enveloping algebra, and the Lusztig canonical bases for the entire
quantum groups, give strong evidence that quantum groups are the
shadows of a much richer structure that we might call a `categorified
quantum group'.

Lusztig's geometric approach produces monoidal categories associated
to quantum groups: categories of perverse sheaves.  Crane and Frenkel
hoped that these categories could be given a combinatorial or
algebraic formulation revealing a Hopf category structure.  Recently
there has been some progress towards fulfilling Crane and Frenkel's
hopes.  In particular, these categories of perverse sheaves have been
reformulated into an algebraic language related to the
categorification of $U_q^+\mathfrak{g}$~\cite{KL,VV}.  The entire quantum
group $U_q\mathfrak{sl}_n$ has been categorified by Khovanov and Lauda
\cite{Lau1,KL3}, and they also gave a conjectural categorification of
the entire quantum group $U_q\mathfrak{g}$ for every simple Lie
algebra $\g$.  Categorified representation theory, or
`2-representation theory', has taken off, thanks largely to the
foundational work of Chuang and Rouquier~\cite{CR,Rou}.

There is much more that needs to be understood.  In particular,
categorification of quantum groups at roots of unity has received only
a little attention \cite{Kh5}, and the Hopf category structure has not
been fully developed.  Furthermore, these approaches have not yet
obtained braided monoidal bicategories of 2-representations of
categorified quantum groups.  Nor have they constructed 4d TQFTs.

\subsection*{Freed (1994)}

In 1994, Freed published an important paper \cite{Freed} which
exhibited how higher-dimensional algebraic structures arise naturally
from the Lagrangian formulation of topological quantum field theory.
Among many other things, this paper clarified the connection between
quasitriangular Hopf algebras and 3d TQFTs.  It also introduced an
informal concept of `2-Hilbert space' categorifying the concept of
Hilbert space.  This was later made precise, at least in the
finite-dimensional case \cite{hda2,Bartlett}, so it is now tempting to
believe that much of the formalism of quantum theory can be
categorified.  The subtleties of analysis involved in understanding
infinite-dimensional 2-Hilbert spaces remain challenging, with close
connections to the theory of von Neumann algebras \cite{BBFW}.

\subsection*{Kontsevich (1994)}

In a lecture at the 1994 International Congress of Mathematicians in
Z\"urich, Kontsevich \cite{Kontsevich2} proposed the `homological
mirror symmetry conjecture', which led to a burst of work relating
string theory to higher categorical structures.  A detailed discussion
of this work would drastically increase the size of this paper.  So,
we content ourselves with a few elementary remarks.

We have already mentioned the concept of an `$A_\infty$ space': a
topological space equipped with a multiplication that is associative
up to a homotopy that satisfies the pentagon equation up to a
homotopy...  and so on, forever, in a manner governed by the Stasheff
polytopes \cite{Stasheff1}.  This concept can be generalized
to any context that allows for a notion of homotopy between maps.  In
particular, it generalizes to the world of `homological algebra',
which is a simplified version of the world of homotopy theory.  In
homological algebra, the structure that takes the place of a
topological space is a \textbf{chain complex}: a sequence of abelian
groups and homomorphisms
\[
\xymatrix{
 V_{0}
&
 V_{1} \ar[l]_{d_{1}}
&
 V_{2} \ar[l]_{d_{2}}
&
 \cdots \ar[l]_{d_{3}}
}
\]
with $d_i d_{i+1} = 0$.  In applications to physics, we focus on the
case where the $V_i$ are vector spaces and the $d_i$ are linear operators.
Regardless of this, we can define maps between chain complexes,
called `chain maps', and homotopies between chain maps, called
`chain homotopies'.

\vskip 2em
\vbox{
\begin{center}
{\small
\begin{tabular}{|c|c|}                    \hline
topological spaces    &  chain complexes         \\  \hline
continuous maps   &  chain maps                \\   \hline
homotopies   &  chain homotopies        \\     \hline
\end{tabular}}
\vskip 1em
Analogy between homotopy theory and homological algebra
\end{center}
}
\vskip 0.5em

\noindent
For a very readable introduction to these matters, see the book by Rotman
\cite{Rotman}; for a more strenuous one that goes further, try the
book with the same title by Weibel \cite{Weibel}.

The analogy between homotopy theory and homological algebra ultimately
arises from the fact that while homotopy types can be seen as
$\infty$-groupoids, chain complexes can be seen as $\infty$-groupoids
that are `strict' and also `abelian'.  The process of turning a
topological space into a chain complex, so important in algebraic
topology, thus amounts to taking a $\infty$-groupoid and simplifying
it by making it strict and abelian.

Since this fact is less widely appreciated than it should be, let us
quickly sketch the basic idea.  Given a chain complex $V$, each
element of $V_0$ corresponds to an object in the corresponding
$\infty$-groupoid.  Given objects $x,y \in V_0$, a
morphism $f \maps x \to y$ corresponds to an element $f \in V_1$ with
\[                 d_1 f + x = y  .\]
Given morphisms $f,g \maps x \to y$, a 2-morphism
$\alpha \maps f \To g$ corresponds to an element $\alpha \in V_2$ with
\[                d_2 \alpha + f = g ,\]
and so on.  The equation $d_i d_{i+1} = 0$ then says that an
$(i+1)$-morphism can only go between two $i$-morphisms that share
the same source and target---just as we expect in the globular
approach to $\infty$-categories.

The analogue of an $A_\infty$ space in the world of chain complexes is
called an `$A_\infty$ algebra' \cite{Stasheff3,Keller,KS}.  More
generally, one can define a structure called an `$A_\infty$ category',
which has a set of objects, a chain complex $\hom(x,y)$ for any pair
of objects, and a composition map that is associative up to a chain
homotopy that satisfies the pentagon identity up to a chain
homotopy... and so on.  Just as a monoid is the same as a category
with one object, an $A_\infty$ algebra is the same as an $A_\infty$
category with one object.

Kontsevich used the language of $A_\infty$ categories to formulate a
conjecture about `mirror symmetry', a phenomenon already studied by
string theorists.  Mirror symmetry refers to the observation that
various pairs of superficially different string theories seem in fact to be
isomorphic.  In Kontsevich's conjecture, each of these theories is
a `open-closed topological string theory'.  We already introduced
this concept near the end of our discussion of Lawrence's 1993 paper.
Recall that such a theory is designed to describe processes involving
open strings (intervals) and closed strings (circles).  The
basic building blocks of such processes are these:
\[
\psset{xunit=.4cm,yunit=.4cm}
\begin{pspicture}[.2](27,3.5)
  \rput(0,1){\multl}
  \rput(4,1){\comultl}
  \rput(7,2){\birthl}
  \rput(9,2.4){\deathl}
  \rput(12,1){\multc}
  \rput(16,1){\comultc}
  \rput(19,2){\birthc}
  \rput(21,1.6){\deathc}
  \rput(23,1){\ctl}
  \rput(26,1){\ltc}
\end{pspicture}
\]
In the simple approach we discussed, the space of states of the open
string is a Frobenius algebra.  The space of states of the closed
string is a commutative Frobenius algebra, typically the center of the
Frobenius algebra for the open string.  In the richer approach
developed by Kontsevich and subsequent authors, notably Costello
\cite{Costello}, states of the open string are instead described by an
$A_\infty$ category with some extra structure mimicking that of a
Frobenius algebra.  The space of states of the closed string is
obtained from this using a subtle generalization of the concept of
`center'.

To get some sense of this, let us ignore the `Frobenius' aspects and
simply regard the space of states of an open string as an algebra.
Multiplication in this algebra describes the process of two open
strings colliding and merging together:
\vskip 1em
\noindent
\[
 \begin{pspicture}[.5](4,2.5)
  \rput(2,0){\multl}
\end{pspicture}
\]
The work in question generalizes this simple idea in two ways.  First,
it treats an algebra as a special case of an $A_\infty$ algebra,
namely one for which only the 0th vector space in its underlying chain
complex is nontrivial.  Second, it treats an $A_\infty$ algebra as a
special case of an $A_\infty$ category, namely an $A_\infty$ category
with just one object.

How should we understand a general $A_\infty$ category as describing
the states of an open-closed topological string?  First, the different
objects of the $A_\infty$ category correspond to different boundary
conditions for an open string.  In physics these boundary conditions
are called `$D$-branes', because they are thought of as membranes in
spacetime on which the open strings begin or end.  The `$D$' stands
for Dirichlet, who studied boundary conditions back in the mid-1800's.
A good introduction to $D$-branes from a physics perspective can be
found in Polchinski's books \cite{Polchinski}.

For any pair of $D$-branes $x$ and $y$, the $A_\infty$ category gives
a chain complex $\hom(x,y)$.  What is the physical meaning of this?
It is the space of states for an open string that starts on the
$D$-brane $x$ and ends on the $D$-brane $y$.  Composition describes a
process where open strings in the states $g \in \hom(x,y)$ and $f \in
\hom(y,z)$ collide and stick together to form an open string in the
state $fg \in \hom(x,z)$.

However, note that the space of states $\hom(x,y)$ is not a mere
vector space.  It is a chain complex---so it is secretly a strict
$\infty$-groupoid!  This lets us talk about states that are not
\textit{equal}, but still \textit{isomorphic}.  In particular,
composition in an $A_\infty$ category is associative only up to
isomorphism: the states $(fg)h$ and $f(gh)$ are not usually equal,
merely isomorphic via an associator:
\[            a_{f,g,h} \maps (fg)h \to f(gh)   .\]
In the language of chain complexes, we write this as follows:
\[            da_{f,g,h} + (fg)h = f(gh) .\]
This is just the first of an infinite list of equations that
are part of the usual definition of an $A_\infty$ category.  The
next one says that the associator satisfies the pentagon identity
up to $d$ of something, and so on.

Kontsevich formulated his homological mirror symmetry conjecture as
the statement that two $A_\infty$ categories are equivalent.  The
conjecture remains unproved in general, but many special cases
are known.  Perhaps more importantly, the conjecture has become
part of an elaborate web of ideas relating gauge theory to
the `Langlands program'---which itself is a vast generalization
of the circle of ideas that gave birth to Wiles' proof of
Fermat's Last Theorem.  For a good introduction to all this,
see the survey by Edward Frenkel \cite{Frenkel}.

\subsection*{Gordon--Power--Street (1995)}

In 1995, Gordon, Power and Street introduced the definition and basic
theory of `tricategories'---or in other words, weak 3-categories \cite{GPS}.
Among other things, they defined a `monoidal bicategory' to be a
tricategory with one object.  They then showed that a monoidal
bicategory with one object is the same as a braided monoidal category.
This is a precise working-out of the categorified Eckmann--Hilton
argument sketched in our discussion of Turaev's 1992 paper.

So, a tricategory with just one object and one morphism is the same as
a braided monoidal category.  There is also, however, a notion of
`strict 3-category': a tricategory where all the relevant laws hold as
equations, not merely up to equivalence.  Not surprisingly, a strict
3-category with one object and one morphism is a braided monoidal
category where all the braiding, associator and unitors are
\textit{identity} morphisms.  This rules out the possibility of
nontrivial braiding, which occurs in categories of braids or tangles.
As a consequence, not every tricategory is equivalent to a strict
3-category.

All this stands in violent contrast to the story one dimension down,
where a generalization of Mac Lane's coherence theorem can be used to
show every bicategory is equivalent to a strict 2-category.  So, while
it was already known in some quarters \cite{JT}, Gordon, Power and
Street's book made the need for weak $n$-categories clear to all: in a
world where all tricategories were equivalent to strict 3-categories,
there would be no knots!

Gordon, Power and Street did, however, show that every tricategory
is equivalent to a `semistrict' 3-category, in which some but not all
the laws hold as equations.  They called these semistrict 3-categories
`Gray-categories', since their definition relies on John Gray's
prescient early work \cite{Gray}.  Constructing a workable theory of
semistrict $n$-categories for all $n$ remains a major challenge.

\subsection*{Baez--Dolan (1995)}

In \cite{hdatqft}, Baez and Dolan outlined a program for
understanding extended TQFTs in terms of $n$-categories.  A key part
of this is the `periodic table of $n$-categories'.  Since this only
involves weak $n$-categories, let us drop the qualifier `weak' for the
rest of this section, and take it as given.  Also, just for the sake
of definiteness, let us take a globular approach to $n$-categories:
\[
\begin{tabular}{|c|c|c|c|c|}
  \hline
   \textbf{objects} & \textbf{morphisms} & \textbf{2-morphisms} & \textbf{3-morphisms} &
   $\cdots$ \\
  \hline \hline
   $\bullet$ & $\xy
  (-8,0)*+{\bullet}="1";
  (0,10)*+{}; 
  (0,-10)*+{}; 
  (8,0)*+{\bullet}="2";
  {\ar "1";"2"};
 \endxy$ & $\xy (-8,0)*+{\bullet}="4"; (8,0)*+{\bullet}="6";
  {\ar@/^1.65pc/^{}"4";"6"};
  {\ar@/_1.65pc/_{} "4";"6"};
{\ar@{=>}(0,3)*{};(0,-3)*{}} ;
\endxy$ & $\xy 0;/r.22pc/:
  (-20,0)*+{\bullet}="1";
  (0,0)*+{\bullet}="2";
  {\ar@/^2pc/ "1";"2"};
  {\ar@/_2pc/ "1";"2"};
   (-10,6)*+{}="A";
   (-10,-6)*+{}="B";
  {\ar@{=>}@/_.7pc/ "A"+(-2,0) ; "B"+(-1,-.8)};
  {\ar@{=}@/_.7pc/ "A"+(-2,0)  ; "B"+(-2,0)};
  {\ar@{=>}@/^.7pc/ "A"+(2,0)  ; "B"+(1,-.8)};
  {\ar@{=}@/^.7pc/ "A"+(2,0)  ; "B"+(2,0)};
  {\ar@3{->} (-12,0)*{}; (-8,0)*{}};
 \endxy$ & Globes  \\
  \hline
\end{tabular}
\]

\noindent
So, in this section `2-category' will mean `bicategory' and
`3-category' will mean `tricategory'.  (Recently this sort of
terminology as been catching on, since the use of Greek prefixes to name
weak $n$-categories becomes inconvenient as the value of $n$ becomes
large.)

We have already seen the beginning of a pattern involving
these concepts:
\begin{itemize}
\item A category with one object is a monoid.
\item A 2-category with one object is a monoidal category.
\item A 3-category with one object is a monoidal 2-category.
\end{itemize}
The idea is that we can take an $n$-category with one object and
think of it as an $(n-1)$-category by ignoring the object, renaming
the morphisms `objects', renaming the 2-morphisms `morphisms', and so
on.  Our ability to compose morphisms in the original $n$-category
gets reinterpreted as an ability to `tensor' objects in the
resulting $(n-1)$-category, so we get a `monoidal' $(n-1)$-category.

However, we can go further: we can consider a monoidal $n$-category
with one object.  We have already looked at two cases of this,
and we can imagine more:
\begin{itemize}
\item A monoidal category with one object is a commutative monoid.
\item A monoidal 2-category with one object is a braided monoidal category.
\item A monoidal 3-category with one object is a braided monoidal 2-category.
\end{itemize}
Here the Eckmann--Hilton argument comes into play, as explained our
discussion of Turaev's 1992 paper.  The idea is that given a monoidal
$n$-category $C$ with one object, this object must be the unit for the
tensor product, $1 \in C$.  We can focus attention on $\hom(1,1)$,
which an $(n-1)$-category.  Given $f,g \in \hom(1,1)$, there
are \text{two} ways to combine them: we can compose them, or tensor
them.  As we have seen, we can visualize these operations as putting
together little squares in two ways: vertically, or horizontally.
\[
  \xy
 (-6,6)*{}="TL";
 (-6,-6)*{}="BL";
 (6,6)*{}="TR";
 (6,-6)*{}="BR";
 (-6,0)*{}="LM";
 (6,0)*{}="LR";
 "TL";"TR" **\dir{-};
 "TL";"BL" **\dir{-};
 "TR";"BR" **\dir{-};
 "BL";"BR" **\dir{-};
  "LM";"LR" **\dir{-};
 (0,-3)*{f};
 (0,3)*{g};
 (0,-9)*{fg};
 \endxy
\qquad \qquad \qquad
 \xy
 (-6,6)*{}="TL";
 (-6,-6)*{}="BL";
 (6,6)*{}="TR";
 (6,-6)*{}="BR";
 (0,6)*{}="MT";
 (0,-6)*{}="MB";
 "TL";"TR" **\dir{-};
 "TL";"BL" **\dir{-};
 "TR";"BR" **\dir{-};
 "BL";"BR" **\dir{-};
 "MT";"MB" **\dir{-};
 (3,0)*{g};
 (-3,0)*{f};
 (0,-9)*{f \tensor  g};
 \endxy
\]
These operations are related by an `interchange' morphism
\[  (f f') \tensor (g g') \to (f \tensor g)(f' \tensor g') ,\]
which is an equivalence (that is, invertible in a suitably weakened
sense).  This allow us to carry out the Eckmann--Hilton argument and
get a braiding on $\hom(1,1)$:
\[        B_{f,g} \maps f \tensor g \to g \tensor f  .\]

Next, consider braided monoidal $n$-categories with
one object.  Here the pattern seems to go like this:
\begin{itemize}
\item A braided monoidal category with one object is a commutative monoid.
\item A braided monoidal 2-category with one object is a symmetric
monoidal category.
\item A braided monoidal 3-category with one object is a sylleptic
monoidal 2-category.
\item A braided monoidal 4-category with one object is a sylleptic
monoidal 3-category.
\end{itemize}
The idea is that given a braided monoidal $n$-category with one
object, we can think of it as an $(n-1)$-category with \textit{three}
ways to combine objects, all related by interchange equivalences.  We
should visualize these as the three obvious ways of putting together
little cubes: side by side, one in front of the other, and one on top
of the other.

In the first case listed above, the third operation doesn't give
anything new.  Just like a monoidal category with one object, a braided
monoidal category with one object is merely a commutative monoid.  In the
next case we get something new: a braided monoidal 2-category with
one object is a \textit{symmetric} monoidal category.  The reason is
that the third monoidal structure allows us to interpolate between the
Eckmann--Hilton argument that gives the braiding by moving $f$ and $g$
around clockwise, and the argument that gives the reverse braiding by
moving them around them counterclockwise.  We obtain the equation
\[ \xy
 \vtwist~{(-5,8)}{(5,8)}{(-5,-8)}{(5,-8)}<>|>>><;
 (-7,5)*{\scriptstyle f};
 (7,5)*{\scriptstyle g};
\endxy
\quad = \quad
 \xy
 \vtwistneg~{(-5,8)}{(5,8)}{(-5,-8)}{(5,-8)}<>|>>><;
 (-7,5)*{\scriptstyle g};
 (7,5)*{\scriptstyle f};
\endxy \]
which characterizes a symmetric monoidal category.

In the case after this, instead of an equation, we obtain an
\textit{2-isomorphism} that describes the \textit{process} of
interpolating between the braiding and the reverse braiding:
\[ s_{f,g} \maps \; \xy
 \vtwist~{(-5,8)}{(5,8)}{(-5,-8)}{(5,-8)}<>|>>><;
 (-7,5)*{\scriptstyle x};
 (7,5)*{\scriptstyle y};
\endxy
\quad \To \quad
 \xy
 \vtwistneg~{(-5,8)}{(5,8)}{(-5,-8)}{(5,-8)}<>|>>><;
 (-7,5)*{\scriptstyle y};
 (7,5)*{\scriptstyle x};
\endxy \]
The reader should endeavor to imagine these pictures as drawn in
4-dimensional space, so that there is room to push the top strand in
the left-hand picture `up into the fourth dimension', slide it behind
the other strand, and then push it back down, getting the right-hand
picture.  Day and Street \cite{DS} later dubbed this 2-isomorphism
$s_{f,g}$ the `syllepsis' and formalized the theory of sylleptic
monoidal 2-categories.  The definition of a fully weak sylleptic
monoidal 2-category was introduced still later by Street's student
McCrudden \cite{McCrudden}.

To better understand the patterns at work here, it is useful to
define a `$k$-tuply monoidal $n$-category' to be an $(n+k)$-category
with just one $j$-morphism for $j < k$.  A chart of these appears
below.  This is called the `periodic table', since like Mendeleyev's
original periodic table it guides us in extrapolating the behavior of
$n$-categories from simple cases to more complicated ones.  It is not
really `periodic' in any obvious way.

\begin{center}
\periodictable
\vskip 1em
The Periodic Table: \break
hypothesized table of $k$-tuply monoidal $n$-categories
\end{center}

The periodic table should be taken with a grain of salt.  For example,
a claim like `2-categories with one object and one morphism are the
same as commutative monoids' needs to be made more precise.  Its truth
may depend on whether we consider commutative monoids as forming a
category, or a 2-category, or a 3-category!  This has been
investigated by Cheng and Gurski \cite{ChGur2}.  There have also been
attempts to craft an approach that avoids such subtleties
\cite{BaezShulman}.

But please ignore such matters for now: just stare at the table.  The
most notable feature is that the $n$th column of the periodic table
seems to stop changing when $k$ reaches $n+2$.  Baez and Dolan called
this the `stabilization hypothesis'.  The idea is that adding extra
monoidal structures ceases to matter at this point.  Simpson later
proved a version of this hypothesis in his approach to $n$-categories
\cite{Simpson4}.  So, let us assume the stabilization hypothesis is
true, and call a $k$-tuply monoidal $n$-category with $k \ge n+2$
a `stable $n$-category'.

In fact, stabilization is just the simplest of the many intricate
patterns lurking in the periodic table.  For example, the reader will
note that the syllepsis
\[  s_{f,g} \maps B_{f,g} \To B_{g,f}^{-1}  \]
is somewhat reminiscent of the braiding itself:
\[  B_{f,g} \maps f \tensor g \to g \tensor f  .\]
Indeed, this is the beginning of a pattern that continues as we
zig-zag down the table starting with monoids.  To go from monoids to
commutative monoids we add the equation $fg = gf$.  To go from
commutative monoids to braided monoidal categories we then replace
this equation by an isomorphism, the braiding $B_{f,g} \maps f \tensor
g \to g \tensor f$.  But the braiding engenders another isomorphism
with the same source and target: the reverse braiding $B_{g,f}^{-1}$.
To go from braided monoidal categories to symmetric monoidal
categories we add the equation $B_{f,g} = B_{g,f}^{-1}$.  To go from
symmetric monoidal categories to sylleptic monoidal 2-categories we
then replace this equation by a 2-isomorphism, the syllepsis $s_{f,g}
\maps B_{f,g} \To B_{g,f}^{-1}$.  But this engenders another
2-isomorphism with same source and target: the `reverse syllepsis'.
Geometrically speaking, this is because we can also deform the left
braid to the right one here:
\[  \xy
 \vtwist~{(-5,8)}{(5,8)}{(-5,-8)}{(5,-8)}<>|>>><;
 (-7,5)*{\scriptstyle x};
 (7,5)*{\scriptstyle y};
\endxy
\quad \To \quad
 \xy
 \vtwistneg~{(-5,8)}{(5,8)}{(-5,-8)}{(5,-8)}<>|>>><;
 (-7,5)*{\scriptstyle y};
 (7,5)*{\scriptstyle x};
\endxy \]
by pushing the top strand \textit{down} into the fourth dimension and
then behind the other strand.  To go from sylleptic monoidal 2-categories
to symmetric ones, we add an equation saying the syllepsis equals
the reverse syllepsis.  And so on, forever!  As we zig-zag down the
diagonal, we meet ways of switching between ways of switching between...
ways of switching things.

This is still just the tip of the iceberg: the patterns that arise
further from the bottom edge of the periodic table are vastly more
intricate.  To give just a taste of their subtlety, consider the
remarkable story told in Kontsevich's 1999 paper \textsl{Operads and
Motives and Deformation Quantization} \cite{Kontsevich3}.  Kontsevich
had an amazing realization: quantization of ordinary classical
mechanics problems can be carried out in a systematic way using ideas
from string theory.  A thorough and rigorous approach to this issue
required proving a conjecture by Deligne.  However, early attempts to
prove Deligne's conjecture had a flaw, first noted by Tamarkin, whose
simplest manifestation---translated into the language of
$n$-categories---involves an operation that first appears for braided
monoidal 6-categories!

For this sort of reason, one would really like to see precisely what
features are being added as we march down any column of the periodic
table.  Batanin's approach to $n$-categories offers a beautiful answer
based on the combinatorics of trees \cite{Batanin3}.  Unfortunately,
explaining this here would take us too far afield.  The slides of a
lecture Batanin delivered in 2006 give a taste of the richness of his
work \cite{Batanin4}.

Baez and Dolan also emphasized the importance of $n$-categories with
duals at all levels: duals for objects, duals for morphisms, ... and
so on, up to $n$-morphisms.  Unfortunately, they were only able to
precisely define this notion in some simple cases.  For example, in
our discussion of Doplicher and Roberts' 1989 paper we defined
monoidal, braided monoidal, and symmetric monoidal categories with
duals---meaning duals for both objects and morphisms.  We noted that
tangles in 3d space can be seen as morphisms in the free braided
monoidal category on one object.  This is part of a larger pattern:
\begin{itemize}
\item
The category of framed 1d tangles in 2d space, $1\Tang_1$, is
the free monoidal category with duals on one object.
\item
The category of framed 1d tangles in 3d space, $1\Tang_2$,
is the free braided monoidal category with duals on one object.
\item
The category of framed 1d tangles in 4d space, $1\Tang_3$,
is the free symmetric monoidal category with duals on one object.
\end{itemize}
A technical point: here we are using `framed' to mean `equipped
with a trivialization of the normal bundle'.  This is how the word
is used in homotopy theory, as opposed to knot theory.  In fact
a framing in this sense determines an orientation, so a `framed
1d tangle in 3d space' is what ordinary knot theorists would
call a `framed oriented tangle'.

Based on these and other examples, Baez and Dolan formulated the
`tangle hypothesis'.  This concerns a conjectured $n$-category
$n\Tang_k$ where:
\begin{itemize}
\item objects are collections of framed points in $[0,1]^k$,
\item morphisms are framed 1d tangles in $[0,1]^{k+1}$,
\item 2-morphisms are framed 2d tangles in $[0,1]^{k+2}$,
\item and so on up to dimension $(n-1)$, and finally:
\item $n$-morphisms are isotopy classes of framed $n$-dimensional
tangles in $[0,1]^{n+k}$.
\end{itemize}
For short, we call the $n$-morphisms `$n$-tangles in $(n+k)$
dimensions'.  Figure 2 may help the reader see how simple these
actually are: it shows a typical
$n$-tangle in $(n+k)$ dimensions for various values of $n$ and
$k$.  This figure is a close relative of the periodic table.
The number $n$ is the \textit{dimension} of the tangle, while
$k$ is its \textit{codimension}: that is, the number of \textit{extra}
dimensions of space.

The tangle hypothesis says that $n\Tang_k$ is the free $k$-tuply
monoidal $n$-category with duals on one object.  As usual, the
one object, $x$, is simply a point.  More precisely,
$x$ can be any point in $[0,1]^k$ equipped with a framing that
makes it positively oriented.

\begin{figure}
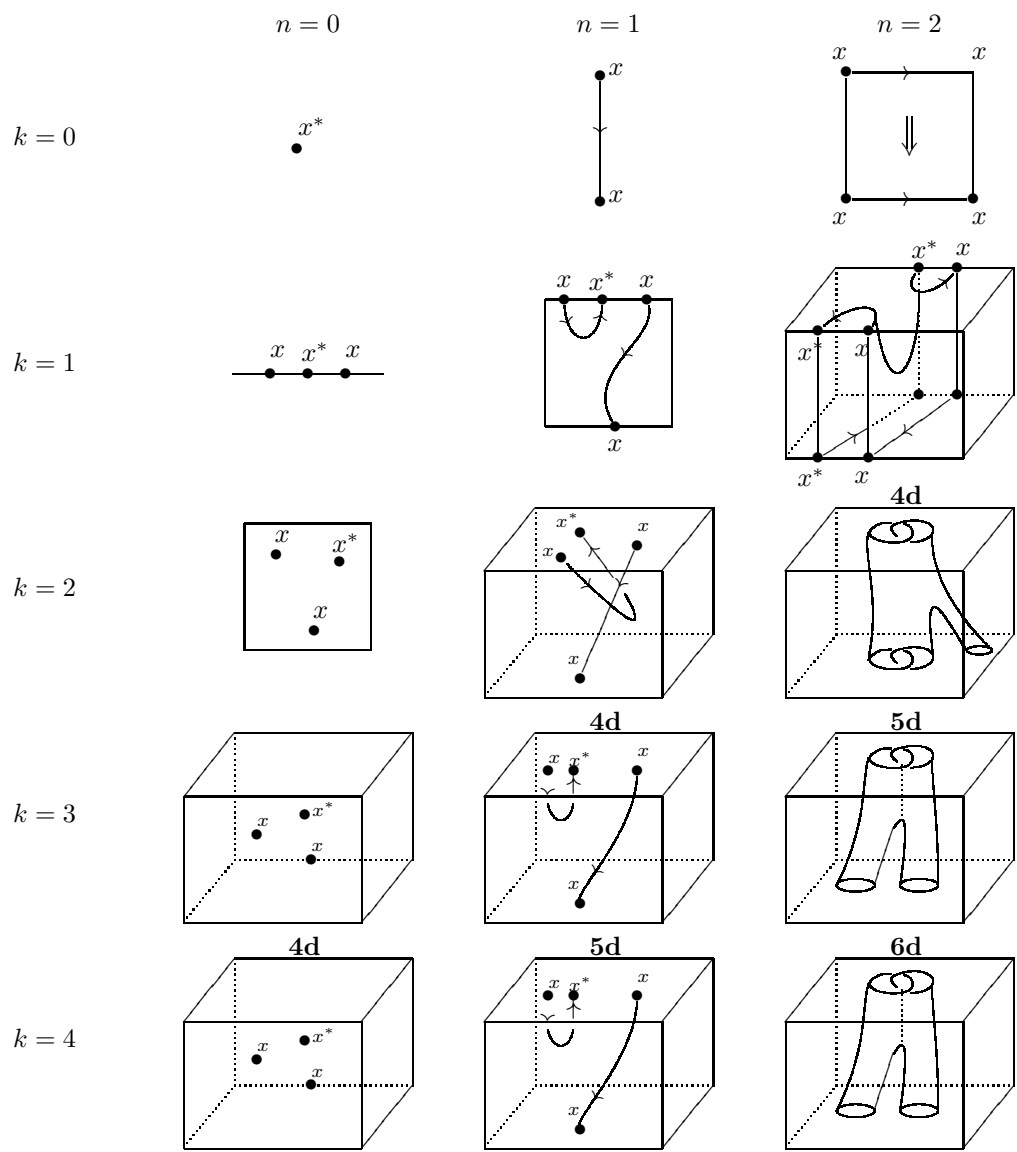

\begin{center}
\periodictableII
\caption{Examples of $n$-tangles in $(n+k)$-dimensional
space.}
\end{center}
\end{figure}

Combining the stabilization hypothesis and the tangle hypothesis, we
obtain an interesting conclusion: the $n$-category $n\Tang_k$
stabilizes when $k$ reaches $n+2$.  This idea is backed up by a
well-known fact in topology: any two embeddings of a compact
$n$-dimensional manifold in $\R^{n+k}$ are isotopic if $k \ge n+2$.
In simple terms: when $k$ is this large, there is enough room to untie
any $n$-dimensional knot!

So, we expect that when $k$ is this large,
the $n$-morphisms in $n\Tang_k$ correspond to `abstract' $n$-tangles,
not embedded in any ambient space.  But this is precisely how we think
of cobordisms.  So, for $k \ge n+2$, we should expect that $n\Tang_k$
is a stable $n$-category where:
\begin{itemize}
\item objects are compact framed 0-dimensional manifolds;
\item morphisms are framed 1-dimensional cobordisms;
\item 2-morphisms are framed 2-dimensional `cobordisms
between cobordisms',
\item 3-morphisms are framed 3-dimensional `cobordisms
between cobordisms between cobordisms',
\end{itemize}
and so on up to dimension $n$, where we take equivalence classes.
Let us call this $n$-category
$n\Cob_n$, since it is a further elaboration of the 2-category
$n\Cob_2$ in our discussion of Lawrence's 1993 paper.

The `cobordism hypothesis' summarizes these ideas: it says that
$n\Cob_n$ is the free stable $n$-category with duals on one object
$x$, namely the positively oriented point.
We have already sketched how `once extended' $n$-dimensional TQFTs
can be treated as symmetric monoidal functors
\[             Z \maps n\Cob_2 \to 2\Vect  .\]
This suggests that fully extended $n$-dimensional TQFTs should
be something similar, but with $n\Cob_n$ replacing $n\Cob_2$.
Similarly, we should replace $2\Vect$ by some sort of $n$-category:
something deserving the name $n\Vect$, or even better, $n\Hilb$.

This leads to the `extended TQFT hypothesis', which says
that a unitary extended TQFT is a map between stable $n$-categories
\[            Z \maps n\Cob_n \to n\Hilb  \]
that preserves all levels of duality.  Since $n\Hilb$ should be a
stable $n$-category with duals, and $n\Cob_n$ should be the free such
thing on one object, we should be able to specify a unitary extended
TQFT simply by choosing an object $H \in n\Hilb$ and saying that
\[             Z(x) = H \]
where $x$ is the positively oriented point.  This is the `primacy
of the point' in a very dramatic form.

What progress has there been on making these hypotheses precise and
proving them?  In 1998, Baez and Langford \cite{hda4} came close to
proving that $2\Tang_2$, the 2-category of 2-tangles in
4d space, was the free braided monoidal 2-category with duals
on one object.  (In fact, they proved a similar result for
oriented but unframed 2-tangles.)  In 2009, Schommer-Pries
\cite{Schommer-Pries} came close to proving that $2\Cob_2$ was the
free symmetric monoidal 2-category with duals on one object.
(In fact, he gave a purely algebraic description of $2\Cob_2$
as a symmetric monoidal 2-category, but not explicitly using the
language of duals.)

But the really exciting development is the paper that Jacob Lurie \cite{Lurie}
put on the arXiv in 2009.  Entitled \textsl{On the Classification of
Topological Field Theories}, this outlines a precise statement and
proof of the cobordism hypothesis for all $n$.

Lurie's version makes use, not of $n$-categories, but of
`$(\infty,n)$-categories'.  These are $\infty$-categories such that
every $j$-morphism is an equivalence for $j > n$.  This helps avoid
the problems with duality that we mentioned in our discussion of
Atiyah's 1988 paper.  There are many approaches to
$(\infty,1)$-categories, including the `$A_\infty$ categories'
mentioned in our discussion of Kontsevich's 1994 lecture.  Prominent
alternatives include Joyal's `quasicategories' \cite{Joyal2}, first
introduced in the early 1970's under another name by Boardmann and
Vogt \cite{BV}, and also Rezk's `complete Segal spaces' \cite{Rezk}.
For a comparison of some approaches, see the survey by Bergner
\cite{Bergner}.  Another good source of material on quasicategories is
Lurie's enormous book on higher topos theory \cite{Lurie2}.  The study
of $(\infty,n)$-categories for higher $n$ is still in its infancy.  At
this moment Lurie's paper is the best place to start, though he
attributes the definition he uses to Barwick, who promises a
two-volume book on the subject \cite{Barwick}.

\subsection*{Khovanov (1999)}

In 1999, Mikhail Khovanov found a way to categorify the Jones
polynomial~\cite{Kh}.   We have already seen a way to
categorify an algebra that has a basis $e^i$ for which
\[
e^i e^j = \sum_{k} m^{ij}_k e^k
\]
where the constants $m^{ij}_k$ are \textit{natural numbers}.
Namely, we can think of these numbers as dimensions of
\textit{vector spaces} $M^{ij}_k$.  Then we can seek a 2-algebra
with a basis of irreducible objects $E^i$ such that
\[
E^i \otimes E^j = \sum_{k} M^{ij}_k \otimes E^k .
\]
We say this 2-algebra categorifies our original algebra: or, more
technically, we say that taking the `Grothendieck group' of the
2-algebra gives back our original algebra.  In this simple example,
taking the Grothendieck group just means forming a vector space with
one basis element $e^i$ for each object $E^i$ in our basis of
irreducible objects.

The Jones polynomial, and other structures related to quantum
groups, present more challenging problems.  Here instead of
natural numbers we have polynomials in $q$ and $q^{-1}$.
Sometimes, as in the theory of canonical bases, these polynomials
have natural number coefficients.  Elsewhere, as in the Jones
polynomial, they have integer coefficients.  How can we generalize
the concept of `dimension' so it can be a polynomial of this sort?

In fact, problems like this were already tackled by Emmy Noether in
the late 1920's, in her work on homological algebra \cite{Dick}.  We
have already defined the concept of a `chain complex', but this term
is used in several slightly different ways, so now let us change our
definition a bit and say that a \textbf{chain complex} $V$ is a
sequence of vector spaces and linear maps
\[
\xymatrix{
\cdots
&
 V_{-1} \ar[l]_{d_{-1}}
&
 V_{0} \ar[l]_{d_{0}}
&
 V_{1} \ar[l]_{d_{1}}
&
 V_{2} \ar[l]_{d_{2}}
&
 \cdots \ar[l]_{d_{3}}
}
\]
with $d_i d_{i+1} = 0$.  If the vector spaces are finite-dimensional
and only finitely many are nonzero, we can define the \textbf{Euler
characteristic} of the chain complex by
\[   \chi(V) = \sum_{i = -\infty}^\infty (-1)^i  \dim(V_i)  .\]
The Euler characteristic is a remarkably robust invariant: we can change
the chain complex in many ways without changing its Euler characteristic.
This explains why the number of vertices minus the number of
edges plus the number of faces is equal to 2 for every convex polyhedron!

We may think of the Euler characteristic as a generalization of
`dimension' which can take on arbitrary \textit{integer} values.  In
particular, any vector space gives a chain complex for which only
$V_0$ is nontrivial, and in this case the Euler characteristic reduces
to the ordinary dimension.  But given any chain complex $V$, we can
`shift' it to obtain a new chain complex $sV$ with
\[         sV_i = V_{i+1}  ,\]
and we have
\[          \chi(sV) = -\chi(V)  .\]
So, shifting a chain complex is like taking its `negative'.

But what about polynomials in $q$ and $q^{-1}$?  For these, we need
to generalize vector spaces a bit further, as indicated here:

\vskip 1em
\begin{center}
{\small
\begin{tabular}{|c|c|}
                               \hline
vector spaces        &  natural numbers                      \\
\hline
chain complexes      &  integers                    \\
\hline
graded               &  polynomials in $q^{\pm 1}$   \\
vector spaces        &  with natural number coefficients   \\
\hline
graded               & polynomials in $q^{\pm 1}$   \\
chain complexes      & with integer coefficients   \\
\hline
\end{tabular}} \vskip 1em
Algebraic structures and the values of their `dimensions'
\end{center}
\vskip 0.5em
\noindent
A \textbf{graded vector space} $W$ is simply a series of vector spaces
$W_i$ where $i$ ranges over all integers.   The
\textbf{Hilbert--Poincar\'e series} $\dim_q(W)$ of a graded vector
space is given by
\[      \dim_q(W) = \sum_{i = -\infty}^\infty \dim(W_i) \, q^i .  \]
If the vector spaces $W_i$ are finite-dimensional and only finitely
many are nonzero, $\dim_q(W)$ is a polynomial in $q$ and $q^{-1}$ with
natural number coefficients.  Similarly, a
\textbf{graded chain complex} $W$ is a series of chain complexes
$W_i$, and its \textbf{graded Euler characteristic} $\chi(W)$
is given by
\[      \chi_q(W) = \sum_{i = -\infty}^\infty \chi(W_i) \, q^i .  \]
When everything is finite enough, this is a polynomial in $q$
and $q^{-1}$ with integer coefficients.

Khovanov found a way to assign a graded chain complex to any link in
such a way that its graded Euler characteristic is the Jones
polynomial of that link, apart from a slight change in normalizations.
This new invariant can distinguish links that have the same Jones
polynomial~\cite{BN}.  Even better, it can be extended to an invariant
of tangles in 3d space, and also \textit{2-tangles in 4d space!}

To make this a bit more precise, note that we can think of a
2-tangle in 4d space as a morphism $\alpha \maps
S \to T$ going from one tangle in 3d space, namely $S$, to another,
namely $T$.  For example:
\[ \alpha \maps
\xy
(5,10)*{}; (-10,-10)*{} **\crv{(6,-2)&(-12,4)}
\POS?(.25)*{\hole}="x" \POS?(.45)*{\hole}="y" \POS?(.6)*{\hole}="z";
"y"+(0,-1); (2,-10)*{} **\crv{}\POS?(.2)*{\hole}="M";
(-10,10)*{}; "z" **\crv{(-9,0)};
"z"; "M" **\crv{};
"M"; "x" **\crv{(5,0)};
"x"; "y" **\crv{(0,7) & (-5,6)};
\endxy
\qquad \to \qquad
\xy
(4,2)*{}="P";
(5,10)*{}; (-10,-10)*{} **\crv{(6,2)&(-12,4)}
\POS?(.25)*{\hole}="x" \POS?(.45)*{\hole}="y" \POS?(.6)*{\hole}="z";
(-8,10)*{}; "z" **\crv{(-9,2)};
"P"; (3,-10)*{} **\crv{(-2,4)} \POS?(.7)*+{}="M";
"P"; "M"+(.5,-.2) **\crv{(8,0) & (5,-5)};
"z"; "M" **\crv{(-9,2)};
\endxy
\]
In its most recent incarnation, Khovanov homology makes use of
a certain monoidal category $C$.  Its precise definition takes
a bit of work \cite{BN2}, but its objects are built using graded chain
complexes, and its morphisms are built using maps between these.
Khovanov homology assigns to each tangle $T$ in 3d space an
object $Z(T) \in C$, and assigns to each 2-tangle in 4d
space $\alpha \maps T \To T'$ a morphism $Z(\alpha) \maps Z(T) \to Z(T')$.

What is especially nice is that $Z$ is a monoidal functor.  This means
we can compute the invariant of a 2-tangle by breaking it into pieces,
computing the invariant for each piece, and then composing and
tensoring the results.  Actually, in the original construction due
to Jacobsson \cite{Jac1} and Khovanov \cite{Kh2}, $Z(\alpha)$ was only
well-defined up to a scalar multiple.  But later, using the
streamlined approach introduced by Bar-Natan~\cite{BN2}, this problem
was fixed by Clark, Morrison, and Walker \cite{CMW}.

So far we have been treating 2-tangles as morphisms.  But in fact
we know they should be 2-morphisms.  There should be a
braided monoidal bicategory $2\Tang_2$ where, roughly speaking:
\begin{itemize}
\item objects are collections of framed points in the square $[0,1]^2$,
\item morphisms are framed oriented tangles in the cube $[0,1]^3$,
\item 2-morphisms are framed oriented 2-tangles in $[0,1]^4$.
\end{itemize}
The tangle hypothesis asserts that $2\Tang_2$ is the
\textit{free} braided monoidal bicategory with duals on one object
$x$, namely the positively oriented point.  Indeed, a version of this
claim ignoring framings is already known to be true \cite{hda4}.

This suggests that Khovanov homology could be defined in a way that
takes advantage of this universal property of $2\Tang_2$.  For this we
would need to see the objects and morphisms of the category $C$ as
morphisms and 2-morphisms of some braided monoidal bicategory with
duals, say $\overline{C}$, equipped with a special object $c$.  Then
Khovanov homology could be seen as the essentially unique braided
monoidal functor preserving duals, say
\[        \overline{Z} \maps 2\Tang_2 \to \overline{C}  ,\]
with the property that
\[          \overline{Z}(x) = c . \]
This would be yet another triumph of `the primacy of the point'.

It is worth mentioning that the authors in this field have chosen to
study higher categories with duals in a manner that does not distinguish
between `source' and `target'.  This makes sense, because duality
allows one to convert input to outputs and vice versa.  In 1999,
Jones introduced `planar algebras' \cite{Jones2}, which can thought
of as a formalism for handling certain categories with duals.
In his work on Khovanov homology, Bar-Natan introduced a structure
called a `canopolis' \cite{BN2}, which is a kind of categorified
planar algebra.   The relation between these ideas and other approaches
to $n$-category theory deserves to be clarified and generalized
to higher dimensions.

One exciting aspect of Khovanov's homology theory is that it breathes
new life into Crane and Frenkel's dream of understanding the special
features of smooth 4-dimensional topology in a purely combinatorial
way, using categorification.  For example, Rasmussen \cite{Ras} has
used Khovanov homology to give a purely combinatorial proof of the
Milnor conjecture---a famous problem in topology that had been solved
earlier in the 1990's using ideas from quantum field theory, namely
Donaldson theory \cite{KM}.  And as the topologist Gompf later pointed
out \cite{Ras2}, Rasmussen's work can also be used to prove the
existence of an exotic $\R^4$.

In outline, the argument goes as follows.  A knot in $\R^3$ is said to
be \textbf{smoothly slice} if it bounds a smoothly embedded disc in
$\R^4$.  It is said to be \textbf{topologically slice} if it bounds a
topologically embedded disc in $\R^4$ and this embedding extends to a
topological embedding of some thickening of the disc.  Gompf had shown
that if there is a knot that is topologically but not smoothly slice,
there must be an exotic $\R^4$.  However, Rasmussen's work can be
used to find such a knot!

Before this, all proofs of the existence of exotic $\R^4$'s had
involved ideas from quantum field theory: either Donaldson theory or
its modern formulation, Seiberg--Witten theory.  This suggests a
purely combinatorial approach to Seiberg--Witten theory is within
reach.  Indeed, Ozsv\'{a}th and Szab\'{o} have already introduced a
knot homology theory called `Heegaard Floer homology' which has a
conjectured relationship to Seiberg-Witten theory~\cite{OS}.  Now that
there is a completely combinatorial description of Heegaard--Floer
homology~\cite{MOS,MOST}, one cannot help but be optimistic that some
version of Crane and Frenkel's dream will become a reality.

In summary: the theory of $n$-categories is beginning to shed light on
some remarkably subtle connections between physics, topology, and
geometry.  Unfortunately, this work has not yet led to concrete
successes in elementary particle physics or quantum gravity.  But
given the profound yet simple ways that $n$-categories unify and
clarify our thinking about mathematics and physics, we can hope that
what we have seen so far is just the beginning.

\subsection*{Acknowledgements}

We thank the denizens of the $n$-Category Caf\'e, including Toby
Bartels, Michael Batanin, David Ben-Zvi, Rafael Borowiecki, Greg Egan,
Alex Hoffnung, Urs Schreiber, and Zoran {\v{S}}koda, for many
discussions and corrections.  JB thanks the Department of the Pure
Mathematics and Mathematical Statistics at the University of Cambridge
for inviting him to give a series of lectures on this topic in July
2004. He also thanks Derek Wise for writing up notes for a course on
this topic at U.\ C.\ Riverside during the 2004-2005 academic year.
AL was partially supported by the NSF grants DMS-0739392 and
DMS-0855713.

\newpage

\newcommand{\hpeprint}[1]{
\href{http://arXiv.org/abs/#1}{arXiv:#1}}

\end{document}